\documentclass{aastex63}

\usepackage{multirow}

\usepackage{amsmath}
\received{\today}
\revised{}
\usepackage[dvipsnames]{xcolor}
\accepted{}
\submitjournal{ApJ}

\shorttitle{Zero-Age Massive Stellar Population of W49A}
\shortauthors{Juárez-Gama et al.}

\begin{document}

\title{The Zero-Age Massive Stellar Population of W49A from VLA Observations}

\correspondingauthor{Mariana Juárez-Gama, Roberto Galván-Madrid}
\email{m.juarez@irya.unam.mx, r.galvan@irya.unam.mx}

\author[0009-0001-4382-7134]{Mariana Juárez-Gama}
\affil{Instituto de Radioastronom\'ia y Astrof\'isica, Universidad Nacional Aut\'onoma de M\'exico, Morelia, Michoac\'an 58089, M\'exico.}
\affil{Instituto Nacional de Astrofísica, Óptica y Electrónica, Luis E. Erro 1, 72840 Tonantzintla, Puebla, México.}

\author[0000-0003-1480-4643]{Roberto Galv\'an-Madrid}
\affil{Instituto de Radioastronom\'ia y Astrof\'isica, Universidad Nacional Aut\'onoma de M\'exico, Morelia, Michoac\'an 58089, M\'exico.}

\author[0000-0002-2011-4924]{Genaro Su\'arez}
\affil{Department of Astrophysics, American Museum of Natural History, Central Park West at 79th Street, NY 10024, USA.}

\author[0000-0003-3115-9359]{Christopher G. De Pree}
\affil{National Radio Astronomy Observatory, 520 Edgemont Rd., Charlottesville, VA 22903, USA.}

\author[0000-0002-6195-0152]{John J. Tobin}
\affil{National Radio Astronomy Observatory, 520 Edgemont Rd., Charlottesville, VA 22903, USA.}

\author[0000-0002-6971-5755]{Gustavo Bruzual}
\affil{Instituto de Radioastronom\'ia y Astrof\'isica, Universidad Nacional Aut\'onoma de M\'exico, Morelia, Michoac\'an 58089, M\'exico.}

\author[0000-0001-6431-9633]{Adam Ginsburg}
\affil{Department of Astronomy, University of Florida, PO Box 112055, USA.}

\author{José E. Mendoza-Torres}
\affiliation{Instituto Nacional de Astrofísica, Óptica y Electrónica, Luis E. Erro 1, 72840 Tonantzintla, Puebla, México.}

\author[0000-0003-2300-2626]{Hauyu Baobab Liu}
\affil{Department of Physics, National Sun Yat-Sen University, No. 70, Lien-Hai Road, Kaohsiung City 80424, Taiwan, ROC.}
\affil{Center of Astronomy and Gravitation, National Taiwan Normal University, Taipei 116, Taiwan, ROC.}

\author[0000-0003-1526-7587]{David Wilner}
\affil{Center for Astrophysics $\vert$ Harvard \& Smithsonian, Cambridge, MA 02138, USA.}

\author[0000-0003-3246-0821]{Thomas Nony}
\affil{Laboratoire d’Astrophysique de Bordeaux, Univ. Bordeaux, CNRS, B18N, allée Geoffroy Saint-Hilaire, 33615 Pessac, France.}
\affil{Instituto de Radioastronom\'ia y Astrof\'isica, Universidad Nacional Aut\'onoma de M\'exico, Morelia, Michoac\'an 58089, M\'exico.}

\author[0009-0001-8553-1227]{Adara Parra-López}
\affil{Facultad de Ciencias de la Tierra y el Espacio, Universidad Autónoma de Sinaloa, Culiacán 80040, Sinaloa, México.}

\author[0000-0002-2162-8441]{Rudy Rivera-Soto}
\affil{Instituto de Radioastronom\'ia y Astrof\'isica, Universidad Nacional Aut\'onoma de M\'exico, Morelia, Michoac\'an 58089, M\'exico.}

\author[0000-0002-5456-523X]{Anna F. McLeod}
\affil{Centre for Extragalactic Astronomy/Institute for Computational Cosmology, Durham University, South Road, Durham DH1 3LE, UK.}

\author[0000-0003-3152-8564]{Nichol Cunningham}
\affil{SKA Observatory, Jodrell Bank, Lower Withington, Macclesfield, SK11 9FT, United Kingdom.}

\author[0000-0003-2619-9305]{Xing Lu}
\affil{Shanghai Astronomical Observatory, Chinese Academy of Sciences, 80 Nandan Road, Shanghai 200030, P. R. China.}

\author[0000-0002-2640-5917]{Eric F. Jiménez-Andrade}
\affil{Instituto de Radioastronom\'ia y Astrof\'isica, Universidad Nacional Aut\'onoma de M\'exico, Morelia, Michoac\'an 58089, M\'exico.}
\affil{National Radio Astronomy Observatory, 520 Edgemont Rd., Charlottesville, VA 22903, USA.}

\begin{abstract}
We use all-configuration VLA data at 3.3 cm with a physical resolution of $\approx$ 2000 AU to infer the embedded zero-age massive stellar population of the W49A protocluster, as traced by its compact, ultracompact (UC), and hypercompact (HC) H\textsc{ii} regions. 
Our method consists of visual source identification, the derivation of stellar ionizing-photon rates from the observed emission measure, and the further derivation of stellar masses and luminosities using state-of-the-art stellar calibrations. Considering the 87 robust detections, maximum-likelihood estimation fitting of the high-mass end ($M>$ 9.5 $M_{\odot}$) of the sample yields power-law slopes $\Gamma \geq$ 2.23 for the logarithmic representation of the stellar initial mass function (IMF): $dN/d(\log M) \propto M^{-\Gamma}$. The result is robust considering different assumptions for the ionizing stellar systems. The slopes remain at  $\Gamma \geq$ 1.89 after correction for optical depth effects at 3.3 cm.  
Therefore, the inferred distribution of stellar masses presents a clear {\it deficit} in the high-mass end as compared to the standard stellar IMF ($\Gamma = 1.35$). 
We propose that this is due to a shorter lifetime of the radio-detected H\textsc{ii} regions produced by higher-mass stars, but evolutionary effects in the mass distribution of star formation within embedded protoclusters cannot be discarded.  
\end{abstract}

\keywords{ISM: individual: W49 --- H II regions --- stars: formation}

\section{Introduction} \label{sec:intro}
The W49A cloud, especially its central clump W49N, is a prototypical example of extreme star formation in the Milky Way. It is located at a distance of 11.1 kpc from the Sun \citep{Zhang2013}, and it is one of the handful of Galactic star formation regions exceeding a bolometric luminosity of $10^7~L_\odot$ 
\citep{Sievers1991,Lin2016}. 
While its gas and dust content has been mapped in detail from sub-pc to $\sim 100$ pc scales \citep{Miyawaki2009,GM13,Rugel19,Barnes2020b,Miyawaki22}, its stellar content remains elusive. A fraction of the stellar population has been identified in the near- and mid-IR \citep{HomeierAlves05,Saral15,BinderPovich2018,DeBuizer2021}, possibly including stars with masses in excess of $100~M_\odot$ \citep{AlvesHomeier2003,Wu2016}. However, a large fraction of the young stellar and protostellar population is still deeply embedded. One way of peering into the deeply embedded (proto)stellar population is through centimeter and (sub)millimeter continuum observations \citep[e.g.,][]{DePree2000,Wilner2001,Depree2020,Nony2024}. 

In recent years, significant efforts have been made to account for the dusty core population of protocluster clouds  \citep[e.g.,][]{Motte18_NatAs,Sanhueza19,Louvet2024,Cheng2024,Coletta2025}. Similar progress is needed in characterizing the population of young, massive stars embedded in those clouds. Multiwavelength censuses of the protostars and  embedded young stars will provide the ultimate test for theories regarding the origin of the stellar initial mass function \citep[for reviews, see][]{Bastian2010,Kroupa2026}. 

The centimeter continuum is a powerful tool for surveying the free-free emission of ultracompact (UC) and hypercompact (HC) H\textsc{ii} regions 
\citep[for reviews, see][]{Churchwell2002,kurtz2005,Hoare07}, which arises  when a massive star, approximately in the Zero-Age Main Sequence (ZAMS), ionizes its own dense core \citep[e.g.,][]{Keto07,Peters10,Tanaka2016}. Interferometric radio continuum studies of massive star-forming regions have been carried out for decades, and the methods to infer the spectral type of the stars ionizing UC H\textsc{ii} regions are well known \citep[e.g.,][]{Garay1993,Kurtz94,RiveraSoto20,Meng2022,GM2024}. However, there is a lack of studies that go further and infer the corresponding distribution of \textit{stellar masses}. To our knowledge, this was first done in the pioneering work of \citet{HoHaschick81}, using early observations with the Karl G. Jansky Very Large Array (VLA), but it has seldom been attempted in later literature. 

In this study, we make use of sensitive, high angular resolution, all-configuration VLA data to infer the masses of a significant part of the young, massive stellar population of W49A in a systematic way. 
This paper is organized as follows. In Section~\ref{sec:data} we describe the observations and data reduction. Section~\ref{sec:sources} presents the identification of the  H\textsc{ii} regions. In Section~\ref{sec:results} we derive their physical properties, infer the stellar masses, and analyze the resulting stellar initial mass function. The implications of our results are discussed in Section~\ref{sec:disc}, and our main conclusions are given in Section~\ref{sec:conc}. 

\section{Data} \label{sec:data}

W49A was observed with the VLA at 3.3 cm in the B, A, D, and then C configurations in 2015–2016 (15A-089), with 3~hr executions in each configuration, resulting in 1.75~hr on W49A for all configurations except C, which had 1.9~hr on source. The A-configuration observations were made on 2015 June 24, the B-configuration on 2015 February 08, the C-configuration on 2016 January 29, and the D-configuration on 2015 October 10. For all observations, 3C48 was used as the flux density calibrator, 3C84 was the bandpass calibrator, and J1925+2106 was the complex gain calibrator. We used the 8-bit samplers, allowing two 1 GHz basebands centered at 8.5 and 9.716 GHz, with eight contiguous sub-bands at each frequency. 
Five sub-bands were dedicated to the continuum, each having a bandwidth of 128 MHz and 128 channels of 1 MHz. Calibrator sources and other details of the observations are listed in Table 1, Columns 3-4 of \citet{Depree2020}.

We calibrated the data using the VLA pipeline within CASA 6.5.4 
\citep[pipeline version 2023.1.0.124,][]{CASA2022}. The pipeline performs bandpass calibration, radio frequency interference (RFI) flagging, flux density bootstrapping, and standard complex gain calibration. We performed additional flagging where necessary to remove any data impacted by strong RFI and uncorrectable gain errors. The data were self-calibrated to improve image quality by correcting for phase and amplitude gain errors that remained after standard calibration. 
For this, version 1.2.0 of \texttt{auto\_selfcal} \citep{auto_selfcal} was used after the standard VLA pipeline.
It performed self-calibration per-spectral window for all solution intervals where it was possible, based-on the S/N of the target and gain solutions, and the data from each configuration were self-calibrated independently. The A, B, C, and D configuration data had phase-only self-calibration applied to solution intervals down to timescales of 33~s, 3~s (1 integration), 3~s (1 integration), and 9~s, respectively. Additionally, amplitude self-calibration was applied following the completion of phase-only self-calibration on timescales that spanned a single scan, using the options \texttt{calmode=`ap'} and \texttt{solnorm=True}. The latter option normalizes the solutions around 1.0 for each antenna, ensuring that the flux density scale is almost unchanged.

The final images were created using the visibilities from all four self-calibrated configurations combined, utilizing a pixel size of 0.042\arcsec, and were truncated at the 0.1 level of the primary-beam response, i.e., with a diameter of 8.38\arcmin~after primary-beam correction. 
We used the \texttt{mtmfs} deconvolver with \texttt{nterms=2}, Briggs weighting \citep{briggs1995} with \texttt{robust=-1.5}, and a manually-drawn clean mask that encompassed the large and small-scale features throughout the image. 
This processing yielded a synthesized beam FWHM of $\theta_\mathrm{PSF} = 0.174\arcsec \times 0.167\arcsec$ (0.01 pc). 
We found that the approximately equal integration times in each configuration and tuning of the \texttt{robust} parameter enabled us to create images of the combined data with a  good beam shape. Multi-configuration data with different integration time ratios between configurations may require their relative weights to be rescaled to obtain a Gaussian beam.

An overview of the continuum emission across the W49A field is shown in Figure~\ref{fig:EVLAW49Azoom}. 
The ICRS coordinates of the image center are $\alpha =19^\mathrm{h}10^\mathrm{m}12.931^\mathrm{s}$ and $\delta = 09^\circ06\arcmin11.883\arcsec$.
Prior to primary-beam correction, the rms noise in regions far from bright emission reaches down to $\sigma_\mathrm{rms} = 7~\mu$Jy beam$^{-1}$, and it increases by up to $\times 3$ closer to the brightest emission in the map. 
The maximum recoverable scale (MRS) of an interferometric array is often estimated from the shortest baselines. At 3.3 cm and for the D-configuration\footnote{\url{https://science.nrao.edu/facilities/vla/docs/manuals/oss/performance/resolution}}, the corresponding MRS would be 194\arcsec (10.4 pc). In practice,  arrays start to filter-out emission at scales $\lesssim \textrm{MRS}$\footnote{\url{https://zenodo.org/records/18793803}}. Nevertheless, the diagonal extent of the largest sources in our catalog is $\lesssim 25\arcsec$, well below the limit where interferometric filtering becomes important.

\section{Identification and Nature of the Sources} \label{sec:sources}

H\textsc{ii} regions are known to exhibit a series of morphologies based on physical criteria \citep[spherical, shell-like, cometary, bipolar, irregular, as well as the unresolved; see][]{Churchwell2002,DePree2005,Peters2010b}. As an initial step for source identification, we utilized source masks identified with the \texttt{astrodendro} package \citep{Rosolowsky2008}. 
However, the non-uniform noise across the image complicates the functioning of the dendrogram algorithm. 
To account for this, we defined eight different zones without emissions across the image to establish the representative local rms noise level $\sigma$.
Our final source identification was conducted through an  extensive visual inspection of the image using CARTA \citep{CARTA2021}, aided by previously defined catalogs for the brightest sources in the central part of W49A \citep{DePree2000}. We note that the 3.3 cm image presented in this paper is the most sensitive to date at comparable wavelengths.

Source boundaries were adjusted visually, taking as an initial reference a $3\sigma$ threshold in uncrowded areas and $\geq 5\sigma$ in more crowded regions, using the local $\sigma$ as defined above.  

In the specific case of two sources (idx 51 and 58) where the contours remained blended, we separated their individual contributions using the parameters reported by \cite{Depree2020}. For this, we derived the ratios of reported peak intensities and areas for each component; we then applied these proportions to our total integrated fluxes and mask areas to estimate the individual properties of each component. This procedure
ensures consistent treatment across regions with varying noise levels and allows complex or irregular morphologies to be properly characterized. 

\begin{figure}[!ht]
\centering    
\includegraphics[width=0.98\linewidth]{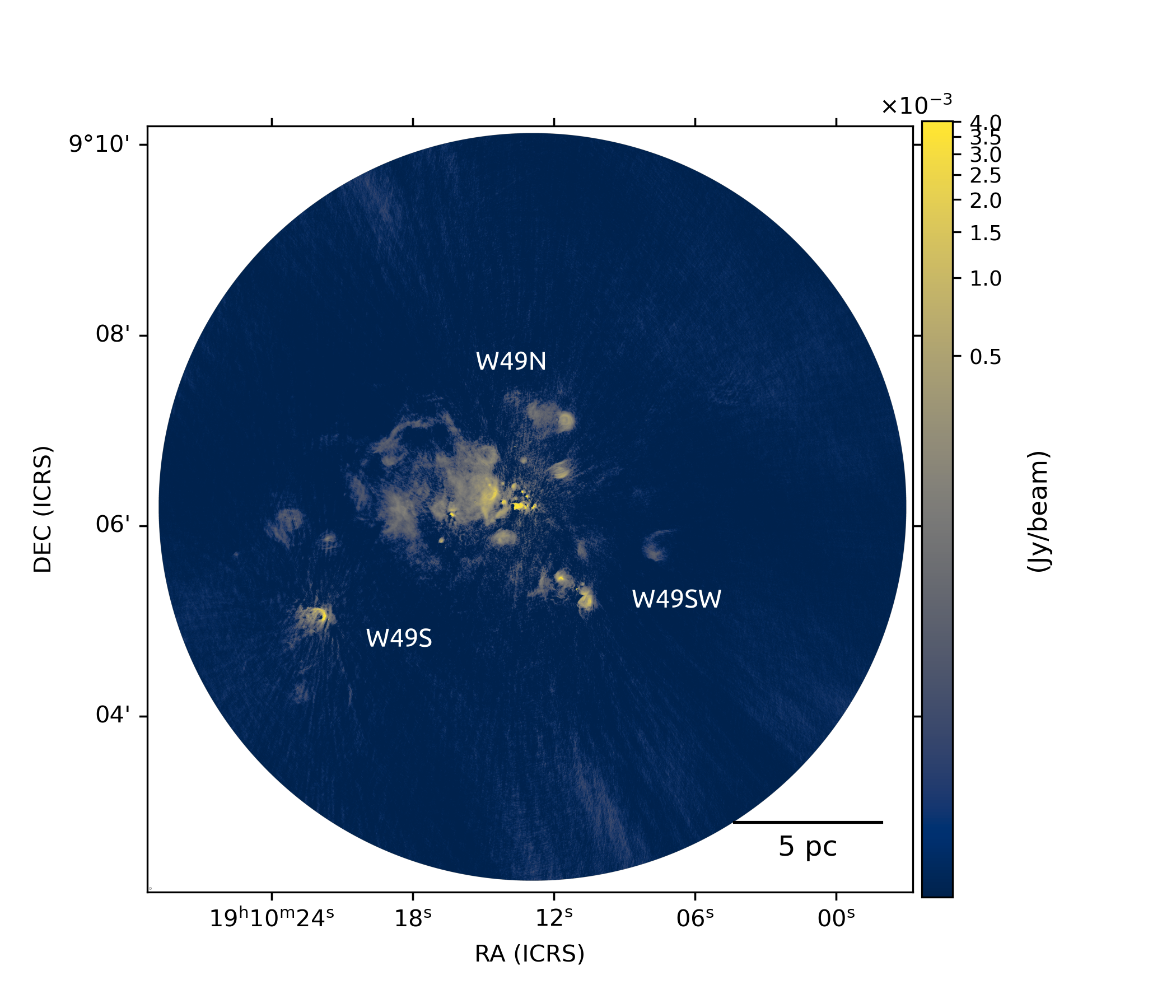}
\caption{
All-configuration VLA 9.1 GHz (3.3 cm) continuum map of W49A. The image is shown prior to primary beam correction. Individual zoomed-in views for the identified H\textsc{ii} regions are provided in Figures \ref{fig:zoom} and \ref{fig:individual}. The FWHM of the synthesized beam is \ensuremath{\theta_{\mathrm{PSF}} = 0.174\arcsec \times 0.167\arcsec} (0.01 pc). The maximum recoverable scale is $\lesssim 194\arcsec$ (10.4 pc). The rms noise far from bright emission is \ensuremath{\sigma_{\mathrm{rms}} \approx 7~\mu\mathrm{Jy}\,\mathrm{beam}^{-1}}, and increases by up to $\times 3$ close to bright emission.}
\label{fig:EVLAW49Azoom}
\end{figure}

\begin{figure*}[!ht]
    \centering
    \includegraphics[width=0.32\textwidth]{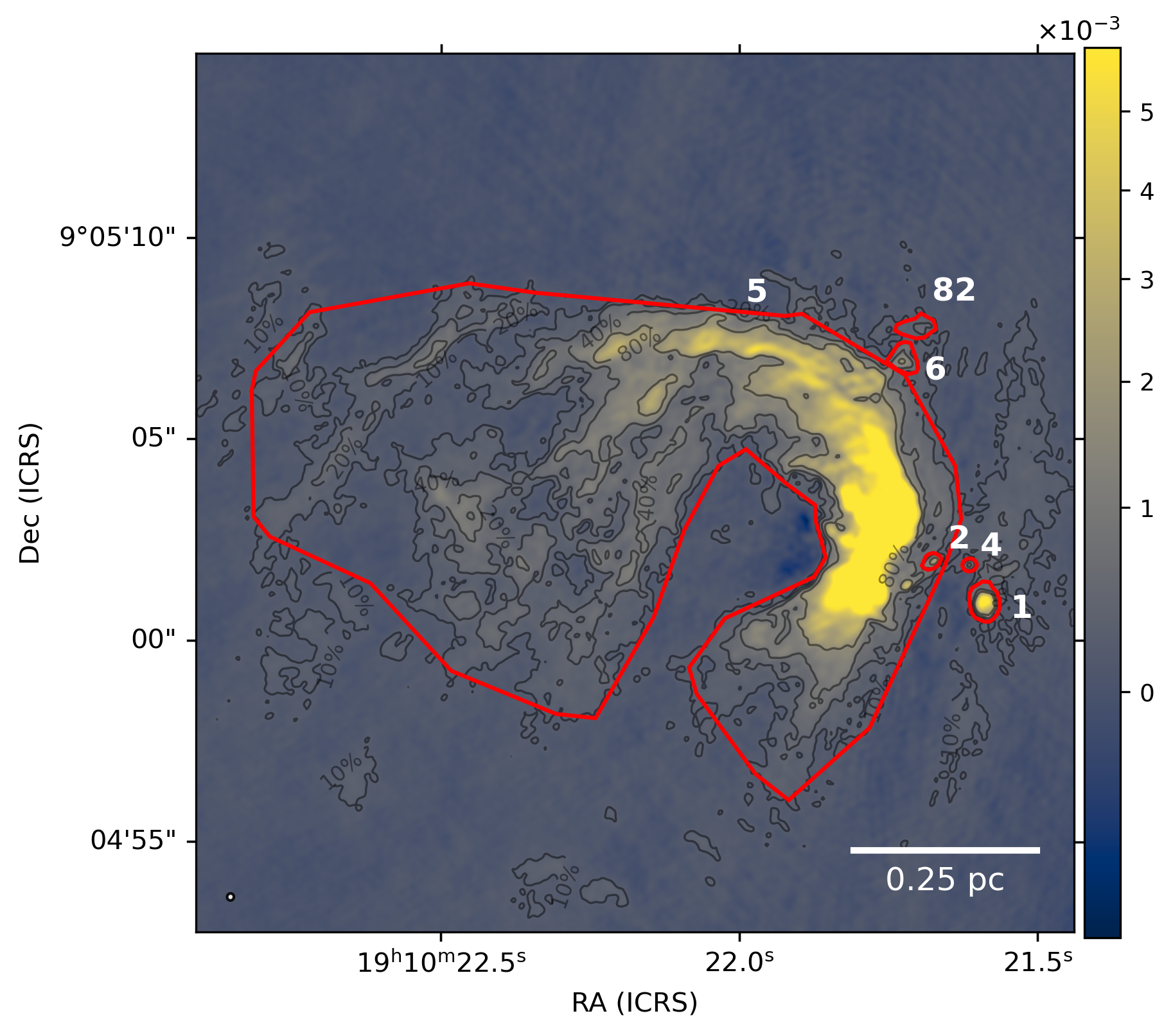}
    \includegraphics[width=0.32\textwidth]{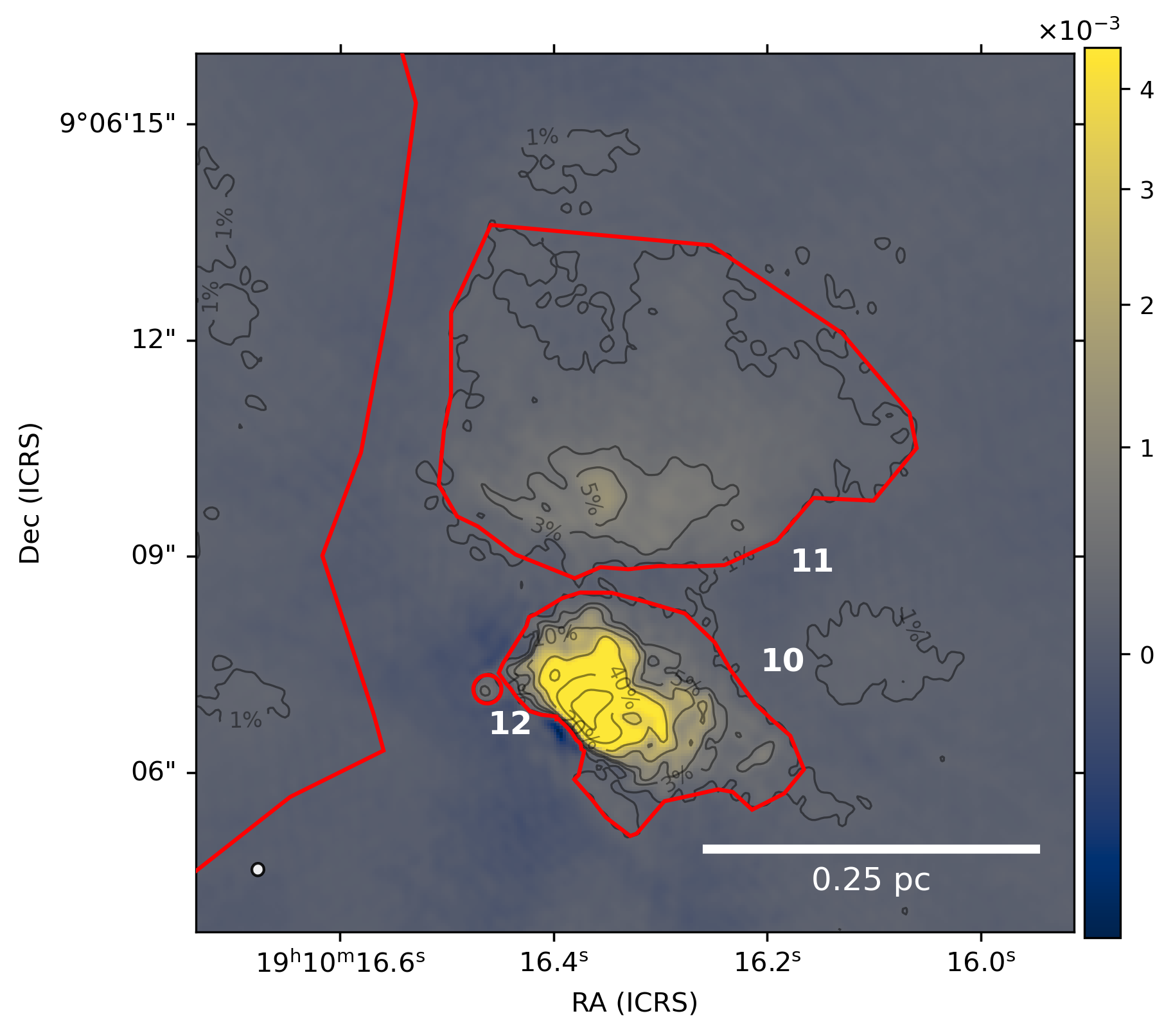}
    \includegraphics[width=0.32\textwidth]{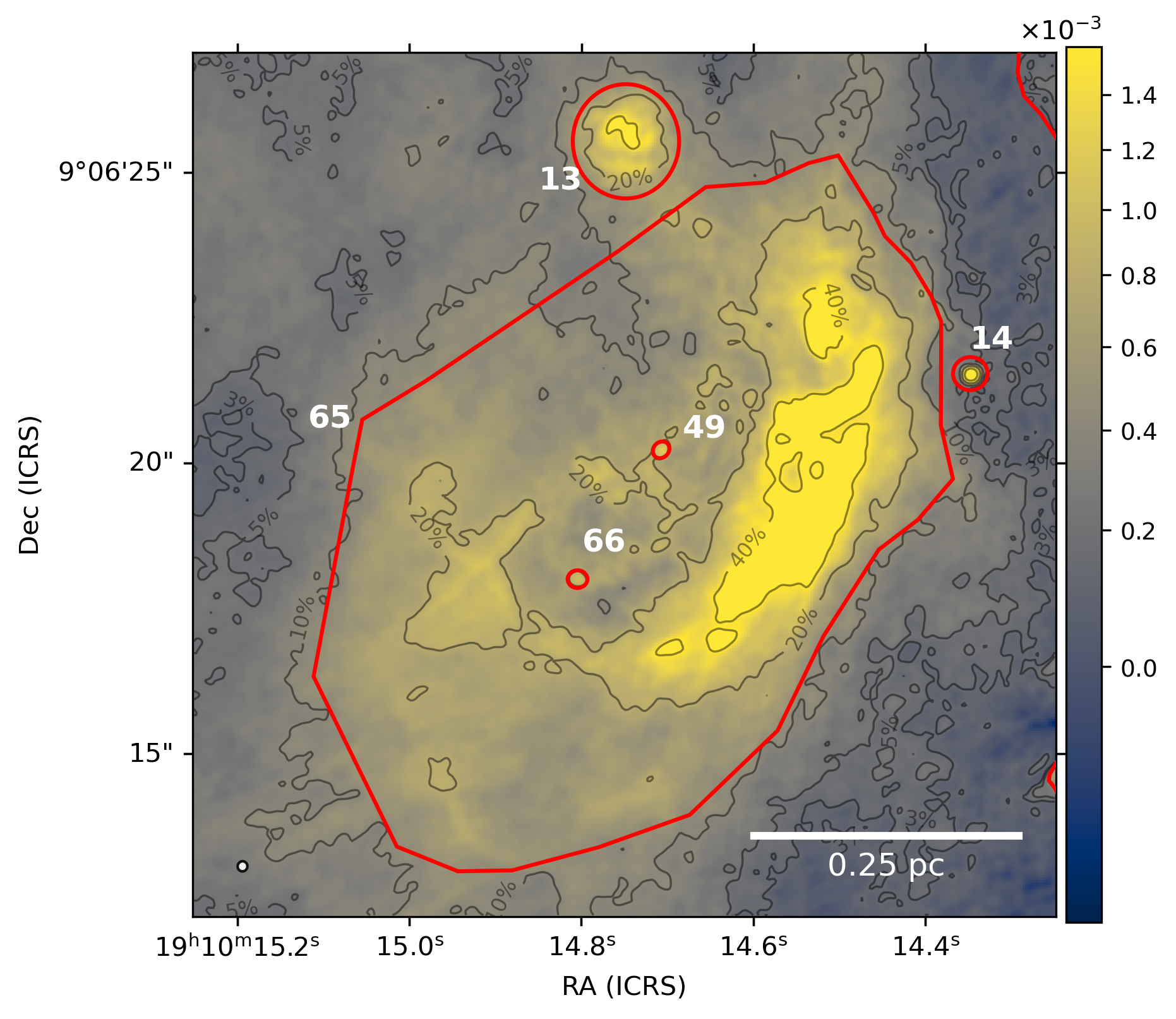}
    \includegraphics[width=0.32\textwidth]{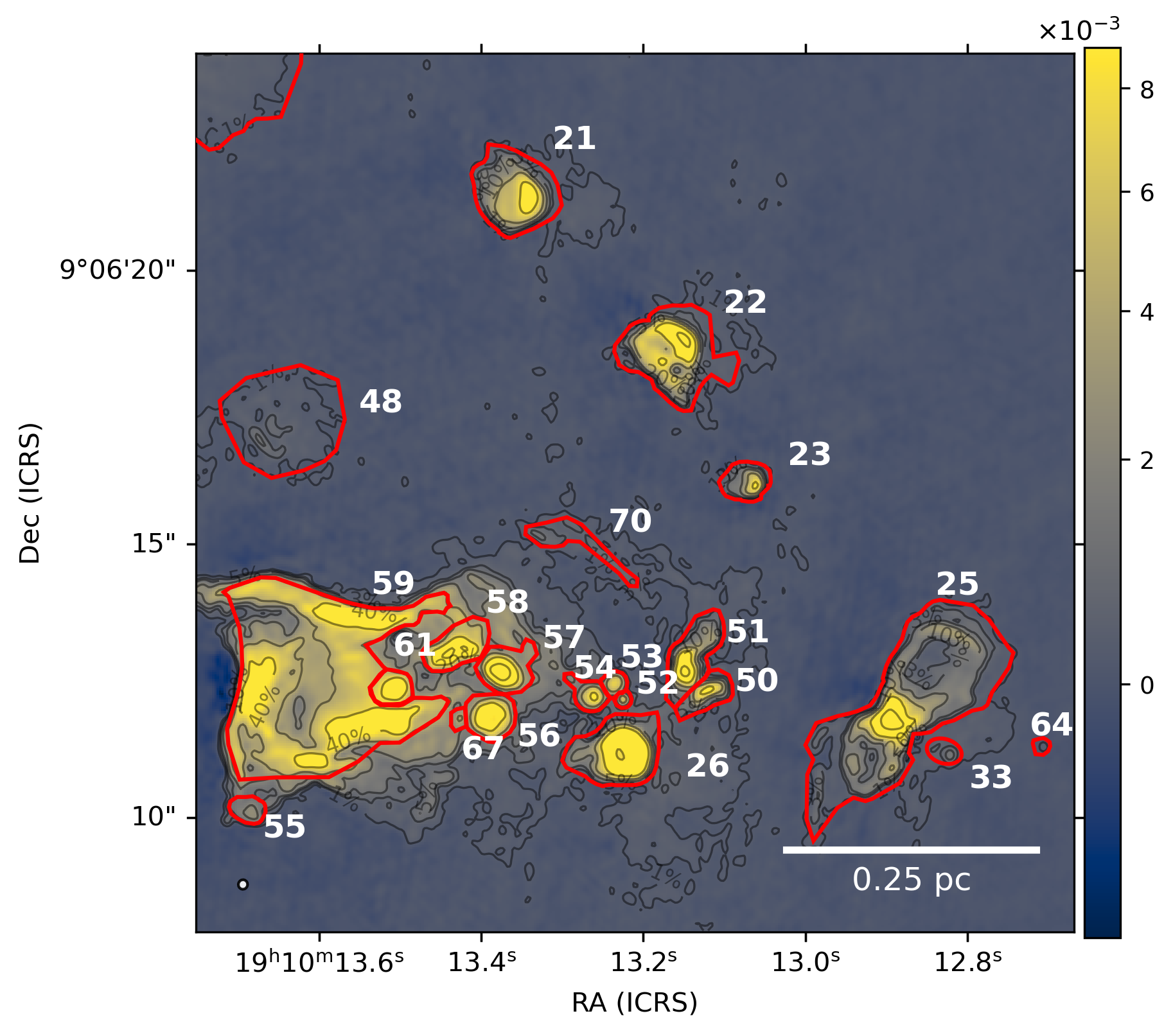}
    \includegraphics[width=0.32\textwidth]{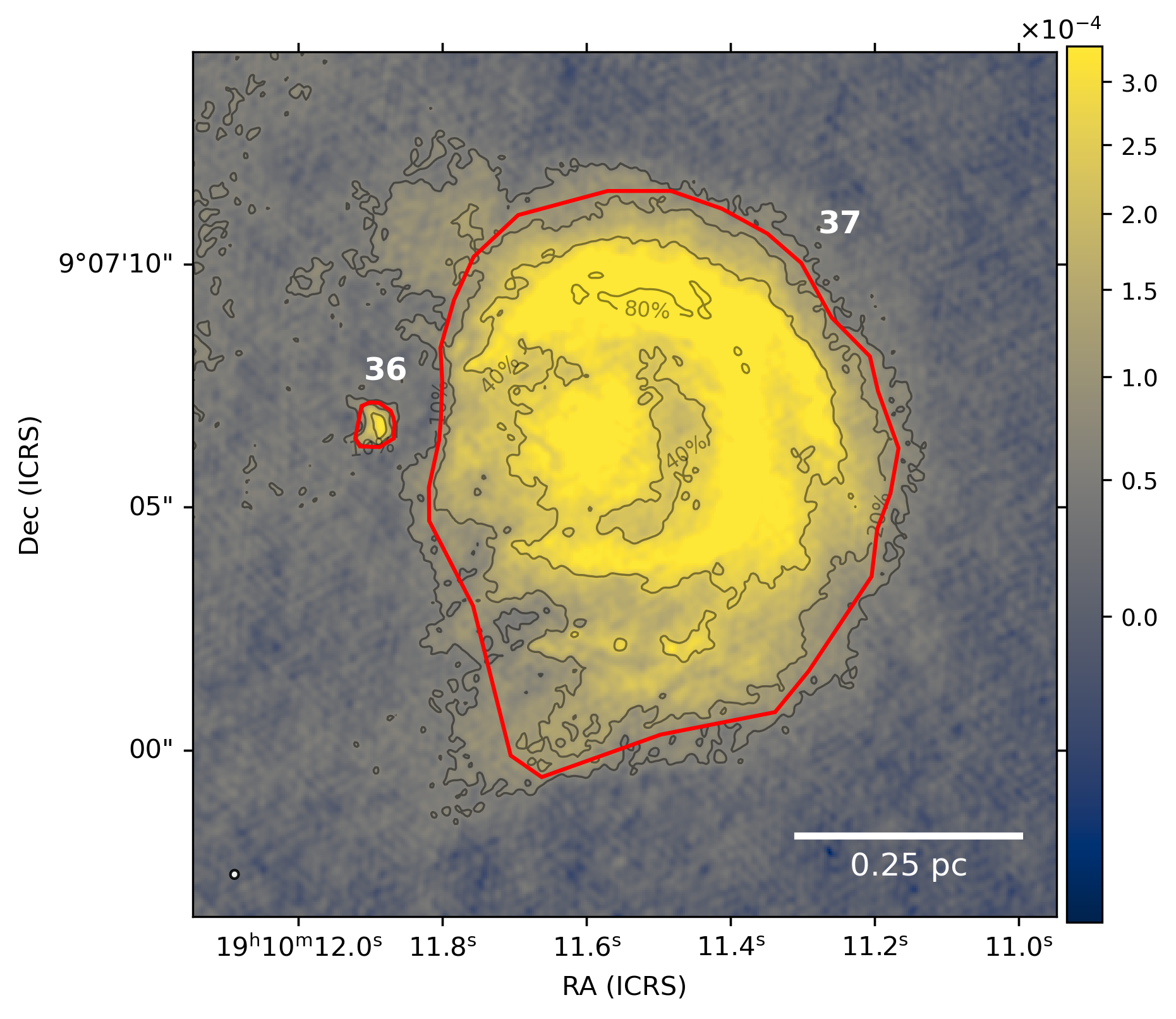}
    \includegraphics[width=0.32\textwidth]{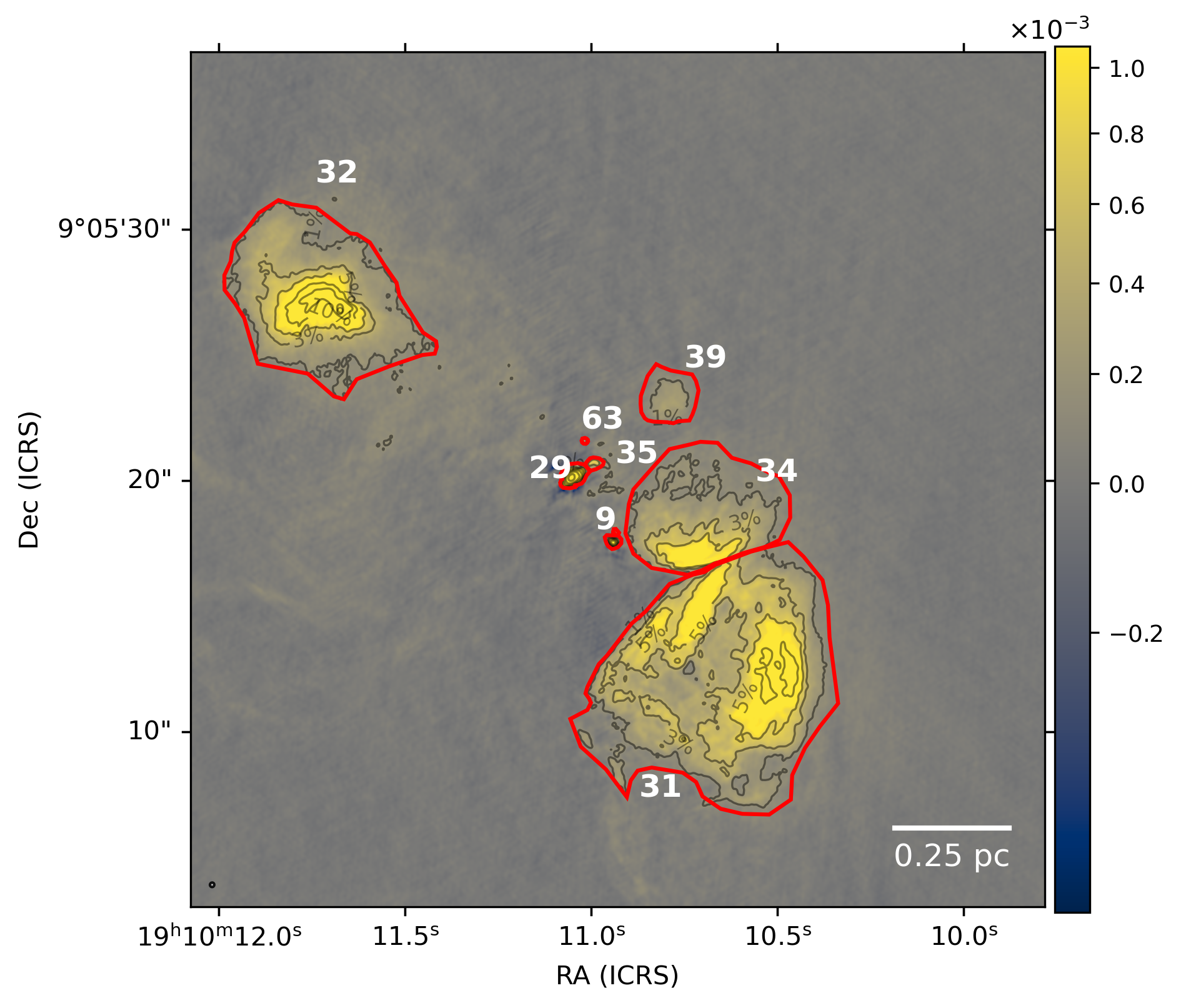}
    \caption{Zoomed-in views of groups of H\textsc{ii} regions. In each panel, a white scale bar in the lower right corner indicates the physical size as labeled. The synthesized beam (FWHM) is shown as a white ellipse in the lower left corner for resolution comparison. Contours are relative percentages of the mean peak intensity of the sources within each panel. 
    } \label{fig:zoom}
\end{figure*}

We performed source identification on a map prior to primary-beam correction to achieve a noise level closer to spatial uniformity, 
and later corrected the photometry of each source for the primary-beam response.
Figure \ref{fig:EVLAW49Azoom} shows the 9.1 GHz (3.3 cm) continuum map used for the identification of the ionized sources. 
Figure \ref{fig:zoom} shows zoomed-in panels for a representative subset of groups of sources, while the complete catalog of individual sources is presented in the Appendix \ref{sec:imaging_sources} (Figure \ref{fig:individual}, with the online figure set containing all 84 images).

From the masks defining each source, we obtained the peak intensity ($I_\mathrm{pk}$), the coordinates of the peak intensity ($\alpha_{\rm pk}$, $\delta_{\rm pk}$), the measured area ($A_\mathrm{m}$), and the flux integrated over the exact mask ($F_\nu$).  
We  defined the measured equivalent diameter as $D_\mathrm{eq,m} = 2\sqrt{A_\mathrm{m}/\pi}$. 
To obtain a deconvolved source diameter $D_\mathrm{eq,d}$, we subtract the equivalent diameter of the gaussian beam $D_\mathrm{eq,b}$ in quadrature, i.e., $D_\mathrm{eq,d}^2=D_\mathrm{eq,m}^2 - D_\mathrm{eq,b}^2$. 
The deconvolved area is thus $A_\mathrm{d}=\pi D_\mathrm{eq,d}^{2}/4$. 
The error in  $A_\mathrm{m}$ for each source was estimated by varying the boundary of the mask by $\pm 1$ pixels.
The error in $F_\nu$ was estimated as the quadratic sum of a 5\% error in the absolute flux scale of the VLA and the photometric flux error, propagating the local rms noise over the source mask. 
We apply  background subtraction, as described in \citet{RiveraSoto20}. The local background for each source was defined using the average intensity in rings exterior to the source's mask, or groups of them in crowded areas, while avoiding any contribution from neighboring sources. 
These properties are listed for each source in Table \ref{tab:prop} in Appendix \ref{sec:indiv_sources}.

We corroborated that the centimeter emission of W49A is mostly thermal by calculating the spectral index between 1.3 and 9.1 GHz, using a reprojection of the publicly available images of the  SARAO MeerKAT 1.3 GHz Galactic Plane Survey \citep[SMGPS, $8\arcsec$ resolution, ][]{Goedhart2024}. The global spectral index between these frequencies is $\alpha \approx -0.13$, corresponding to optically-thin free-free emission.
Additionally, we verified that the level of contamination from background extragalactic sources is negligible. Using the prescription in the appendix of \citet{Anglada1998} at 9.1 GHz for the $\approx 6\arcmin \times 5\arcmin$ area of our detections, the expected number of background sources with flux $> 0.1$ mJy (the faintest in our sample) is $N_\mathrm{bg} \approx 1$. Recently, deep cosmological surveys at centimeter wavelengths have provided updated measurements. We use the VLA X-band results of the GOODS-N survey presented in \citet{Jimenez2024}. The resulting number of background sources with flux $> 0.1$ mJy  is $N_\mathrm{bg} \approx 2$ to 3. 
Furthermore, any background contaminant is expected to be very compact or point-like. We have verified that only four sources in our sample have $D_\mathrm{eq,d} < D_\mathrm{eq,b}$. These are naturally excluded 
for the derivation of the high end of the stellar mass function presented in Section \ref{sec:MF} due to their low fluxes.  

\section{Results} \label{sec:results}
\subsection{Physical Properties of Young H\textsc{ii} Regions} \label{sec:phys-prop}

\begin{figure}[!t]  
\centering 
\includegraphics[width=0.7\linewidth]{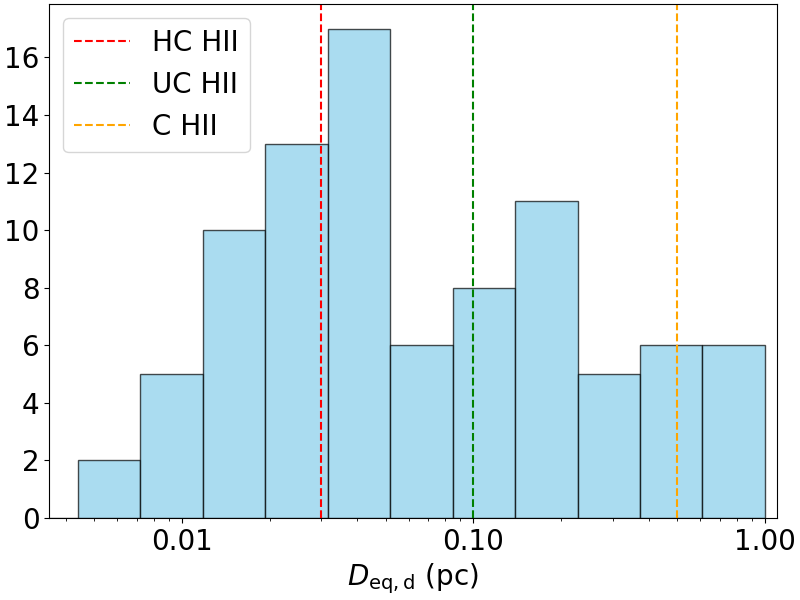}
\caption{
Distribution of deconvolved equivalent diameters ($D_{\mathrm{eq,d}}$) for the 87 identified H\textsc{ii} regions. The sample spans from 0.0053 to 0.99 pc. Vertical dashed lines indicate the classification boundaries following \citet{kurtz2005}: hypercompact (red, $D_{\mathrm{eq,d}}\leq$0.03 pc), ultracompact (green, 0.03$<D_{\mathrm{eq,d}}\leq$0.1 pc), and compact (yellow, 0.1$<D_{\mathrm{eq,d}}\leq$0.5 pc). Sources with $D_{\mathrm{eq,d}}>$0.5 pc are classified as classical H\textsc{ii} regions.}
\label{fig:Deq_histo}
\end{figure}

We identified a total of 87 sources with deconvolved sizes $D_\mathrm{eq,d}$ ranging from 0.0053 to 0.99 pc ($\sim 1.1\times10^3$ to $2.0\times10^5$ au), with the smallest values being consistent with high-resolution interferometric reports of compact structures in the region \citep[e.g.,][]{Depree2020}.
Following the classification scheme of \citet{kurtz2005}, 28 of these are considered hypercompact  (HC, $D_\mathrm{eq,d} \leq 0.03$ pc), 24 ultracompact  (UC, $0.03 < D_\mathrm{eq,d} \leq 0.1$ pc), 26 compact ($0.1 < D_\mathrm{eq,d} \leq 0.5$ pc), and 9 classical H\textsc{ii} regions ($D_\mathrm{eq,d} > 0.5$ pc). Figure \ref{fig:Deq_histo} shows the distribution of sizes in our sample. 

We used the background-subtracted flux ($F_\nu$) and deconvolved area ($A_\mathrm{d}$) of each unique source to estimate their physical properties, such as the emission measure (EM) and the amount of Lyman continuum photons from the ionizing star ($Q_\mathrm{0}$) \citep[see, e.g.,][]{Kurtz94, RiveraSoto20}. The properties are listed in Table \ref{tab:propderiv} in Appendix \ref{sec:indiv_sources}. We derived the Rayleigh-Jeans brightness temperature using $T_\mathrm{b} =c^2F_{\nu}/2k\nu^2\Omega_\mathrm{d}$, where $F_\nu$ is the integrated flux density of each source, $\nu$ is the observing frequency, and $\Omega_\mathrm{d} = A_\mathrm{d}/d^2$ is the deconvolved solid angle computed from the deconvolved area $A_\mathrm{d}$ listed in Table~\ref{tab:prop}, assuming a distance of $d = 11.1$ kpc. 

We derived the continuum optical depth from $\tau_\nu=-\ln (1-T_\mathrm{b}/T_\mathrm{e})$, assuming an electron temperature of $T_\mathrm{e}=10000$~K, which is consistent with the LTE electronic temperatures derived from radio recombination lines by \cite{depree1997}.
A few sources approach the limit $T_\mathrm{b} \lesssim T_\mathrm{e}$ at their peaks, indicating large optical depths at their centers at 3.3 cm. To correct for these, we also made use of the available photometry at 7 mm from \citet{DePree2000}. 

From the optical depth, we then derived the emission measure $\mathrm{EM} \equiv \int n_\mathrm{e}^2 ds$ (units of pc cm$^{-6}$) using $\mathrm{EM} =  (1/8.235\times10^{-2}) T_\mathrm{e}^{1.35}\nu^{2.1}\tau_{\nu}$, where $\nu$ is in GHz \citep[see, e.g.,][]{Tools_6thed}. We further estimated the electron density from $n_\mathrm{e}=\sqrt{\mathrm{EM}/s}$, where we take the line-of-sight depth $s$ of the H\textsc{ii} region to be equal to $D_\mathrm{eq,d}$. Finally, we calculated the the ionizing photon rate from $Q_\mathrm{0}= (1.471\pi/6)\, \mathrm{EM}\,\alpha_\mathrm{t}\,D_\mathrm{eq,d}^2$ \citep{MezgerHenderson1967}. We consider the case-B recombination coefficient $\alpha_\mathrm{t} = \alpha_\mathrm{B} \approx 2.6 \times 10^{-13}$  cm$^3$ s$^{-1}$ due to the high densities in UC H\textsc{ii} regions \citep{Osterbrock2006}. For the photometric errors ($\sigma_\mathrm{p}$) in the derived quantities, we propagated the errors in the flux ($\sigma_\mathrm{F}$), area ($\sigma_\mathrm{A}$) and also considered a 10\% error in the assumed temperature $\sigma_{T_\mathrm{e}} = 0.1T_\mathrm{e}$.

\subsection{Inferred Stellar Masses} \label{sec:stellar-masses}

\begin{figure}[!t]
    \centering
    \includegraphics[width=0.49\textwidth]{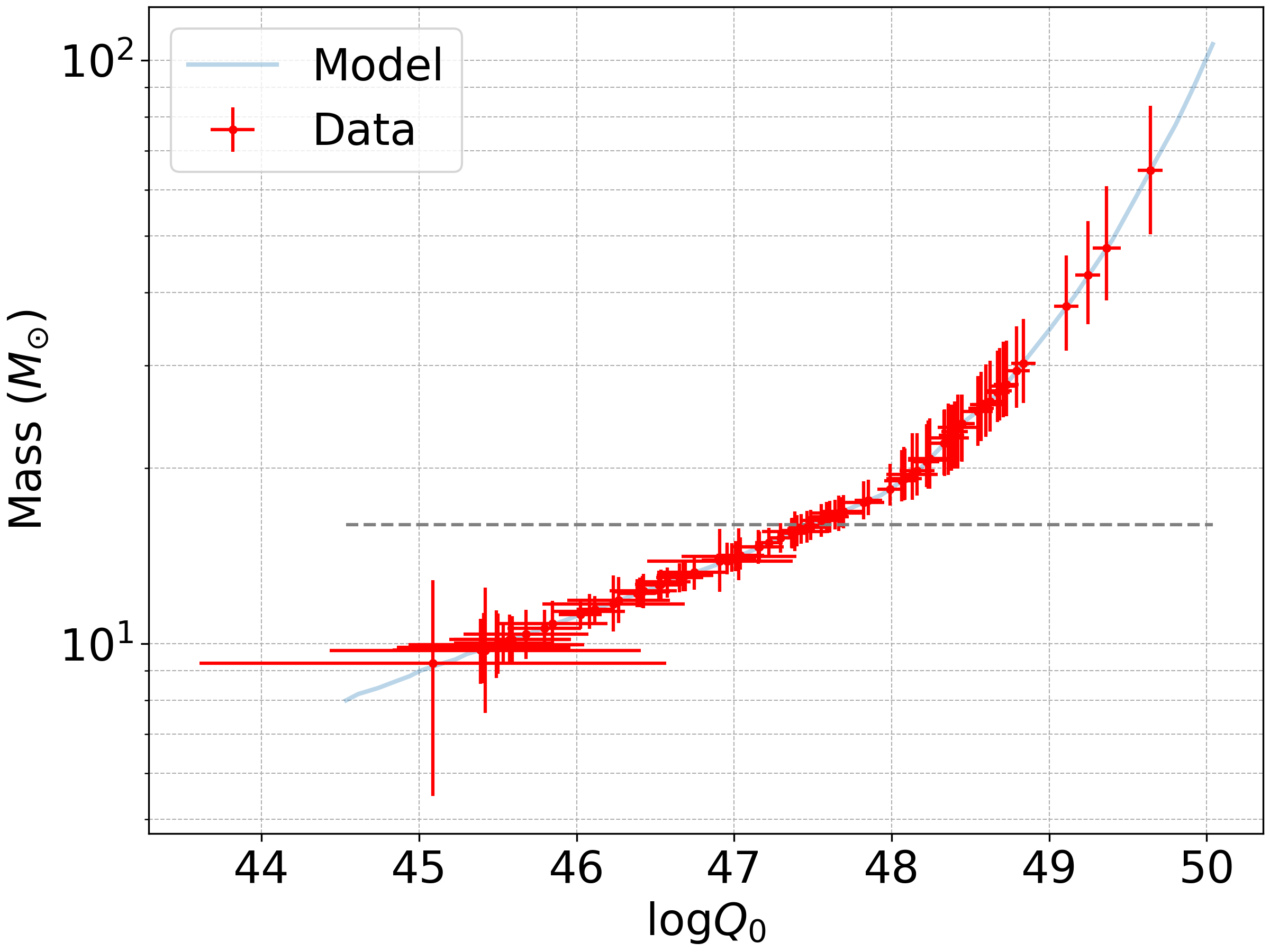}
    \includegraphics[width=0.49\textwidth]{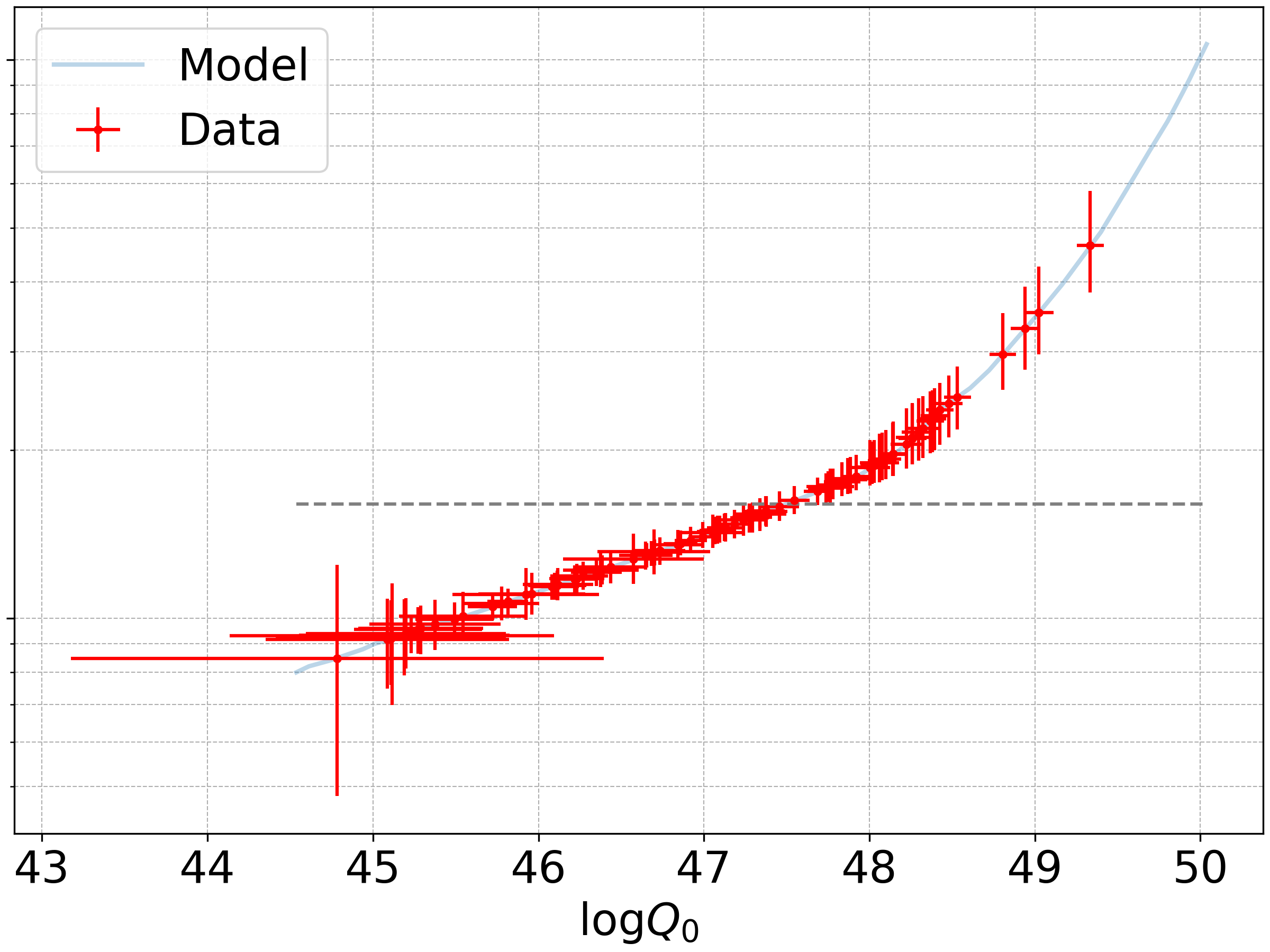}
    \caption{Relationship between stellar mass $M$ and $\log Q_{0}$, derived using the CB19 calibration for ZAMS stars of solar metallicity, shown by the blue line. Red dots correspond to our observed H\textsc{ii} regions, assuming  a single-star scenario ({\it left panel}) and an equal-mass binary scenario ({\it right panel}). The horizontal gray dashed line at $M = 16~M_{\odot}$ marks the approximate transition between the O-type and B-type stellar regimes.} 
    \label{fig:calib}
\end{figure}

We estimate the mass of the ionizing stars responsible for the observed H\textsc{ii} regions using the $Q_\mathrm{0}$\,vs.\,$M$ stellar calibration from the revised version of the \citet[][BC03]{BruzualCharlot2003} stellar population synthesis models introduced by \citet[][CB19 models hereafter]{Plat2019}. The CB19 models represent a major revision of the BC03 models. The CB19 models follow the \texttt{PARSEC} evolutionary tracks \citep{Marigo2013,Chen2015} for 16 different chemical compositions. Tables 8 to 12 of \cite{Sanchez2022} list the different metallicities and describe in detail the theoretical and empirical stellar spectral libraries used to build the CB19 models.
In this work, we use the CB19 calibration for solar metallicity $Z=0.017$, restricted to ZAMS stars, thus ensuring consistency with the assumption of young, massive ionizing sources. 
This relationship,  shown as a {\it blue line} in Figure \ref{fig:calib}, is built as follows.
The \texttt{PARSEC} tracks in the theoretical H-R diagram provide the effective temperature and bolometric luminosity of stars of mass $M_k$ on the ZAMS. To each of these stars, we assign a spectrum from the \texttt{Tlusty} grid of stellar model atmospheres for O- and B-type stars computed by \cite{Lanz2003,Lanz2007}, respectively, interpolating their solar metallicity grids to the physical parameters of star $M_k$, and then scaling the model spectrum to its bolometric luminosity. 
From the scaled spectrum we compute $Q_\mathrm{0}(M_k)$.
The line shown in Figure \ref{fig:calib} ranges from $M_k\,=$
8 to 100 $M_\odot$. To facilitate the mass conversion, we binned the stellar parameters into mass intervals and computed the mean values and associated standard errors for the ionizing photon production rate $\log Q_0$ and the bolometric luminosity $L$. The resulting calibration values are summarized in Table \ref{tab:cb19_zams_calibration}.
\texttt{Tlusty} is a non-LTE code which predicts the intensity of thousand of atomic lines assuming plane-parallel stellar atmospheres in hydrostatic and radiative equilibrium. The non-LTE treatment is fundamental for hot stars, for which LTE models fail and, in particular, the ionizing flux is significantly affected by line-blanketing. \cite{Lanz2003} compare the values of $Q_\mathrm{0}$ derived from the \texttt{Tlusty} models to the values derived from models computed with the non-LTE, line-blanketed, expanding model atmospheres \texttt{WM-Basic} code
\citep{pauldrach2001}, concluding that both sets of $Q_\mathrm{0}$ values agree within 0.1 dex, except for a few models with non-solar metallicity.
Thus, we do not expect that the use of the $Q_\mathrm{0}$ values derived from the \texttt{Tlusty} models introduces larger systematic errors.
The results obtained with this procedure are almost identical to those obtained using the O-type stellar calibration from \citet{Martins2005}, but the proposed procedure also allows us to infer stellar masses $M<16~M_\odot$ (B-type).

We obtained the total error in our determination of $M$ through Monte Carlo simulations. For each source, we generated a set of random values of $\log Q_\mathrm{0}$, assuming a Gaussian distribution that varied concurrently due to two independent sources of uncertainty: the photometric error ($\sigma_\mathrm{p}$) and the error in the assumed stellar model ($\sigma_\mathrm{c}$). The latter term represents the intrinsic dispersion in the $\log Q_\mathrm{0}$--$M$ relation within the CB19 model grid. 
Although this dispersion varies with stellar mass, we adopt a conservative approach and use the maximum value measured across the full mass range, $\sigma_\mathrm{c} \simeq 0.21$ dex, for all sources.
We implemented a nested sampling approach: we generated $N_{\sigma_\mathrm{p}}=10^{3}$ random values based on the photometric error, and for each of these, we generated $N_{\sigma_\mathrm{c}}=10^{3}$ additional values based on the stellar calibration error. This resulted in a total of $N_{\sigma_\mathrm{p}}N_{\sigma_\mathrm{c}}=10^6$ simulations per source. The reported $M$ is the median of the resulting mass distribution. The upper and lower limits of the associated error are defined by the 16th and 84th percentiles of the distribution. The resulting values of $M$ are listed in Table \ref{tab:propderiv}.

These mass estimates primarily assume a single-star scenario for ionization.
However, we also performed the same procedure under the alternative assumption of an equal-mass binary system ionizing the H\textsc{ii} region.
This binary scenario also serves as a simplified way to account for unresolved multiplicity, whereby a single detected H\textsc{ii} region may be powered by more than one ionizing source. 
In the binary scenario, each star contributes $F_\nu/2$ to the observed flux.  
Figure \ref{fig:calib} shows the calculated masses and associated errors, based on the CB19 models, for both scenarios.

We obtained stellar masses ranging from 9.3 to 64.8 $M_\odot$ in the individual-star scenario, while in the binary-system scenario, the stellar mass range is from 8.5 to 46.5 $M_{\odot}$.
The sum of the stellar masses of all identified sources is $M_{\rm tot,single} \simeq $\,1.58$\times10^3\,M_\odot$ in the single-star scenario and $M_{\rm tot,binary} \simeq $2.75$\,\times10^3\,M_\odot$ in the equal-mass binary scenario.
Since the stellar mass is dominated by the lower-mass population, the total stellar mass in the cluster should be one order of magnitude higher (see Section \ref{sec:disc_incomplete}).

\label{fig:mass_q0}
We perform an a-posteriori consistency check to verify that our interpretation of the identified sources as H\textsc{ii} regions powered by young massive stars is compatible with the known bolometric luminosity of the W49A cloud. Although stellar luminosities are not a primary focus of this work, they can be estimated self-consistently from the same CB19 stellar models used to infer stellar masses. 
The total far-IR luminosity of W49A has been estimated to be $L_{\rm bol} \simeq 3.7\times10^{7}\,L_\odot$ \citep{Lin2016}. Summing the stellar luminosities corresponding to the inferred masses of all identified sources, we obtain a total luminosity of $L_{\rm tot,single} \simeq $4.69$\times10^{6}\,L_\odot$ in the single-star scenario and $L_{\rm tot,binary} \simeq $6.12$\times10^{6}\,L_\odot$ in the equal-mass binary scenario. 
In both cases, the inferred stellar luminosity remains below the observed bolometric luminosity of the cloud. This is expected since there should be other sources of radiation in the cloud besides the ionizing stars of the H\textsc{ii} region population, namely, massive protostars prior to the formation of an H\textsc{ii} region \citep{Nony2024}, the more evolved  massive stars not detected as radio sources \citep{DeBuizer2021}, and their low-mass counterparts \citep{Saral15}.  

\subsection{Inferred Stellar Mass Function} \label{sec:MF}

For the entire sample, we performed a fit of a non-discretized power-law distribution of the form:
\begin{equation}\label{eq:pwplfit}
\frac{dN}{d(\log M)} \propto M^{-(b-1)} = M^{-\Gamma}.
\end{equation}
We carried out this fit with the Python package \texttt{plfit}, which is based on the work of \cite{Clauset09} to obtain the slope $\Gamma$ through the combination of maximum likelihood fitting methods and goodness-of-fit tests based on the Kolmogorov-Smirnov (KS) statistic and likelihood ratios.
We evaluated the fit at each possible value of the minimum fitted mass $M_{\rm min}$, starting at $M_{\rm min} = $9.3$~M_\odot$ (8.5$~M_\odot$ considering binaries) to utilize all data. For each case, we calculated the respective KS statistic and $p$-value.
We emphasize that this does not imply the sample is complete down to $M_{\rm min}$; rather, the IMF is fitted only for masses above this threshold, where the sample is considered reliable.
We examined the resulting $\Gamma$–KS relations and selected the best-fit solution as the one minimizing the KS statistic while retaining at least $\sim 40$ sources above $M_{\rm min}$ to avoid biases due to small-number statistics. We considered fits with $p$-values greater than 0.1 acceptable, i.e., those not rejecting a power law as a valid model. 

For the originally calculated distribution of masses under the single-star assumption, the maximum-likelihood fit yields a slope of $\Gamma \approx$ 2.60. 
To propagate the uncertainties in the inferred stellar masses, we performed Monte Carlo realizations of the mass distribution, perturbing the individual stellar masses in $\log M$ according to their error. For each realization, we re-fitted the IMF slope following the same procedure used for the unperturbed mass distribution, including the adopted criteria for $M_{\rm min}$.
The resulting distribution of slopes has a median value of $\Gamma = $ 2.60, with uncertainties defined by the 16th and 84th percentiles of the Monte Carlo realizations (see Table~\ref{tab:propfits}). We adopt this interval as our final uncertainty estimate, as it incorporates the effect of observational errors in the inferred stellar masses.

\begin{figure}[!t]
    \centering
    \includegraphics[width=0.48\textwidth]{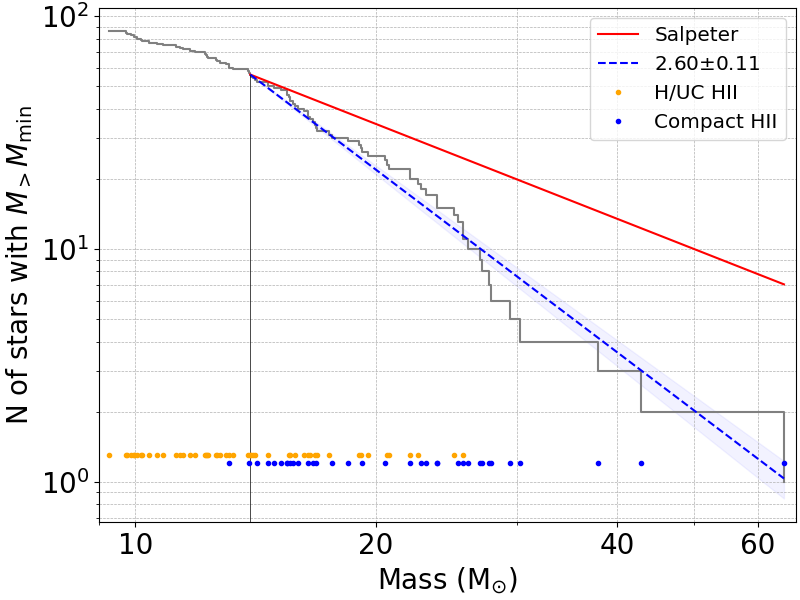}
    \includegraphics[width=0.48\textwidth]{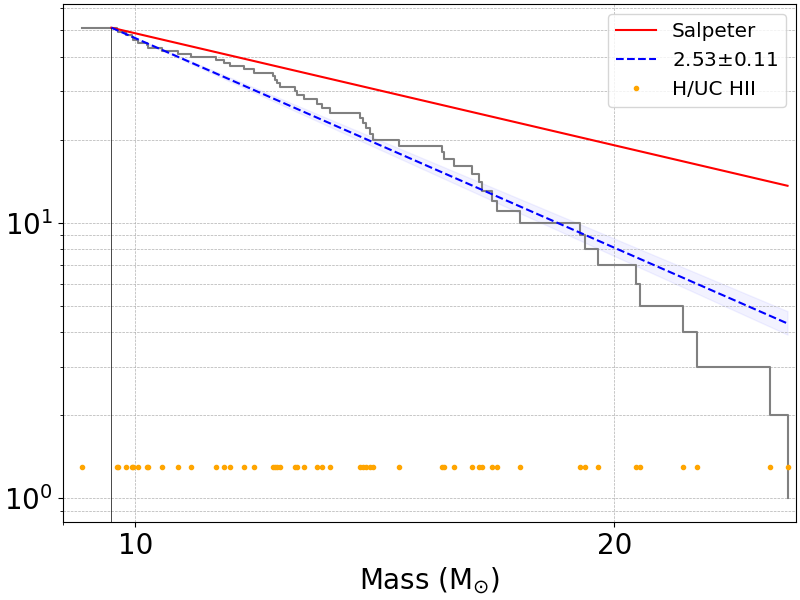}
    \includegraphics[width=0.48\textwidth]{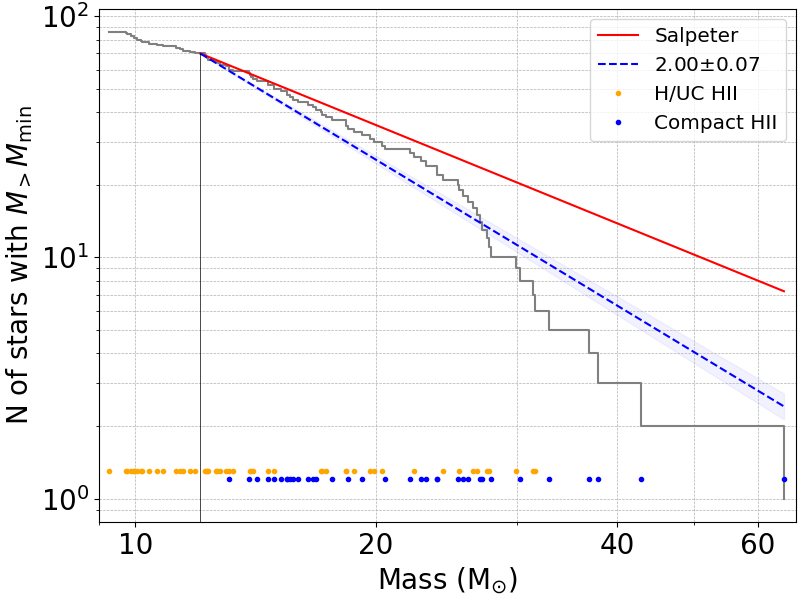}
    \includegraphics[width=0.48\textwidth]{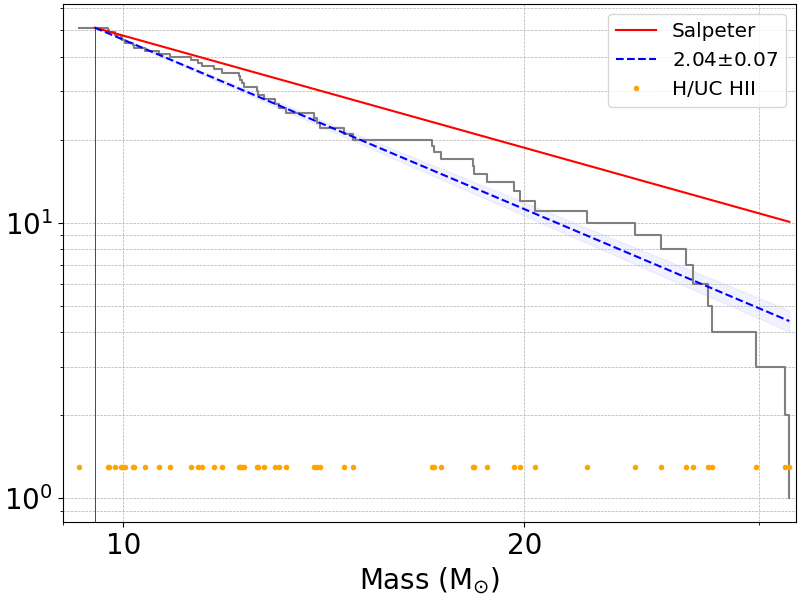}
    \caption{
    Logarithmic distributions of the inferred stellar masses, shown as complementary cumulative distribution functions (CCDFs). The {\it top row} presents the masses inferred from 3.3~cm continuum flux densities. The {\it bottom row} shows the results after applying optical depth corrections using the 7~mm data. The gray line shows the CCDF assuming a single-star scenario for the full sample ({\it left}), and the non-elongated UC and HC H\textsc{ii} regions ({\it right}). The blue dashed line indicates the best-fit power law for $M > M_{\text{min}}$. For the 3.3 cm data ({\it top}) $M_{\text{min}} = 13.9~M_{\odot}$ and $M_{\text{min}} = $9.7$~M_{\odot}$, while for the corrected 7~mm data ({\it bottom}),  $M_{\text{min}} = $12.1$~M_{\odot}$ and $M_{\text{min}} = $9.5$~M_{\odot}$ for the left and right panels, respectively. The red line represents the high-mass Salpeter slope, $\Gamma = 1.35$. Orange and blue points at the bottom mark individual stellar masses from the most compact (UC and HC) and less compact H\textsc{ii} regions, respectively.
    }
    \label{fig:imfplfit}
\end{figure}

Table \ref{tab:propfits} summarizes the best-fitting parameters for each of the considered physical scenarios. Additionally, results are shown for the full sample and for a subset that includes only the most compact sources (UC and HC H\textsc{ii} regions) and excludes a single visually identified elongated source (Source 70, characterized by an axis ratio $b/a<$0.5).  
In this subsample, we excluded the larger or elongated sources  because they are more likely to trace unresolved blends of multiple sources, represent regions that deviate significantly from the assumed geometry, or be externally photoionized extended structures.
In all cases prior to optical-depth correction (see below), the inferred power-law indices are consistently steeper than $\Gamma =$ 2.42 after considering $1\sigma$ errors.
Figure \ref{fig:imfplfit} presents the distribution of the inferred stellar masses for the single-star scenario, both for the full and restrictive samples. 
The best-fit power-law slopes above $M_{\rm min}$ are overplotted and compared with the Salpeter IMF slope
\citep[$\Gamma_{\rm Salp} = 1.35$,][]{Salpeter1955}, which is taken as an exact benchmark. 
The fitted slopes are inconsistent with the Salpeter slope by $11\sigma$. In the binary scenario, the differences are even larger, reaching  $\mathbf{13\sigma}$ to $\mathbf{16\sigma}$.

\begin{table}
\caption{Best-fit power-law parameters obtained with \protect\texttt{plfit} for the inferred stellar mass distributions. Results are shown for single-star and binary assumptions, considering both the full sample and the subset of non-elongated UC and HC H\protect\textsc{ii} regions (see text). Rows labeled as ``$\tau$-corr Single'' indicate the results after applying optical depth corrections to selected sources using 7 mm data. The bottom row shows the result of a robustness test after excluding the maximum number of potentially dubious sources (diffuse, non-discrete, or marginally resolved). The minimum fitted mass $M_{\rm min}$, number of sources above $M_{\rm min}$, power-law index $\Gamma$, and the fraction of Monte Carlo simulations that produce a $p_{\rm KS}$ value $>0.1$.}
\label{tab:propfits}
\centering
\begin{tabular}{cccccc}
\hline
\multicolumn{2}{c}{Scenario}  &  $M_{\rm min}$ (M$_{\odot}$) & $N$($>M_{\rm min}$)&$\Gamma$ & $p_{\rm KS}>0.1$ \\
\hline \hline

\multirow{2}{4em}{Single} & All sources   & 13.9 &56& 2.60 $\pm_{0.11}^{0.13}$ & 99.4 \\
                             & Compact sources & 9.7& 50& 2.53 $\pm_{0.11}^{0.10}$  & 31.8 \\\hline
\multirow{2}{4em}{Binary} & All sources & 13.9 & 98& 3.37$\pm_{0.15}^{0.13}$ & 98.8 \\
                           & Compact sources & 10.5 &76& 3.53 $\pm_{0.14}^{0.14}$ & 48.2 \\
\hline
\multirow{2}{4em}{$\tau$-corr Single} &All sources& 12.6&65&2.12 $\pm_{0.08}^{0.09}$ &93.0\\
                            &Compact&9.5&50&2.04 $\pm_{0.08}^{0.07}$&82.0\\
\hline
Single&\multirow{2}{4em}{Excluding dubious sources}& 12.2&58&2.23$\pm_{0.10}^{0.12}$&90.4 \\
$\tau$-corr Single&& 11.8&59&1.89$\pm$0.07&94.8\\
&\\
\hline
\end{tabular}
\end{table}

\begin{figure}[!t]  
\centering 
\includegraphics[width=0.8\linewidth]{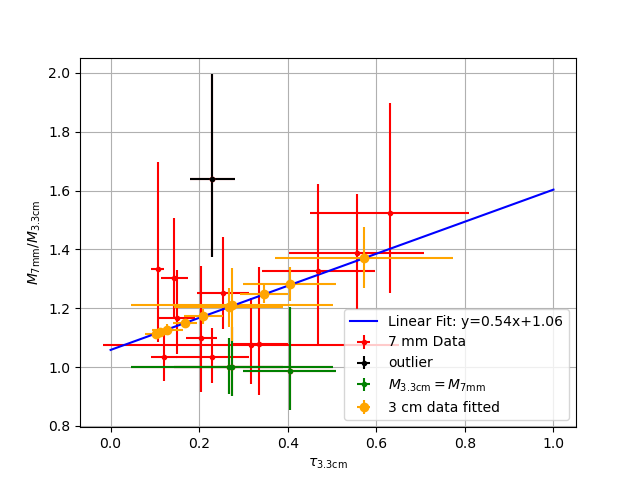}
\caption{
Correlation between the mean optical depth at 3.3 cm ($\tau_\mathrm{3.3cm}$) and the mass ratio ($M_{7\mathrm{mm}}/M_{3.3,\mathrm{cm}}$). The sources with 7 mm measurements used for the fit are plotted in red. The outlier that was excluded from the regression is shown in black. The sources that are likely affected by interferometric filtering at 7 mm are shown in green. The best‑fit linear relation is drawn in blue, while the yellow points represent the remaining sources with $\tau_{3.3\mathrm{cm}}>0.1$.
        }
\label{fig:pearson}
\end{figure}

The steep slope inferred from the 3.3 cm data might be partly influenced by optical-depth effects in the densest H\textsc{ii} regions. As mentioned in section \ref{sec:phys-prop}, a fraction of the sources exhibit peak brightness temperatures approaching the electron temperature of $10,000$ K, indicating that the 3.3 cm emission at the center of these sources is partially optically thick, which would lead to an underestimation of their stellar masses. 
In the 3.3 cm data, we identify a total of 23 sources with mean $\tau_{\rm 3.3cm}> 0.1$.  
Since the optical depth of free-free emission scales with frequency as $\tau_\mathrm{ff} \propto \nu^{-2.1}$ \citep{Tools_6thed}, to correct their inferred masses, we obtained photometry for 16 out of these 23 sources that were clearly detected within the field of view of the 7 mm maps of the central region of W49 presented in \citep{depree1997,DePree2000}.  
Of these sources, three (idx 53, 56, and 67) have approximately the same derived mass at 7 mm and 3.3 cm, contrary to expectations. This likely results from the loss of extended emission due to interferometric filtering in the higher resolution, lower-quality 7 mm map. Consequently, these sources were treated as having underestimated masses, similar to those lacking 7 mm data. 
Using the results from these 13 sources at 7 mm, we found a correlation (Pearson coefficient $r=$0.60, $p$-value of 0.04) between the 3.3 cm mean optical depth and the mass ratio $M_{7\mathrm{mm}}/M_{3.3\mathrm{cm}}$ (Figure \ref{fig:pearson}).
This trend indicates that increasing optical depth at 3.3 cm leads to a progressively stronger underestimation of stellar masses when inferred solely from centimeter-wavelength data.  We use this empirical relation to estimate the optical-depth-corrected stellar masses for the remaining 10 sources with $\tau_{3.3\mathrm{cm}} > 0.1$ that lack 7 mm mass measurements.

To account for the uncertainties in the estimated masses, we performed a two-step error propagation analysis. First, we performed a Monte Carlo simulation to determine the error of the correction factor $f_M = M_{7\,\mathrm{mm}}/M_{3.3\,\mathrm{cm}}$ for each source. For every measurement, we generated $10^3$ random samples of $\tau_{3.3\,\mathrm{cm}}$ following a Gaussian distribution centered on the observed value, with a standard deviation equal to its error. Subsequently, the final error of the corrected mass was calculated by propagating the errors from both the estimated correction factor $f_M$ and the original 3.3 cm mass measurement $M_{3.3\,\mathrm{cm}}$. The corrected mass was then used to obtain the corrected luminosity and its corresponding correction factor $f_L = L_{\mathrm{corr}}/L_{3.3\,\mathrm{cm}}$ following the calibration described in Section \ref{sec:stellar-masses}. Both factors are presented in Table \ref{tab:propderiv} of Appendix \ref{sec:indiv_sources}.

Applying this correction and recomputing the stellar mass distribution and fits under the single-star assumption results in flatter power-law slopes compared to the uncorrected case, $\Gamma = $2.12$ \pm_{0.08}^{0.09}$ (all sources), and $\Gamma = $2.04$_{-0.08}^{+0.07}$ (compact sources). These are still significantly steeper than the standard slope of 1.35 by $9\sigma$ to 10$\sigma$. This test confirms that optical-depth effects at 3.3 cm can partially bias the inferred mass function slope but are insufficient to reconcile the observed distribution with the standard slope of the stellar IMF at high masses. Furthermore, integrating the resulting stellar masses and luminosities (incorporating the corrected values for the 27 sources) yields a total mass of $M_{\rm tot} \simeq $1.67$\times10^{3}\,M_\odot$ and a total luminosity of $L_{\rm tot} \simeq $5.48$\times10^{6}\,L_\odot$, which remains below the total bolometric luminosity of the cloud ($L_{\rm bol} \simeq 3.7\times10^{7}\,L_\odot$).

To further test the robustness of our results against potential contaminants, we performed an additional fit excluding the maximum number of potentially dubious sources (i.e., those identified as diffuse, non-discrete, associated with larger structures, or marginally resolved). The resulting IMF fit yields a slope of $\Gamma = 2.23\pm_{0.10}^{0.12}$ (without correction) and $\Gamma = 1.89\pm0.07$ (with correction). Although these values are slightly lower than our main results, they remain consistently steeper than the Salpeter slope, with $\Gamma \geq 1.89$ under all tested scenarios.

\begin{figure}[!t]  
\centering 
\includegraphics[width=0.8\linewidth]{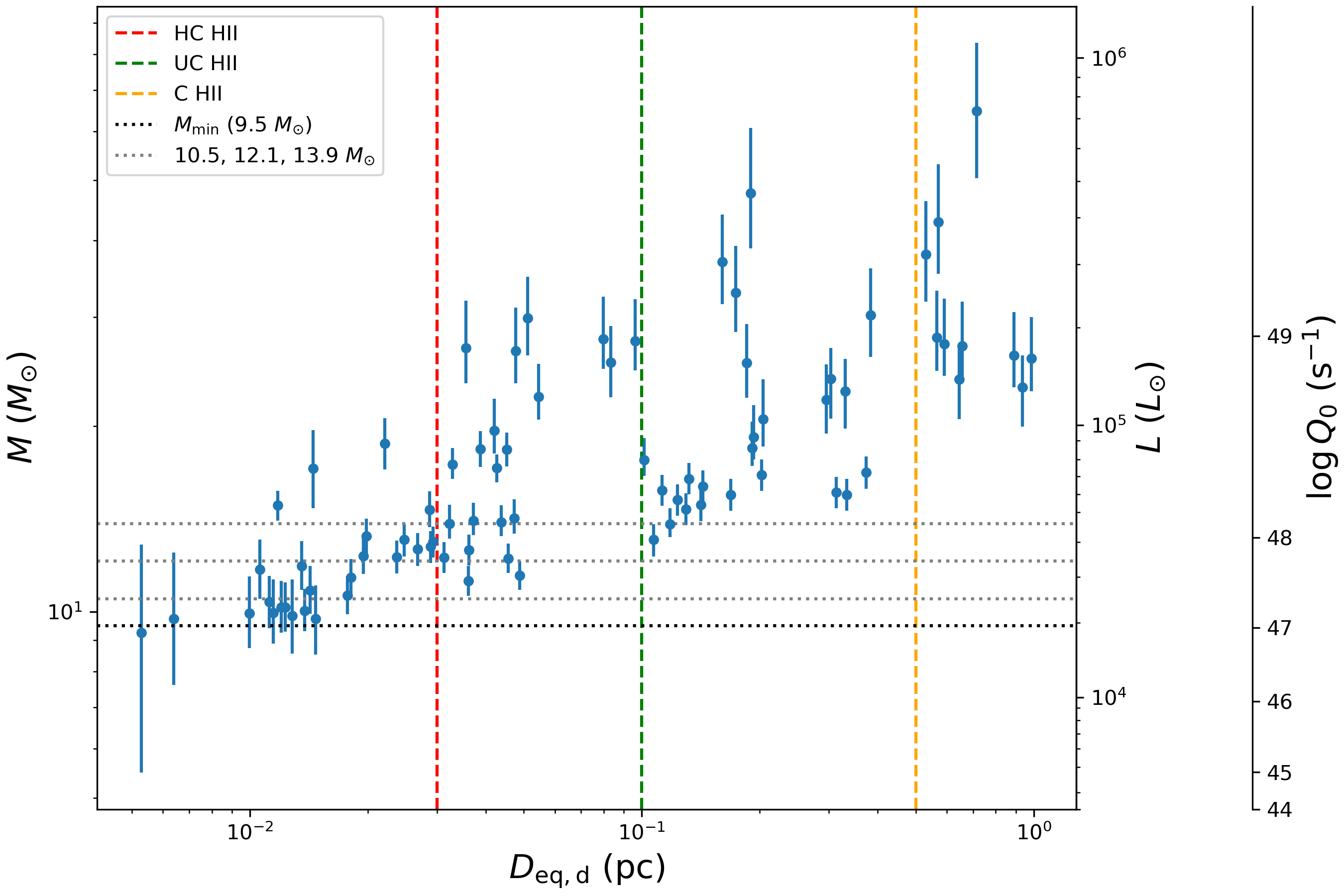}
\caption{
Stellar mass $M$ as a function of equivalent diameter $D_\mathrm{eq,d}$ of the source sample. Dashed vertical lines indicate the classification for hypercompact (red), ultracompact (green) and compact (yellow) H\textsc{ii} regions. The horizontal black dashed line corresponds to the minimum mass threshold ($M_\mathrm{min} = $9.5$~M_{\odot}$) of the fits, , while the gray dotted lines represent alternative minimum-mass cuts across the various scenarios. The right-hand axes display the corresponding luminosity ($L$, inner right) and ionizing photon rate ($\log Q_0$, outer right).
        }
\label{fig:DeqvsM}
\end{figure}

Additionally, we inspect the equivalent diameter ($D_\mathrm{eq,d}$) versus the corrected masses of the sources (see Figure \ref{fig:DeqvsM}). Given that the sources with the smallest physical size correspond to the lowest-mass objects, the adopted minimum-mass cuts across the various scenarios exclude them from the fitting range (indicated by the dotted gray lines, alongside the primary cut in black), ensuring that the high-mass end is determined independently.

\section{Discussion}\label{sec:disc}

We used 3.3 cm radio data of the W49A protocluster at an unprecedented combination of high angular resolution, spatial dynamic range, and sensitivity to conduct a deep census of the H\textsc{ii} regions in one of the most luminous star formation regions in the Galaxy. We used the 87 detected sources to infer the corresponding ionizing-photon fluxes in the standard way and adopted an updated version of the \citet{BruzualCharlot2003} stellar population models to derive the mass of the ionizing stars. Given that radio-detected H\textsc{ii} regions trace ionizing stars essentially in the ZAMS, this method can, in principle, be used to directly infer the stellar IMF in protoclusters. 
The inferred IMF is significantly steeper than the standard IMF power-law slope of $\Gamma_{\rm Salp}=1.35$, with measured slopes $\Gamma \geq 2.23$ for single-star or equal-mass binary assumptions, as well as for the entire sample, a subsample of the most compact, non-elongated sources, and a further subset excluding the maximum number of potentially dubious sources. The inferred IMF slopes remain $\Gamma\geq 1.89$ after optical-depth corrections. 
In this analysis, we do not consider the possibility of  dust attenuation within the H\textsc{ii} regions. While the presence of small amounts of dust inside more evolved H\textsc{ii} regions is well established \citep[e.g.,][]{Paladini2012,Yoo2026}, the possible presence and impact of dust inside UC and HC H\textsc{ii} regions is undetermined. Some models suggest that dust absorption inside the youngest H\textsc{ii} regions could be important \citep{Arthur2004}.

To explore the robustness of our results, we performed several tests varying the source definitions and blending criteria. These experiments show that our derived properties and overall conclusions remain robust against these methodological variations, with changes in the slopes $\Gamma$ of at most $15\%$.
Therefore, it is clear that the young massive stellar population of W49A, traced by its H\textsc{ii} regions, departs from a canonical IMF, having a relative deficit of the most massive stars. Below, we provide what we consider to be the two most likely explanations for our findings. 

\subsection{Shortened Lifetimes of H\textsc{ii} Regions for the More Massive Stars}\label{sec:disc_lifetime}
\smallskip
The relative underabundance of observed stars in the higher-mass end could be explained if the observable lifetime of radio-detected H\textsc{ii} regions becomes shorter for more massive stars, probably due to the increased relative importance of stellar feedback. 
Specifically, the origin of the steep inferred IMF could be the large dependence of the total and ionizing stellar luminosities on stellar mass, coupled with the small size ($< 0.1$ pc) of the molecular cores that become ionized. 
If the observable lifetime of a small H\textsc{ii} region ionized by a star of mass $M$ is $t_\mathrm{HII} \propto M^{-a}$,  where $a$ is a positive constant, the observed mass  distribution derived from H\textsc{ii} regions is: 

\begin{equation}
\frac{dN_\mathrm{HII}}{d (\log M)} \propto M^{-\Gamma_\mathrm{obs}},     
\end{equation}

\noindent
where $\Gamma_\mathrm{obs} = \Gamma_\mathrm{IMF} + a$. 
If we additionally assume that the stellar population is forming with a Salpeter mass spectrum  $\Gamma_\mathrm{IMF} = \Gamma_\mathrm{Salp} = 1.35$, and take the best power law results for the H\textsc{ii} region sample after optical-depth corrections $\Gamma_\mathrm{obs} \geq 1.89$, then $a \geq 0.54$. 
The physical origin of this possible dependence of the lifetime of H\textsc{ii} regions on the ionizing-star mass will require further investigation. 

Observational surveys typically retrieve a single value for the lifetime of UC and HC \textsc{Hii} regions \citep[e.g.,][]{WC89,Kalcheva18,Ginsburg20}, which are then used in observational studies of single clouds and theoretical work \citep[e.g.,][]{Peters10,LiuHonli21,Nony2024}. Using a large sample of \textsc{Hii} regions across the Galaxy, \citet{Mottram2011} concluded that the lifetime of their ``compact'' H\textsc{ii} regions could be slightly decreasing with luminosity: $t_\mathrm{CHII} \propto L^{-0.13\pm0.16}$. However, their sample contains, on average, somewhat larger \textsc{Hii} regions that fall between the definitions of ``compact'' and ``ultra-compact'' (see Fig. \ref{fig:Deq_histo}), and their observations are less sensitive. 

There is evidence from both observations and simulations that radiation pressure and other effects, such as stellar winds, are important in young H\textsc{ii} regions 
\citep{Geen2020,Barnes2020a,Olivier21}. If it is the case that feedback effects play a relatively larger role at higher masses, then the most massive stars would disperse their surrounding ionized gas more rapidly; their \textsc{Hii} regions may cease to appear as compact centimeter sources while the stars themselves remain present in the cluster. 
This is consistent with our observation that the lower-mass ionizing sources in W49A are preferentially associated with the smaller HC and UC H\textsc{ii} regions, suggesting that the more massive stars have already transitioned out of the most compact stages.
This is also consistent with the fact that the identified compact sources account for $\sim 61$\% of the total 3.3 cm flux measured in the map ($F_{\rm compact}=115.5$ Jy out of $F_{\rm tot}=25.4$~Jy). The remaining $\sim 39$\% of the emission is distributed in extended structures that are not decomposed into compact components. Such diffuse free-free emission likely traces more evolved H\textsc{ii} regions whose ionizing stars are no longer associated with well-defined H\textsc{ii} regions in the radio map. 
The presence of a radio-undetected  population of OB stars was also proposed from the results of cm-wavelength surveys in the W51A protocluster, which shares many characteristics with W49A \citep{Ginsburg2016}. 
Identifying such potentially missing stellar populations requires observations at complementary wavelengths. In particular, deep near- and mid-infrared observations with the JWST \citep[e.g.,][]{Yoo2026} may reveal the massive stars that are not associated with radio UC and HC H\textsc{ii} regions, allowing for a more complete census of the cluster populations.

\subsection{An Evolving High-Mass Population}\label{sec:disc_incomplete}
An alternative and potentially complementary explanation comes from the fact that W49A is still actively forming stars \citep[e.g.,][]{GM13,Saral15,Miyawaki2022,Nony2024}, such that the inferred stellar mass distribution represents the current snapshot rather than the final outcome of star formation in the cluster. In this scenario, the apparent deficit of the most massive stars could disappear in the future as star formation proceeds.
Recently, \citet{PadoanGieles2026} reported that a relatively steep luminosity function of H\textsc{ii} regions with respect to the final stellar population is an indication of a relatively slow massive star formation process, as proposed in their inertial-inflow model of star formation in clusters \citep{Padoan2020}.
This interpretation would require that, somehow, massive stars form over timescales as long as a few Myr.

We estimate the amount of missing stars in the high end of the mass distribution by calculating 100 realizations of a cluster following the Kroupa IMF \citep{Kroupa2001} with the same number of stars above $M > 13~M_\odot$ as in our observations,  $N_{>13M_\odot} = 61$. These model clusters have a mass of $M_\mathrm{cl} \approx 1.4\times10^4~M_\odot$. On average, the clusters with the standard IMF have $N_{>20M_ \odot} = 41$  and $N_{>30M_ \odot} = 22$, whereas in W49A we observe $N_{>20M_ \odot} = 29$ and $N_{>30M_ \odot} = 8$. Therefore, in order for this scenario to be plausible, about a dozen objects in our sample that currently have a stellar mass $M > 13~M_\odot$ ought to still be accreting and growing in mass. Accretion rates $\dot{M} > 3\times10^{-5}~M_\odot$ yr$^{-1}$ are needed to increase their mass by $> 10~M_\odot$ in the typical UC H\textsc{ii} lifetime of $\sim3\times10^5$ yr.

That accretion can proceed through the early stages of HC and UC H\textsc{ii} regions has been shown in analytical models \citep{Keto2003,Keto07}, as well as in simulations \citep{Peters10,KuiperHosokawa2018}. However, searches for clearcut examples have proven to be challenging \citep[e.g.,][]{KZK2008,Klaassen2018,Ginsburg20}. The UC H\textsc{ii} at the center of G10.6--0.4 is the clearest known case \citep{Sollins2005,KetoWood06}, but this object is ionized by a compact cluster that renders the radius of accretion of the ionized gas large enough to be analyzed in detail with current facilities \citep{GM23}. In W49A, a few UC/HC H\textsc{ii} regions have been reported to exhibit changes in their centimeter continuum and recombination lines \citep{DePree2018,Depree2020,Rodriguez2020}, which can be linked to accretion events \citep{GM2011}. 
Notwithstanding the previously mentioned evidence, it has also been shown theoretically that if accretion occurs at rates higher than $\dot{M} \geq 10^{-4}~M_\odot$ yr$^{-1}$, the massive protostar can become ``bloated'', with a colder photosphere incapable of producing sufficient ionizing photons to create observable H\textsc{ii} regions \citep{Hosokawa2009,HosokawaOmukai2010}. 

Detailed theoretical predictions and deep mid-IR observations with the JWST would also be helpful in testing the 
hypotheses regarding the existence of a significant population of accreting UC/HC H\textsc{ii} regions, bloated massive protostars, or both.

\section{Conclusions}\label{sec:conc}

We conducted a deep census of compact, ultracompact, and hypercompact H\textsc{ii} regions in W49A using high-resolution, high-dynamic-range VLA 3.3 cm observations. We inferred the masses of their ionizing stars based on updated stellar calibrations. The resulting high-mass slope derived from the compact-source population is significantly steeper than a canonical Salpeter IMF, and this result remains robust under different assumptions regarding binarity or source selection, as well as after corrections for optical depth effects.

We propose two possible explanations for the steep high-mass slope inferred from the H\textsc{ii} region census in W49A. The first is that the radio-selected sample is biased against the most massive stars due to shorter radio H\textsc{ii} region lifetimes. If stellar feedback becomes increasingly efficient with stellar mass, the most massive stars may disperse their surrounding ionized gas more rapidly, shortening the observable lifetime of embedded, radio-detected H\textsc{ii} regions at higher masses.  
The second possibility is that W49A is still assembling its most massive members, such that the observed mass distribution represents an instantaneous snapshot of an evolving population that will ultimately have a mass distribution consistent with the Salpeter IMF slope at high masses. 
We do not discard this possibility, but we point out that it would require about a dozen stars with UC and HC H\textsc{ii} regions to still be actively accreting at a rate large enough to further increase several solar masses, but not so large as to bloat the stellar surface to the point that there are not sufficient ionizing photons to produce an observable H\textsc{ii} region. 

Distinguishing between intrinsic IMF variations, observational biases, and continued mass assembly will require further multiwavelength studies and theoretical work.

\acknowledgments 

The authors thank the anonymous referees for their useful reviews. 
The National Radio Astronomy Observatory is a facility of the National Science Foundation operated under a cooperative agreement by Associated Universities, Inc. 
M.J.G. and R.G.M acknowledge support from the UNAM-DGAPA-PAPIIT project IN105225. 

The 3.3 cm VLA data used in this work is on Zenodo under the DOI: 10.5281/zenodo.18839616
\facilities{VLA}

\software{
\texttt{astrodendro} \citep{Rosolowsky2008},
\textsc{CARTA} \citep{CARTA2021}, 
\textsc{imf} \citep{Richardson2026}, 
\texttt{plfit} (\url{https://github.com/keflavich/plfit})}.  
\bibliography{stars_VLA_W49}{}
\bibliographystyle{aasjournal}



\appendix 

\section{Measurements for individual sources} \label{sec:indiv_sources}

In this appendix, we provide the source catalogs used in this work. Table \ref{tab:prop} lists the basic observational properties, and Table \ref{tab:propderiv} presents the physical and stellar properties derived from the catalog of 92 unique H\textsc{ii} regions (see Sections~\ref{sec:phys-prop} and~\ref{sec:stellar-masses}).

\begin{longtable}{llllccccc} 
\caption{\raggedright Observational properties of the detected sources. From left to right: cross-ID of the sources, as referenced in \citep{depree1997,Depree2020}, source index (idx), ICRS right ascension and declination of the emission peak ($\alpha_{\rm pk}$, $\delta_{\rm pk}$), deconvolved area ($A_{\rm d}$), deconvolved equivalent diameter ($D_{\rm eq,d}$) expressed in parsecs and astronomical units, integrated flux density ($F_\nu$), and peak intensity ($I_\mathrm{pk}$). }
\label{tab:prop} \\
\hline
CROSS ID & idx & $\alpha_{\rm pk}$ & $\delta_{\rm pk}$ & $A_\mathrm{d}$ & $D_\mathrm{eq,d}$ & $D_\mathrm{eq,d}$ & $F_\nu$ & $I_\mathrm{pk}$ \\ 
 & & ($19^h$ $10^m$ ) & (9$^{\circ}$) & arcsec$^2$ & pc & AU & mJy & mJy beam$^{-1}$ \\
\hline \hline
\endfirsthead

\multicolumn{9}{c}{Table \ref{tab:prop} -- \textit{Continued}} \\
\hline
CROSS ID & idx & $\alpha_{\rm pk}$ & $\delta_{\rm pk}$ & $A_\mathrm{d}$ & $D_\mathrm{eq,d}$ & $D_\mathrm{eq,d}$ & $F_\nu$ & $I_\mathrm{pk}$ \\ 
 & & ($19^h$ $10^m$ ) & (9$^{\circ}$) & arcsec$^2$ & pc & AU & mJy & mJy/beam \\
\hline \hline
\endhead

\hline
\endfoot

\endlastfoot

W49S-1 & 1 & 21.59s & 5'0.98" & 0.56$\pm$0.10 & 0.045 & 9341 & 32.2$\pm$1.8 & 7.87 \\
\hline
\multirow{5}{2em}{W49S} 
  & 2  & 21.67s & 5'1.99"  & 0.106$\pm$0.048 & 0.020 & 4085   & 4.69$\pm$0.27 & 2.34 \\
  & 4  & 21.61s & 5'1.86"  & 0.055$\pm$0.037 & 0.014 & 2943   & 0.61$\pm$0.24 & 0.98 \\
  & 5  & 21.79s & 5'1.99"  & 138.9$\pm$2.4   & 0.716 & 147628 & 3744$\pm$210 & 16.67 \\
  & 6  & 21.73s & 5'6.94"  & 0.374$\pm$0.088 & 0.037 & 7665   & 8.74$\pm$0.52 & 1.88 \\
  & 82 & 21.70s & 5'7.70"  & 0.355$\pm$0.092 & 0.036 & 7464   & 2.88$\pm$0.22 & 0.57 \\
\hline
P & 7 & 16.80s & 5'49.96" & 9.93$\pm$0.42 & 0.191 & 39472 & 84.7$\pm$4.3 & 1.19 \\
- & 8 & 16.62s & 5'46.05" & 0.353$\pm$0.078 & 0.036 & 7445  & 0.92$\pm$0.13 & 0.14 \\
- & 9 & 10.94s & 5'17.53" & 0.283$\pm$0.090 & 0.032 & 6660  & 7.70$\pm$0.40 & 5.33 \\
O & 10 & 16.36s & 6'7.01" & 8.20$\pm$0.43 & 0.174 & 35858 & 515$\pm$26 & 17.63 \\
O2 & 11 & 16.35s & 6'9.78" & 23.72$\pm$0.71 & 0.296 & 60998 & 187$\pm$11 & 1.39 \\
-  & 12 & 16.46s & 6'7.14" & 0.089$\pm$0.042 & 0.018 & 3731  & 1.047$\pm$0.077 & 0.64 \\
M  & 13 & 14.74s & 6'25.62" & 2.79$\pm$0.22 & 0.102 & 20939 & 61.1$\pm$4.0 & 1.98 \\
-  & 14 & 14.35s & 6'21.54" & 0.233$\pm$0.067 & 0.029 & 6051  & 3.89$\pm$0.29 & 2.99 \\
-  & 15 & 14.03s & 6'23.60" & 0.103$\pm$0.046 & 0.019 & 4017  & 2.27$\pm$0.13 & 2.10 \\
J1 & 16 & 14.13s & 6'24.94" & 0.166$\pm$0.060 & 0.025 & 5108  & 4.20$\pm$0.22 & 2.31 \\
-  & 17 & 13.96s & 6'25.87" & 0.152$\pm$0.055 & 0.024 & 4887  & 2.26$\pm$0.13 & 0.81 \\
\hline
\multirow{2}{2em}{J2}
  & 18 & 14.15s & 6'27.84" & 0.085$\pm$0.042 & 0.018 & 3656 & 0.543$\pm$0.059 & 0.52 \\
  & 19 & 14.17s & 6'26.79" & 4.71$\pm$0.31 & 0.132 & 27189 & 34.0$\pm$2.1 & 0.86 \\
\hline
I  & 20 & 13.78s & 6'25.07" & 11.29$\pm$0.49 & 0.204 & 42085 & 144.3$\pm$8.7 & 1.47 \\
F  & 21 & 13.34s & 6'21.37" & 1.89$\pm$0.19 & 0.083 & 17218 & 174.7$\pm$9.0 & 11.39 \\
C  & 22 & 13.15s & 6'18.77" & 2.51$\pm$0.26 & 0.096 & 19839 & 279$\pm$14 & 18.98 \\
C1 & 23 & 13.06s & 6'16.08" & 0.494$\pm$0.099 & 0.043 & 8807  & 33.4$\pm$1.7 & 9.28 \\
J  & 24 & 14.19s & 6'15.28" & 9.29$\pm$0.44 & 0.185 & 38175 & 314$\pm$18 & 6.26 \\
\hline
\multirow{2}{2em}{A} 
  & 25 & 12.89s & 6'11.76" & 6.99$\pm$0.48 & 0.161 & 33113 & 423$\pm$22 & 20.82 \\
  & 33 & 12.82s & 6'11.13" & 0.195$\pm$0.064 & 0.027 & 5525  & 2.96$\pm$0.23 & 1.31 \\
\hline
D & 26 & 13.21s & 6'11.42" & 1.73$\pm$0.20 & 0.080 & 16462 & 295$\pm$15 & 15.61 \\
\hline
\multirow{2}{2em}{GG} 
  & 27 & 13.67s & 6'49.72" & 0.521$\pm$0.097 & 0.044 & 9040 & 8.35$\pm$0.43 & 1.42 \\
  & 28 & 13.61s & 6'49.09" & 0.039$\pm$0.034 & 0.012 & 2484 & 0.326$\pm$0.040 & 0.30 \\
\hline
\multirow{2}{2em}{R} 
  & 29 & 11.05s & 5'20.14" & 0.81$\pm$0.13 & 0.055 & 11249 & 91.1$\pm$4.6 & 16.09 \\
  & 35 & 10.99s & 5'20.73" & 0.226$\pm$0.065 & 0.029 & 5959  & 3.24$\pm$0.18 & 0.79 \\
\hline
N  & 30 & 15.37s & 6'14.95" & 0.408$\pm$0.095 & 0.039 & 8000   & 29.4$\pm$1.6 & 13.77 \\
Q  & 31 & 10.47s & 5'12.96" & 76.3$\pm$1.3 & 0.530 & 109398 & 1112$\pm$58 & 2.23 \\
S  & 32 & 11.67s & 5'26.82" & 39.88$\pm$0.93 & 0.383 & 79099  & 592$\pm$34 & 8.14 \\
R3 & 34 & 10.72s & 5'17.03" & 24.98$\pm$0.71 & 0.303 & 62599  & 241$\pm$13 & 1.46 \\
\hline
\multirow{2}{2em}{CC} 
  & 36 & 11.88s & 7'6.69" & 0.56$\pm$0.11 & 0.046 & 9400   & 2.11$\pm$0.18 & 0.35 \\
  & 37 & 11.51s & 7'9.30" & 86.7$\pm$1.3 & 0.565 & 116605 & 465$\pm$31 & 0.62 \\
\hline
HH & 38 & 14.31s & 5'51.09" & 117.3$\pm$1.6 & 0.658 & 135678 & 408$\pm$25 & 0.49 \\
R2 & 39 & 10.81s & 5'22.95" & 4.12$\pm$0.30  & 0.123 & 25438  & 17.2$\pm$1.0 & 0.33 \\
KK & 40 & 21.64s & 5'52.09" & 5.45$\pm$0.35  & 0.142 & 29229  & 14.5$\pm$1.4 & 0.38 \\
EE & 41 & 13.30s & 6'40.53" & 11.07$\pm$0.46 & 0.202 & 41672  & 38.0$\pm$2.7 & 0.36 \\
DD & 42 & 11.68s & 6'31.08" & 94.8$\pm$1.6   & 0.591 & 121940 & 421$\pm$25 & 0.76 \\
FF & 43 & 12.53s & 5'26.69" & 4.56$\pm$0.42  & 0.130 & 26732  & 12.6$\pm$1.2 & 0.29 \\
JJ & 44 & 18.38s & 6'11.29" & 264.7$\pm$2.5  & 0.988 & 203761 & 346$\pm$38 & 0.25 \\
O3 & 45 & 16.78s & 6'13.35" & 113.3$\pm$1.5 & 0.646 & 133304 & 243$\pm$19 & 0.27 \\
AA & 46 & 7.63s  & 5'38.28" & 30.19$\pm$0.86 & 0.334 & 68823  & 20.2$\pm$2.2 & 0.10 \\
II & 47 & 14.88s & 6'38.80" & 215.4$\pm$2.2  & 0.891 & 183810 & 367$\pm$38 & 0.36 \\
HH & 48 & 13.67s & 6'16.92" & 3.44$\pm$0.26  & 0.113 & 23214  & 23.2$\pm$1.7 & 1.15 \\
\hline
\multirow{3}{2em}{L} 
  & 49 & 14.71s & 6'20.24" & 0.031$\pm$0.030 & 0.011 & 2188   & 1.438$\pm$0.098 & 1.24 \\
  & 65 & 14.53s & 6'19.90" & 88.5$\pm$1.3  & 0.571 & 117830 & 1520$\pm$110 & 3.92 \\
  & 66 & 14.81s & 6'18.01" & 0.050$\pm$0.035 & 0.014 & 2798   & 1.57$\pm$0.11 & 1.02 \\
\hline
B1 (2020) & 50 & 13.12s & 6'12.34" & 0.48$\pm$0.11 & 0.042 & 8681 & 40.1$\pm$2.1 & 10.05 \\
\hline
B2 & \multirow{2}{2em}{51} & \multirow{2}{2em}{13.15s} & \multirow{2}{2em}{6'12.68"} & 0.76$\pm$0.12 & 0.053 & 10937 & 93.7$\pm$4.8 & \multirow{2}{2em}{19.72} \\
B3 & & & & 0.254$\pm$0.041 & 0.031 & 6314 & 20.4$\pm$1.0 &  \\
\hline
E1 & 52 & 13.23s & 6'12.18" & 0.038$\pm$0.034 & 0.012 & 2428 & 6.09$\pm$0.32 & 5.50 \\
E2 & 53 & 13.23s & 6'12.47" & 0.133$\pm$0.051 & 0.022 & 4565 & 18.60$\pm$0.96 & 7.17 \\
E3 & 54 & 13.26s & 6'12.22" & 0.293$\pm$0.088 & 0.033 & 6784 & 36.0$\pm$1.9 & 9.37 \\
-  & 55 & 13.68s & 6'10.08" & 0.225$\pm$0.069 & 0.029 & 5935 & 11.90$\pm$0.64 & 2.61 \\
\hline
\multirow{2}{2em}{G1S} 
  & 56 & 13.40s & 6'11.92" & 0.62$\pm$0.11 & 0.048 & 9846  & 122.9$\pm$6.2 & 13.47 \\
  & 57 & 13.38s & 6'12.64" & 0.71$\pm$0.13 & 0.051 & 10564 & 159.5$\pm$8.1 & 20.23 \\
\hline
G2a & \multirow{3}{2em}{58} & \multirow{3}{2em}{13.44s} & \multirow{3}{2em}{6'12.97"} & 0.404$\pm$0.069 & 0.039 & 7966 & 113.2$\pm$5.7 & \multirow{3}{2em}{18.12} \\
G2b & & & & 0.166$\pm$0.028 & 0.025 & 5105 & 22.3$\pm$1.1 &  \\
G2c & & & & 0.174$\pm$0.030 & 0.025 & 5229 & 44.5$\pm$2.3 &  \\
\hline
G3,G4,G5 & 59 & 13.50s & 6'11.80" & 9.75$\pm$0.65 & 0.190 & 39113 & 1712$\pm$87 & 16.44 \\
G3a & 61 & 13.50s & 6'12.39" & 0.345$\pm$0.081 & 0.036 & 7352  & 89.8$\pm$4.5 & 14.27 \\
-   & 62 & 14.09s & 6'21.16" & 0.265$\pm$0.072 & 0.031 & 6449  & 2.18$\pm$0.16 & 0.48 \\
-   & 63 & 11.02s & 5'21.57" & 0.011$\pm$0.025 & 0.006 & 1320  & 0.226$\pm$0.030 & 0.29 \\
-   & 64 & 12.71s & 6'11.29" & 0.045$\pm$0.035 & 0.013 & 2646  & 0.22$\pm$0.22 & 0.33 \\
-   & 67 & 13.43s & 6'11.80" & 0.057$\pm$0.041 & 0.014 & 2990  & 8.19$\pm$0.42 & 4.55 \\
BB  & 68 & 10.99s & 5'43.53" & 26.58$\pm$0.84 & 0.313 & 64579 & 21.8$\pm$2.2  & 0.11 \\
MM  & 69 & 23.06s & 6'4.65"  & 237.8$\pm$2.4 & 0.936 & 193161 & 219$\pm$21 & 0.24 \\
B1 (1997) & 70* & 13.33s & 6'15.16" & 0.61$\pm$0.18 & 0.047 & 9747 & 9.44$\pm$0.67 & 1.18 \\
-   & 71 & 14.33s & 6'8.48"  & 29.59$\pm$0.87 & 0.330 & 68130 & 208$\pm$15 & 0.64 \\
-   & 72 & 16.82s & 6'1.34"  & 0.65$\pm$0.11 & 0.049 & 10068 & 1.14$\pm$0.13 & 0.15 \\
-   & 74 & 16.22s & 6'1.51"  & 3.78$\pm$0.29 & 0.118 & 24343 & 7.88$\pm$0.74 & 0.20 \\
-   & 75 & 14.66s & 6'3.73"  & 7.70$\pm$0.42 & 0.168 & 34744 & 20.3$\pm$2.5  & 0.34 \\
-   & 76 & 14.91s & 6'29.44" & 5.56$\pm$0.34 & 0.143 & 29535 & 26.6$\pm$3.3  & 0.57 \\
-   & 83 & 19.12s & 6'41.11" & 37.92$\pm$0.87 & 0.374 & 77124 & 41.3$\pm$6.0  & 0.13 \\
-   & 84 & 14.98s & 6'9.49"  & 10.12$\pm$0.49 & 0.193 & 39835 & 105.2$\pm$8.7 & 0.73 \\
LL  & 88 & 24.09s & 4'25.78" & 0.034$\pm$0.030 & 0.011 & 2311  & 0.413$\pm$0.041 & 0.41 \\
-   & 89 & 23.09s & 4'14.36" & 0.052$\pm$0.035 & 0.014 & 2847  & 0.298$\pm$0.042 & 0.19 \\
-   & 90 & 17.71s & 3'59.37" & 0.008$\pm$0.023 & 0.005 & 1091  & 0.11$\pm$0.16 & 0.15 \\
-   & 91 & 25.50s & 5'41.92" & 3.12$\pm$0.25 & 0.107 & 22140 & 4.2$\pm$1.4   & 0.17 \\
-   & 92 & 17.97s & 8'30.06" & 0.027$\pm$0.028 & 0.010 & 2058  & 0.27$\pm$0.19 & 0.33 \\
-   & 93 & 18.02s & 8'28.72" & 0.036$\pm$0.032 & 0.011 & 2370  & 0.28$\pm$0.20 & 0.29 \\
-   & 94 & 8.10s  & 7'6.19"  & 0.059$\pm$0.037 & 0.015 & 3036  & 0.21$\pm$0.24 & 0.21 \\
-   & 95 & 8.65s  & 4'30.70" & 0.041$\pm$0.034 & 0.012 & 2539  & 0.340$\pm$0.057 & 0.40 \\
\hline

\end{longtable}
\noindent\textit{Note.} The source marked with an asterisk (*) corresponds to a UC H \textsc{ii} region with an elongated morphology; this source is included in the full-sample analysis but excluded from the compact-only power-law fits to the stellar mass distributions. Reported values are rounded according to the two significant figures of their associated uncertainties.

\clearpage
\begin{table*}[ht]
\centering
\caption{Mean stellar parameters and associated standard errors as a function of ZAMS mass, derived from the CB19 stellar population synthesis models at solar metallicity ($Z = 0.017$). Columns list the stellar mass ($M$), the mean logarithm of the ionizing photon production rate ($\log Q_0$), and the mean bolometric luminosity ($L$).}
\label{tab:cb19_zams_calibration}
\begin{tabular}{ccc@{\hskip 0.5in}ccc}
\hline\hline
$M_\star$ & $\log Q_0$ & $\log(L/L_\odot)$ & $M_\star$ & $\log Q_0$ & $\log(L/L_\odot)$ \\
($M_\odot$) & ($\mathrm{s}^{-1}$) & & ($M_\odot$) & ($\mathrm{s}^{-1}$) & \\
\hline
8.0   & $44.537 \pm 0.069$ & $3.510^{+0.047}_{-0.052}$ & 21.9  & $48.325 \pm 0.016$ & $4.821^{+0.030}_{-0.032}$ \\
8.2   & $44.613 \pm 0.075$ & $3.562^{+0.054}_{-0.062}$ & 23.8  & $48.444 \pm 0.038$ & $4.913^{+0.034}_{-0.037}$ \\
8.4   & $44.741 \pm 0.076$ & $3.597^{+0.056}_{-0.064}$ & 25.8  & $48.610 \pm 0.017$ & $4.982^{+0.033}_{-0.036}$ \\
8.6   & $44.840 \pm 0.098$ & $3.630^{+0.054}_{-0.062}$ & 27.8  & $48.723 \pm 0.032$ & $5.067^{+0.037}_{-0.040}$ \\
8.8   & $44.940 \pm 0.082$ & $3.664^{+0.056}_{-0.064}$ & 29.7  & $48.810 \pm 0.033$ & $5.128^{+0.041}_{-0.045}$ \\
9.0   & $45.011 \pm 0.097$ & $3.682^{+0.049}_{-0.055}$ & 34.5  & $49.000 \pm 0.021$ & $5.277^{+0.038}_{-0.041}$ \\
9.2   & $45.118 \pm 0.074$ & $3.713^{+0.049}_{-0.055}$ & 39.5  & $49.162 \pm 0.017$ & $5.392^{+0.027}_{-0.028}$ \\
9.4   & $45.230 \pm 0.082$ & $3.743^{+0.049}_{-0.055}$ & 49.1  & $49.399 \pm 0.013$ & $5.589^{+0.021}_{-0.022}$ \\
9.6   & $45.302 \pm 0.100$ & $3.772^{+0.049}_{-0.055}$ & 58.6  & $49.556 \pm 0.010$ & $5.739^{+0.018}_{-0.018}$ \\
9.8   & $45.410 \pm 0.097$ & $3.801^{+0.049}_{-0.055}$ & 68.6  & $49.693 \pm 0.010$ & $5.855^{+0.010}_{-0.011}$ \\
10.0  & $45.522 \pm 0.124$ & $3.829^{+0.049}_{-0.055}$ & 77.5  & $49.802 \pm 0.010$ & $5.9616^{+0.0034}_{-0.0035}$ \\
11.0  & $45.938 \pm 0.079$ & $3.961^{+0.051}_{-0.058}$ & 82.2  & $49.847 \pm 0.010$ & $6.0056^{+0.0030}_{-0.0030}$ \\
12.0  & $46.319 \pm 0.098$ & $4.076^{+0.047}_{-0.053}$ & 87.6  & $49.896 \pm 0.010$ & $6.045^{+0.037}_{-0.041}$ \\
14.0  & $46.995 \pm 0.076$ & $4.278^{+0.047}_{-0.053}$ & 92.3  & $49.935 \pm 0.010$ & $6.083^{+0.037}_{-0.041}$ \\
16.0  & $47.495 \pm 0.208$ & $4.444^{+0.049}_{-0.055}$ & 97.1  & $49.971 \pm 0.010$ & $6.120^{+0.037}_{-0.041}$ \\
18.0  & $47.936 \pm 0.072$ & $4.586^{+0.047}_{-0.053}$ & 106.6 & $50.038 \pm 0.010$ & $6.187^{+0.037}_{-0.041}$ \\
20.0  & $48.184 \pm 0.010$ & $4.695^{+0.037}_{-0.041}$ & --    & -- & -- \\
\hline
\end{tabular}
\end{table*}

{
\selectfont\small
\setlength\tabcolsep{2.5pt}
\hspace*{-1cm}

\begin{longtable}{cccccccccccccc}
\caption{ Derived physical and stellar properties of the identified H\textsc{ii} regions assuming a single ionizing source. From left to right, the table lists the source index, average brightness temperature ($T_\mathrm{b}$), peak brightness temperature ($T_{b,\mathrm{pk}}$), average optical depth at 3.3 cm ($\tau_{\rm 3cm}$), emission measure (EM), electron density ($n_\mathrm{e}$), logarithm of the ionizing photon rate ($\log Q_\mathrm{0}$), inferred stellar mass at 3 cm ($M_{\rm 3cm}$), mass correction factor ($f_M$), inferred stellar luminosity at 3 cm ($L_{\rm 3cm}$) and luminosity correction factor ($f_L$).}
 \label{tab:propderiv} \\
 
\hline
idx   &      $T_\mathrm{b}$ &      $T_\mathrm{b,pk}$ &        $\tau_{\rm 3cm}$         &  EM                           &   $n_\mathrm{e}$    &         log$Q_\mathrm{0}$   &  $M_{\rm 3cm}$ & $f_M$  & $L_{\rm 3cm}$ &$f_L$\\  
      &      K  &      K  &                           &$10^7$ pc cm$^{-6}$  &   $10^4$ cm$^{-3}$ & photons s$^{-1}$    &       $M_{\odot}$ && 10$^4$ $L_{\odot}$ &\\   
\hline \hline
\endfirsthead

\multicolumn{8}{c}{Table \ref{tab:propderiv}}
\\
\hline
idx   &      $T_\mathrm{b}$ &      $T_\mathrm{b,pk}$ &        $\tau_{\rm 3cm}$         &  EM                           &   $n_\mathrm{e}$    &         log$Q_\mathrm{0}$   &  $M_{\rm 3cm}$ & $f_M$  & $L_{\rm 3cm}$ &$f_L$\\  
      &      K  &      K  &                           &$10^7$ pc cm$^{-6}$  &   $10^4$ cm$^{-3}$ & photons s$^{-1}$    &       $M_{\odot}$ && 10$^4$ $L_{\odot}$ &\\   
\hline \hline
\endhead

\hline
\endfoot
\endlastfoot

5 & 450$\pm$26 & 8441.4$\pm$8.8 & 0.0461$\pm$0.0055 & 1.45$\pm$0.26 & 0.450$\pm$0.041 & 49.641$\pm$0.078 & 65$_{-14}^{+19}$ &  & 65$_{-24}^{+39}$ &  \\
59 & 2933$\pm$247 & 8324.6$\pm$8.8 & 0.347$\pm$0.054 & 10.9$\pm$2.3 & 2.40$\pm$0.25 & 49.365$\pm$0.090 & 48$_{-9}^{+13}$ & & 37$_{-13}^{+22}$ &  \\
65 & 286$\pm$20 & 1983.4$\pm$8.8 & 0.0290$\pm$0.0036 & 0.92$\pm$0.17 & 0.400$\pm$0.037 & 49.245$\pm$0.080 & 43$_{-8}^{+10}$ &  & 30$_{-10}^{+16}$ &  \\
31 & 244$\pm$13 & 1130.5$\pm$7.5 & 0.0247$\pm$0.0028 & 0.78$\pm$0.14 & 0.383$\pm$0.034 & 49.110$\pm$0.077 & 37.9$_{-6.1}^{+8.4}$ &  & 23$_{-7}^{+12}$ &  \\
32 & 248$\pm$15 & 4119.1$\pm$8.8 & 0.0251$\pm$0.0030 & 0.79$\pm$0.14 & 0.454$\pm$0.041 & 48.836$\pm$0.078 & 30.2$_{-4.4}^{+5.8}$ &  & 14.0$_{-4.4}^{+6.6}$ &  \\
10 & 1050$\pm$77 & 8924.9$\pm$8.8 & 0.111$\pm$0.015 & 3.50$\pm$0.66 & 1.42$\pm$0.13 & 48.794$\pm$0.082 & 29.4$_{-4.0}^{+5.6}$ & 1.1 & 13.1$_{-3.8}^{+6.4}$ & 1.3 \\
37 & 89.7$\pm$6.0 & 311.8$\pm$7.5 & 0.0090$\pm$0.0011 & 0.284$\pm$0.052 & 0.224$\pm$0.020 & 48.728$\pm$0.079 & 27.8$_{-3.3}^{+5.3}$ &  & 11.7$_{-3.0}^{+5.6}$ &  \\
25 & 1012$\pm$87 & 10542.8$\pm$8.8 & 0.107$\pm$0.015 & 3.36$\pm$0.65 & 1.45$\pm$0.14 & 48.708$\pm$0.084 & 27.7$_{-3.2}^{+5.3}$ & 1.3 & 11.6$_{-2.9}^{+5.6}$ & 1.9 \\
42 & 74.2$\pm$4.7 & 384$\pm$12 & 0.00745$\pm$0.00088 & 0.235$\pm$0.042 & 0.199$\pm$0.018 & 48.684$\pm$0.078 & 27.2$_{-3.0}^{+5.0}$ &  & 11.0$_{-2.6}^{+5.2}$ &  \\
38 & 58.1$\pm$3.6 & 248.2$\pm$5.4 & 0.00583$\pm$0.00069 & 0.184$\pm$0.033 & 0.167$\pm$0.015 & 48.670$\pm$0.078 & 26.9$_{-3.0}^{+4.9}$ &  & 10.8$_{-2.5}^{+5.0}$ &  \\
47 & 28.5$\pm$2.9 & 182.5$\pm$8.8 & 0.00285$\pm$0.00041 & 0.090$\pm$0.018 & 0.1004$\pm$0.0099 & 48.623$\pm$0.086 & 26.0$_{-2.9}^{+4.5}$ &  & 9.8$_{-2.2}^{+4.6}$ &  \\
44 & 21.8$\pm$2.4 & 129.0$\pm$8.8 & 0.00219$\pm$0.00033 & 0.069$\pm$0.014 & 0.0835$\pm$0.0084 & 48.598$\pm$0.088 & 25.7$_{-3.0}^{+4.3}$ &  & 9.5$_{-2.2}^{+4.3}$ &  \\
26 & 2851$\pm$357 & 7901.5$\pm$8.8 & 0.336$\pm$0.064 & 10.6$\pm$2.5 & 3.64$\pm$0.44 & 48.60$\pm$0.10 & 25.7$_{-3.1}^{+4.4}$ & 1.1 & 9.5$_{-2.3}^{+4.4}$ & 1.2 \\
24 & 565$\pm$42 & 3167.3$\pm$8.8 & 0.0582$\pm$0.0075 & 1.83$\pm$0.34 & 0.996$\pm$0.094 & 48.568$\pm$0.081 & 25.3$_{-3.1}^{+3.9}$ &  & 9.2$_{-2.3}^{+3.8}$ &  \\
22 & 1855$\pm$216 & 9612.1$\pm$8.8 & 0.205$\pm$0.035 & 6.5$\pm$1.4 & 2.59$\pm$0.29 & 48.547$\pm$0.094 & 25.0$_{-3.2}^{+3.7}$ & 1.1 & 9.0$_{-2.5}^{+3.5}$ & 1.3 \\
34 & 161.2$\pm$9.7 & 741.3$\pm$7.5 & 0.0163$\pm$0.0019 & 0.512$\pm$0.092 & 0.411$\pm$0.037 & 48.444$\pm$0.078 & 23.8$_{-3.3}^{+2.9}$ &  & 8.2$_{-2.7}^{+2.4}$ &  \\
45 & 35.9$\pm$2.8 & 138.9$\pm$5.4 & 0.00360$\pm$0.00046 & 0.113$\pm$0.021 & 0.132$\pm$0.012 & 48.445$\pm$0.081 & 23.8$_{-3.3}^{+2.9}$ &  & 8.2$_{-2.8}^{+2.4}$ &  \\
69 & 15.4$\pm$1.5 & 119.3$\pm$8.8 & 0.00154$\pm$0.00022 & 0.0486$\pm$0.0094 & 0.0720$\pm$0.0070 & 48.399$\pm$0.084 & 23.1$_{-3.2}^{+2.9}$ &  & 7.6$_{-2.7}^{+2.2}$ &  \\
71 & 117.6$\pm$9.3 & 323.9$\pm$8.8 & 0.0118$\pm$0.0015 & 0.373$\pm$0.069 & 0.336$\pm$0.031 & 48.379$\pm$0.081 & 22.8$_{-3.0}^{+2.9}$ &  & 7.3$_{-2.5}^{+2.2}$ &  \\
57 & 3746$\pm$704 & 10241.0$\pm$8.8 & 0.47$\pm$0.13 & 14.8$\pm$4.5 & 5.37$\pm$0.85 & 48.36$\pm$0.13 & 22.5$_{-3.0}^{+3.3}$ & 1.3 & 7.1$_{-2.5}^{+2.4}$ & 1.9 \\
21 & 1545$\pm$176 & 5769.1$\pm$8.8 & 0.168$\pm$0.028 & 5.3$\pm$1.1 & 2.52$\pm$0.28 & 48.336$\pm$0.093 & 22.1$_{-2.7}^{+3.2}$ & 1.1 & 6.8$_{-2.1}^{+2.4}$ & 1.4 \\
11 & 131.4$\pm$8.7 & 702.5$\pm$7.5 & 0.0132$\pm$0.0016 & 0.417$\pm$0.076 & 0.376$\pm$0.034 & 48.332$\pm$0.079 & 22.1$_{-2.6}^{+3.1}$ &  & 6.8$_{-2.1}^{+2.4}$ &  \\
G2a & 4677$\pm$832 & 9176.2$\pm$8.8 & 0.63$\pm$0.18 & 19.9$\pm$6.3 & 7.2$\pm$1.2 & 48.24$\pm$0.14 & 20.8$_{-2.3}^{+3.6}$ & 1.5 & 5.6$_{-1.5}^{+2.9}$ & 2.7 \\
56 & 3325$\pm$621 & 6822.0$\pm$8.8 & 0.40$\pm$0.11 & 12.7$\pm$3.7 & 5.17$\pm$0.79 & 48.23$\pm$0.13 & 20.6$_{-2.2}^{+3.5}$ & 1.3 & 5.5$_{-1.4}^{+2.9}$ & 1.8 \\
20 & 214$\pm$16 & 743.7$\pm$8.8 & 0.0216$\pm$0.0027 & 0.68$\pm$0.13 & 0.578$\pm$0.054 & 48.222$\pm$0.080 & 20.5$_{-2.0}^{+3.3}$ &  & 5.4$_{-1.3}^{+2.7}$ &  \\
61 & 4358$\pm$1050 & 7223.1$\pm$8.8 & 0.57$\pm$0.20 & 18.0$\pm$6.8 & 7.1$\pm$1.4 & 48.13$\pm$0.16 & 19.5$_{-1.9}^{+3.5}$ & 1.4 & 4.7$_{-1.0}^{+2.8}$ & 2.2 \\
84 & 174$\pm$17 & 371.6$\pm$8.8 & 0.0175$\pm$0.0025 & 0.55$\pm$0.11 & 0.535$\pm$0.052 & 48.084$\pm$0.084 & 19.2$_{-1.5}^{+2.4}$ &  & 4.5$_{-0.8}^{+1.9}$ &  \\
B2 & 2053$\pm$344 & 9982.2$\pm$8.8 & 0.230$\pm$0.050 & 7.2$\pm$1.9 & 3.70$\pm$0.50 & 48.08$\pm$0.11 & 19.2$_{-1.6}^{+2.6}$ & 1.6 & 4.5$_{-0.9}^{+2.0}$ & 3.4 \\
29 & 1887$\pm$308 & 8146.4$\pm$8.8 & 0.209$\pm$0.045 & 6.6$\pm$1.7 & 3.48$\pm$0.46 & 48.06$\pm$0.11 & 19.0$_{-1.5}^{+2.5}$ & 1.2 & 4.4$_{-0.8}^{+1.8}$ & 1.6 \\
7 & 142.4$\pm$9.4 & 601$\pm$13 & 0.0143$\pm$0.0017 & 0.452$\pm$0.082 & 0.486$\pm$0.044 & 47.989$\pm$0.079 & 18.4$_{-1.2}^{+1.9}$ &  & 4.1$_{-0.6}^{+1.2}$ &  \\
13 & 365$\pm$38 & 1002.4$\pm$8.8 & 0.0372$\pm$0.0055 & 1.17$\pm$0.23 & 1.08$\pm$0.11 & 47.852$\pm$0.087 & 17.6$_{-1.0}^{+1.5}$ &  & 3.65$_{-0.54}^{+0.82}$ &  \\
G2c & 4262$\pm$758 & 9176.2$\pm$8.8 & 0.56$\pm$0.15 & 17.5$\pm$5.3 & 8.3$\pm$1.3 & 47.82$\pm$0.13 & 17.5$_{-1.1}^{+1.5}$ & 1.4 & 3.56$_{-0.60}^{+0.84}$ & 2.4 \\
50 & 1393$\pm$320 & 5089.9$\pm$8.8 & 0.150$\pm$0.041 & 4.7$\pm$1.4 & 3.35$\pm$0.54 & 47.69$\pm$0.13 & 16.9$_{-1.1}^{+1.1}$ & 1.2 & 3.24$_{-0.56}^{+0.60}$ & 1.5 \\
83 & 18.2$\pm$2.7 & 68$\pm$51 & 0.00182$\pm$0.00032 & 0.057$\pm$0.013 & 0.124$\pm$0.014 & 47.674$\pm$0.097 & 16.8$_{-1.0}^{+1.0}$ &  & 3.22$_{-0.52}^{+0.55}$ &  \\
54 & 2053$\pm$626 & 4743.4$\pm$8.8 & 0.230$\pm$0.083 & 7.2$\pm$2.8 & 4.69$\pm$0.97 & 47.66$\pm$0.17 & 16.8$_{-1.1}^{+1.2}$ & 1.0 & 3.19$_{-0.58}^{+0.64}$ & 1.1 \\
41 & 57.4$\pm$4.7 & 183.0$\pm$7.5 & 0.00576$\pm$0.00075 & 0.182$\pm$0.034 & 0.300$\pm$0.028 & 47.640$\pm$0.081 & 16.7$_{-1.0}^{+1.0}$ &  & 3.14$_{-0.49}^{+0.54}$ &  \\
23 & 1128$\pm$232 & 4699.1$\pm$8.8 & 0.120$\pm$0.029 & 3.8$\pm$1.0 & 2.97$\pm$0.44 & 47.61$\pm$0.12 & 16.5$_{-1.0}^{+1.1}$ & 1.0 & 3.06$_{-0.50}^{+0.58}$ & 1.1 \\
1 & 967$\pm$186 & 3986$\pm$13 & 0.102$\pm$0.023 & 3.21$\pm$0.85 & 2.66$\pm$0.37 & 47.59$\pm$0.12 & 16.5$_{-1.0}^{+1.1}$ & 1.1 & 3.02$_{-0.48}^{+0.56}$ & 1.3 \\
19 & 121$\pm$11 & 436.7$\pm$8.8 & 0.0121$\pm$0.0016 & 0.383$\pm$0.073 & 0.539$\pm$0.052 & 47.593$\pm$0.083 & 16.4$_{-0.9}^{+1.0}$ &  & 3.01$_{-0.45}^{+0.54}$ &  \\
30 & 1203$\pm$288 & 6973.1$\pm$8.8 & 0.128$\pm$0.036 & 4.0$\pm$1.2 & 3.23$\pm$0.53 & 47.55$\pm$0.13 & 16.3$_{-1.0}^{+1.1}$ & 1.1 & 2.92$_{-0.47}^{+0.59}$ & 1.4 \\
76 & 80$\pm$11 & 286.5$\pm$8.8 & 0.0080$\pm$0.0014 & 0.253$\pm$0.055 & 0.421$\pm$0.046 & 47.485$\pm$0.094 & 16.0$_{-0.9}^{+1.0}$ &  & 2.76$_{-0.40}^{+0.54}$ &  \\
G2b & 2245$\pm$399 & 9176.2$\pm$8.8 & 0.254$\pm$0.059 & 8.0$\pm$2.2 & 5.69$\pm$0.80 & 47.46$\pm$0.12 & 15.8$_{-0.9}^{+1.1}$ & 1.3 & 2.71$_{-0.42}^{+0.55}$ & 1.8 \\
48 & 113$\pm$12 & 580.7$\pm$8.8 & 0.0113$\pm$0.0017 & 0.358$\pm$0.071 & 0.564$\pm$0.057 & 47.426$\pm$0.086 & 15.72$_{-0.88}^{+0.96}$ &  & 2.66$_{-0.39}^{+0.49}$ &  \\
B3 & 1343$\pm$225 & 9982.2$\pm$8.8 & 0.144$\pm$0.030 & 4.5$\pm$1.1 & 3.85$\pm$0.50 & 47.40$\pm$0.11 & 15.63$_{-0.92}^{+1.00}$ & 1.3 & 2.62$_{-0.41}^{+0.50}$ & 2.1 \\
68 & 13.7$\pm$1.4 & 54.3$\pm$8.8 & 0.00137$\pm$0.00020 & 0.0432$\pm$0.0085 & 0.117$\pm$0.012 & 47.396$\pm$0.086 & 15.61$_{-0.88}^{+0.95}$ &  & 2.61$_{-0.39}^{+0.47}$ &  \\
53 & 2340$\pm$909 & 3630.6$\pm$8.8 & 0.27$\pm$0.12 & 8.4$\pm$4.0 & 6.2$\pm$1.6 & 47.38$\pm$0.21 & 15.6$_{-1.2}^{+1.3}$ & 1.2 & 2.60$_{-0.52}^{+0.64}$ & 1.7 \\
75 & 44.1$\pm$5.9 & 170.5$\pm$8.8 & 0.00442$\pm$0.00074 & 0.139$\pm$0.030 & 0.288$\pm$0.031 & 47.367$\pm$0.093 & 15.48$_{-0.90}^{+0.95}$ &  & 2.55$_{-0.40}^{+0.46}$ &  \\
46 & 11.2$\pm$1.3 & 49.6$\pm$7.5 & 0.00112$\pm$0.00017 & 0.0353$\pm$0.0071 & 0.103$\pm$0.010 & 47.364$\pm$0.088 & 15.48$_{-0.89}^{+0.94}$ &  & 2.55$_{-0.39}^{+0.45}$ &  \\
39 & 69.7$\pm$6.5 & 168.7$\pm$5.4 & 0.00699$\pm$0.00096 & 0.220$\pm$0.042 & 0.423$\pm$0.041 & 47.295$\pm$0.084 & 15.20$_{-0.88}^{+0.89}$ &  & 2.43$_{-0.39}^{+0.40}$ &  \\
40 & 44.5$\pm$5.1 & 190.8$\pm$8.8 & 0.00446$\pm$0.00068 & 0.141$\pm$0.029 & 0.315$\pm$0.032 & 47.221$\pm$0.088 & 14.91$_{-0.88}^{+0.88}$ &  & 2.30$_{-0.39}^{+0.39}$ &  \\
43 & 46.2$\pm$6.1 & 144.7$\pm$8.8 & 0.00463$\pm$0.00077 & 0.146$\pm$0.031 & 0.336$\pm$0.037 & 47.159$\pm$0.093 & 14.67$_{-0.84}^{+0.89}$ &  & 2.19$_{-0.35}^{+0.39}$ &  \\
55 & 886$\pm$275 & 1319.5$\pm$8.8 & 0.093$\pm$0.032 & 2.9$\pm$1.1 & 3.19$\pm$0.64 & 47.15$\pm$0.16 & 14.7$_{-0.9}^{+1.0}$ &  & 2.18$_{-0.39}^{+0.45}$ &  \\
70* & 260$\pm$78 & 596.2$\pm$8.8 & 0.0264$\pm$0.0085 & 0.83$\pm$0.29 & 1.33$\pm$0.25 & 47.04$\pm$0.15 & 14.2$_{-0.8}^{+1.0}$ &  & 1.98$_{-0.30}^{+0.45}$ &  \\
67 & 2402$\pm$1715 & 2302.7$\pm$7.5 & 0.27$\pm$0.23 & 8.7$\pm$7.3 & 7.7$\pm$3.5 & 47.03$\pm$0.37 & 14.1$_{-1.2}^{+1.7}$ & 1.2 & 1.94$_{-0.45}^{+0.74}$ & 1.7 \\
6 & 390$\pm$95 & 953$\pm$13 & 0.040$\pm$0.011 & 1.25$\pm$0.38 & 1.84$\pm$0.30 & 47.01$\pm$0.13 & 14.04$_{-0.72}^{+0.97}$ &  & 1.92$_{-0.26}^{+0.43}$ &  \\
27 & 268$\pm$52 & 717.0$\pm$8.8 & 0.0272$\pm$0.0060 & 0.86$\pm$0.22 & 1.40$\pm$0.19 & 46.99$\pm$0.11 & 13.97$_{-0.69}^{+0.92}$ &  & 1.88$_{-0.24}^{+0.40}$ &  \\
9 & 455$\pm$147 & 2700.4$\pm$8.8 & 0.047$\pm$0.016 & 1.47$\pm$0.54 & 2.13$\pm$0.43 & 46.95$\pm$0.16 & 13.9$_{-0.8}^{+1.0}$ &  & 1.86$_{-0.27}^{+0.43}$ &  \\
74 & 34.9$\pm$4.2 & 104$\pm$12 & 0.00349$\pm$0.00055 & 0.110$\pm$0.023 & 0.306$\pm$0.032 & 46.956$\pm$0.090 & 13.86$_{-0.66}^{+0.84}$ &  & 1.85$_{-0.23}^{+0.36}$ &  \\
52 & 2710$\pm$2422 & 2784.0$\pm$8.8 & 0.32$\pm$0.33 & 10$\pm$11 & 9.2$\pm$5.3 & 46.91$\pm$0.46 & 13.8$_{-1.6}^{+1.9}$ & 1.1 & 1.84$_{-0.55}^{+0.83}$ & 1.2 \\
2 & 737$\pm$332 & 1183$\pm$13 & 0.077$\pm$0.037 & 2.4$\pm$1.2 & 3.49$\pm$0.95 & 46.75$\pm$0.22 & 13.26$_{-0.87}^{+0.90}$ &  & 1.63$_{-0.31}^{+0.33}$ &  \\
16 & 421$\pm$154 & 1168.9$\pm$8.8 & 0.043$\pm$0.017 & 1.36$\pm$0.56 & 2.34$\pm$0.52 & 46.69$\pm$0.18 & 13.10$_{-0.80}^{+0.79}$ &  & 1.58$_{-0.28}^{+0.28}$ &  \\
91 & 22.3$\pm$7.9 & 88$\pm$51 & 0.00223$\pm$0.00082 & 0.070$\pm$0.028 & 0.256$\pm$0.051 & 46.68$\pm$0.17 & 13.08$_{-0.78}^{+0.78}$ &  & 1.57$_{-0.28}^{+0.27}$ &  \\
14 & 278$\pm$83 & 1516.4$\pm$8.8 & 0.0282$\pm$0.0090 & 0.89$\pm$0.31 & 1.74$\pm$0.33 & 46.65$\pm$0.15 & 13.00$_{-0.74}^{+0.75}$ &  & 1.54$_{-0.26}^{+0.26}$ &  \\
35 & 240$\pm$70 & 398.8$\pm$8.8 & 0.0243$\pm$0.0076 & 0.76$\pm$0.26 & 1.63$\pm$0.30 & 46.57$\pm$0.15 & 12.76$_{-0.76}^{+0.76}$ &  & 1.46$_{-0.27}^{+0.27}$ &  \\
33 & 255$\pm$85 & 663.4$\pm$7.5 & 0.0258$\pm$0.0091 & 0.81$\pm$0.31 & 1.74$\pm$0.36 & 46.54$\pm$0.16 & 12.63$_{-0.76}^{+0.78}$ &  & 1.42$_{-0.26}^{+0.28}$ &  \\
82 & 136$\pm$37 & 286.8$\pm$8.8 & 0.0137$\pm$0.0040 & 0.43$\pm$0.14 & 1.09$\pm$0.19 & 46.52$\pm$0.14 & 12.60$_{-0.72}^{+0.73}$ &  & 1.40$_{-0.24}^{+0.26}$ &  \\
15 & 370$\pm$166 & 1062.0$\pm$8.8 & 0.038$\pm$0.018 & 1.19$\pm$0.58 & 2.47$\pm$0.66 & 46.42$\pm$0.21 & 12.32$_{-0.81}^{+0.86}$ &  & 1.31$_{-0.25}^{+0.30}$ &  \\
17 & 248$\pm$90 & 410.7$\pm$8.8 & 0.0251$\pm$0.0096 & 0.79$\pm$0.32 & 1.83$\pm$0.40 & 46.42$\pm$0.18 & 12.27$_{-0.74}^{+0.78}$ &  & 1.29$_{-0.22}^{+0.28}$ &  \\
62 & 138$\pm$39 & 244.8$\pm$8.8 & 0.0139$\pm$0.0042 & 0.44$\pm$0.14 & 1.18$\pm$0.21 & 46.40$\pm$0.14 & 12.24$_{-0.68}^{+0.74}$ &  & 1.28$_{-0.21}^{+0.26}$ &  \\
36 & 63$\pm$13 & 178.1$\pm$7.5 & 0.0063$\pm$0.0015 & 0.198$\pm$0.053 & 0.659$\pm$0.094 & 46.38$\pm$0.12 & 12.20$_{-0.64}^{+0.70}$ &  & 1.26$_{-0.19}^{+0.25}$ &  \\
66 & 526$\pm$374 & 514.1$\pm$8.8 & 0.054$\pm$0.040 & 1.7$\pm$1.3 & 3.5$\pm$1.5 & 46.27$\pm$0.33 & 11.9$_{-1.0}^{+1.1}$ &  & 1.15$_{-0.27}^{+0.39}$ &  \\
49 & 787$\pm$776 & 625.5$\pm$7.5 & 0.082$\pm$0.085 & 2.6$\pm$2.7 & 4.9$\pm$2.8 & 46.23$\pm$0.45 & 11.7$_{-1.2}^{+1.4}$ &  & 1.11$_{-0.32}^{+0.47}$ &  \\
72 & 29.6$\pm$6.0 & 74.8$\pm$7.5 & 0.00296$\pm$0.00067 & 0.093$\pm$0.025 & 0.437$\pm$0.060 & 46.12$\pm$0.11 & 11.46$_{-0.60}^{+0.62}$ &  & 1.04$_{-0.16}^{+0.18}$ &  \\
12 & 197$\pm$95 & 322.3$\pm$8.8 & 0.0199$\pm$0.0099 & 0.63$\pm$0.32 & 1.86$\pm$0.53 & 46.08$\pm$0.22 & 11.37$_{-0.77}^{+0.81}$ &  & 1.02$_{-0.20}^{+0.24}$ &  \\
8 & 44$\pm$11 & 72$\pm$13 & 0.0044$\pm$0.0012 & 0.138$\pm$0.043 & 0.62$\pm$0.10 & 46.02$\pm$0.14 & 11.23$_{-0.61}^{+0.64}$ &  & 0.98$_{-0.16}^{+0.18}$ &  \\
4 & 185$\pm$144 & 495$\pm$51 & 0.019$\pm$0.015 & 0.59$\pm$0.47 & 2.03$\pm$0.88 & 45.85$\pm$0.35 & 10.8$_{-0.9}^{+1.0}$ &  & 0.88$_{-0.22}^{+0.28}$ &  \\
18 & 107$\pm$54 & 261.9$\pm$8.8 & 0.0107$\pm$0.0056 & 0.34$\pm$0.18 & 1.38$\pm$0.41 & 45.80$\pm$0.23 & 10.64$_{-0.73}^{+0.79}$ &  & 0.83$_{-0.17}^{+0.21}$ &  \\
88 & 203$\pm$180 & 206.9$\pm$8.8 & 0.020$\pm$0.018 & 0.65$\pm$0.59 & 2.4$\pm$1.2 & 45.68$\pm$0.40 & 10.4$_{-1.0}^{+1.1}$ &  & 0.77$_{-0.21}^{+0.27}$ &  \\
28 & 139$\pm$119 & 152.3$\pm$8.8 & 0.014$\pm$0.012 & 0.44$\pm$0.39 & 1.91$\pm$0.94 & 45.57$\pm$0.38 & 10.2$_{-0.9}^{+1.0}$ &  & 0.72$_{-0.19}^{+0.26}$ &  \\
95 & 138$\pm$115 & 202$\pm$13 & 0.014$\pm$0.012 & 0.44$\pm$0.38 & 1.89$\pm$0.89 & 45.59$\pm$0.37 & 10.17$_{-0.88}^{+0.99}$ &  & 0.72$_{-0.18}^{+0.24}$ &  \\
89 & 96$\pm$67 & 96.8$\pm$8.8 & 0.0097$\pm$0.0068 & 0.31$\pm$0.22 & 1.49$\pm$0.59 & 45.53$\pm$0.31 & 10.04$_{-0.74}^{+0.87}$ &  & 0.68$_{-0.15}^{+0.21}$ &  \\
93 & 129$\pm$149 & 145$\pm$51 & 0.013$\pm$0.015 & 0.41$\pm$0.48 & 1.9$\pm$1.2 & 45.50$\pm$0.51 & 10.0$_{-1.1}^{+1.3}$ &  & 0.67$_{-0.20}^{+0.32}$ &  \\
92 & 166$\pm$211 & 168$\pm$51 & 0.017$\pm$0.021 & 0.53$\pm$0.68 & 2.3$\pm$1.6 & 45.49$\pm$0.56 & 10.0$_{-1.2}^{+1.5}$ &  & 0.66$_{-0.21}^{+0.36}$ &  \\
64 & 84$\pm$105 & 165$\pm$51 & 0.008$\pm$0.011 & 0.27$\pm$0.34 & 1.44$\pm$0.96 & 45.41$\pm$0.55 & 9.9$_{-1.3}^{+1.4}$ &  & 0.65$_{-0.23}^{+0.35}$ &  \\
94 & 61$\pm$77 & 105$\pm$51 & 0.0061$\pm$0.0078 & 0.19$\pm$0.25 & 1.14$\pm$0.76 & 45.39$\pm$0.56 & 9.8$_{-1.2}^{+1.3}$ &  & 0.62$_{-0.21}^{+0.30}$ &  \\
63 & 340$\pm$758 & 145.1$\pm$8.8 & 0.035$\pm$0.079 & 1.1$\pm$2.5 & 4.1$\pm$5.2 & 45.42$\pm$0.99 & 9.7$_{-2.1}^{+2.7}$ &  & 0.62$_{-0.38}^{+0.74}$ &  \\
90 & 233$\pm$786 & 75$\pm$51 & 0.024$\pm$0.081 & 0.7$\pm$2.5 & 3.8$\pm$7.0 & 45.1$\pm$1.5 & 9.3$_{-3.8}^{+3.6}$ &  & 0.53$_{-0.73}^{+0.96}$ &  \\

\hline
\end{longtable}
\noindent\textit{Note.} The source marked with an asterisk (*) corresponds to a UC H \textsc{ii} region with an elongated morphology; this source is included in the full-sample analysis but excluded from the compact-only power-law fits to the stellar mass distributions. Reported values are rounded according to the two significant figures of their associated uncertainties.

\clearpage

\section{Zoom-in Views of the Detected H\textsc{ii} regions} \label{sec:imaging_sources}

\begin{figure*}[!ht]
    \centering
    \includegraphics[width=0.32\textwidth]{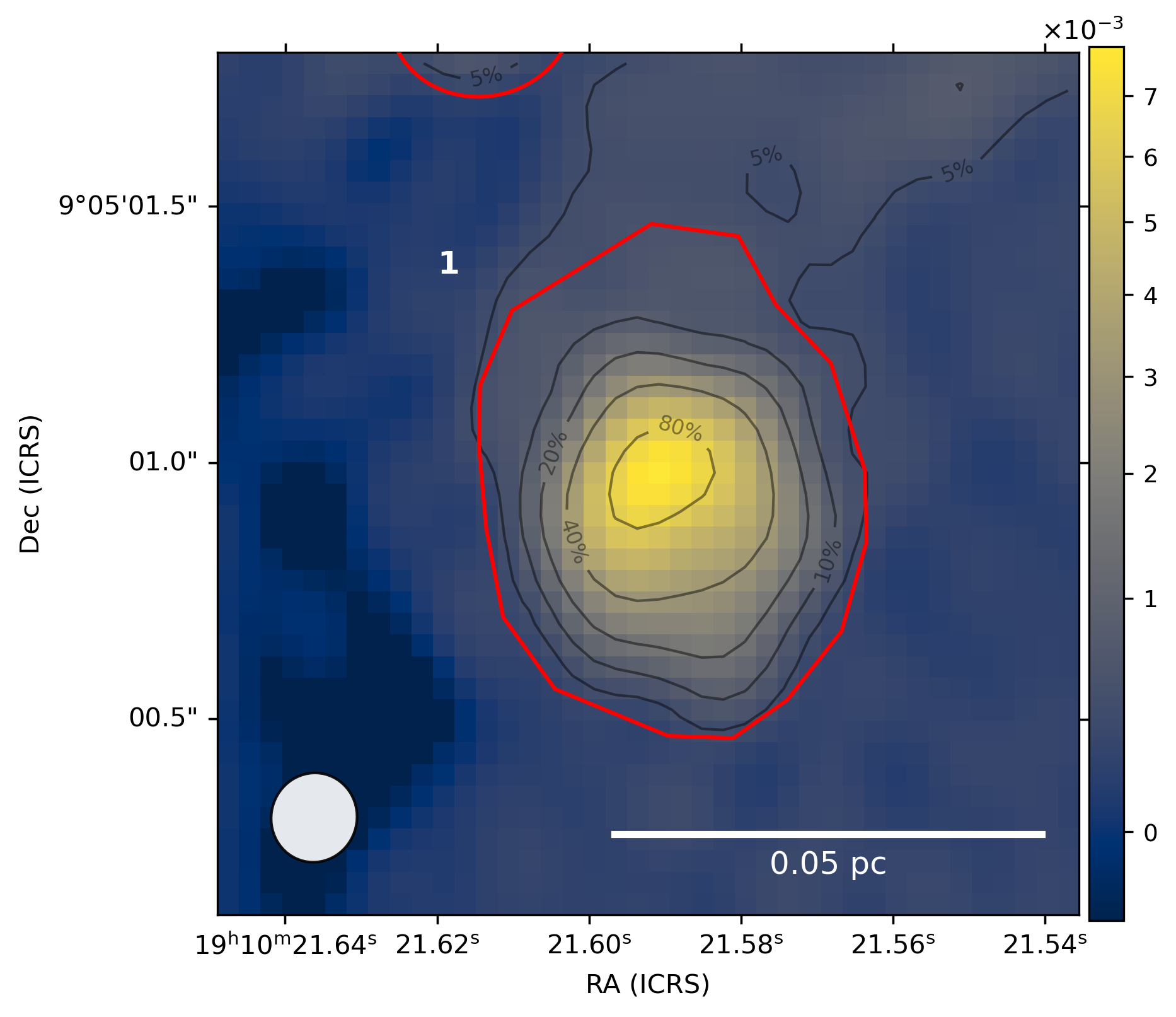}
    \includegraphics[width=0.32\textwidth]{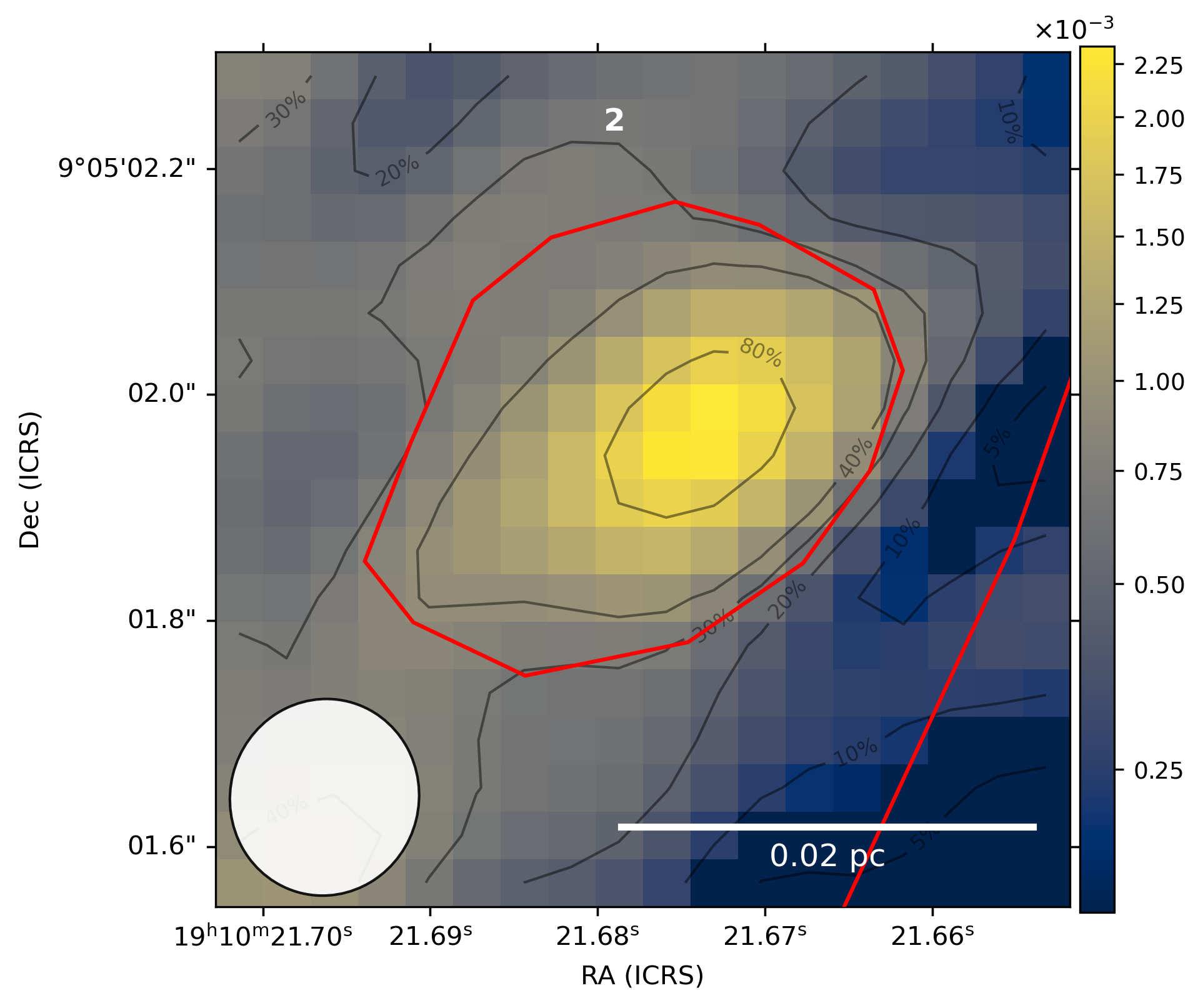}
    \includegraphics[width=0.32\textwidth]{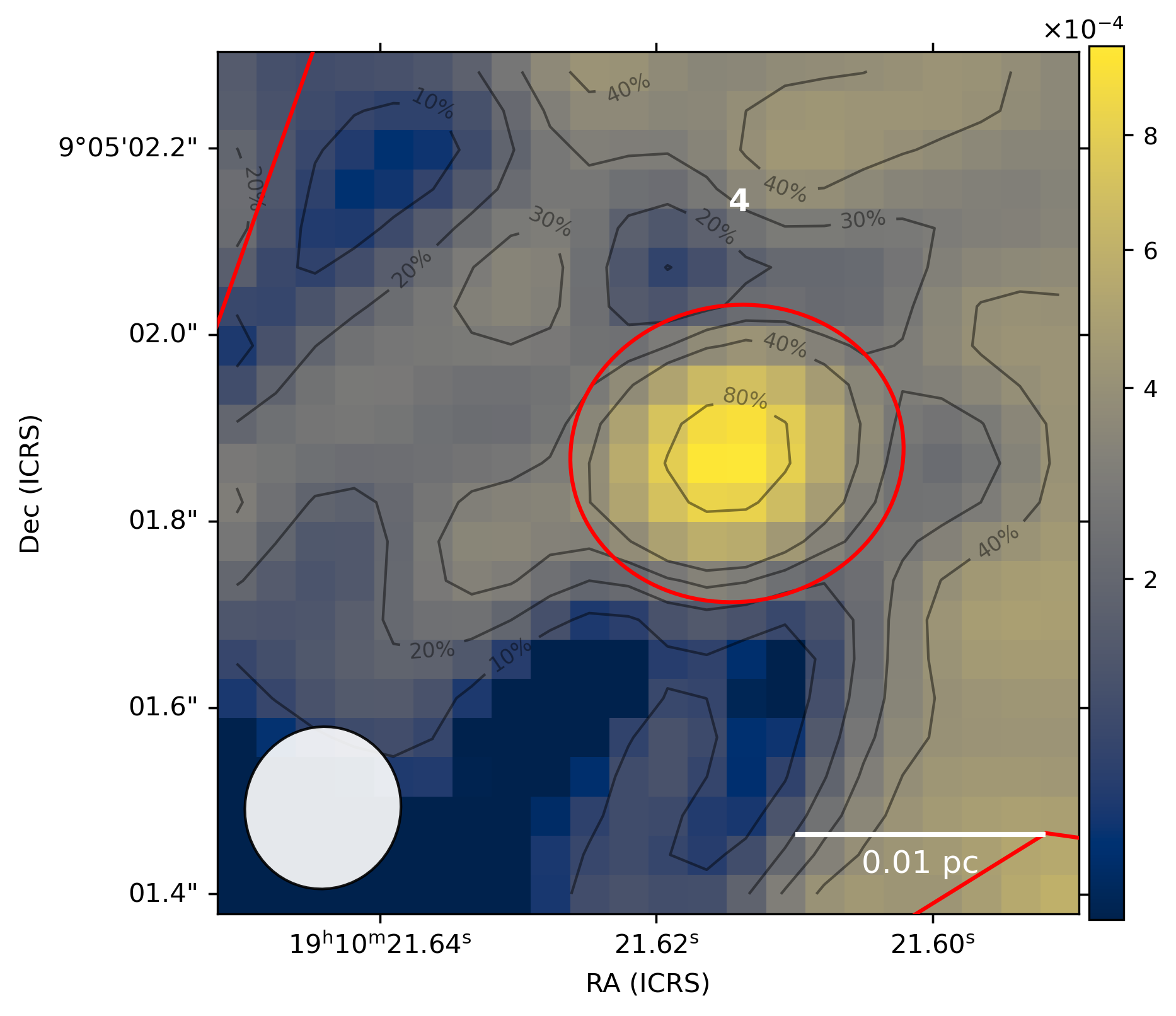}
    \includegraphics[width=0.32\textwidth]{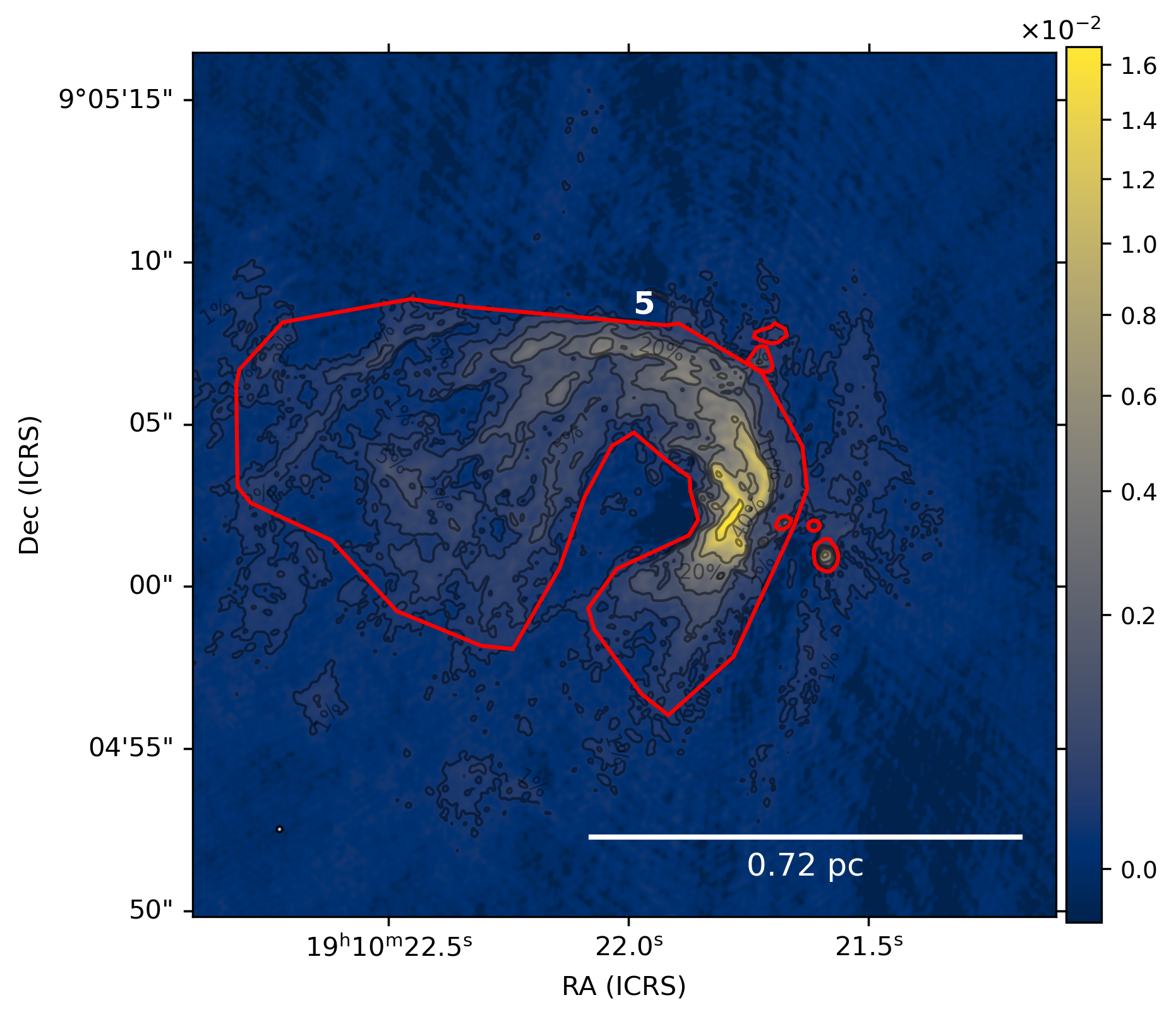}
    \includegraphics[width=0.32\textwidth]{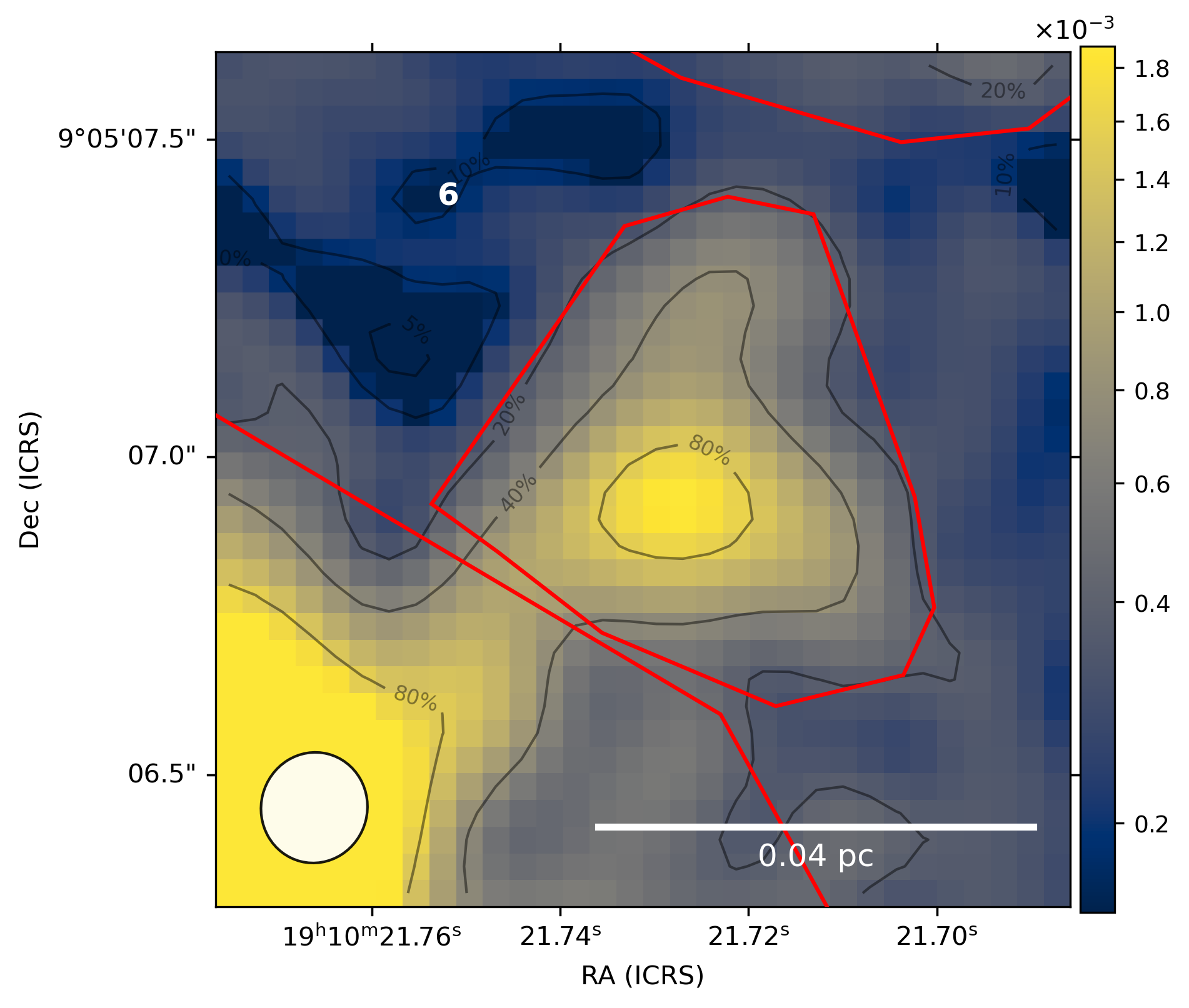}
    \includegraphics[width=0.32\textwidth]{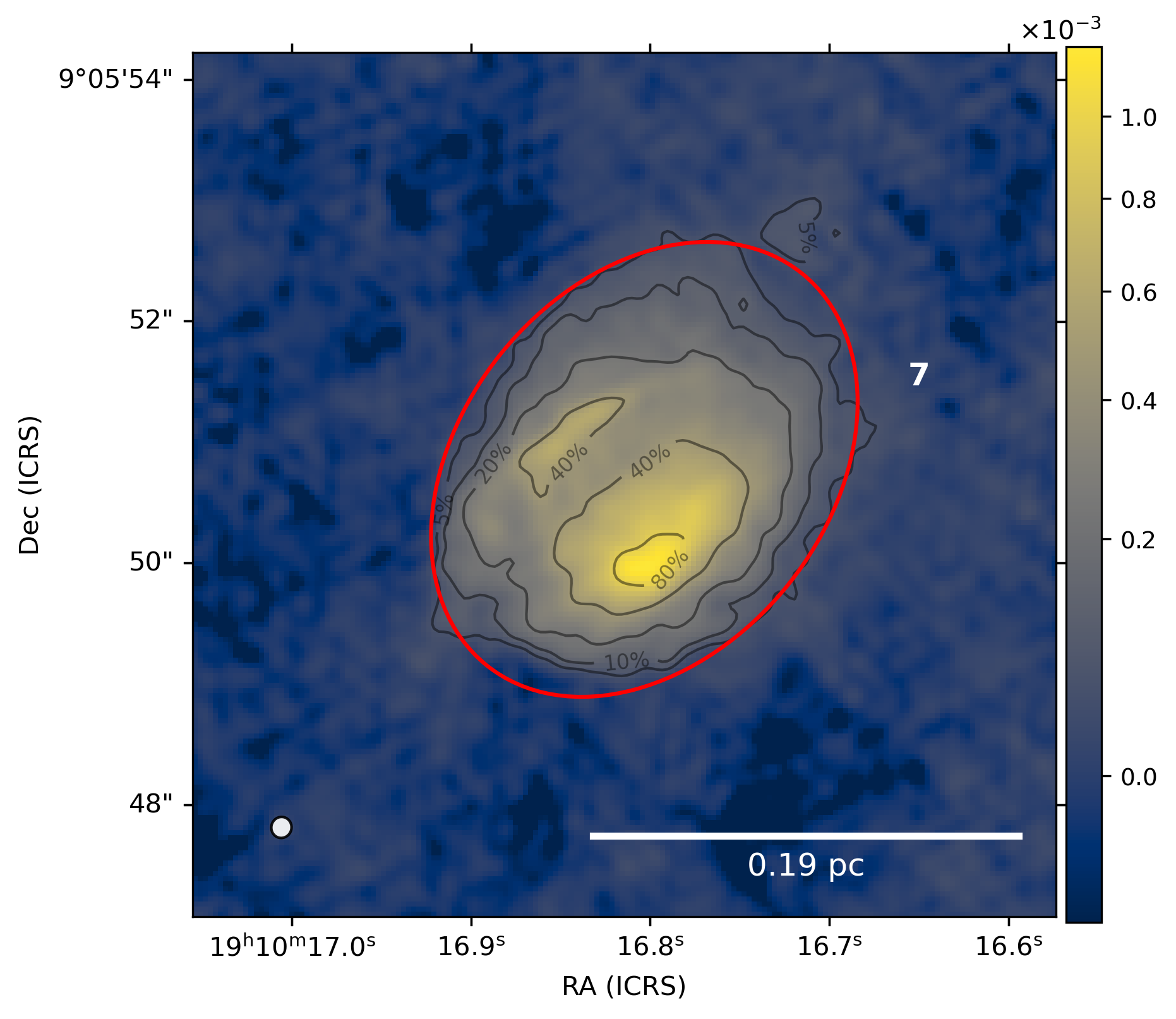}
    \includegraphics[width=0.32\textwidth]{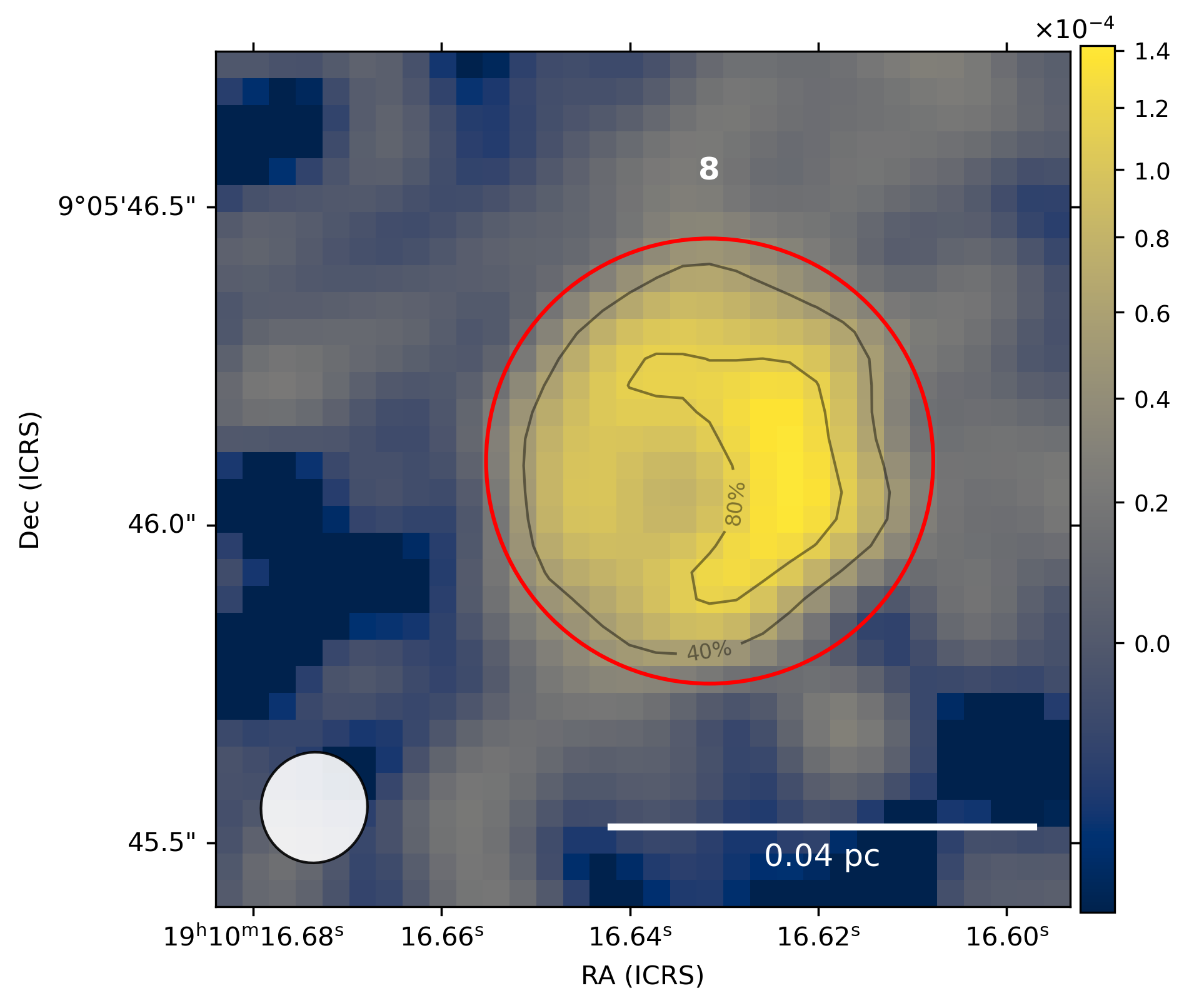}
    \includegraphics[width=0.32\textwidth]{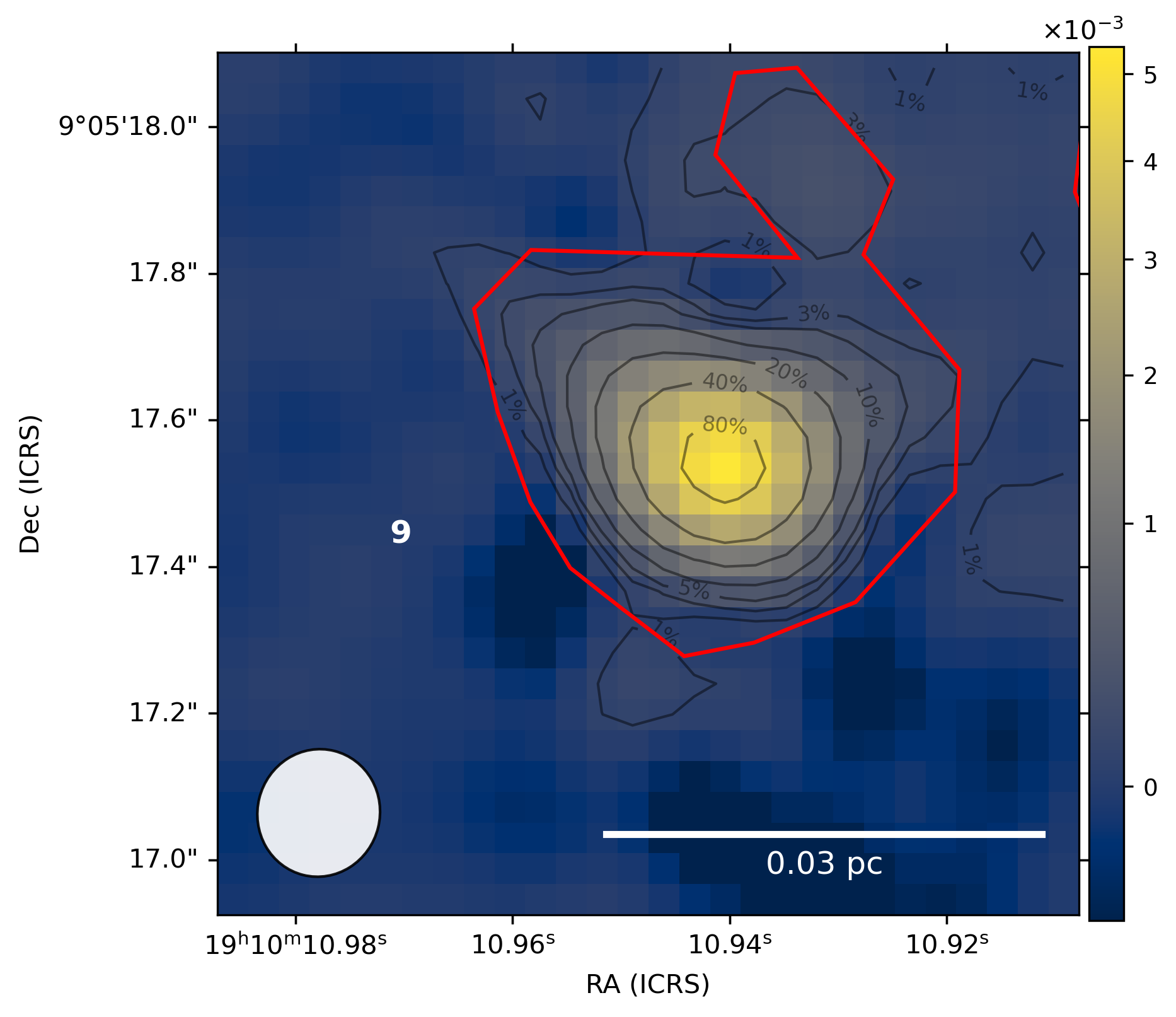}
    \includegraphics[width=0.32\textwidth]{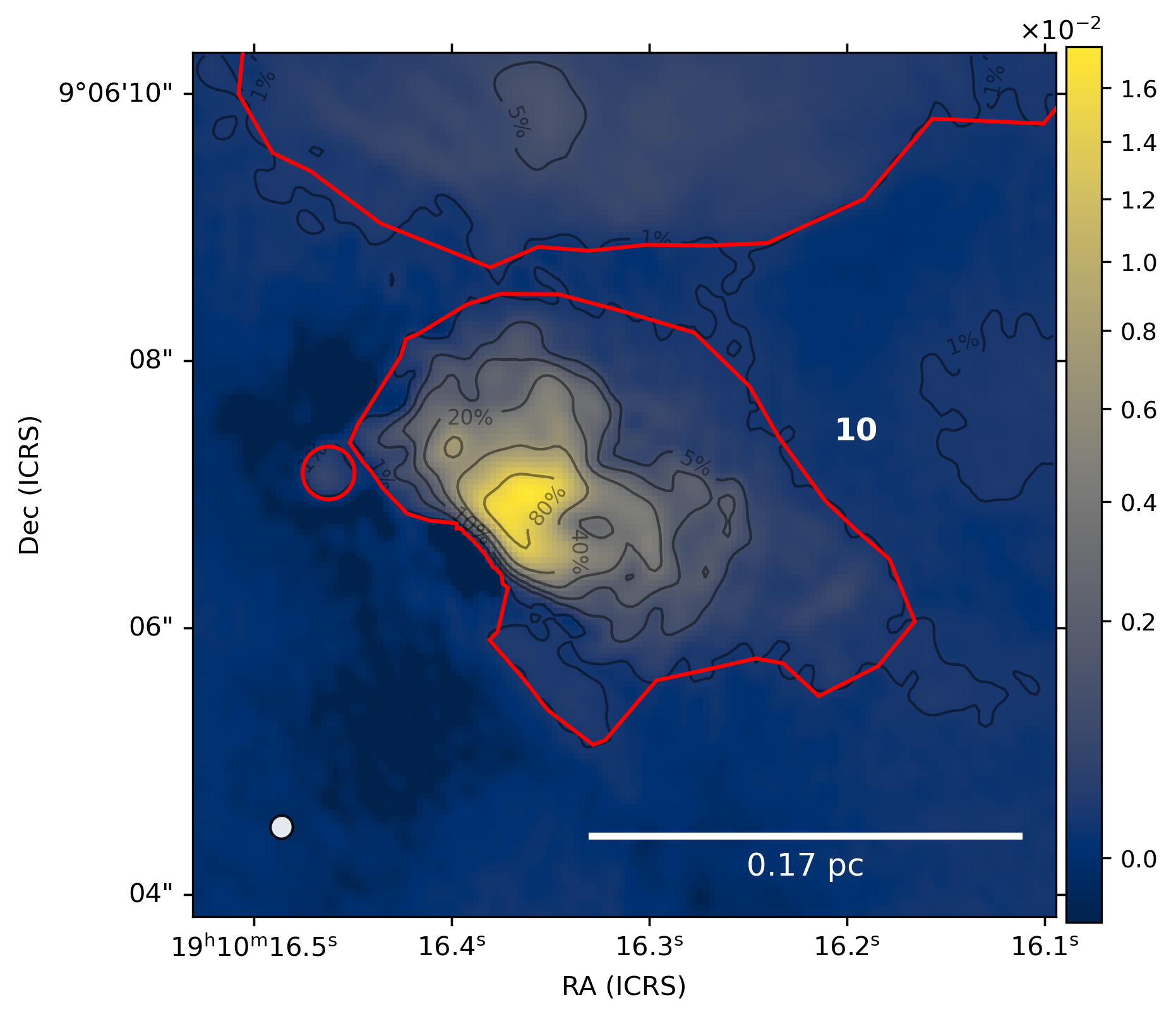}
    \includegraphics[width=0.32\textwidth]{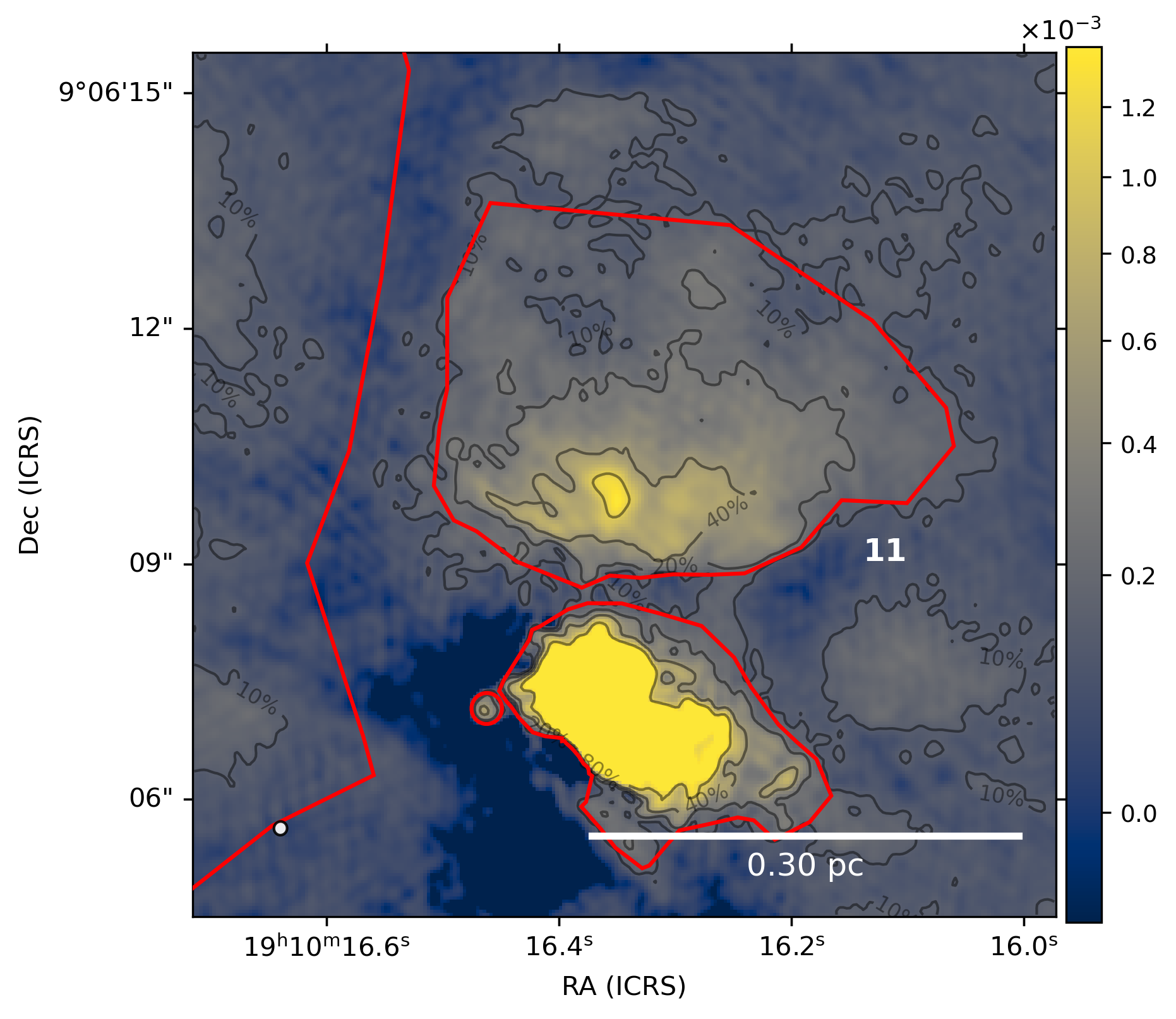}
    \includegraphics[width=0.32\textwidth]{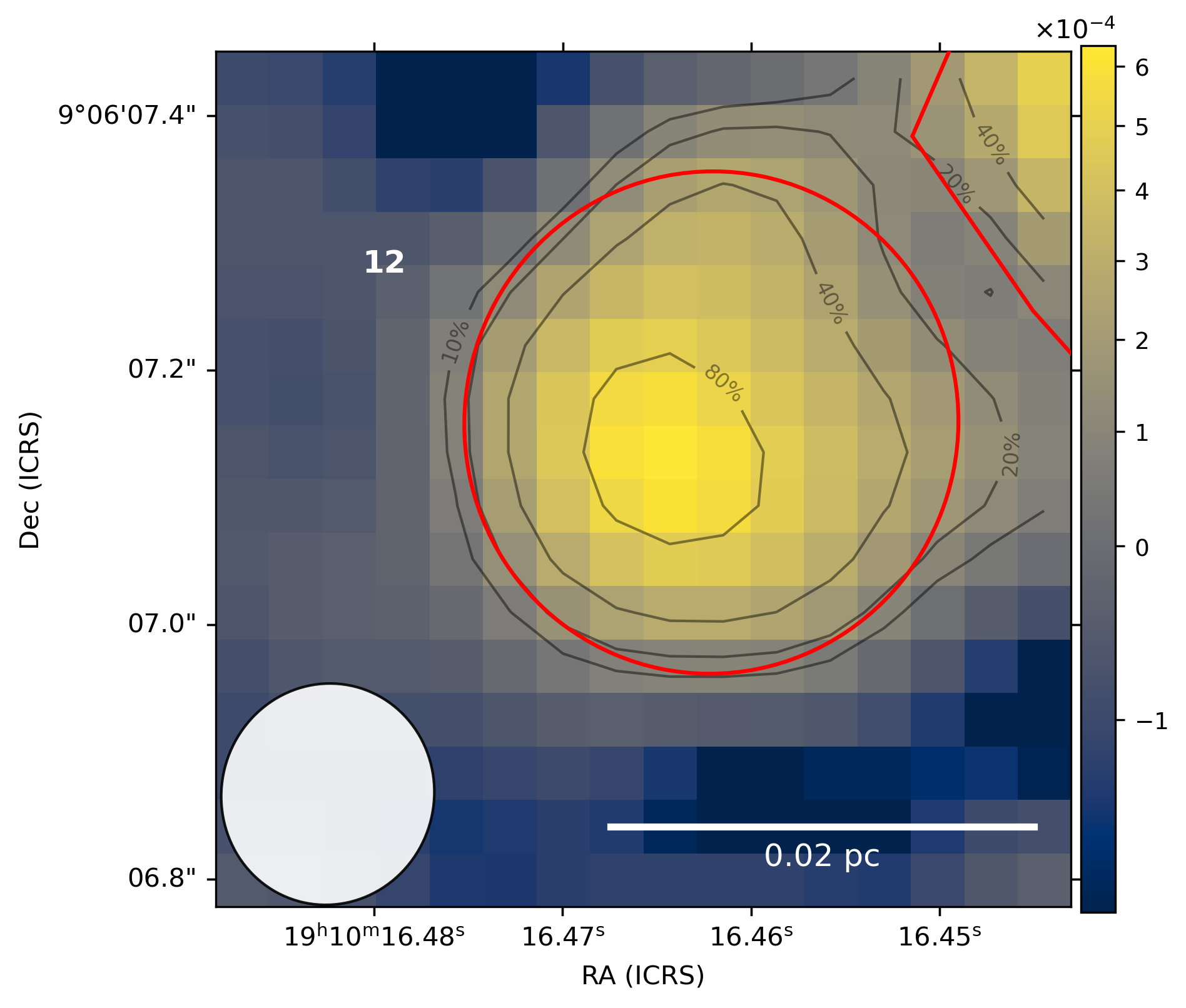}
    \includegraphics[width=0.32\textwidth]{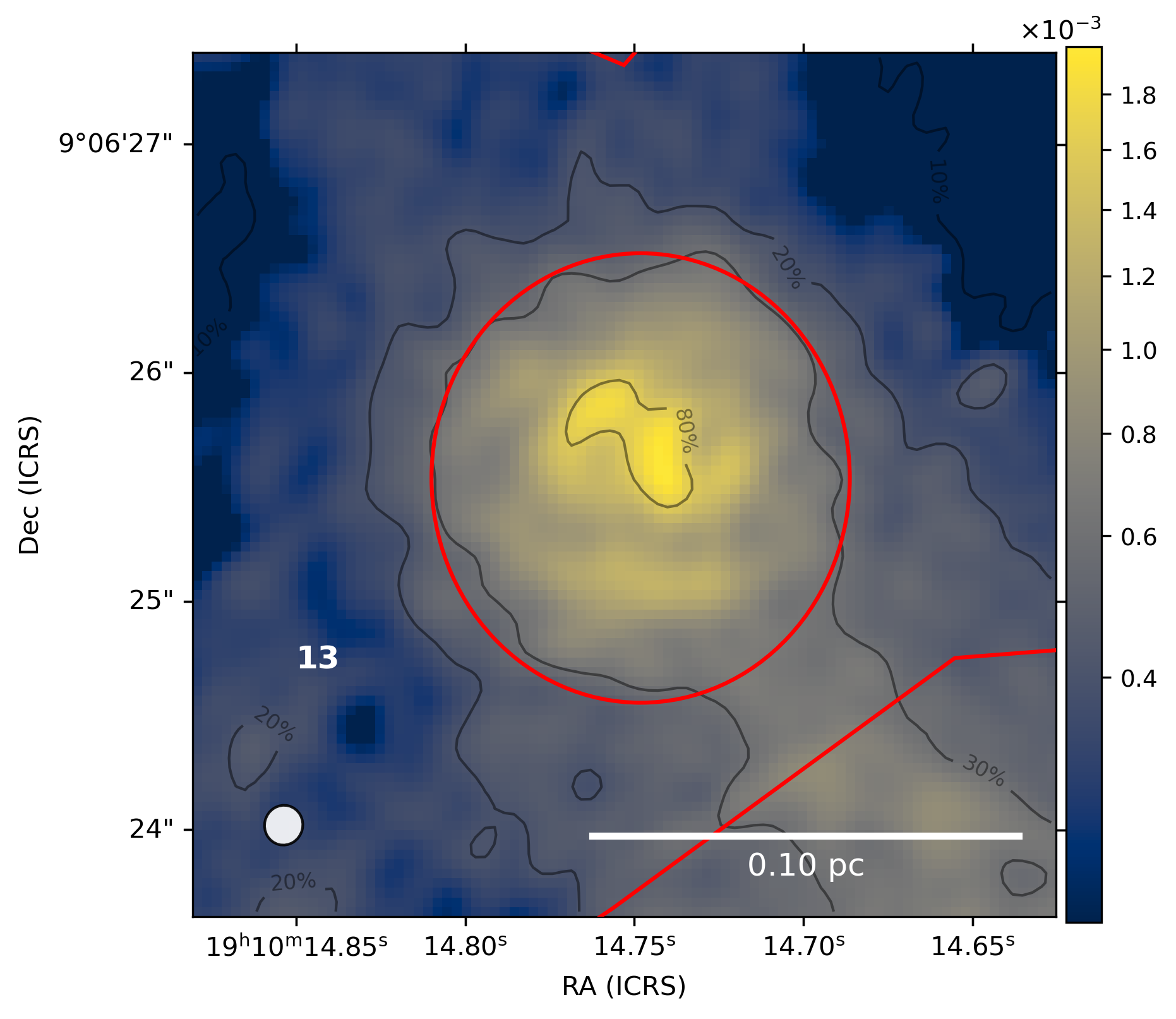}
    \caption{Zoomed-in views of individual H\textsc{ii} regions. In each panel, a white scale bar in the lower right corner indicates the physical size as labeled. The synthesized beam (FWHM) is shown as a white ellipse in the lower left corner for resolution comparison. Contours represent percentages of the peak intensity of each individual source. The color scale is in units of Jy beam\(^{-1}\). The complete figure set (84 images) is available in the online journal.} 
     \label{fig:individual}
\end{figure*}

\begin{figure*}
    \centering
    \includegraphics[width=0.32\textwidth]{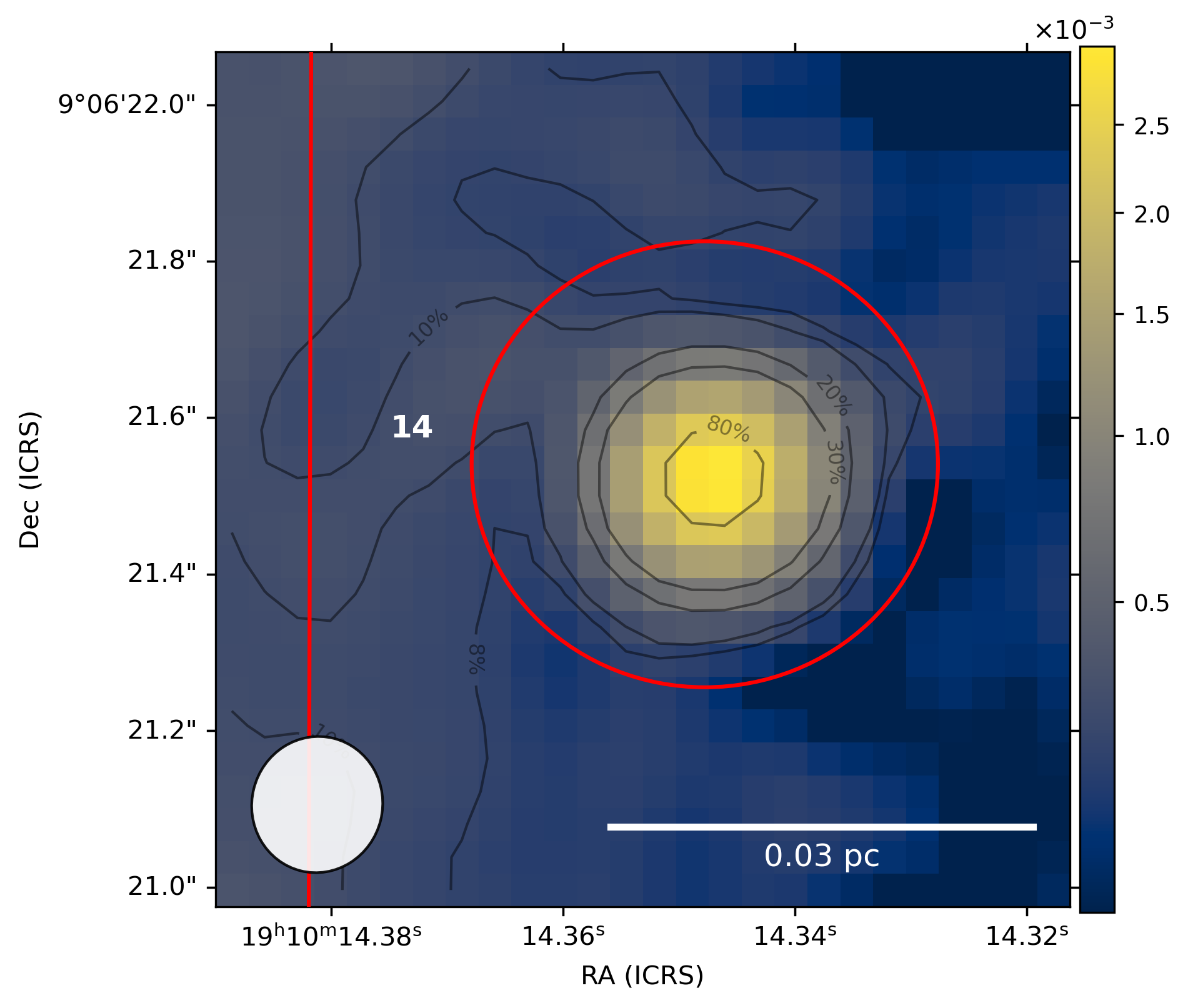}
    \includegraphics[width=0.32\textwidth]{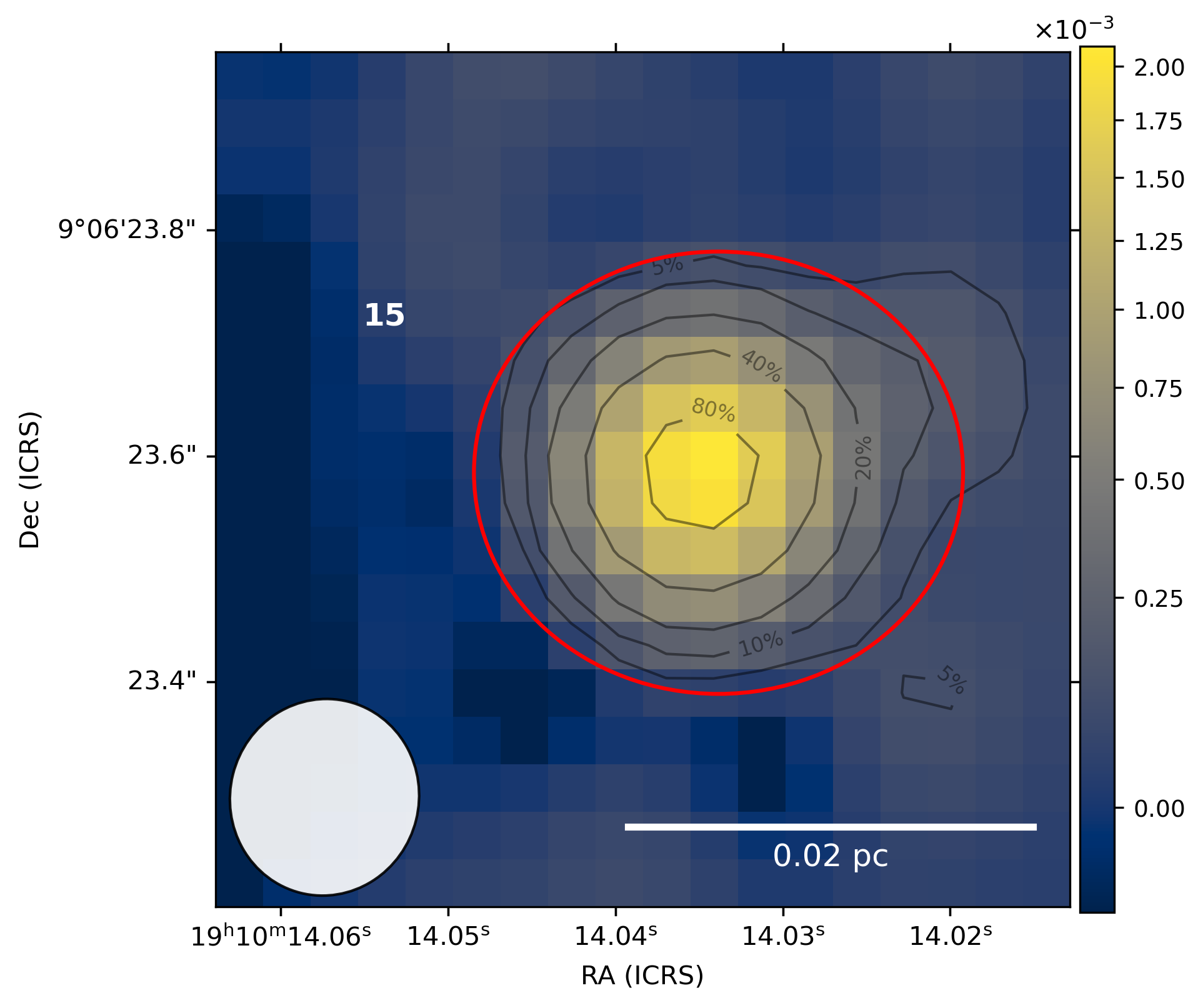}
    \includegraphics[width=0.32\textwidth]{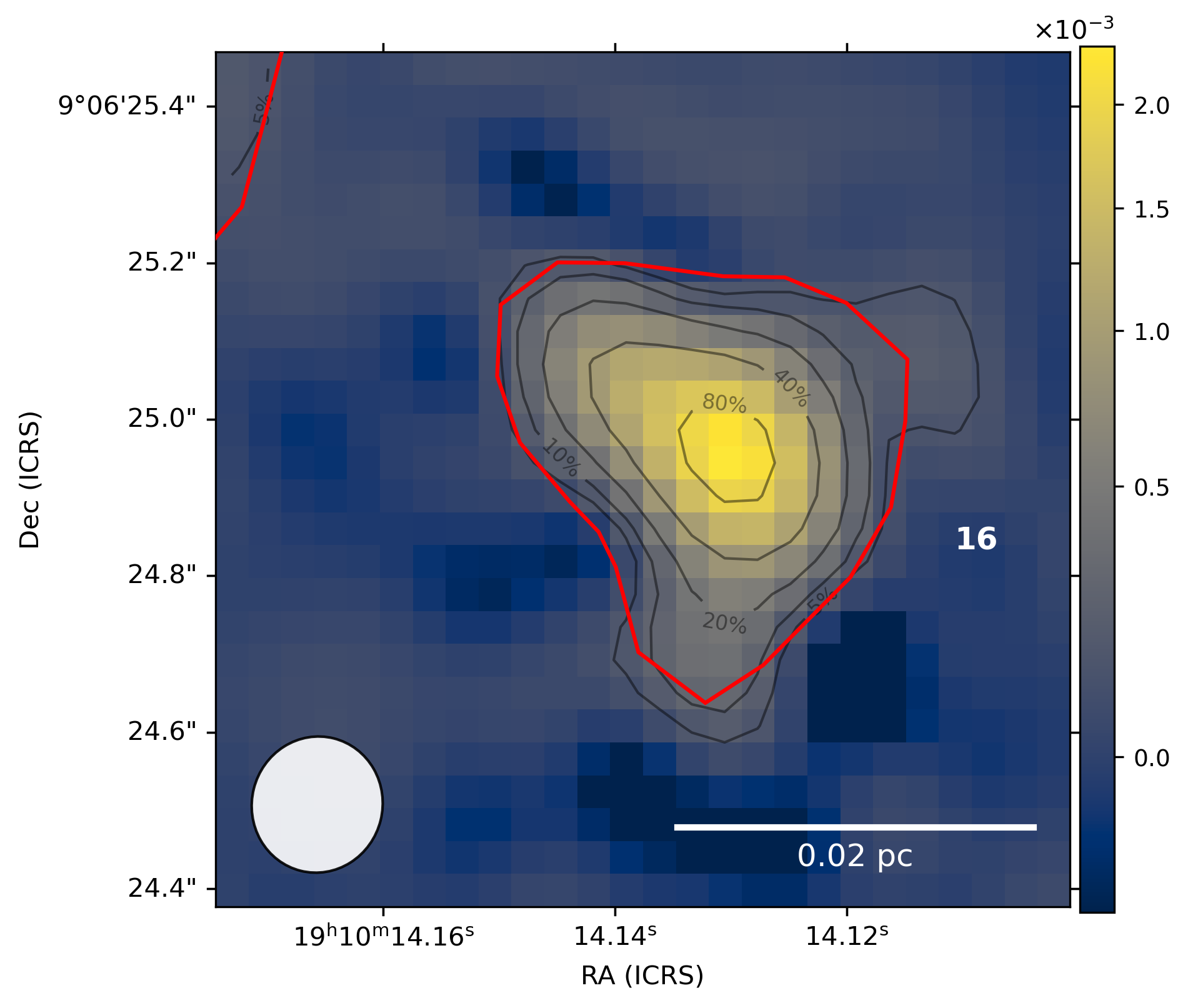}
    \includegraphics[width=0.32\textwidth]{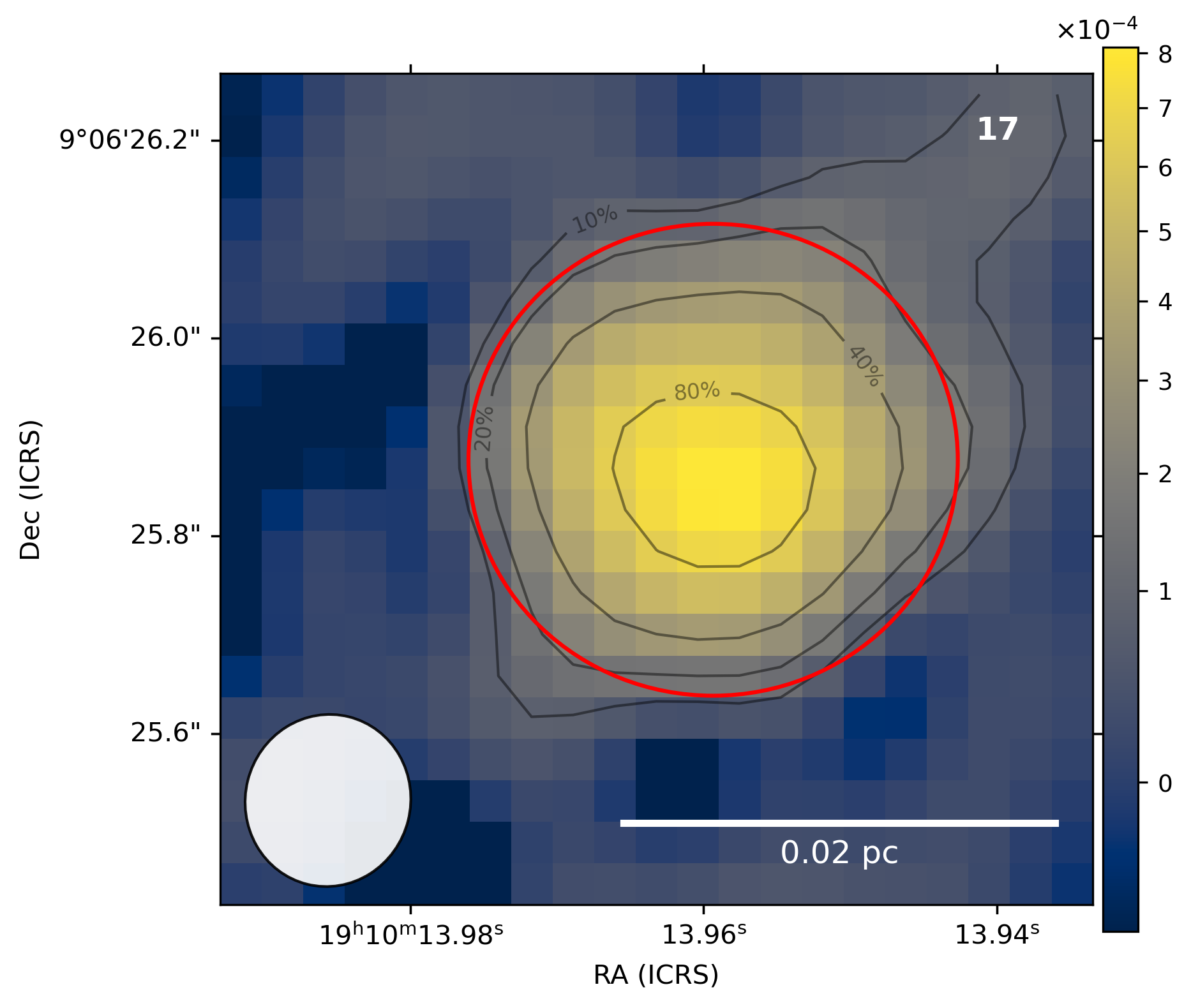}
    \includegraphics[width=0.32\textwidth]{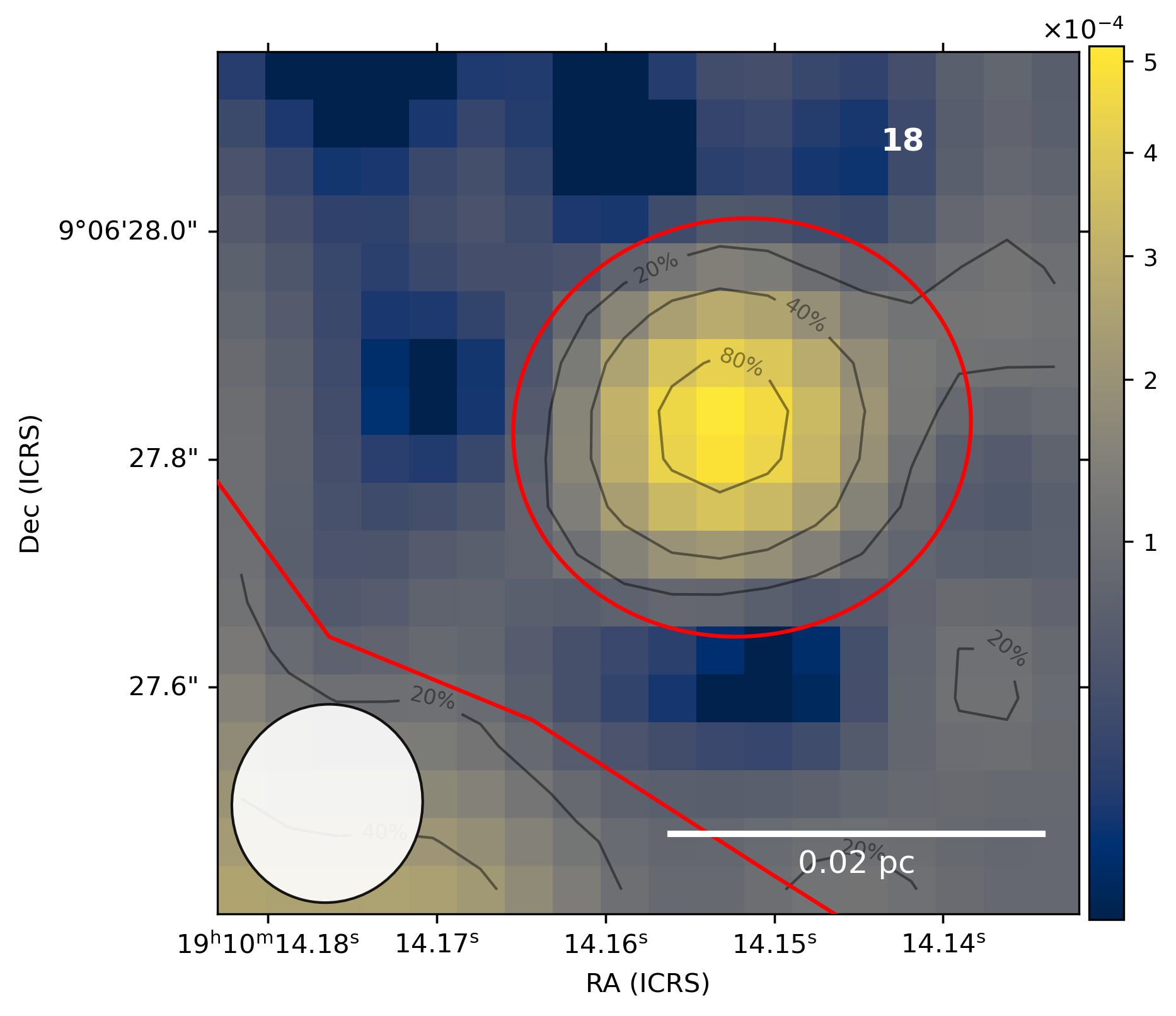}
    \includegraphics[width=0.32\textwidth]{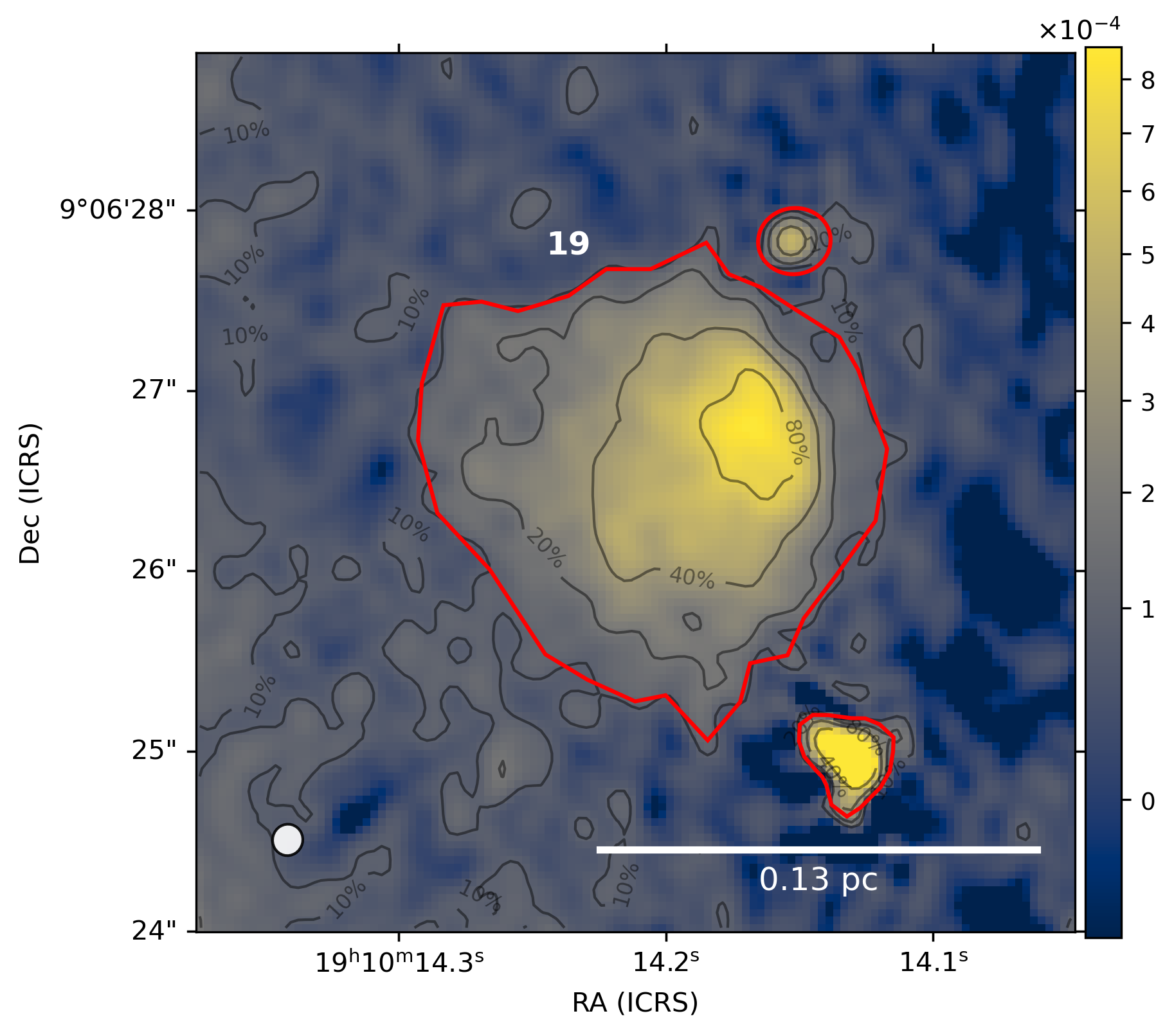}
    \includegraphics[width=0.32\textwidth]{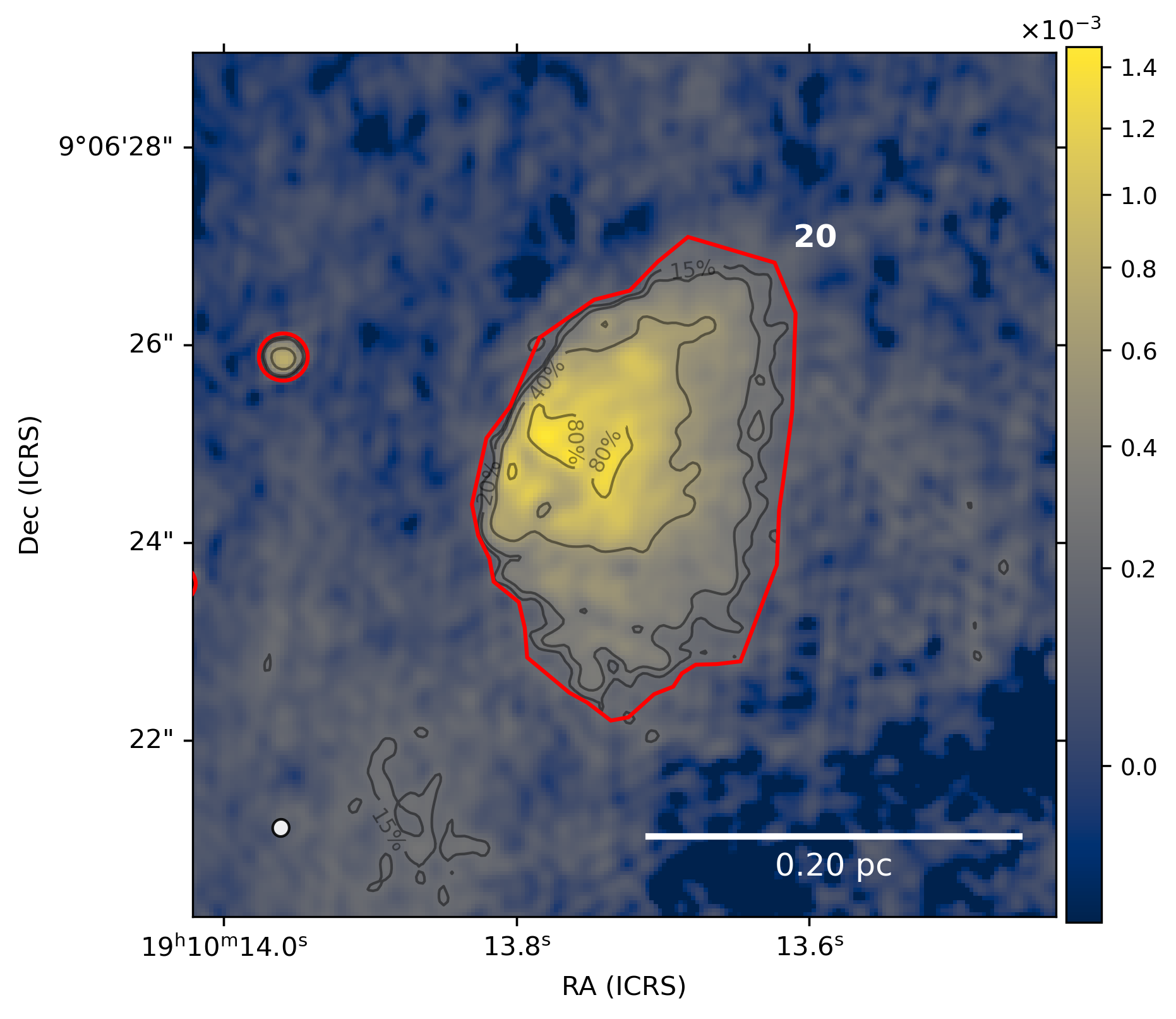}
    \includegraphics[width=0.32\textwidth]{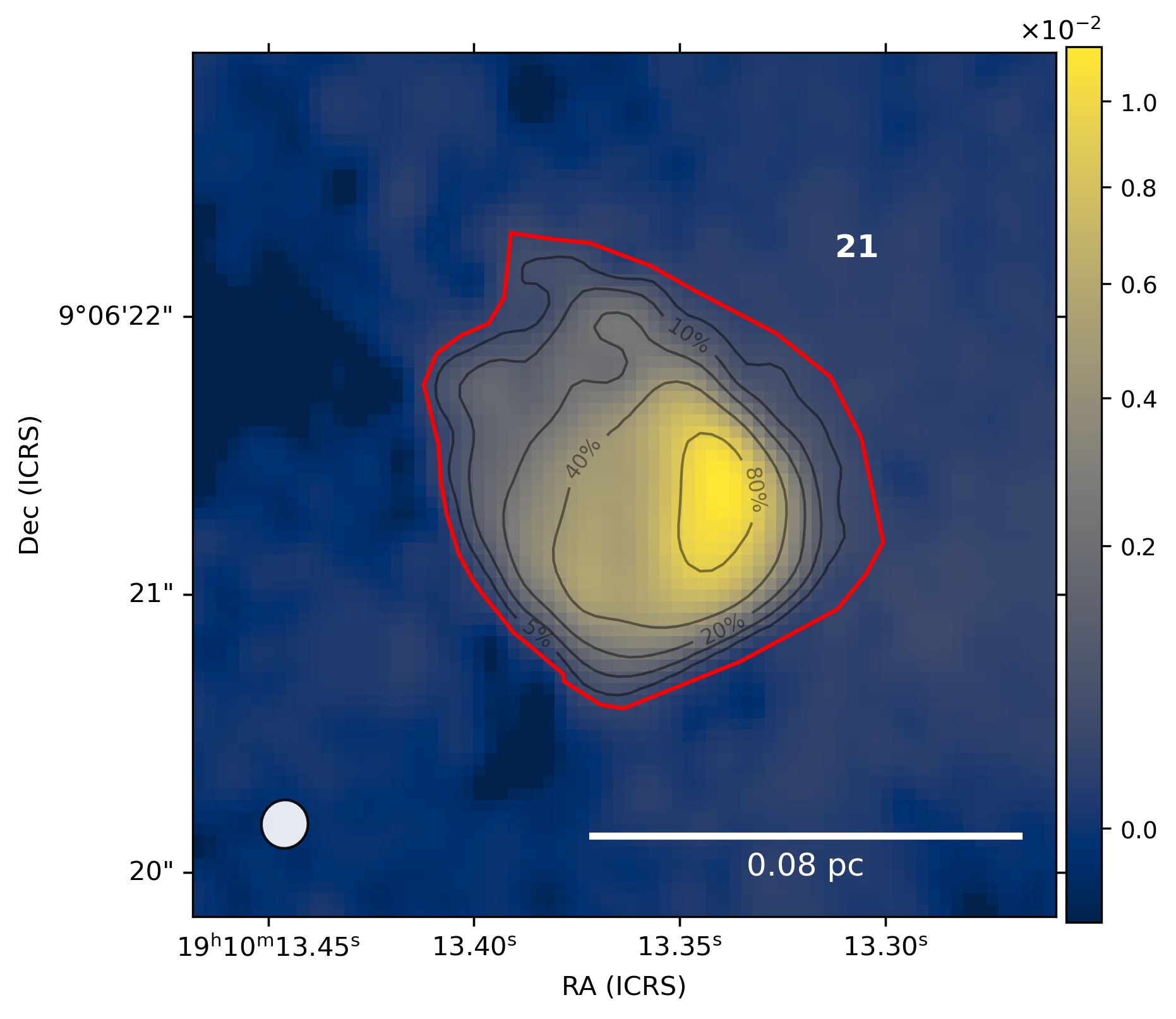}
    \includegraphics[width=0.32\textwidth]{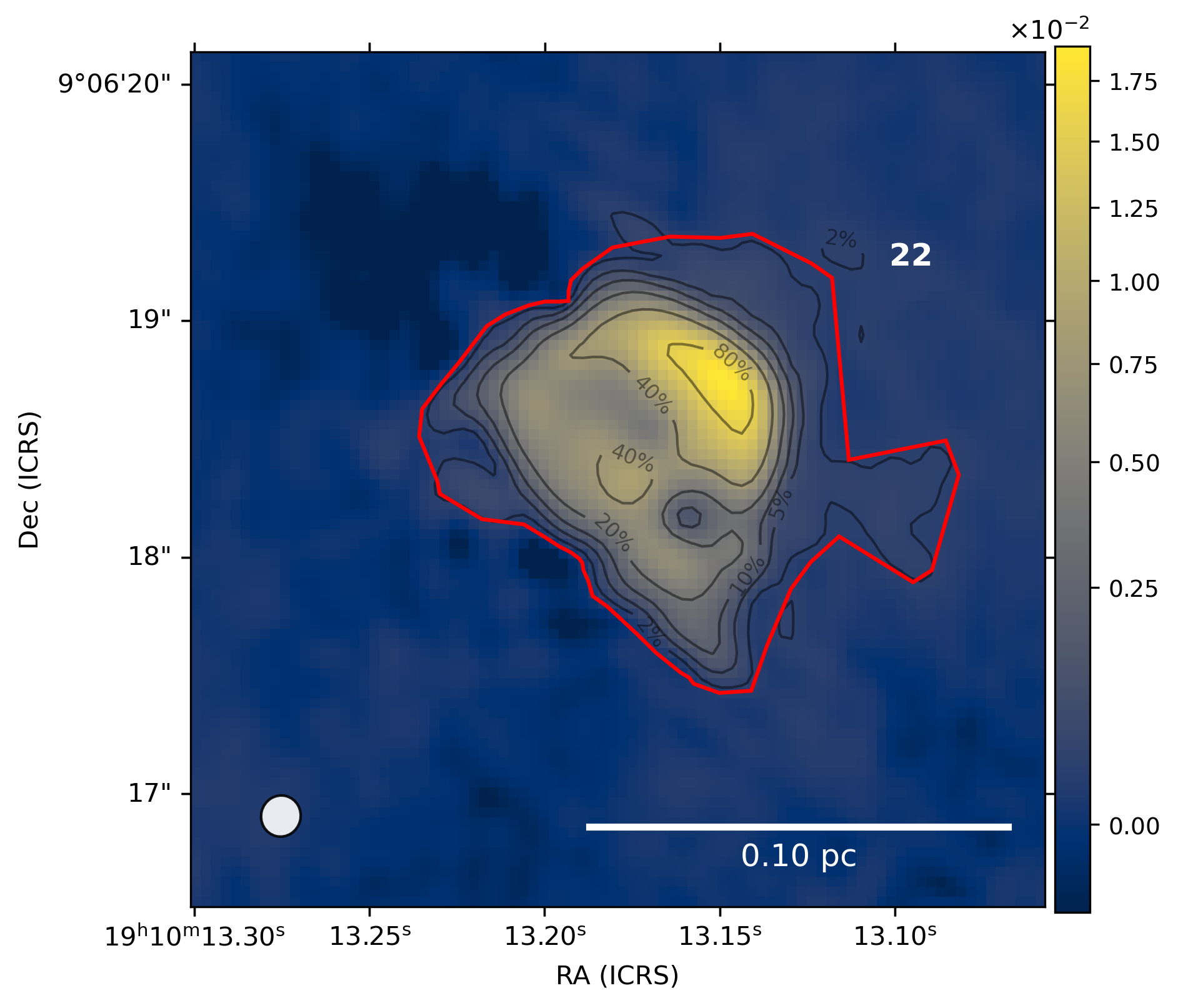}
    \includegraphics[width=0.32\textwidth]{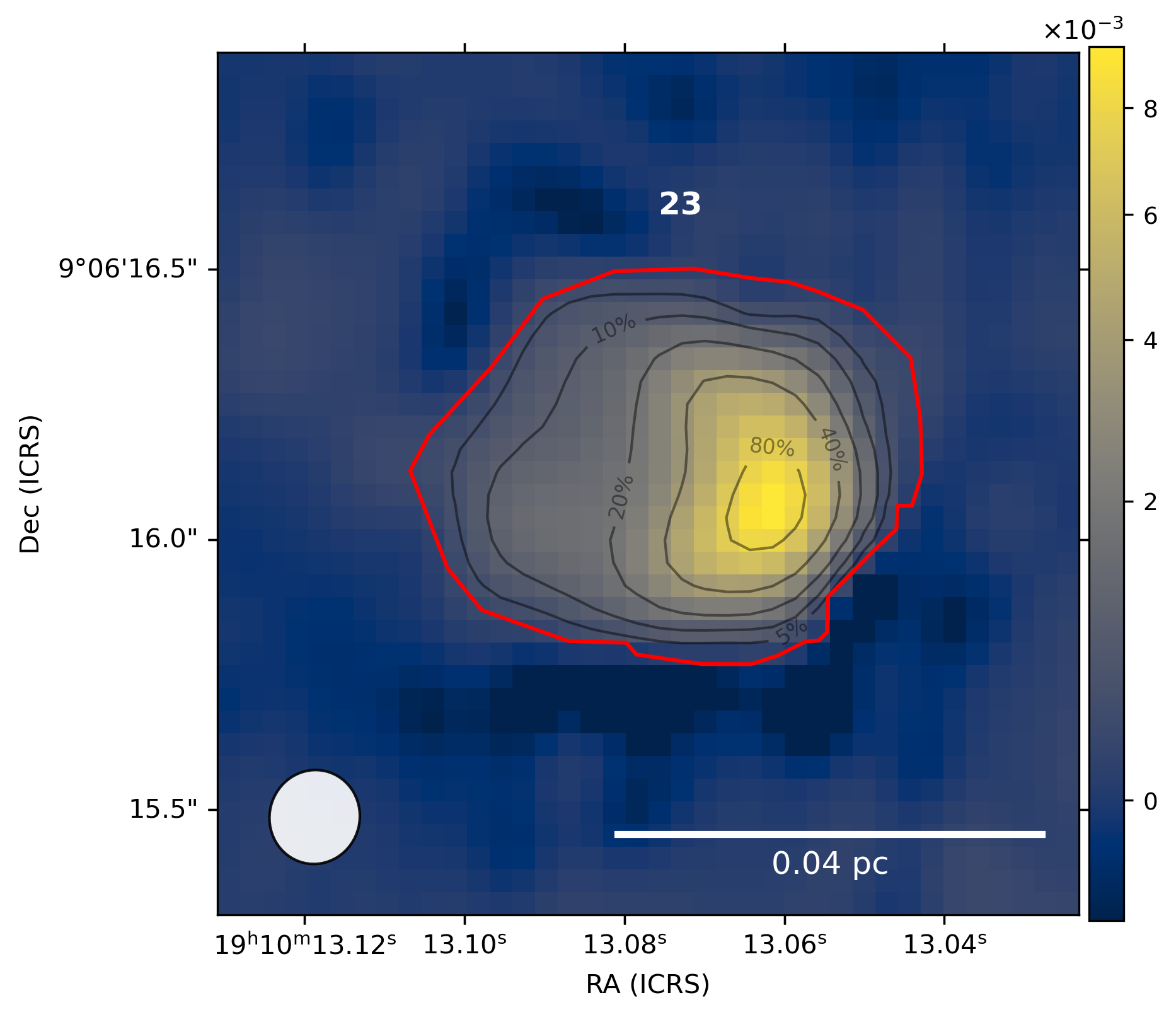}
    \includegraphics[width=0.32\textwidth]{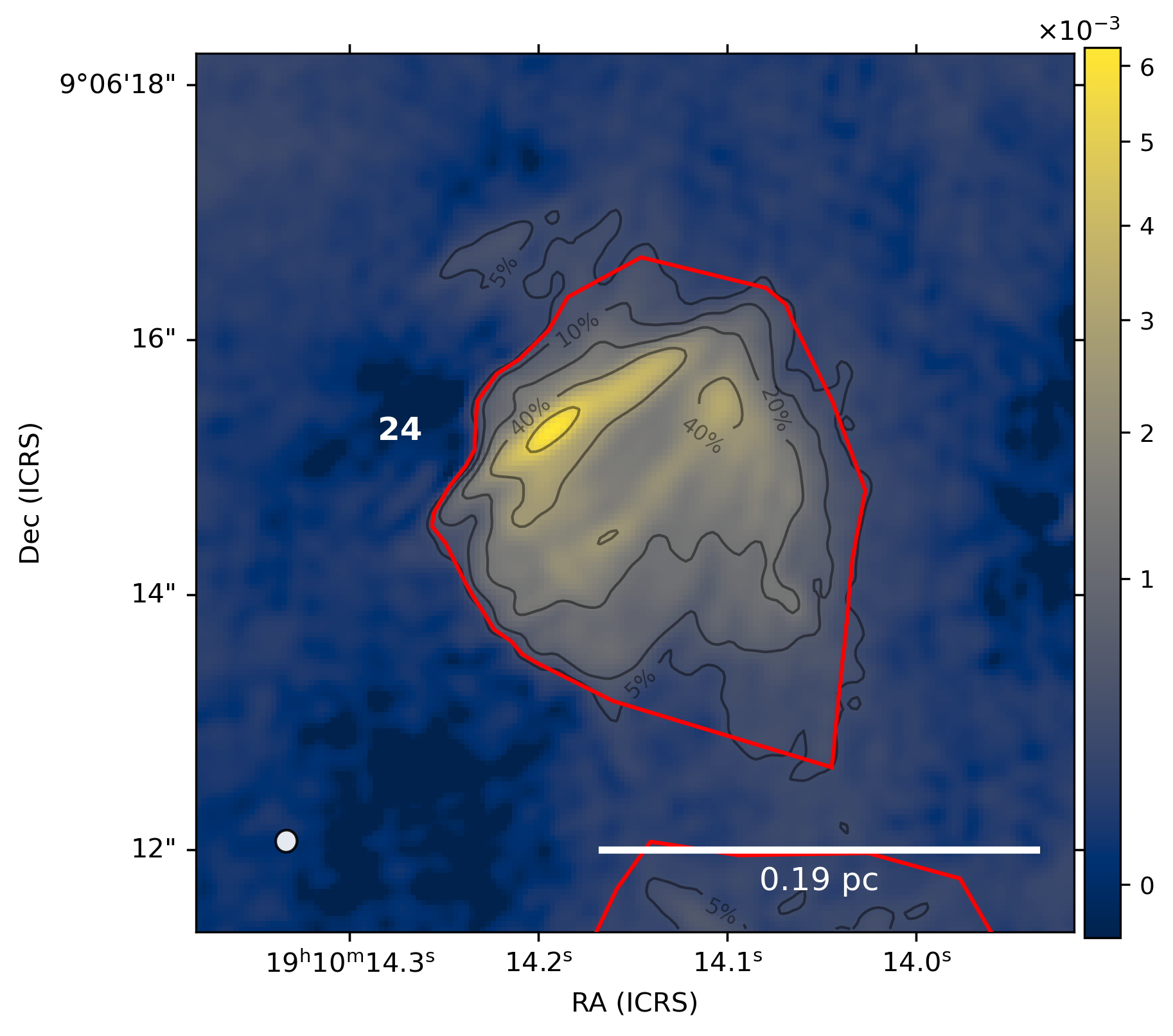}
    \includegraphics[width=0.32\textwidth]{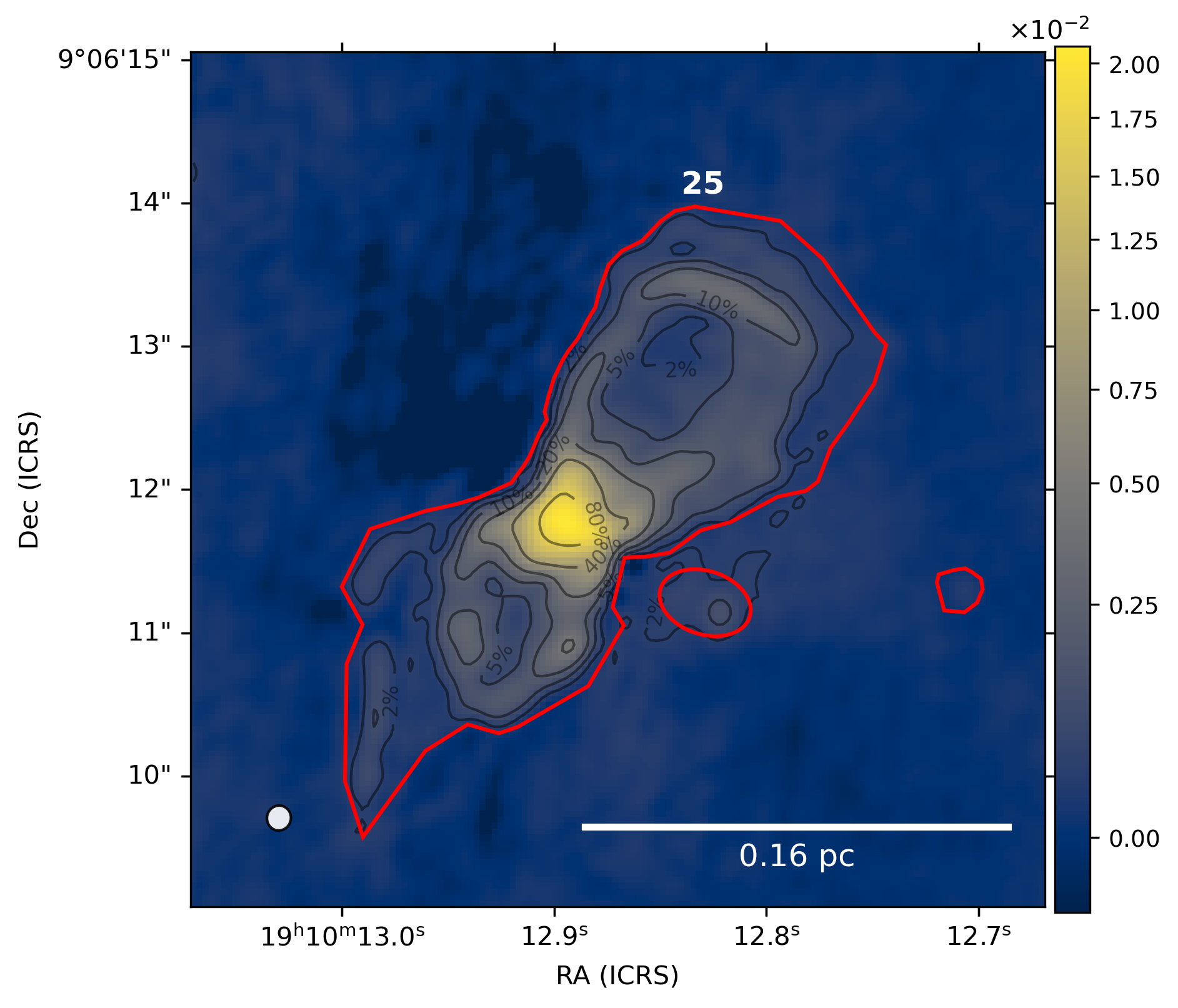}
    \caption{Zoomed-in views (continued).}
\end{figure*}

\begin{figure*}
    \centering
    \includegraphics[width=0.32\textwidth]{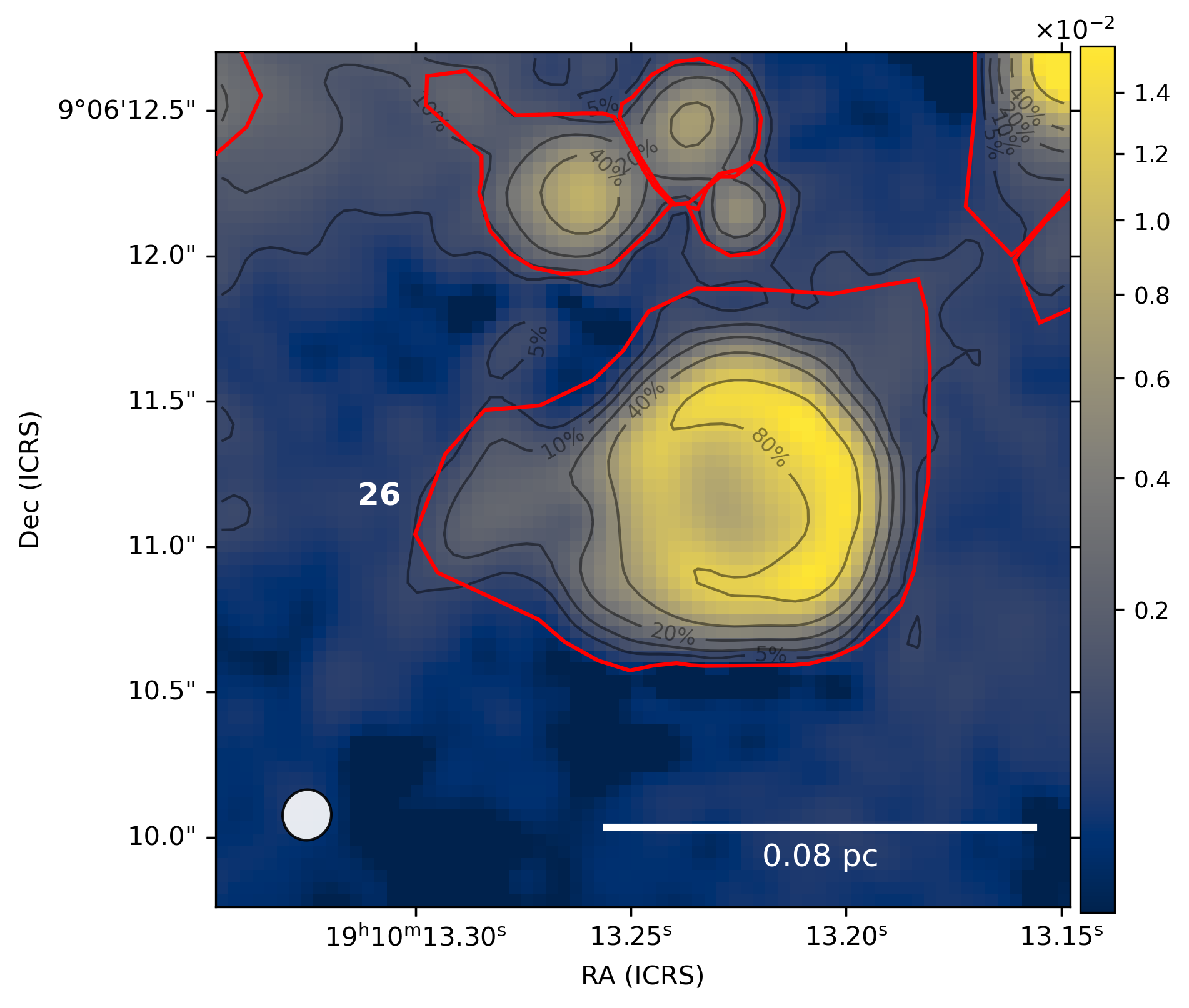}
    \includegraphics[width=0.32\textwidth]{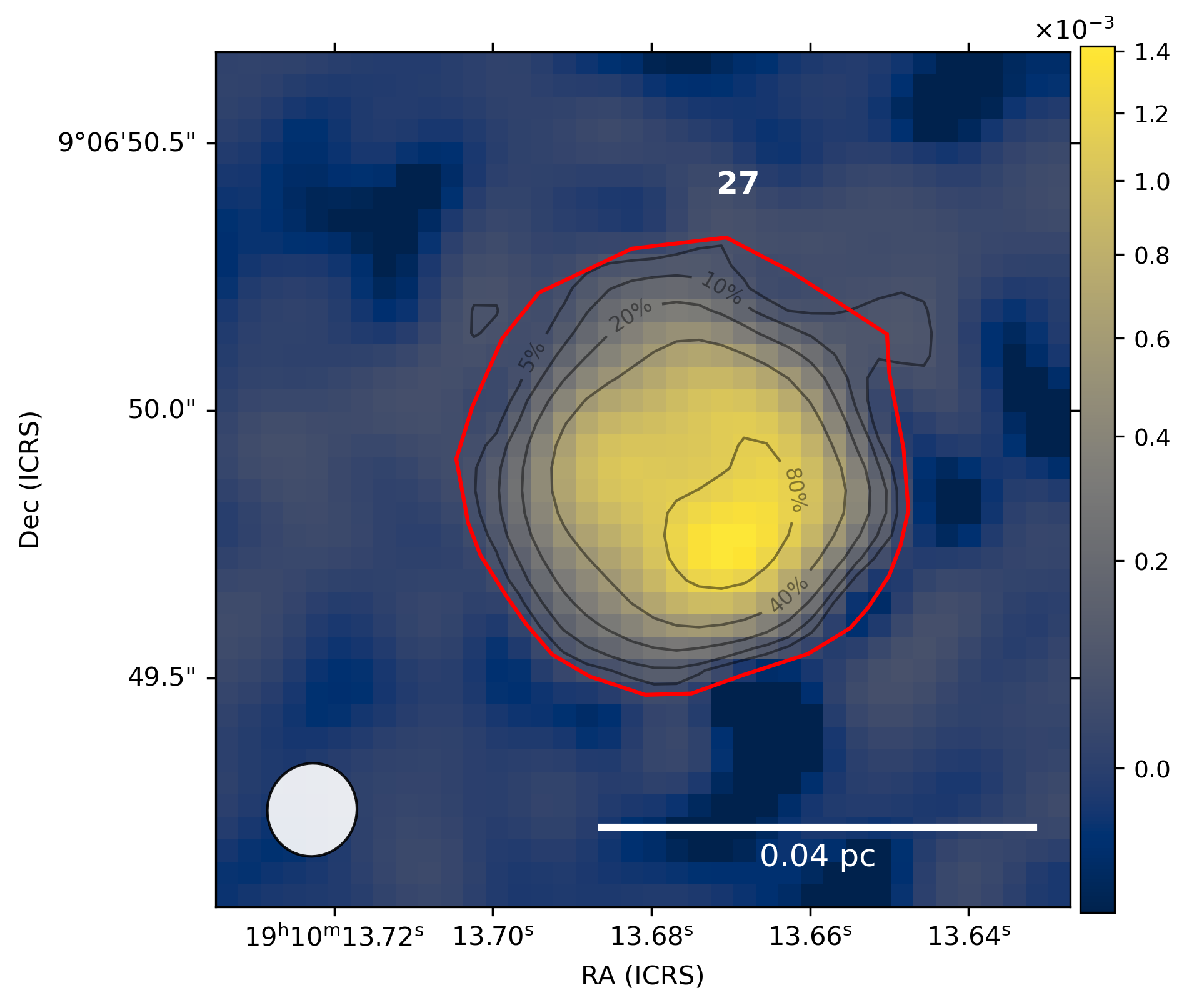}
    \includegraphics[width=0.32\textwidth]{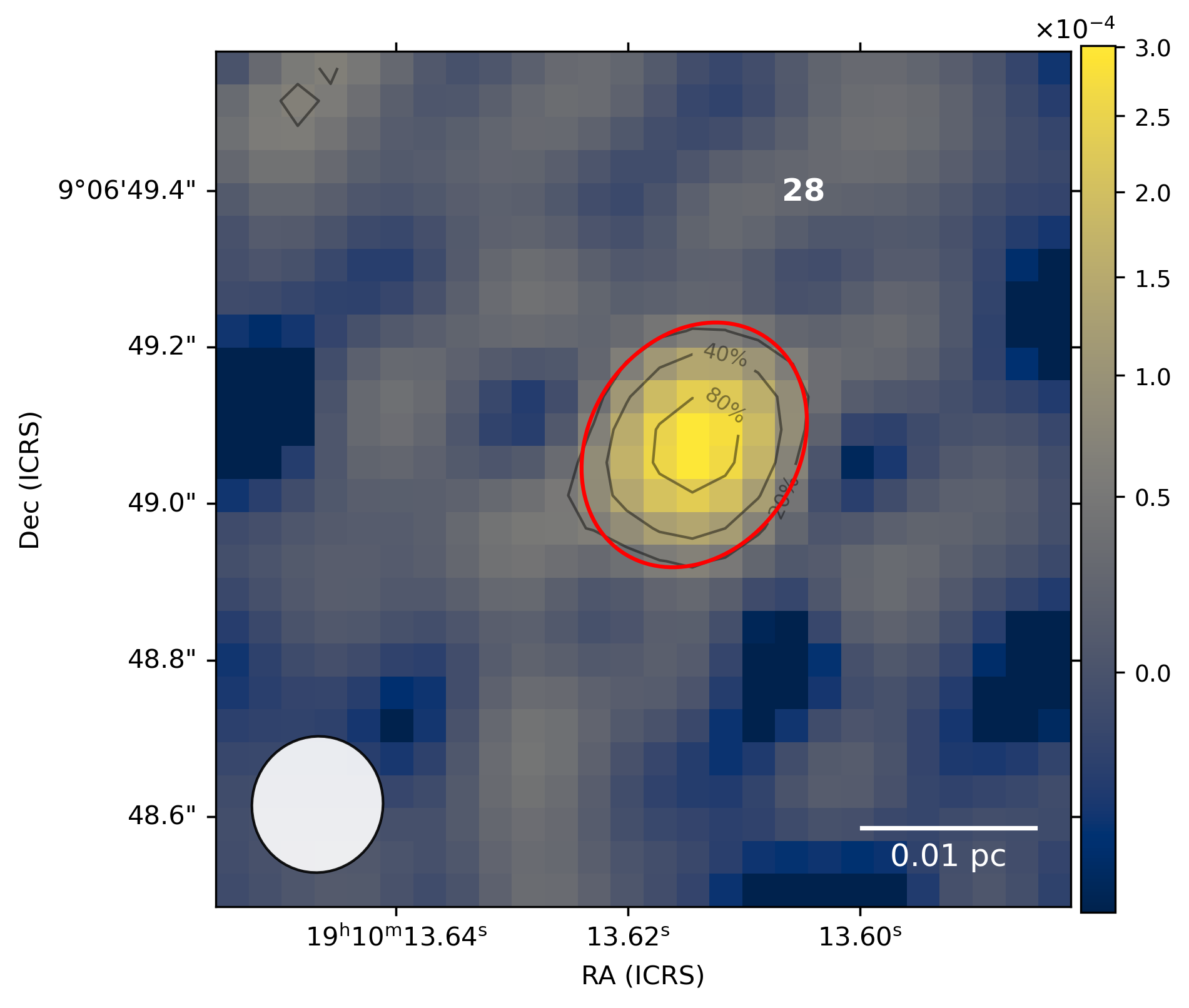}
    \includegraphics[width=0.32\textwidth]{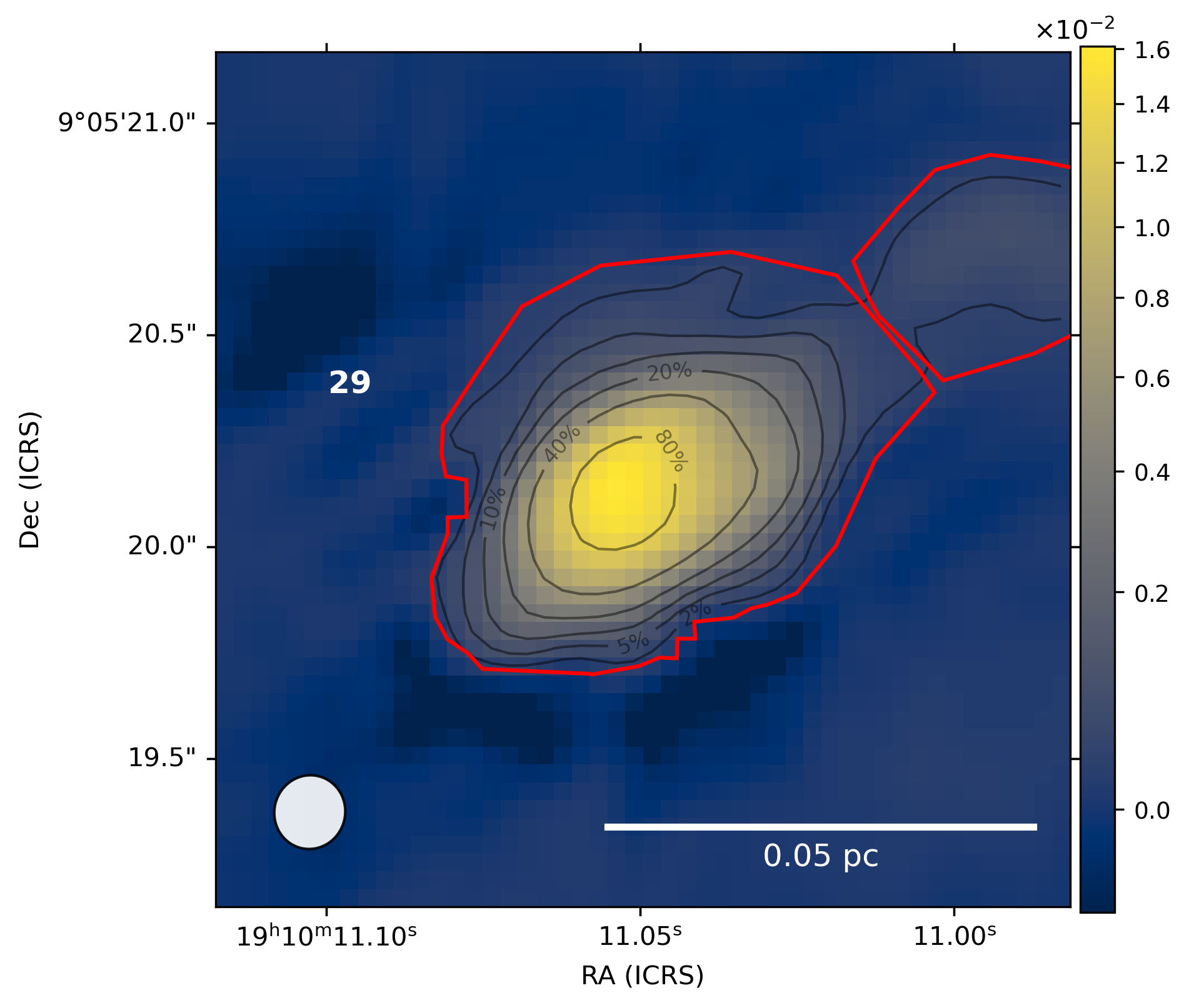}
    \includegraphics[width=0.32\textwidth]{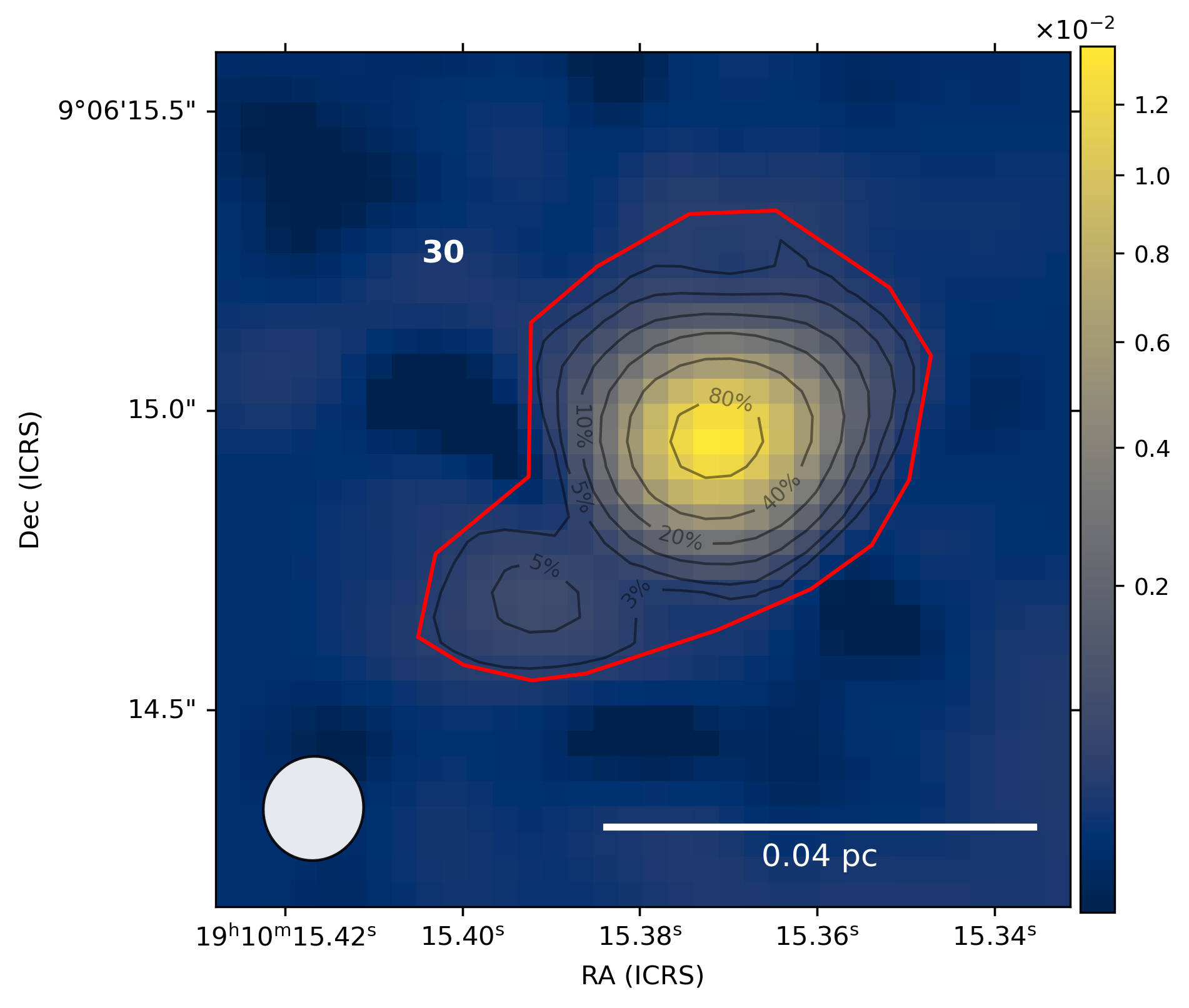}
    \includegraphics[width=0.32\textwidth]{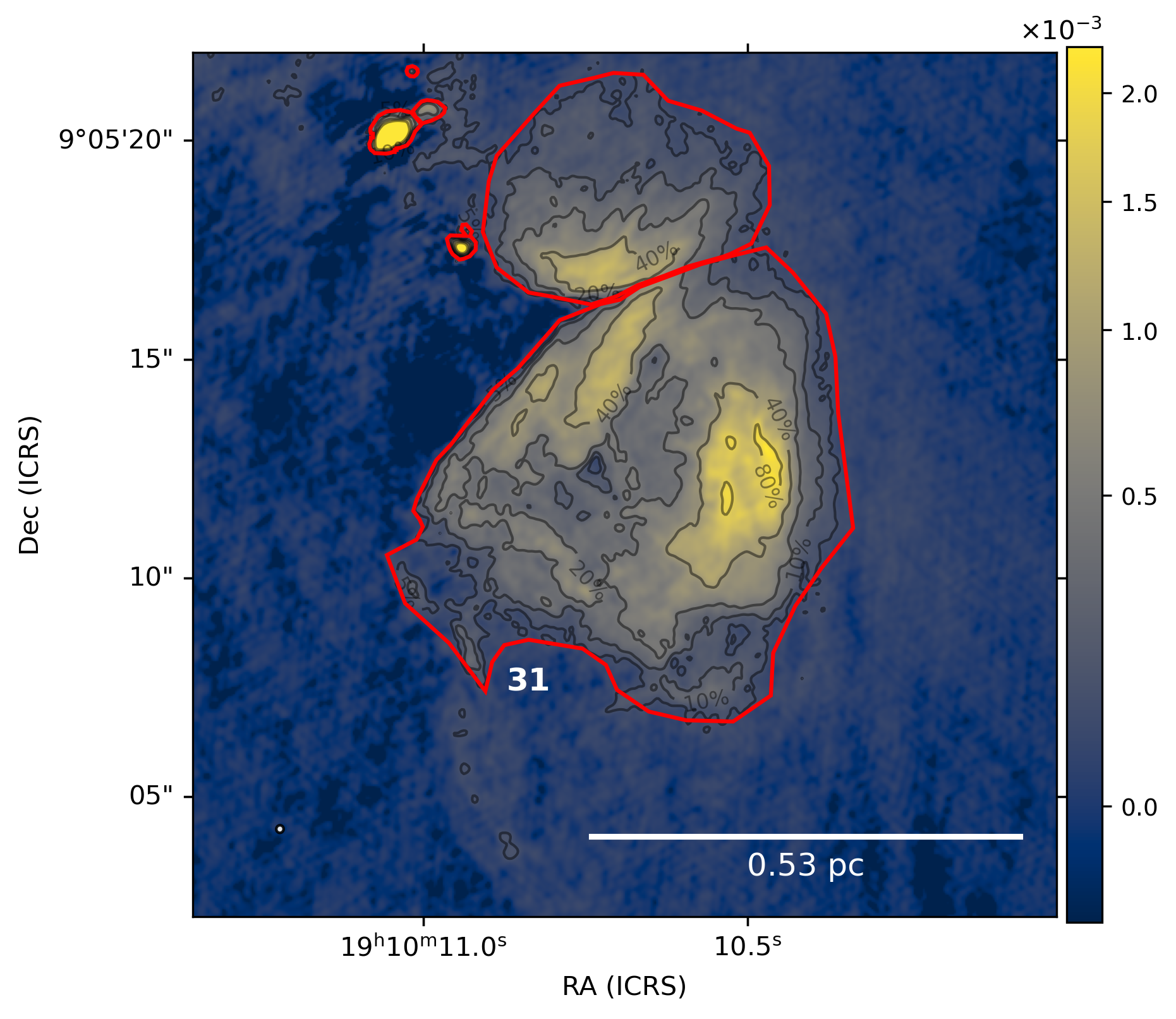}
    \includegraphics[width=0.32\textwidth]{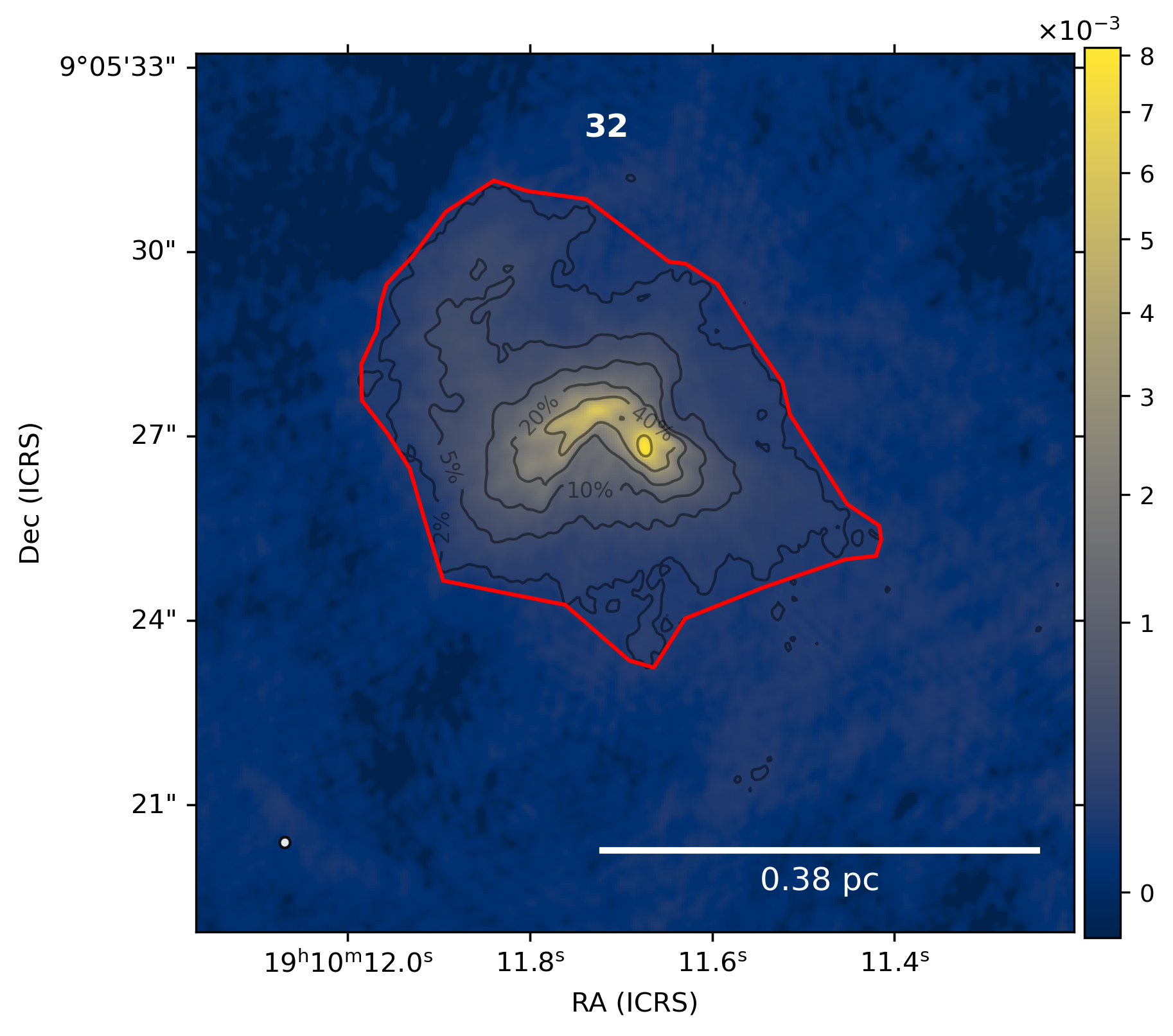}
    \includegraphics[width=0.32\textwidth]{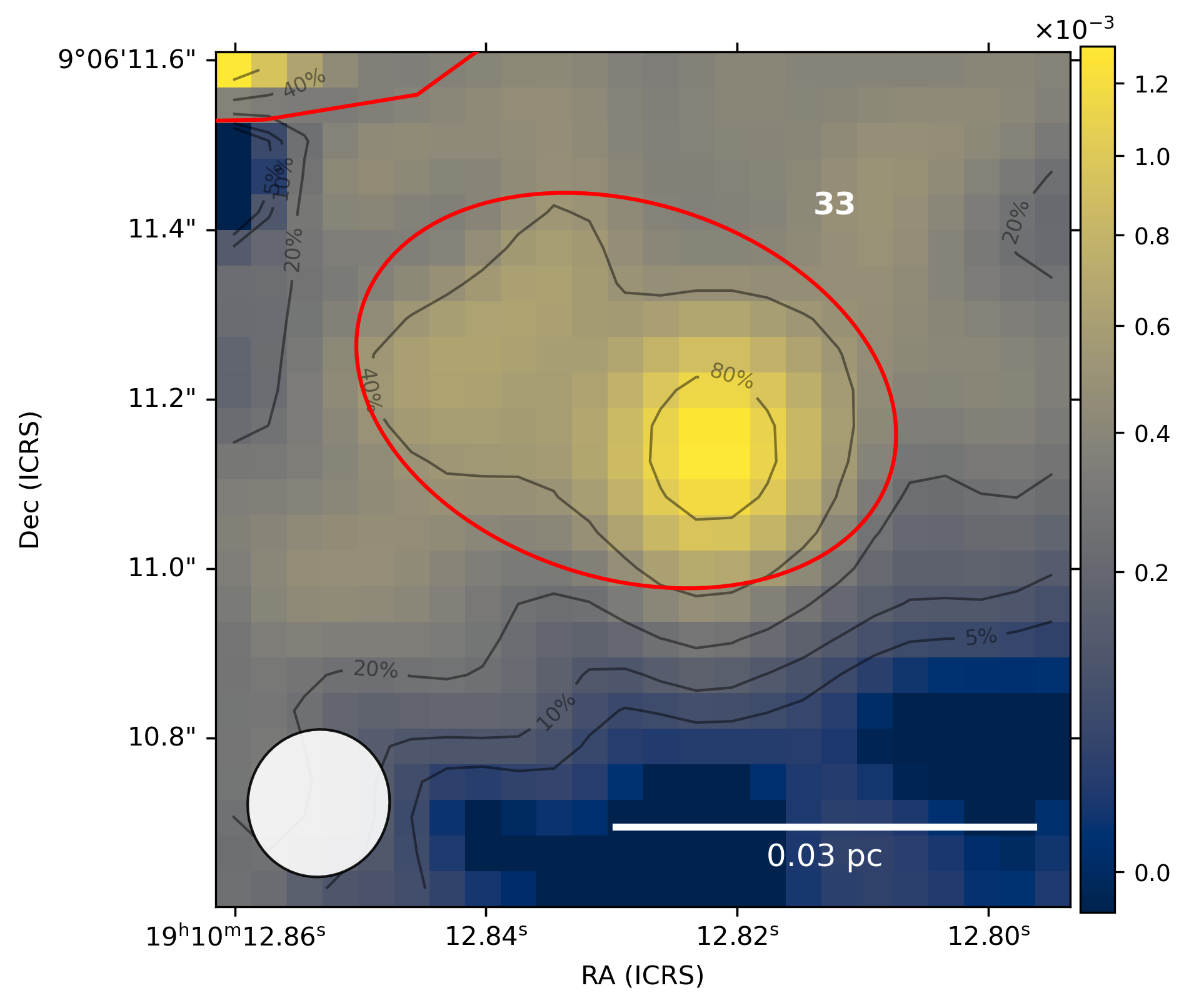}
    \includegraphics[width=0.32\textwidth]{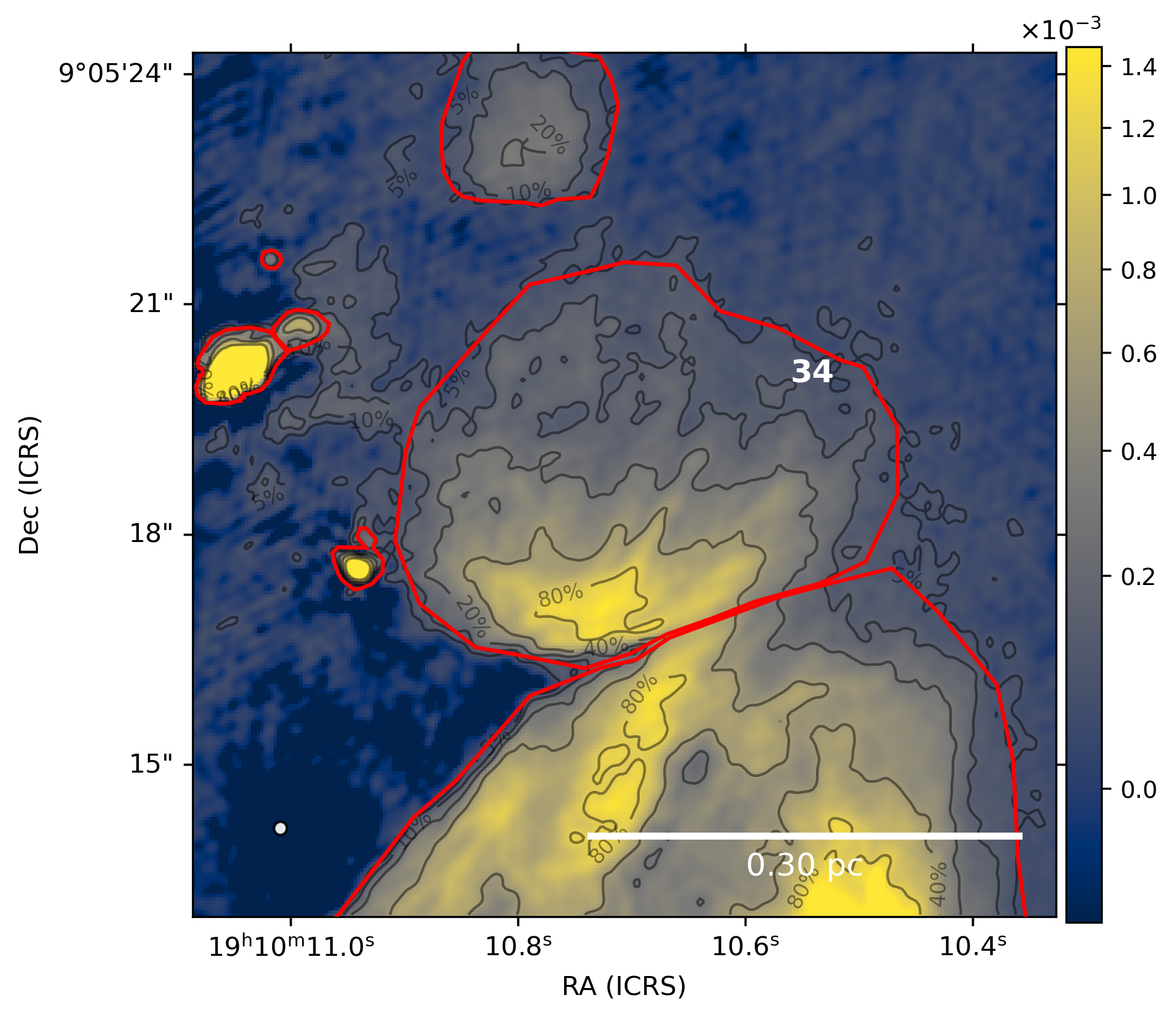}
    \includegraphics[width=0.32\textwidth]{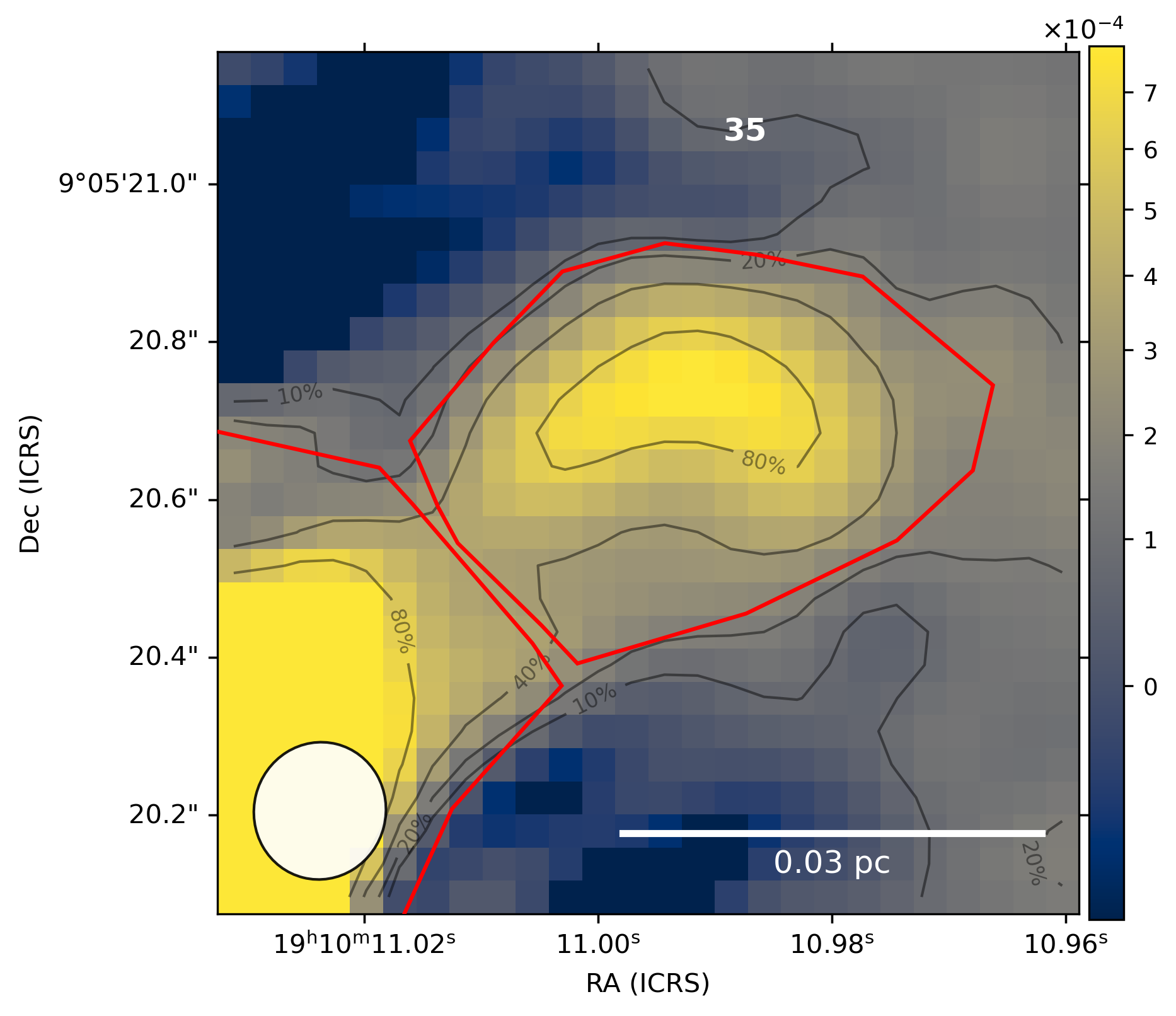}
    \includegraphics[width=0.32\textwidth]{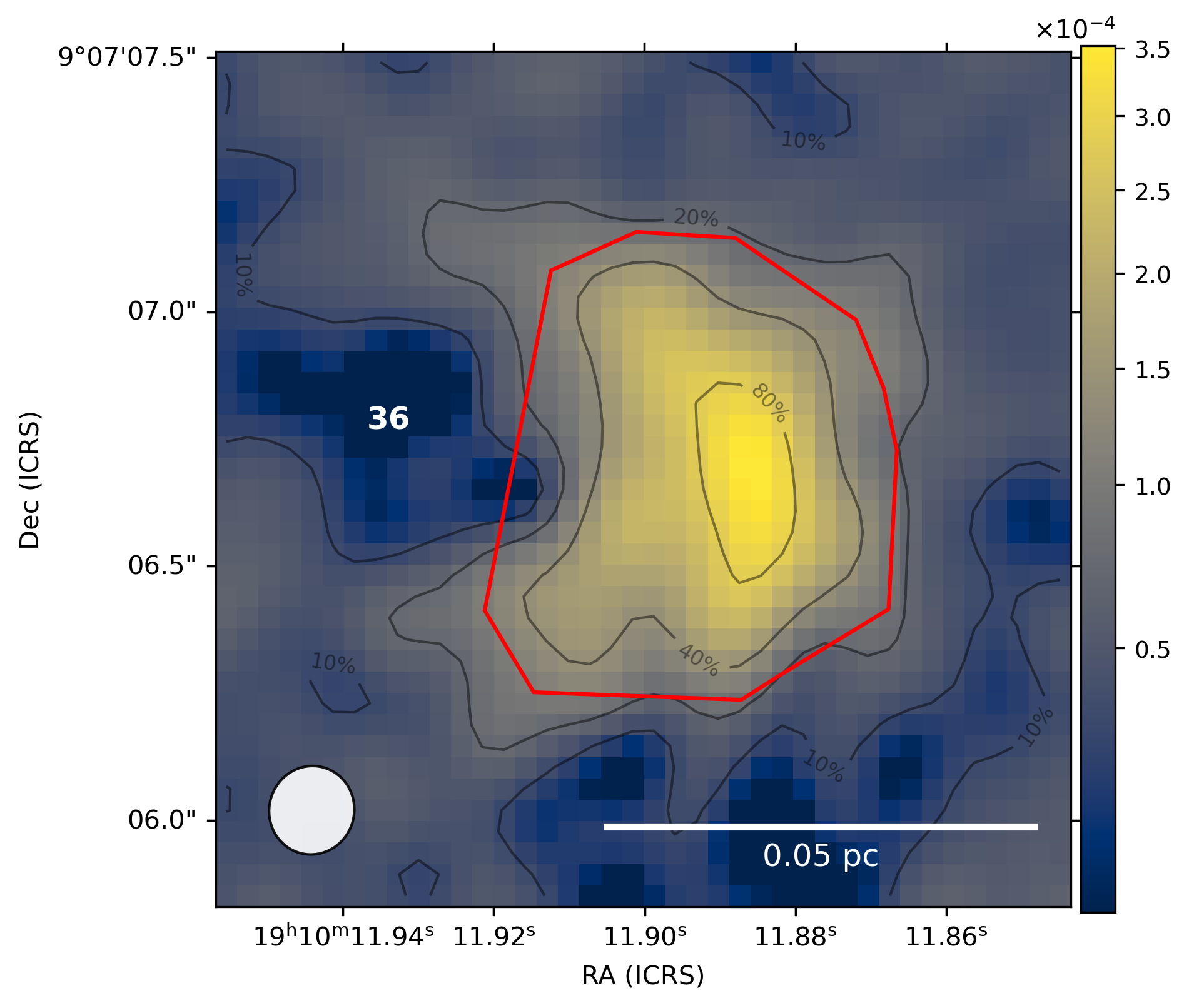}
    \includegraphics[width=0.32\textwidth]{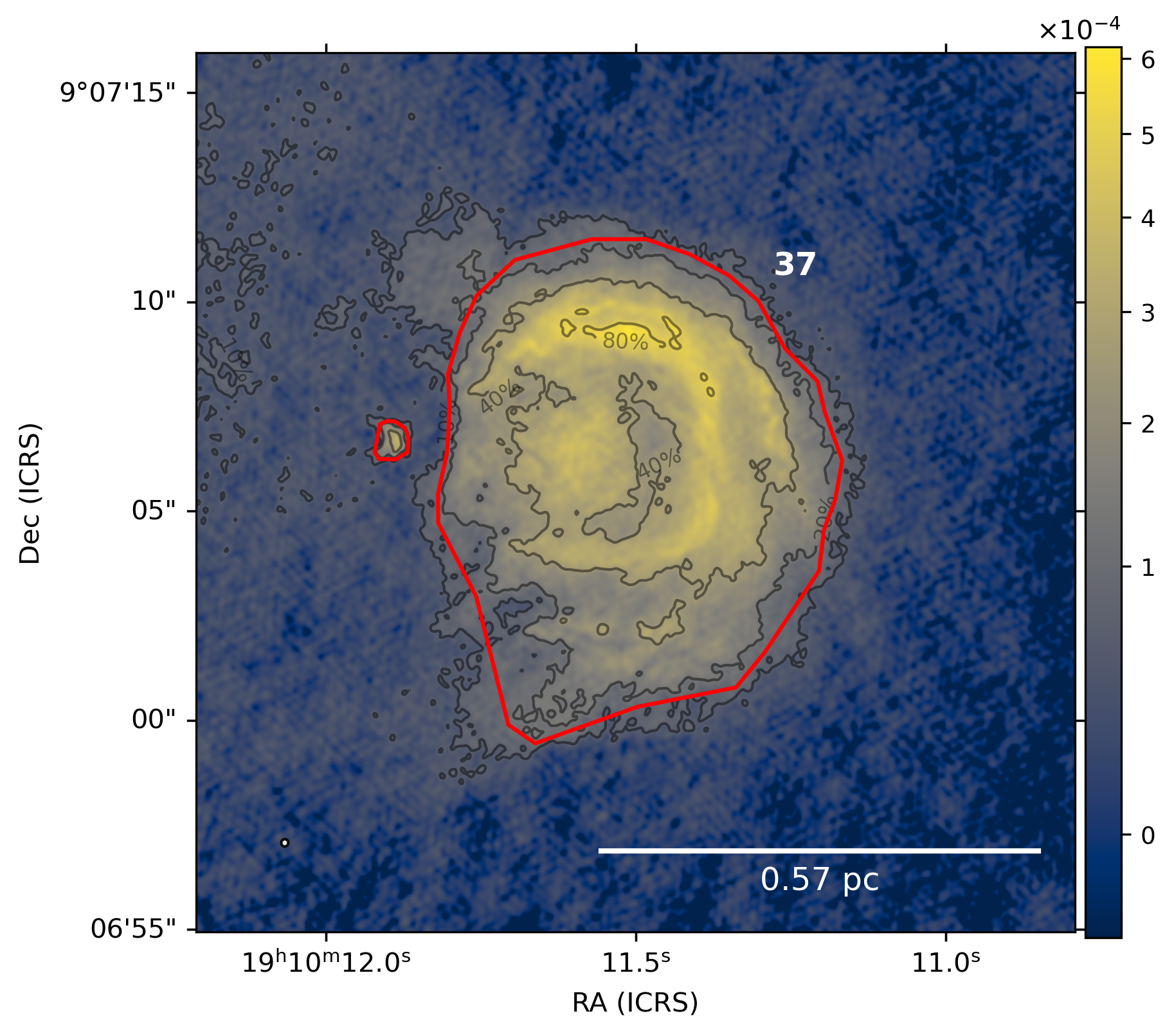}
    \caption{Zoomed-in views (continued).}
\end{figure*}

\begin{figure*}
    \centering
    \includegraphics[width=0.32\textwidth]{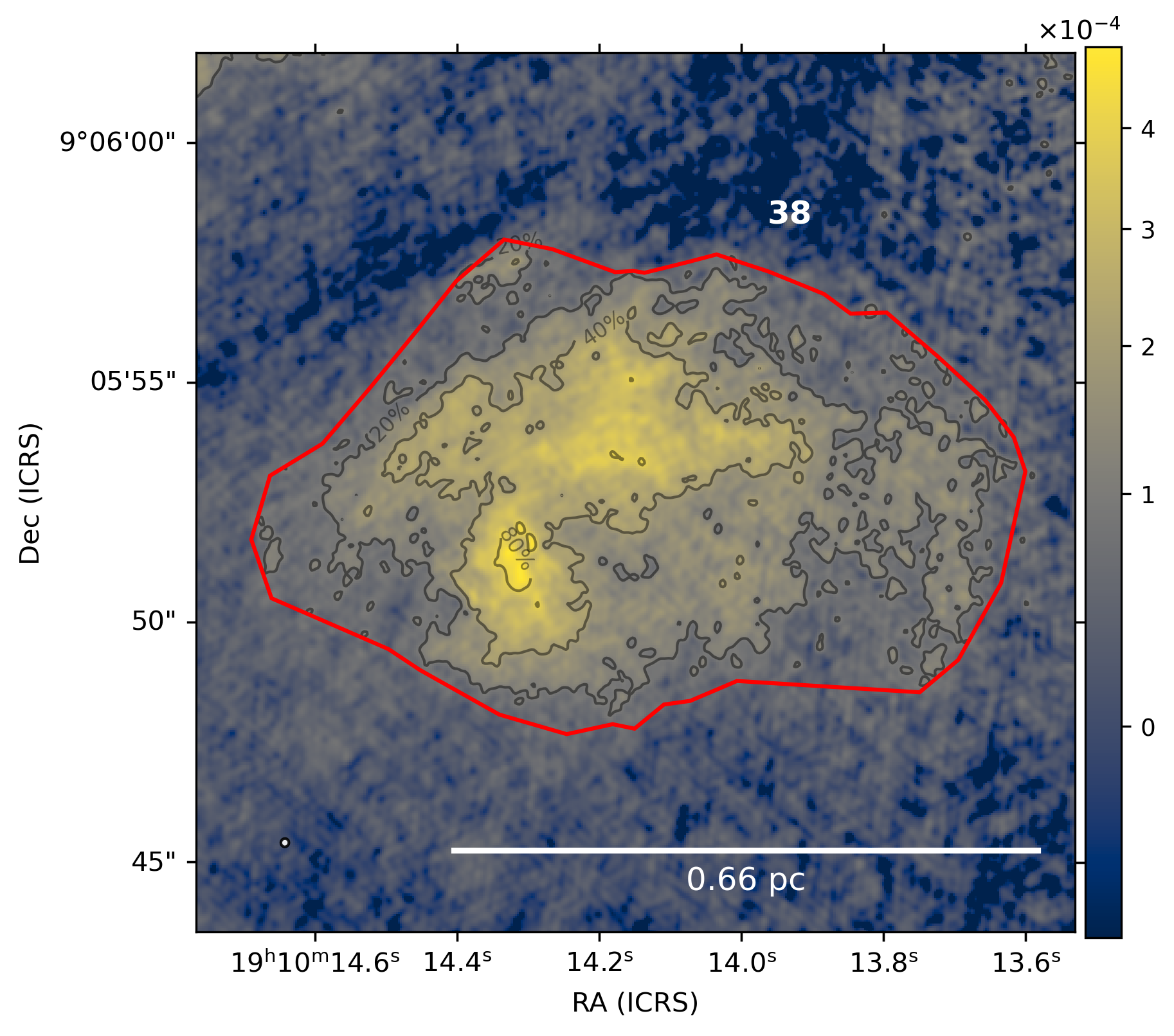}
    \includegraphics[width=0.32\textwidth]{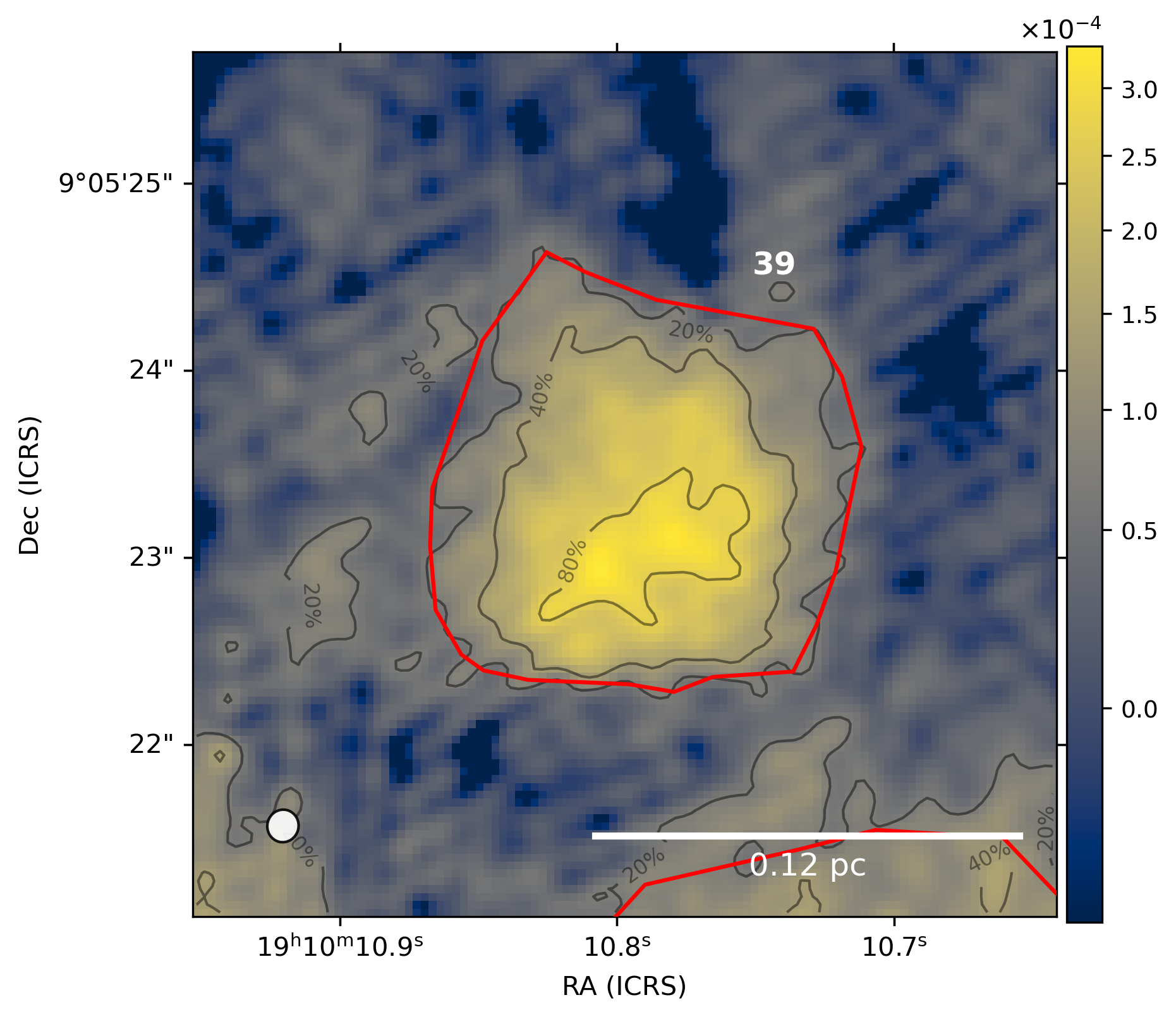}
    \includegraphics[width=0.32\textwidth]{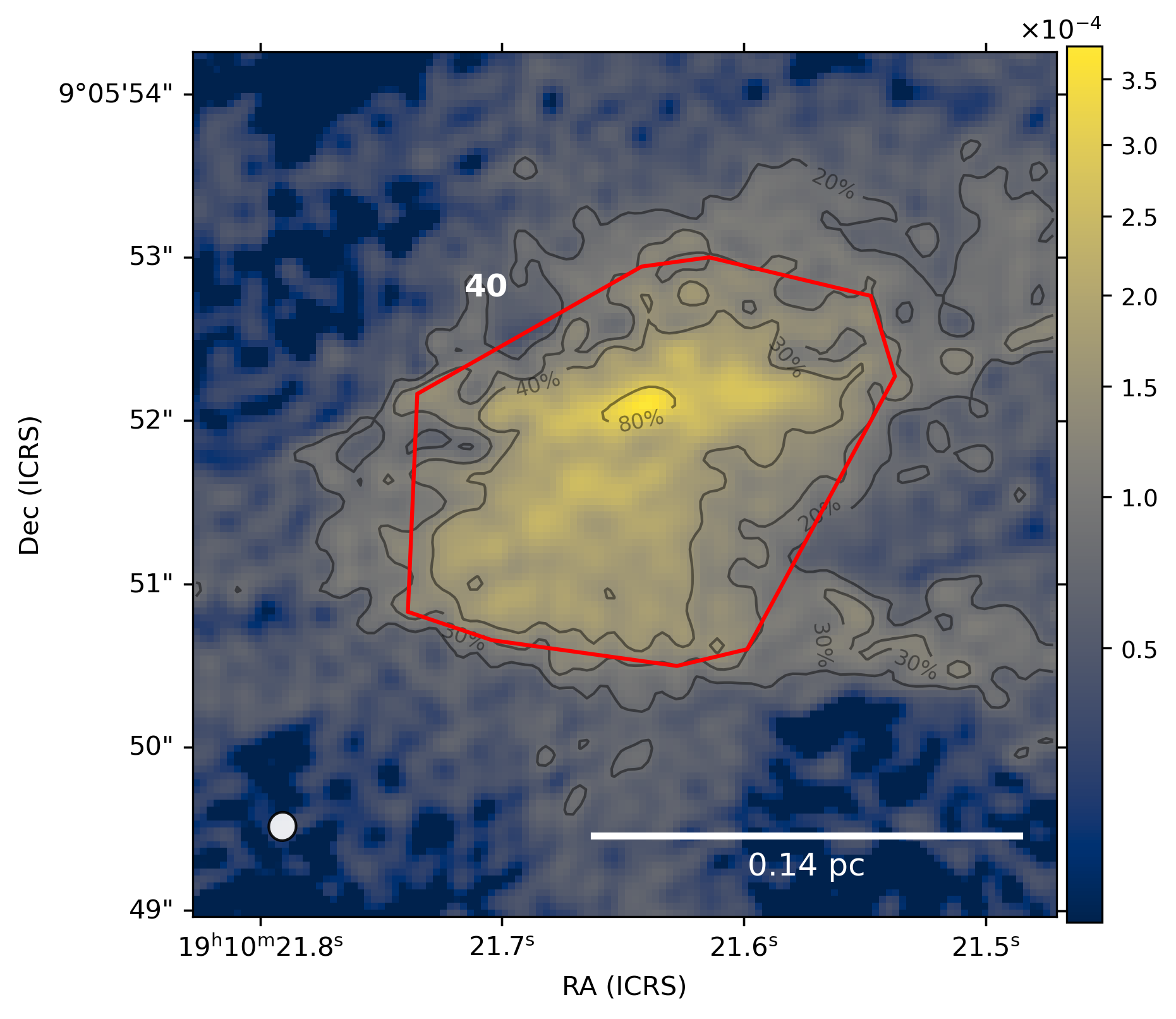}
    \includegraphics[width=0.32\textwidth]{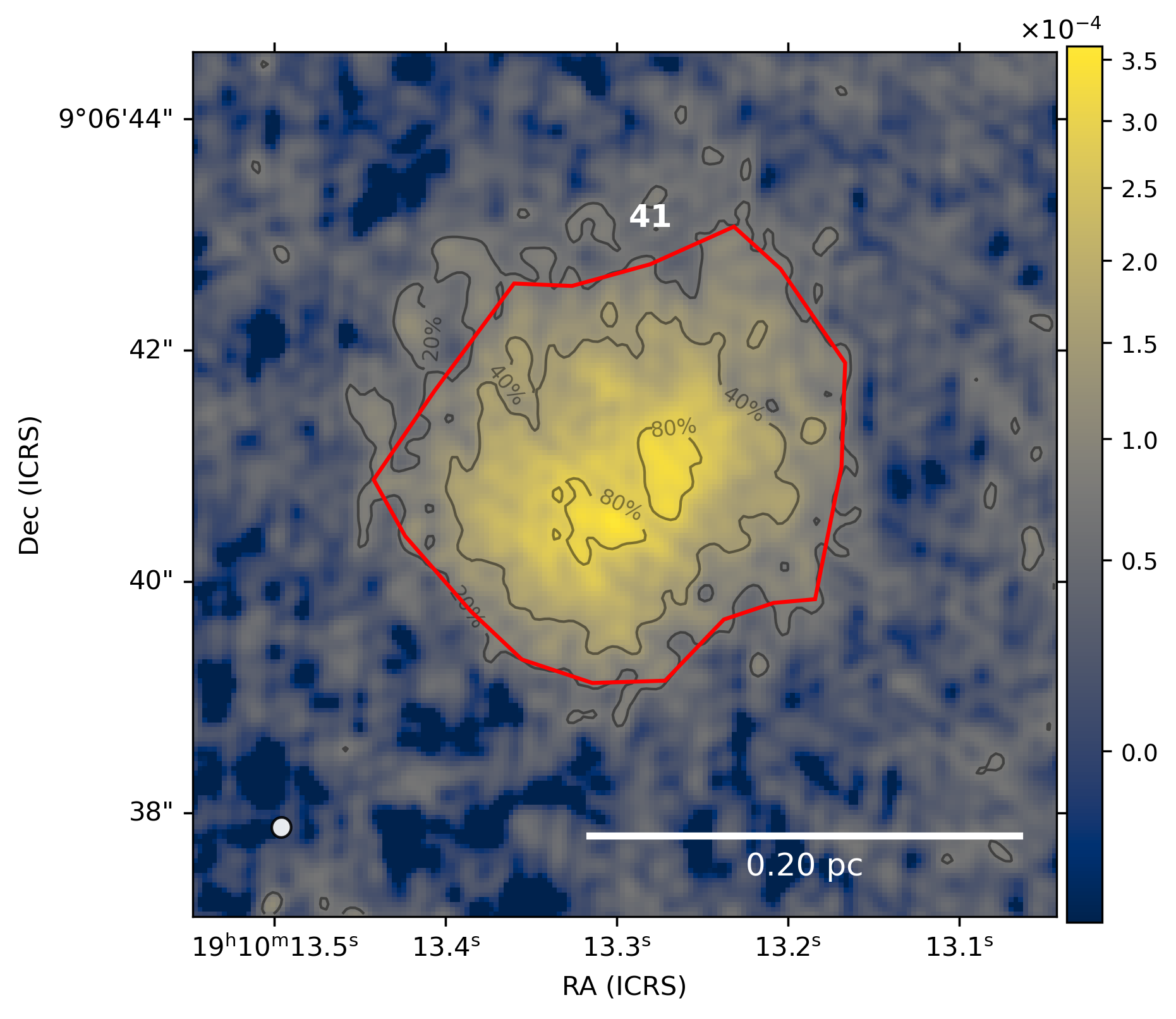}
    \includegraphics[width=0.32\textwidth]{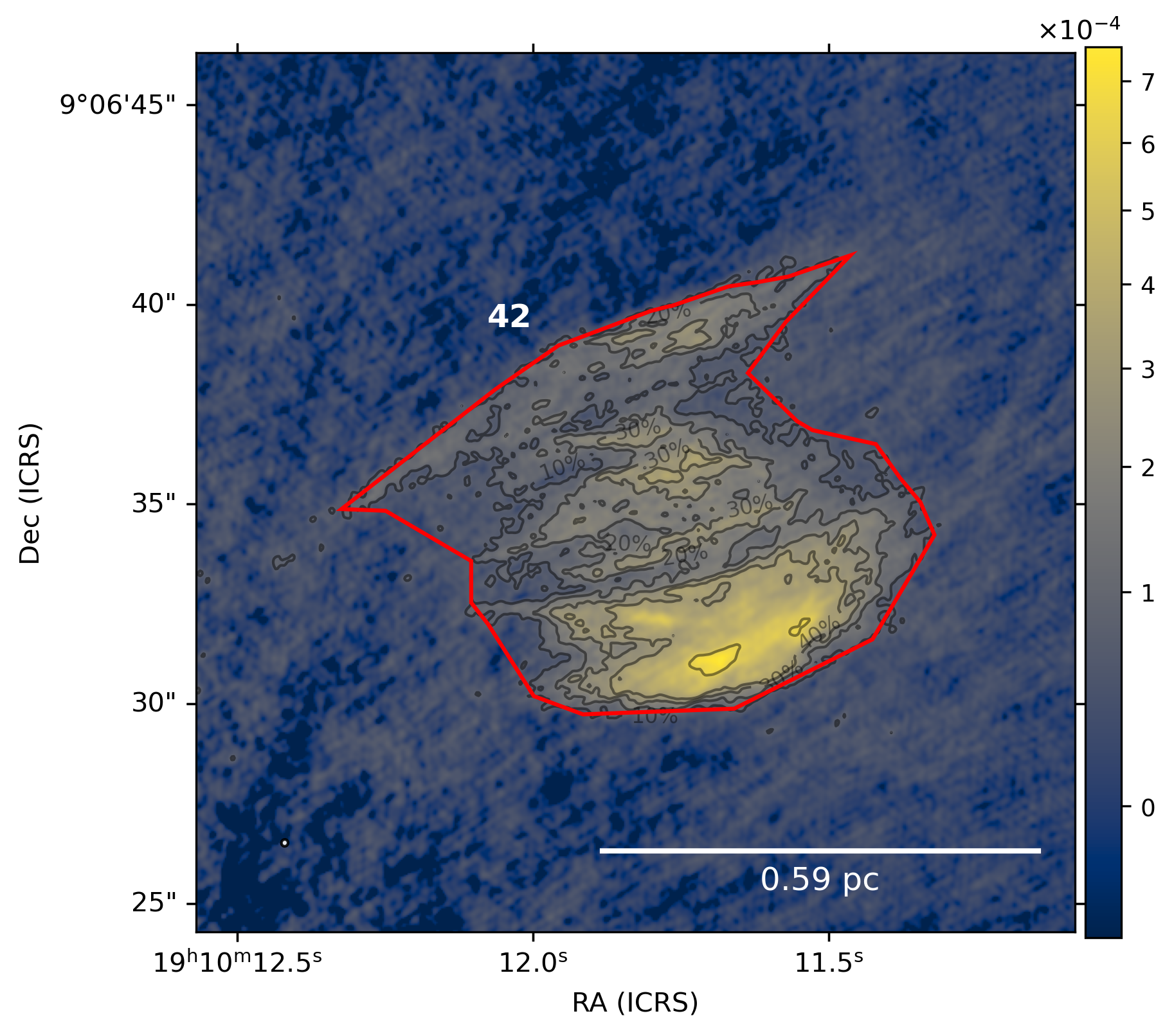}
    \includegraphics[width=0.32\textwidth]{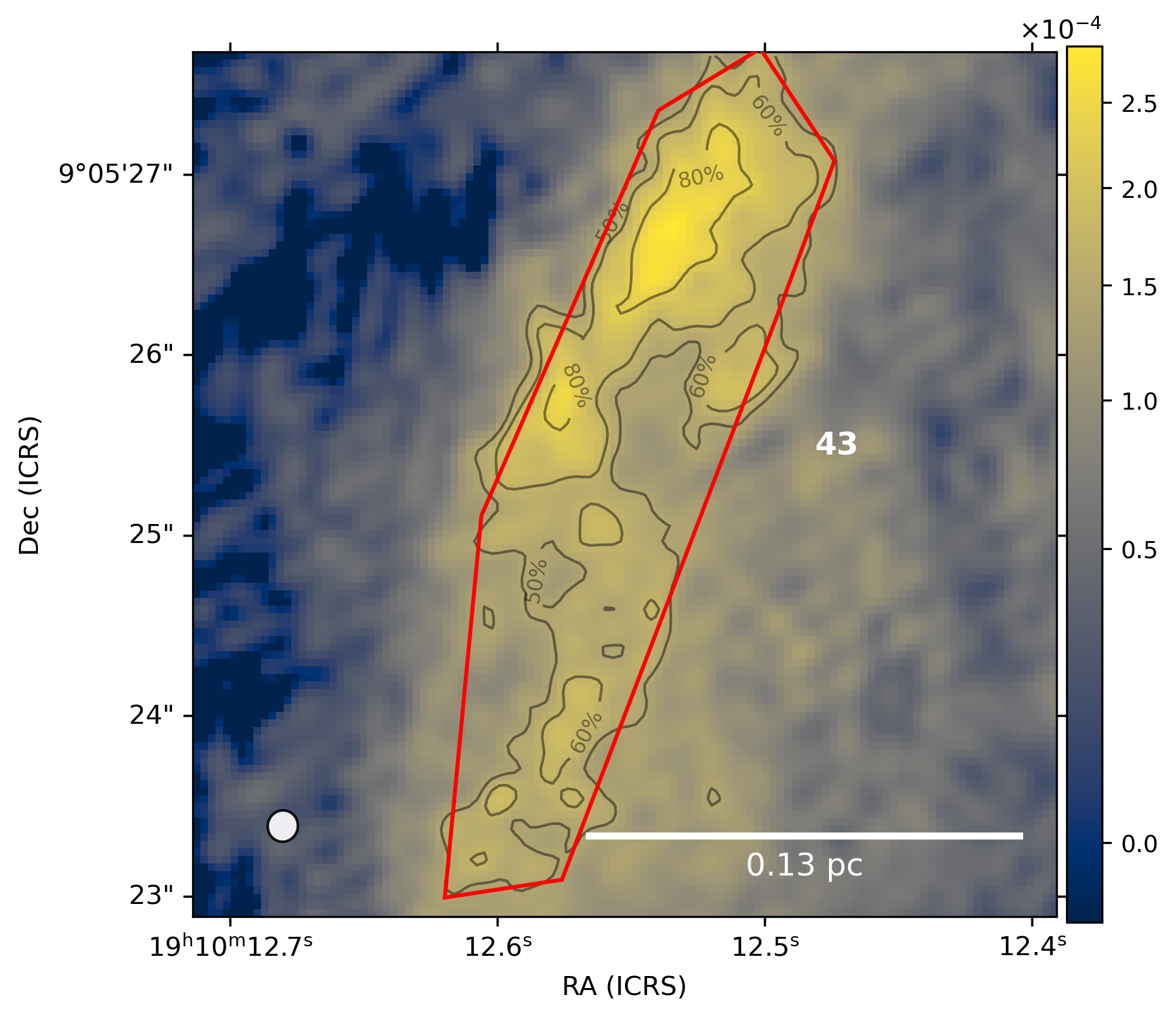}
    \includegraphics[width=0.32\textwidth]{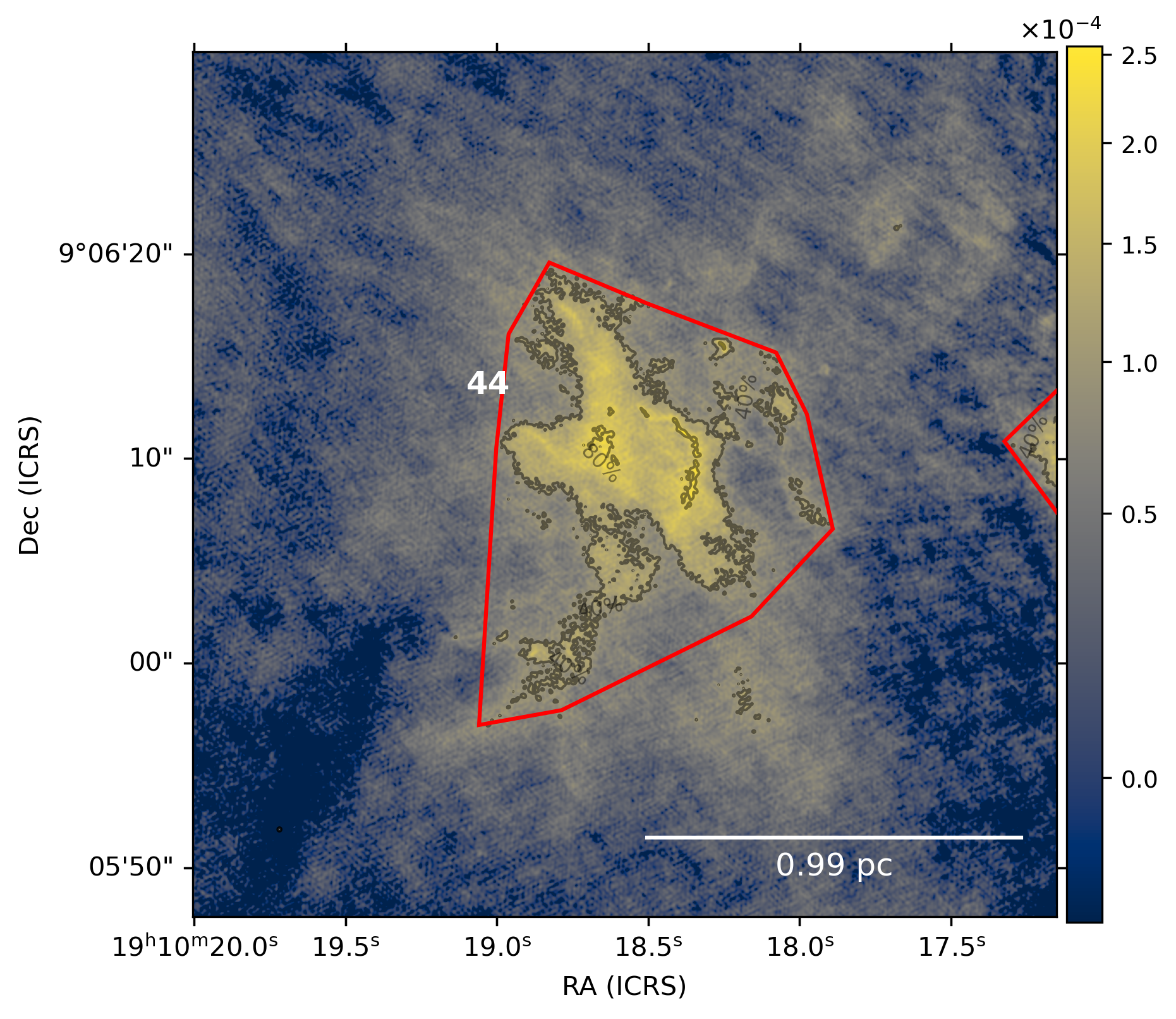}
    \includegraphics[width=0.32\textwidth]{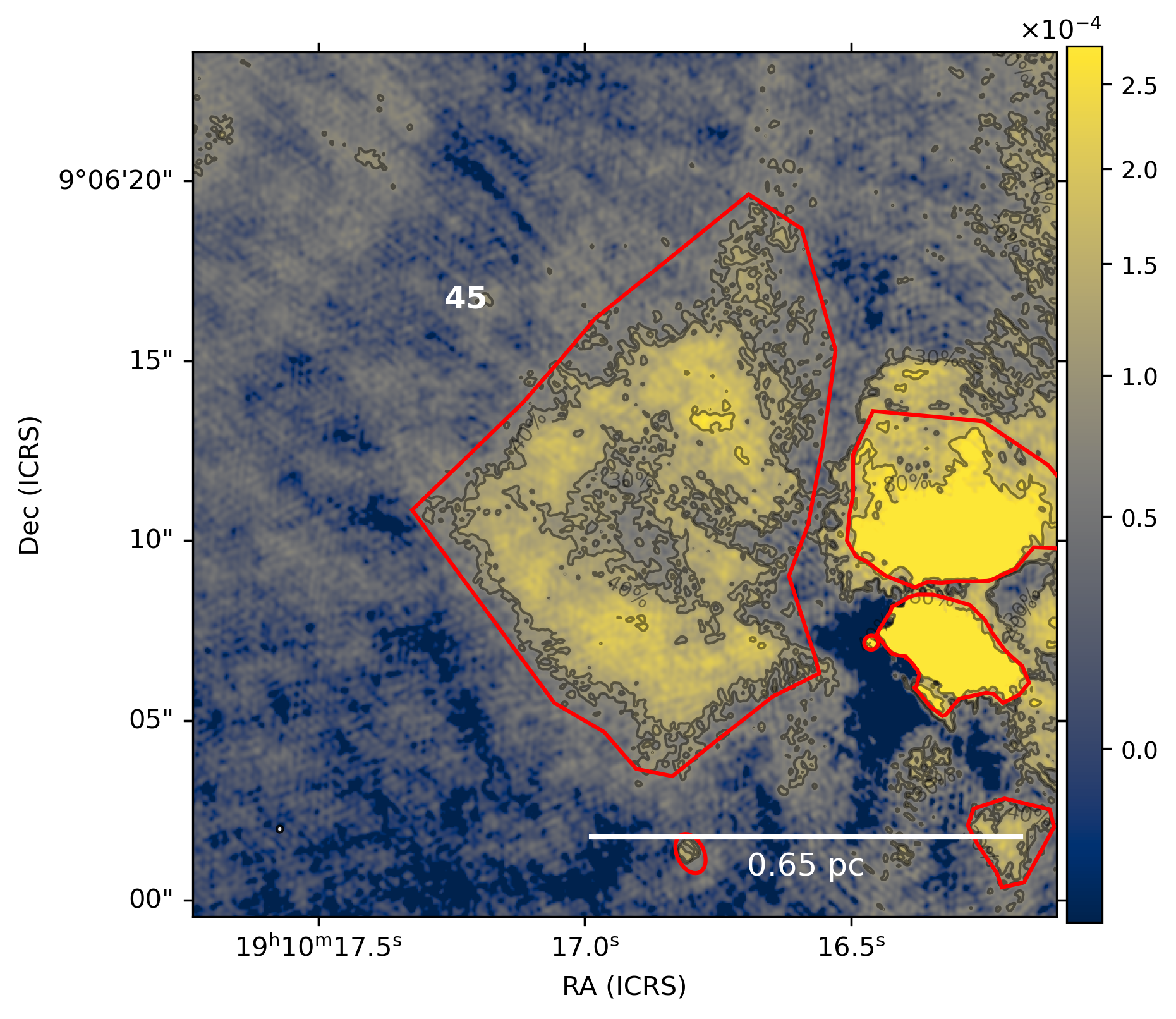}
    \includegraphics[width=0.32\textwidth]{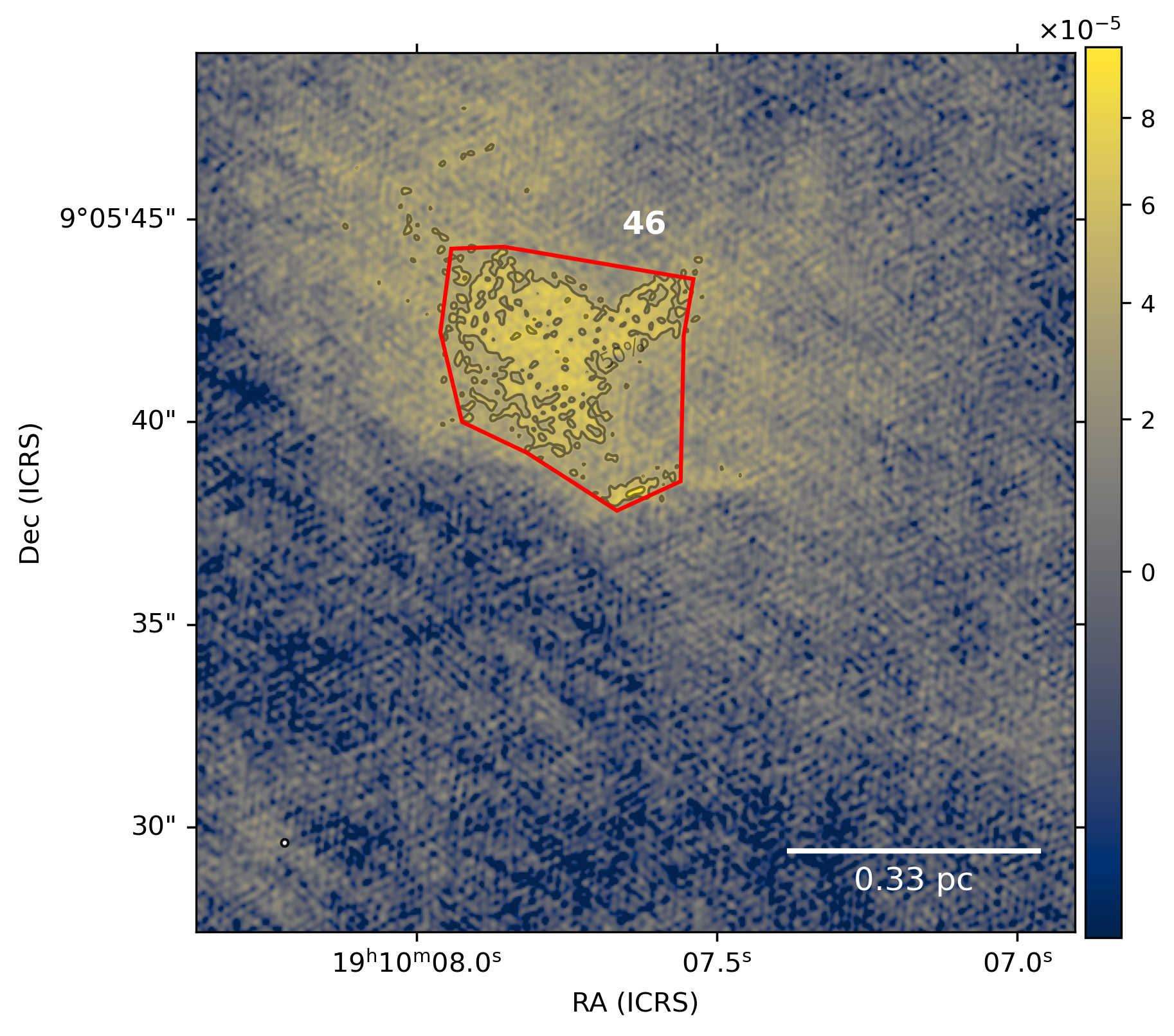}
    \includegraphics[width=0.32\textwidth]{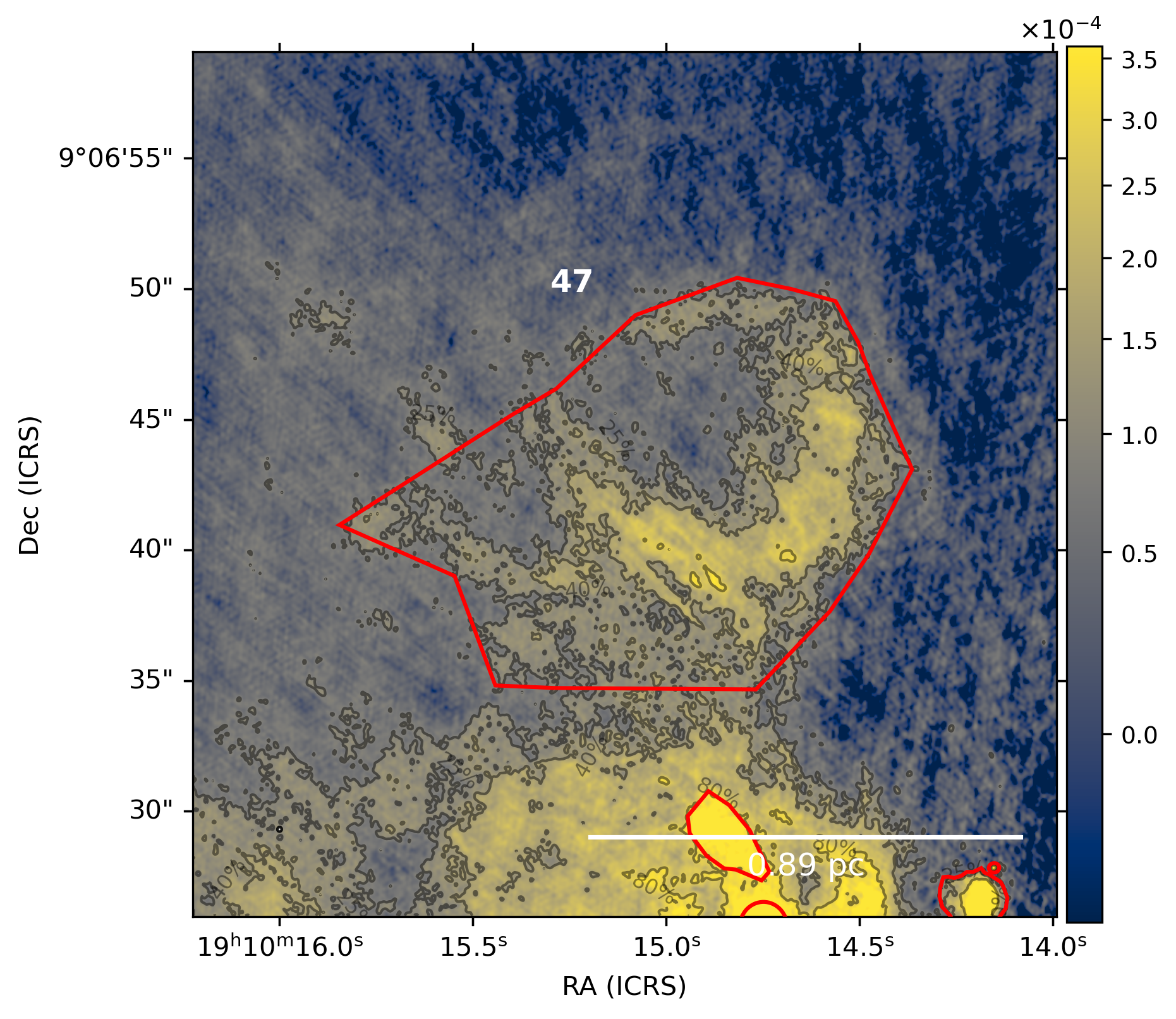}
    \includegraphics[width=0.32\textwidth]{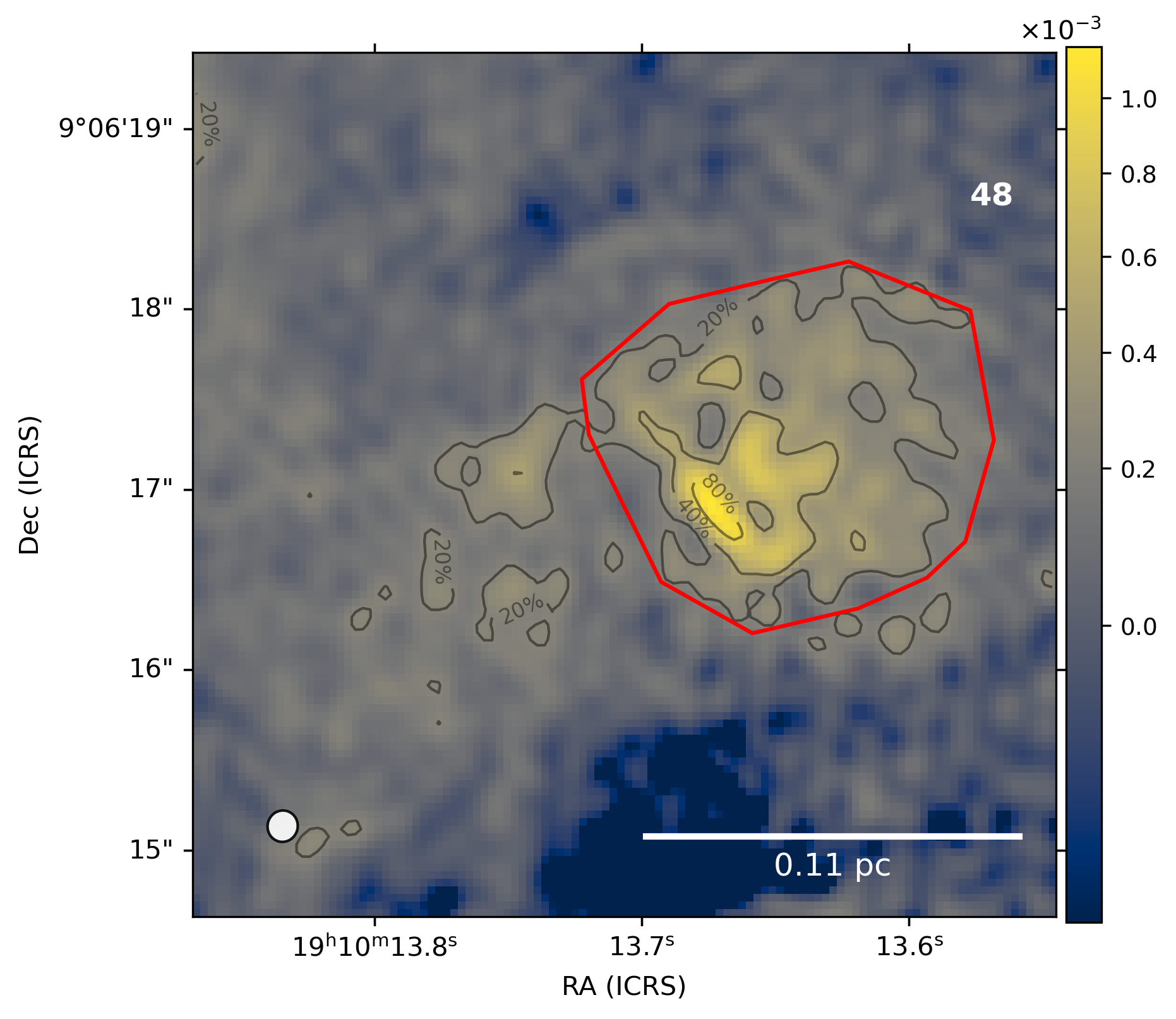}
    \includegraphics[width=0.32\textwidth]{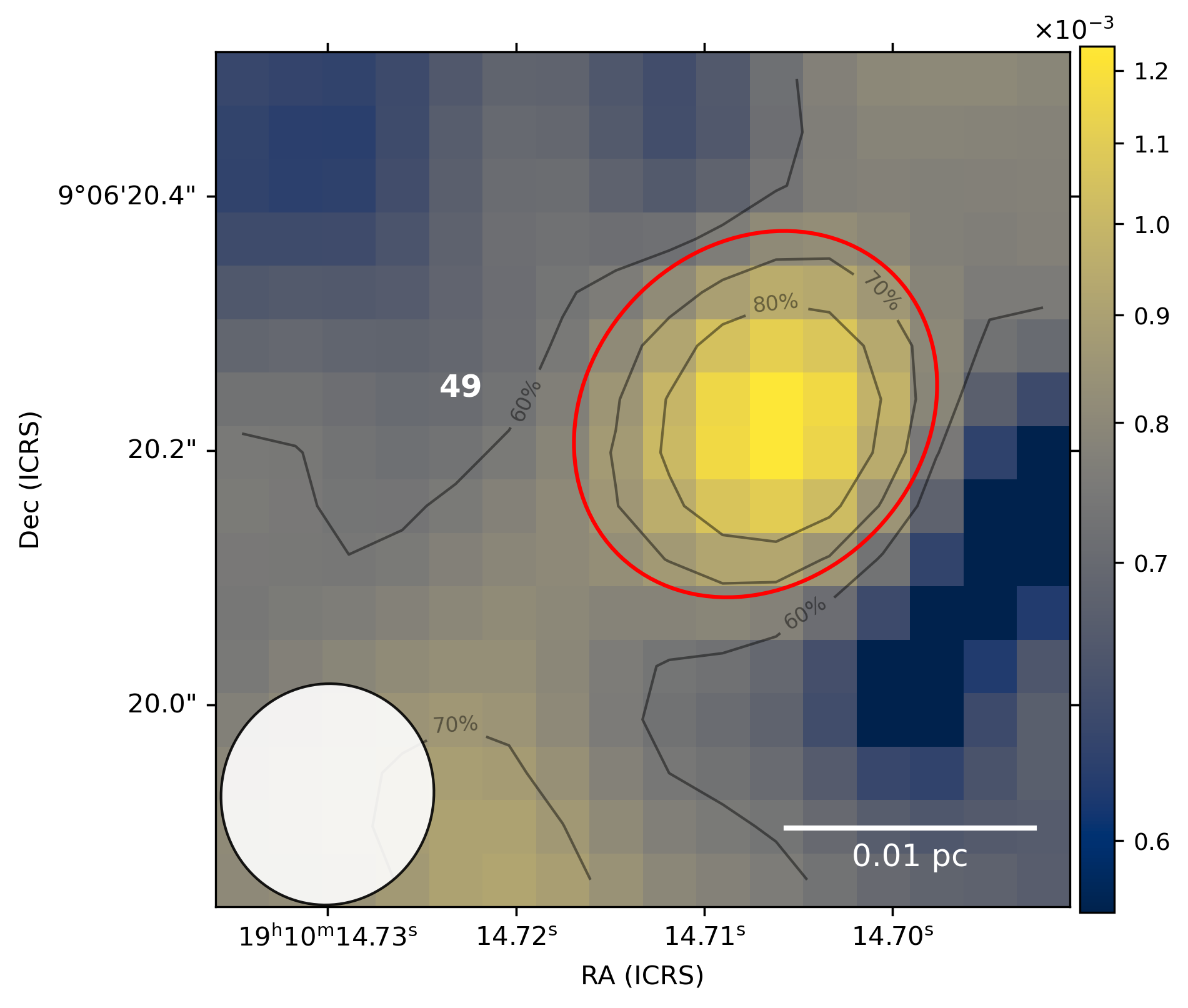}
  \caption{Zoomed-in views (continued).}
\end{figure*}

\begin{figure*}
    \centering
    \includegraphics[width=0.32\textwidth]{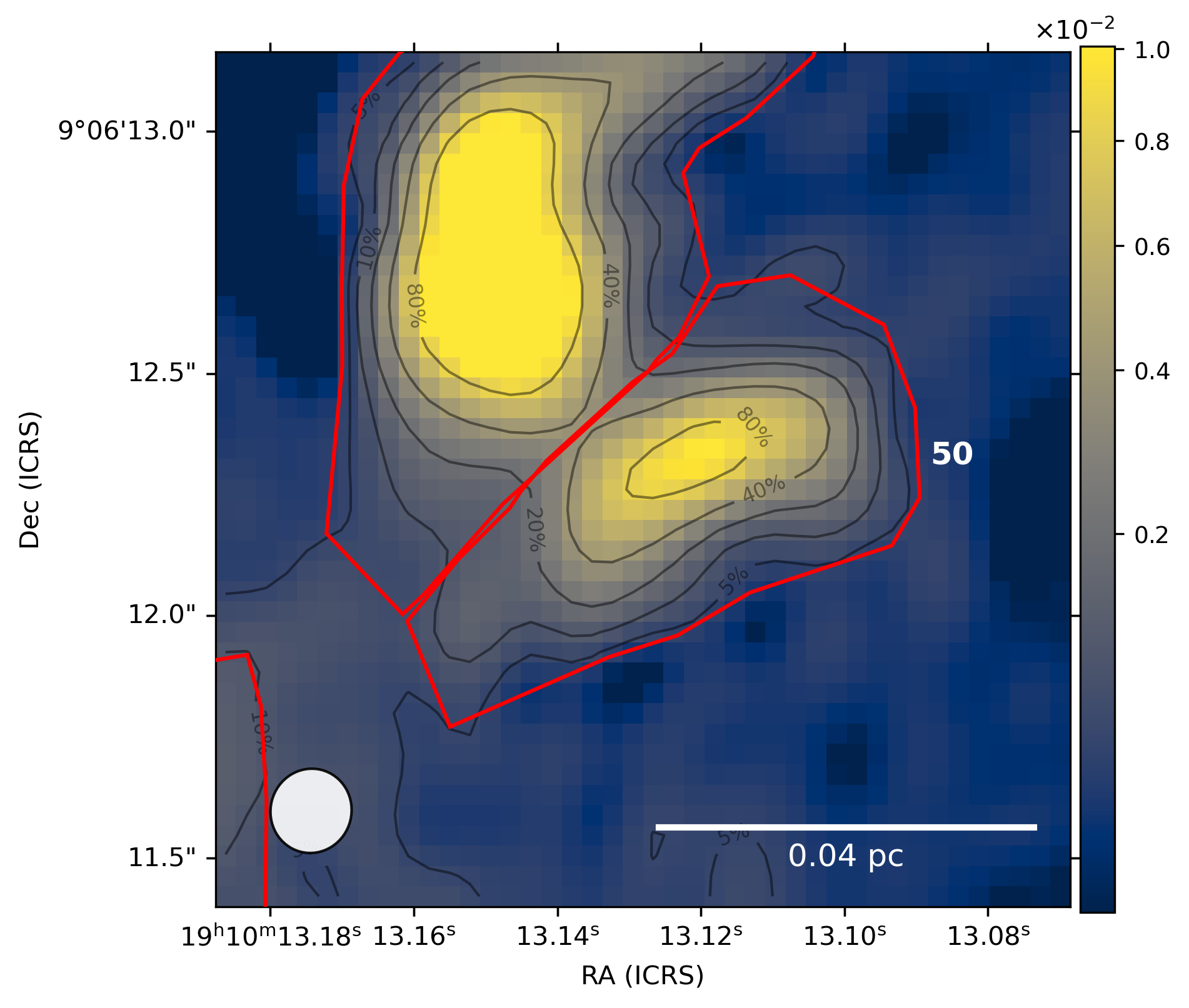}
    \includegraphics[width=0.32\textwidth]{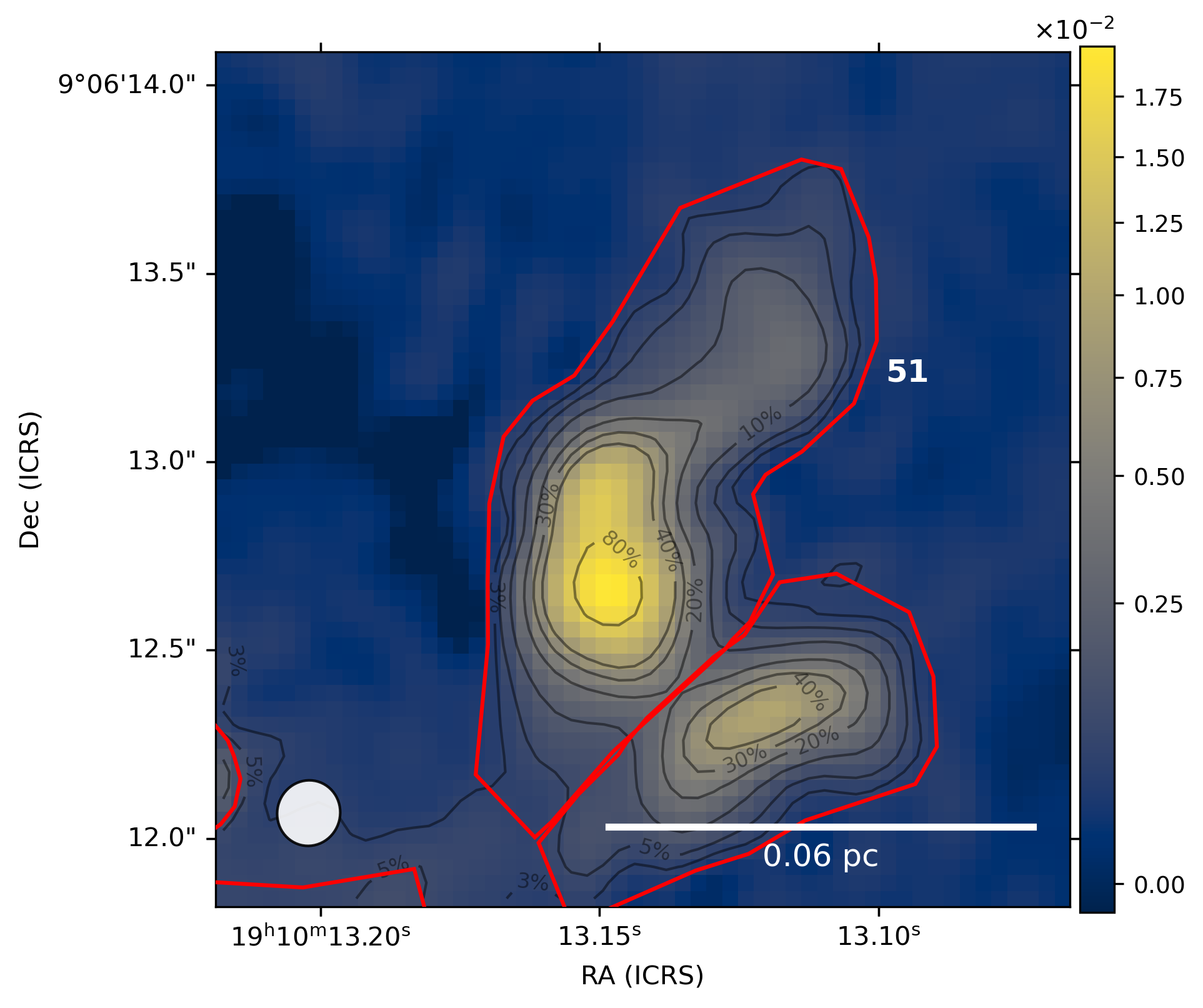}
    \includegraphics[width=0.32\textwidth]{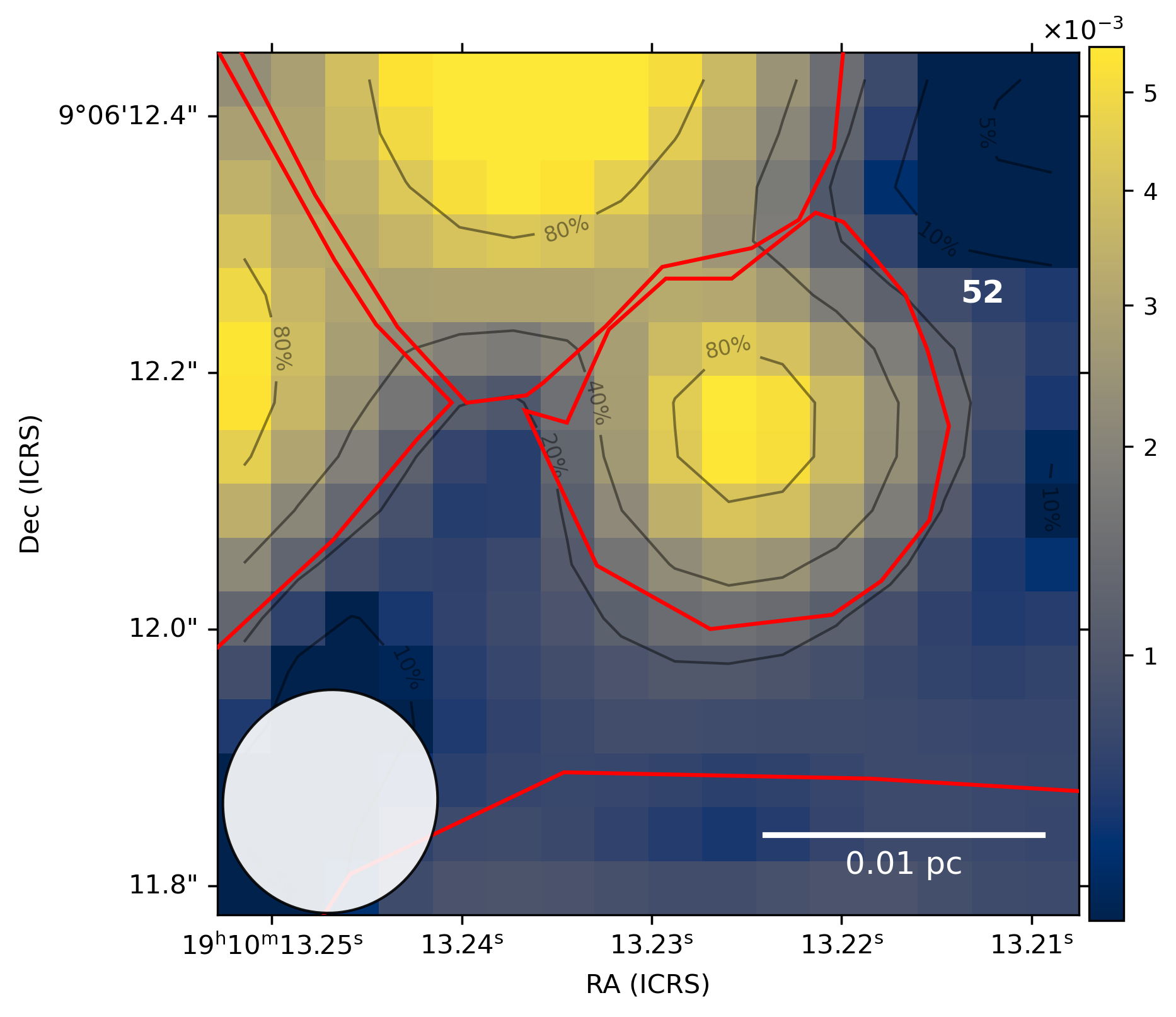}
    \includegraphics[width=0.32\textwidth]{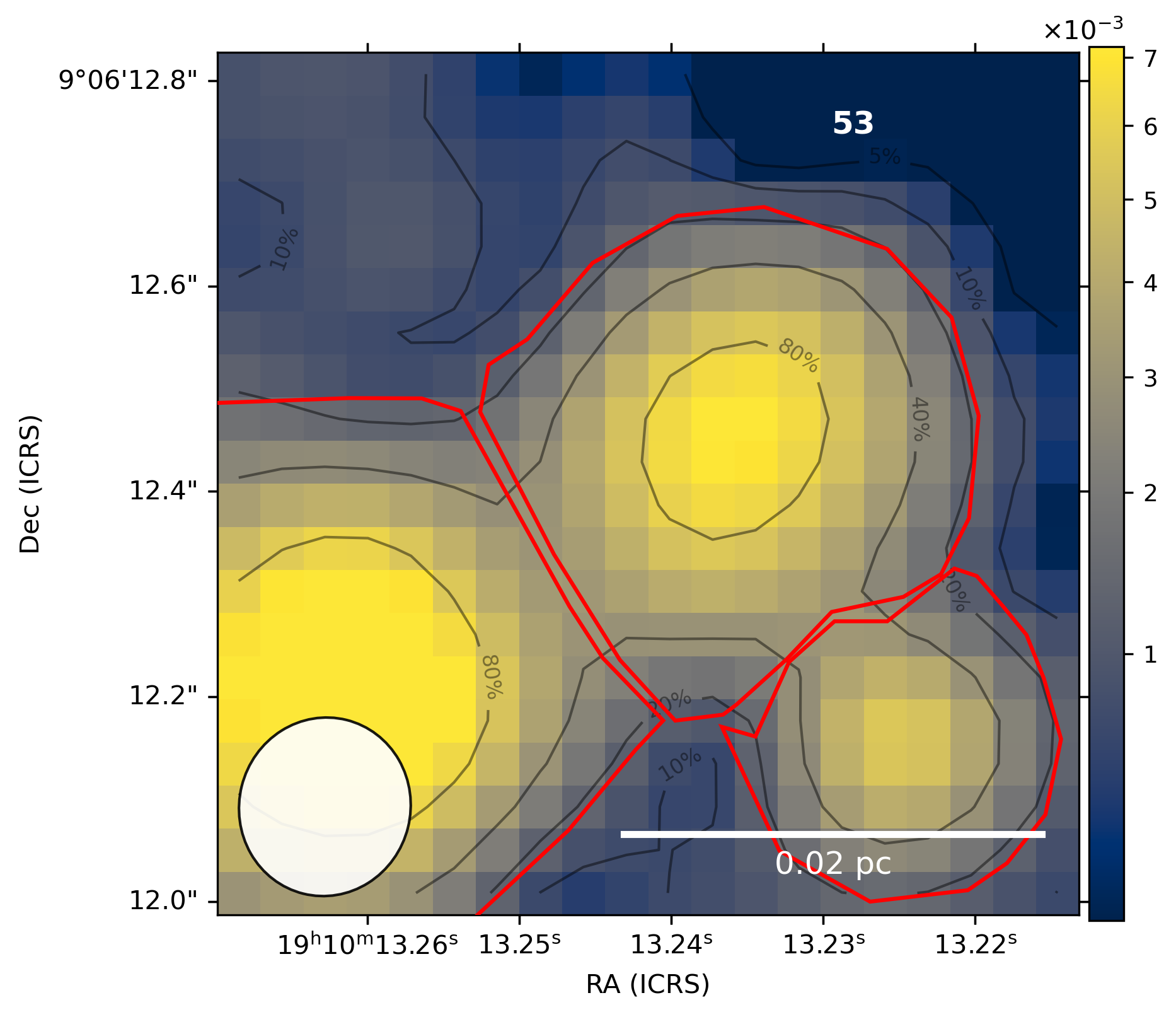}
    \includegraphics[width=0.32\textwidth]{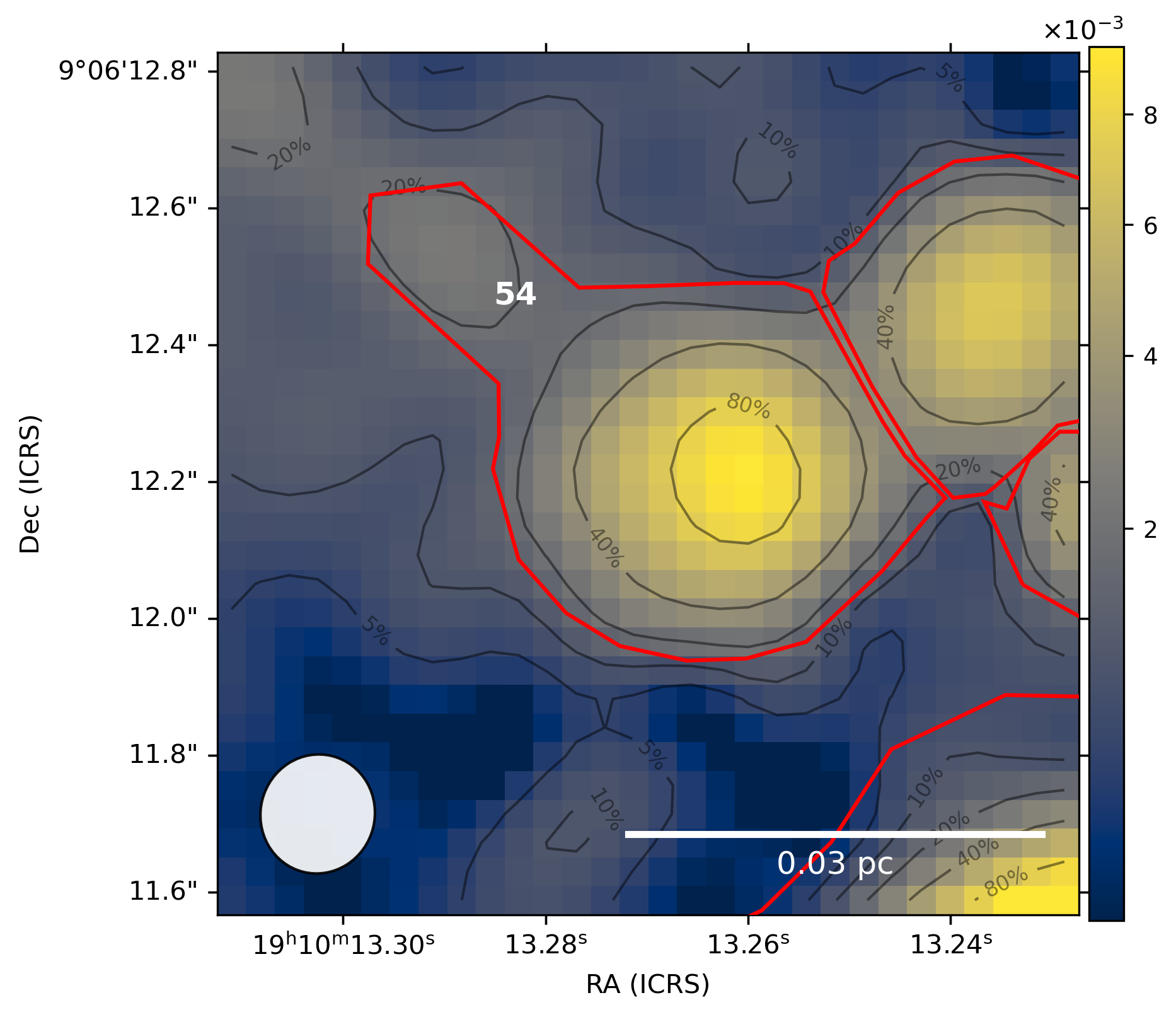}
    \includegraphics[width=0.32\textwidth]{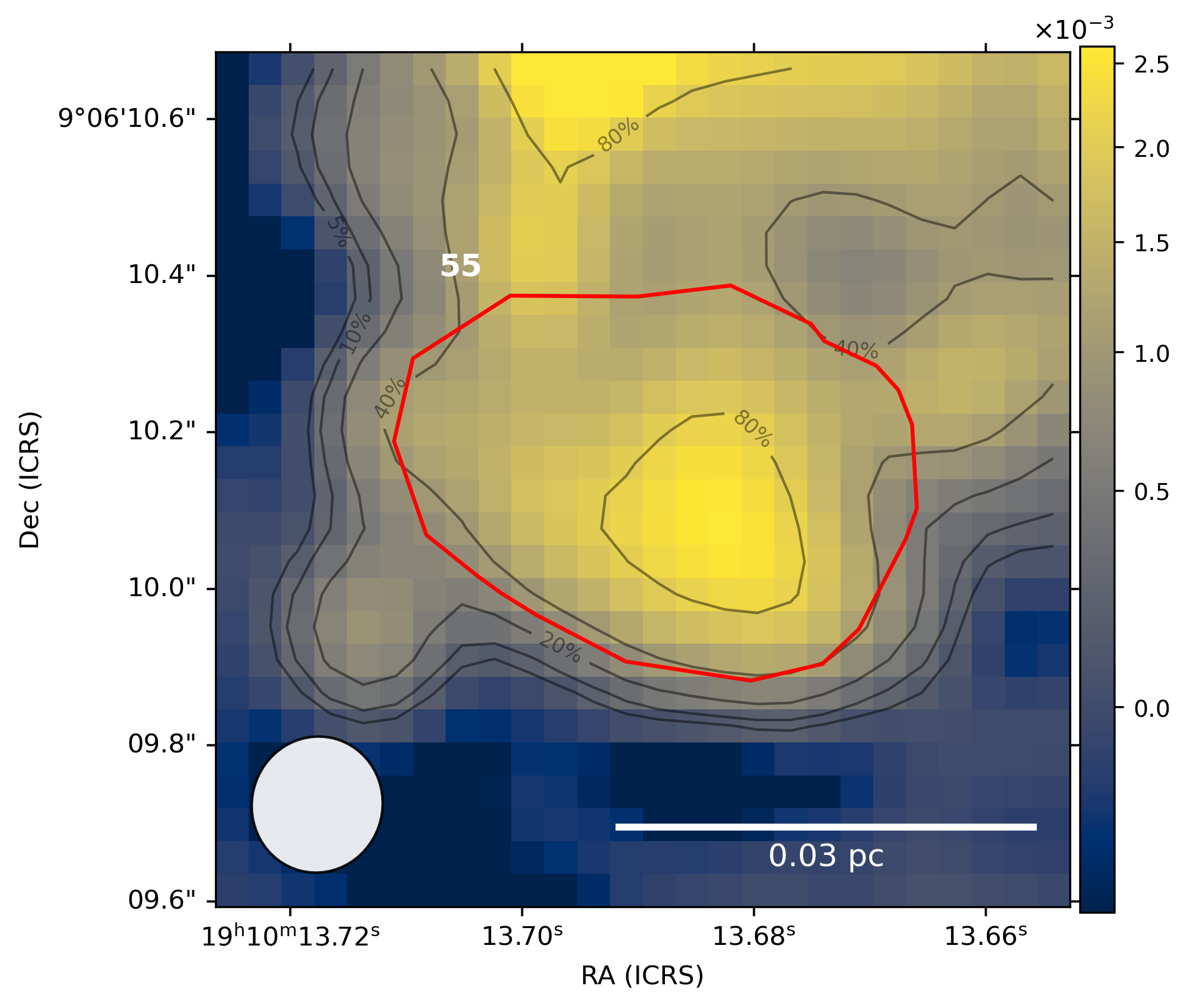}
    \includegraphics[width=0.32\textwidth]{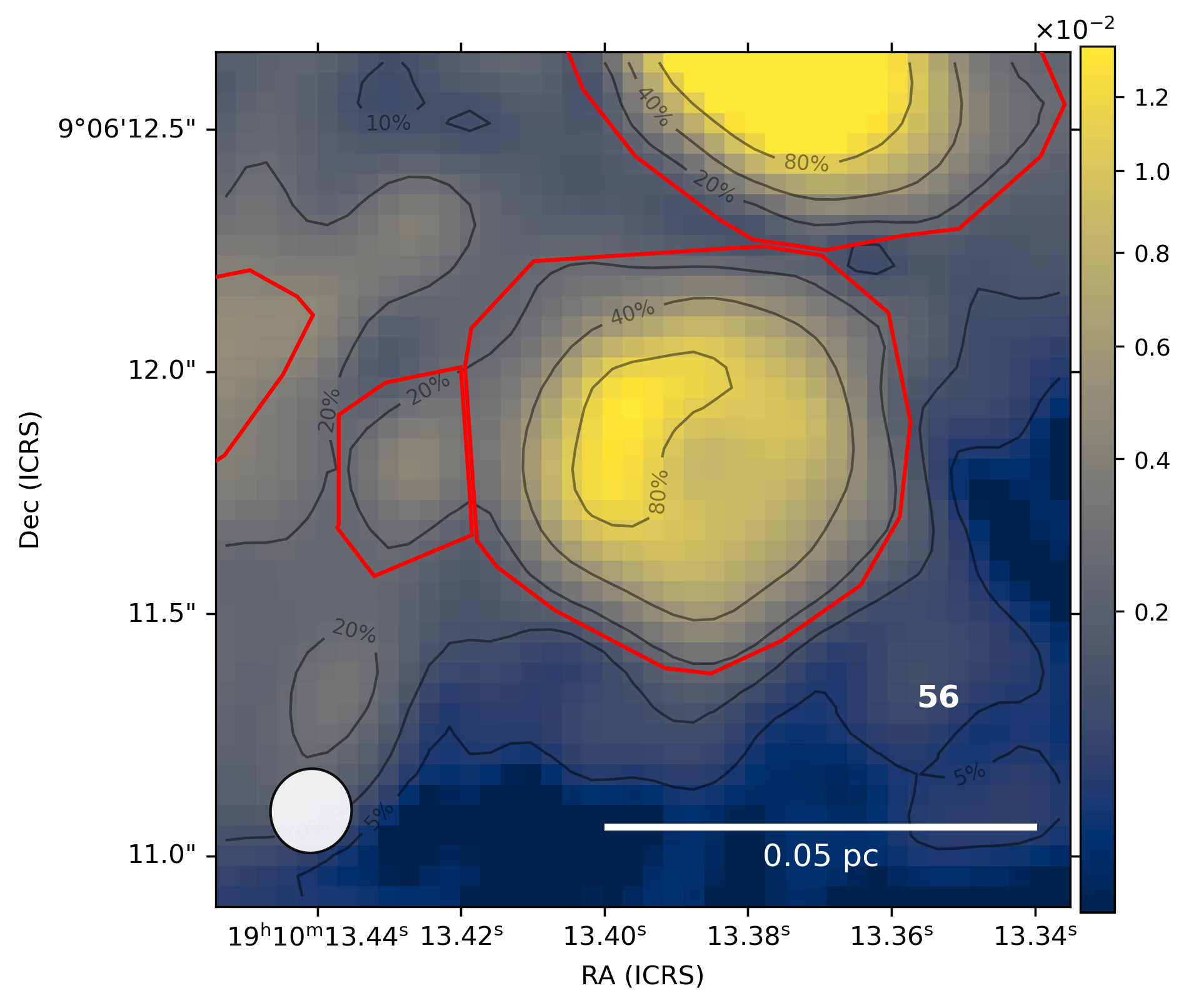}
    \includegraphics[width=0.32\textwidth]{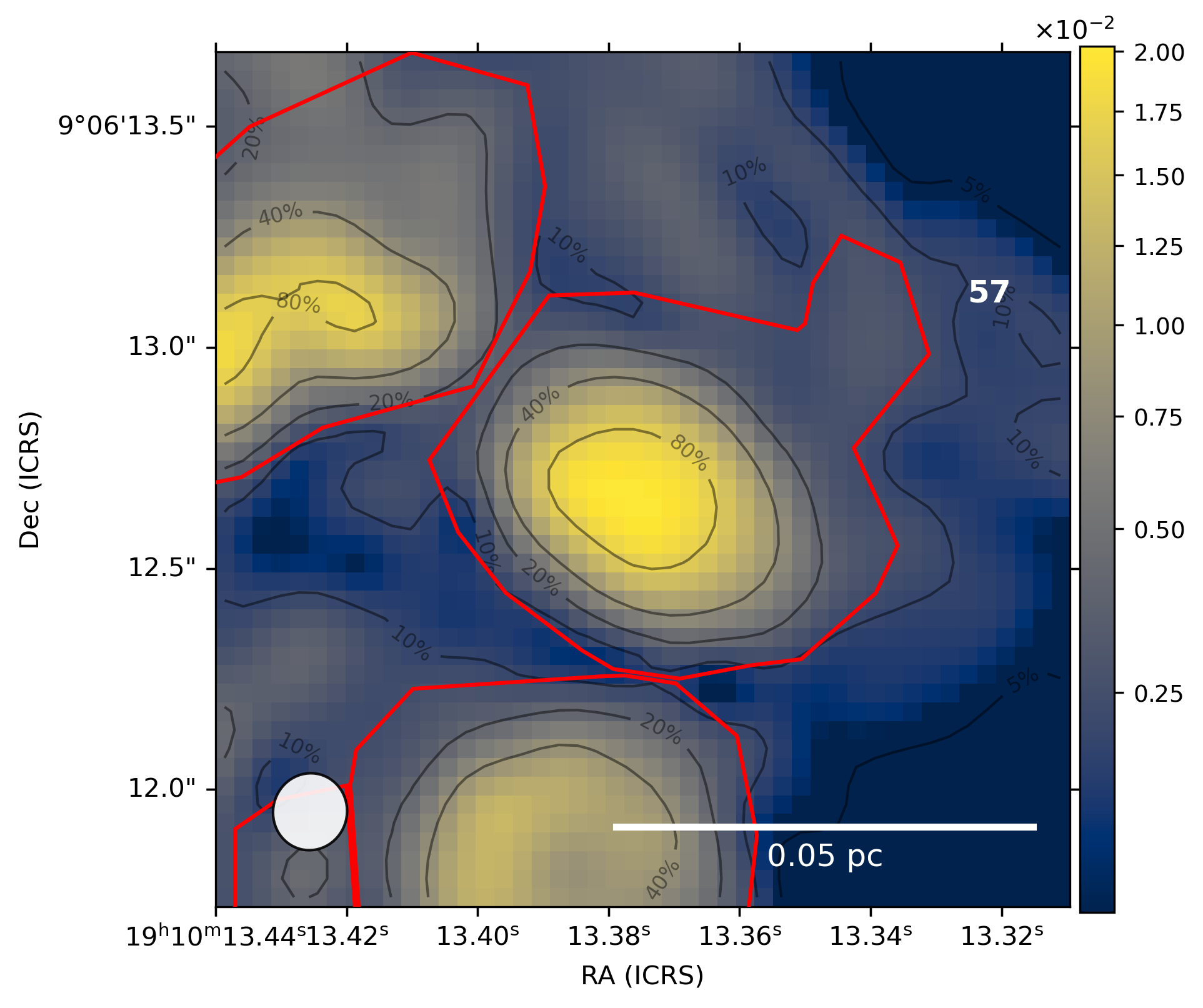}
    \includegraphics[width=0.32\textwidth]{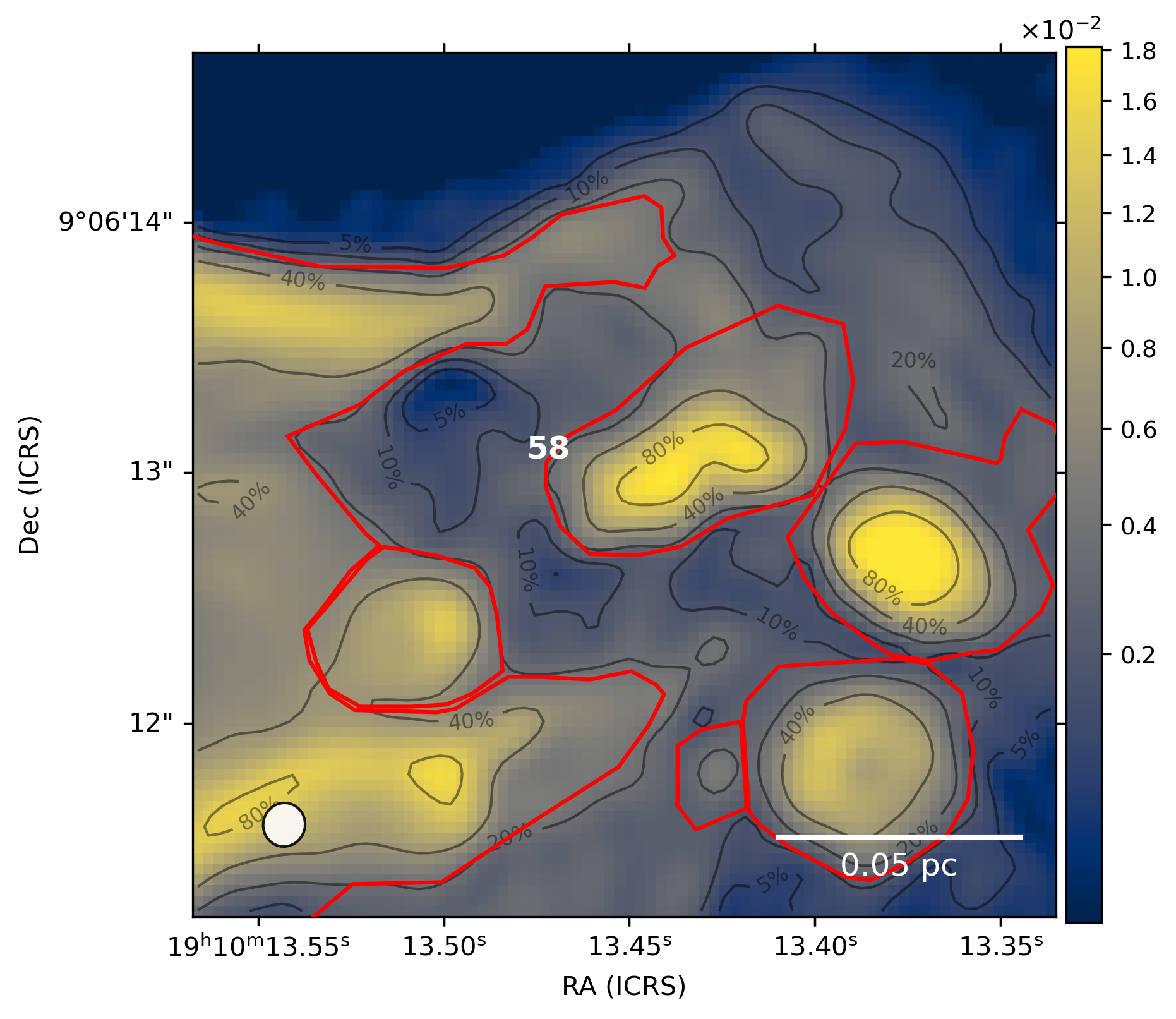}
    \includegraphics[width=0.32\textwidth]{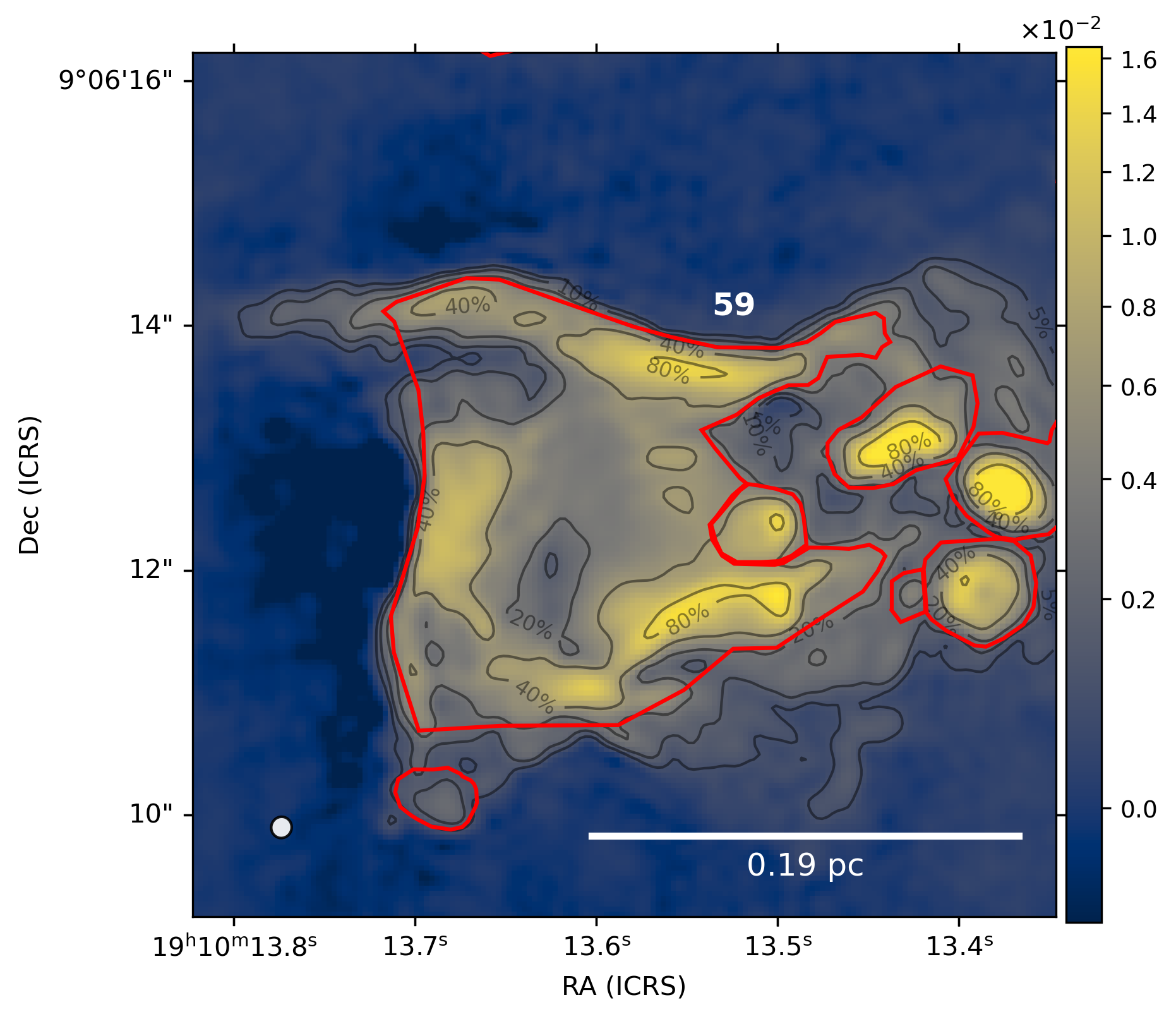}
    \includegraphics[width=0.32\textwidth]{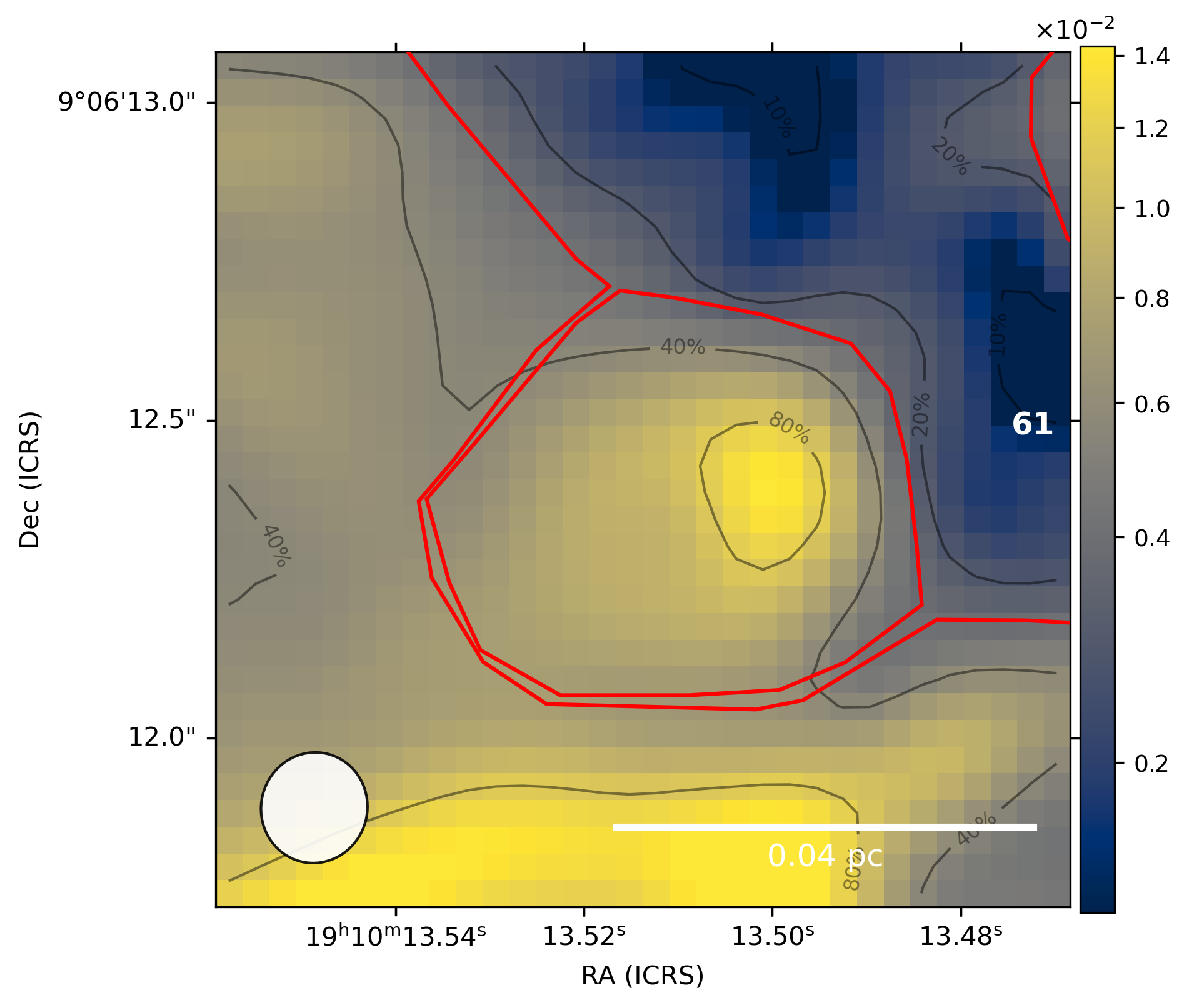}
    \includegraphics[width=0.32\textwidth]{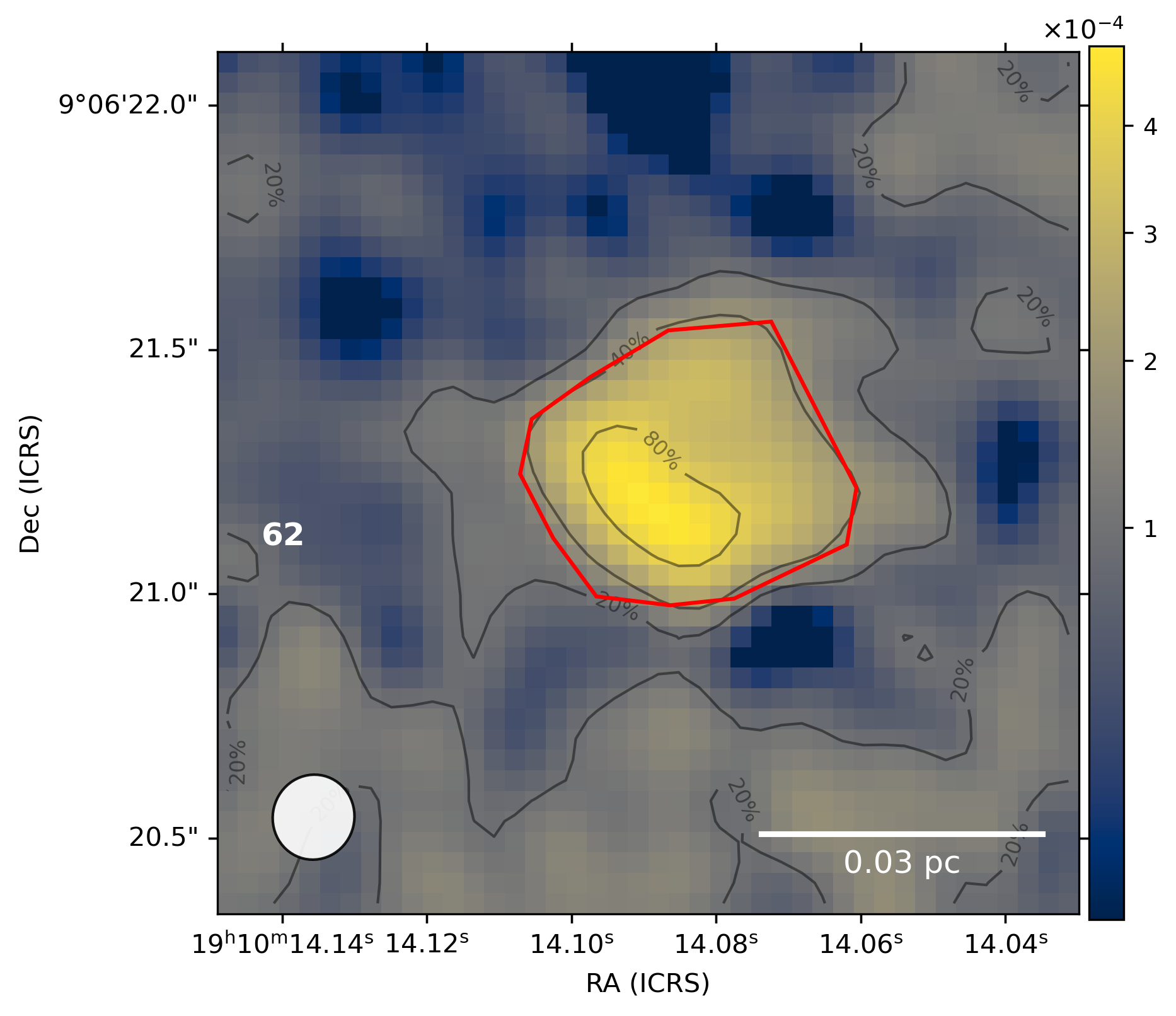}
  \caption{Zoomed-in views (continued).}
\end{figure*} 

\begin{figure*}
    \centering
    \includegraphics[width=0.32\textwidth]{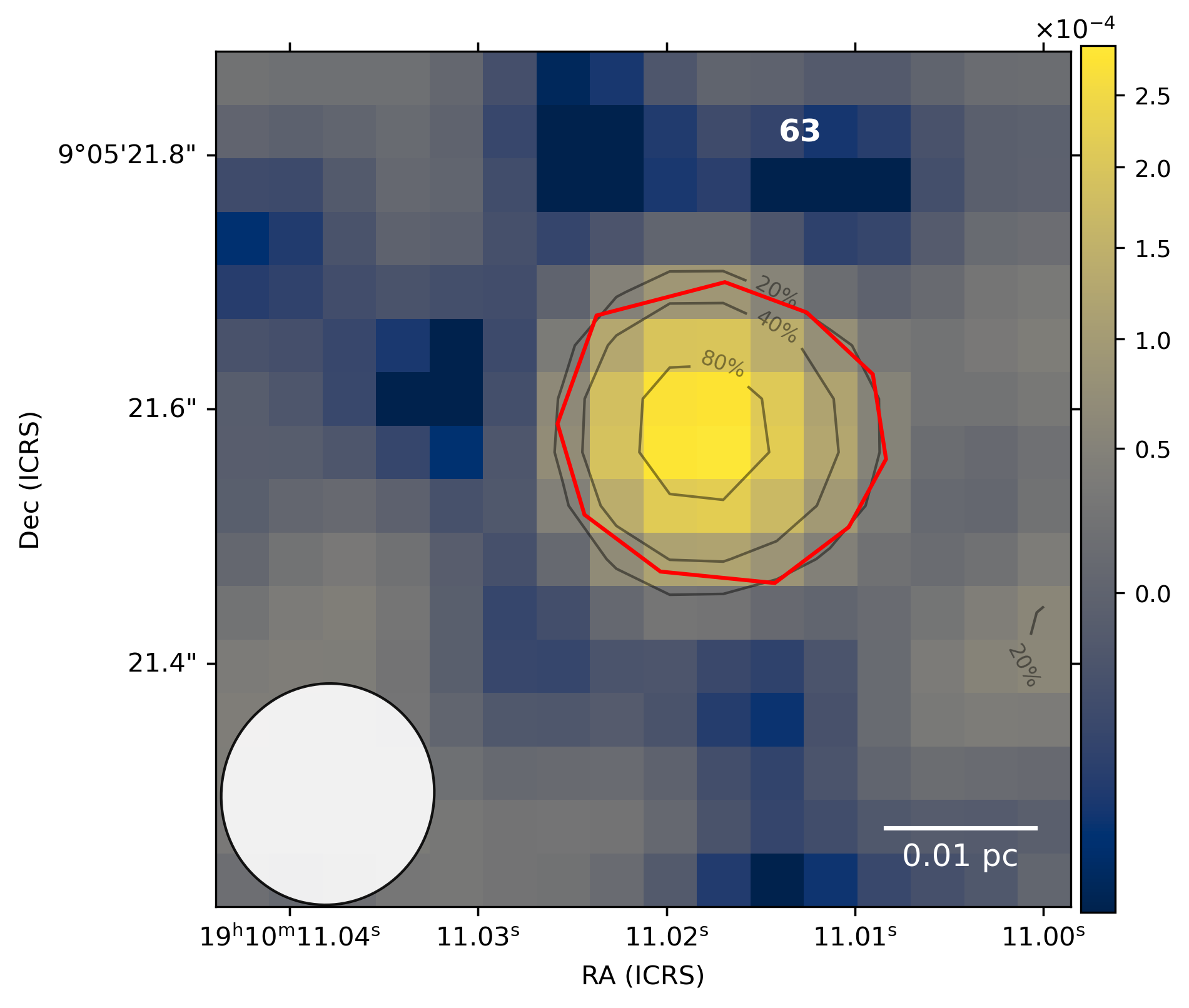}
    \includegraphics[width=0.32\textwidth]{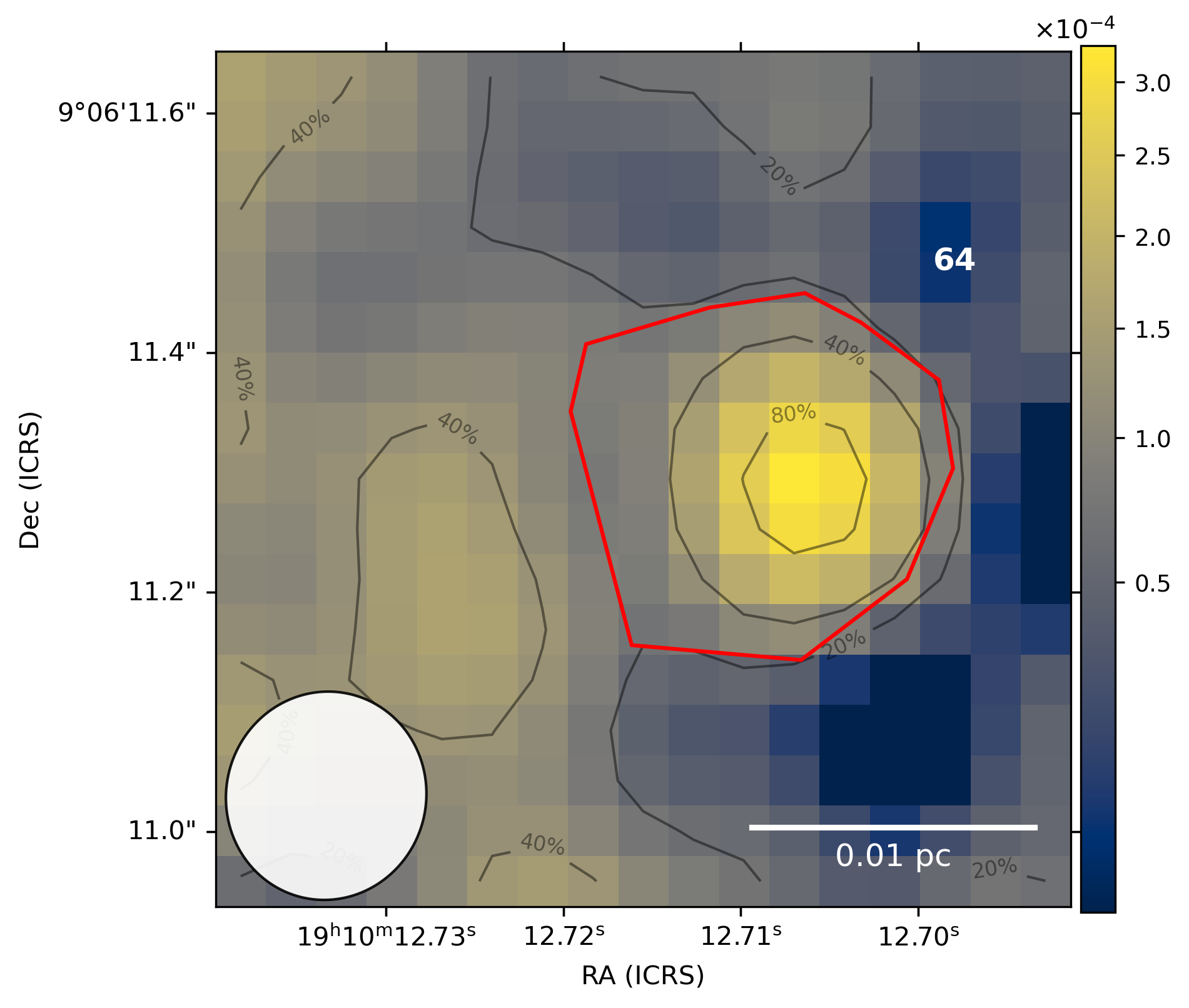}
    \includegraphics[width=0.32\textwidth]{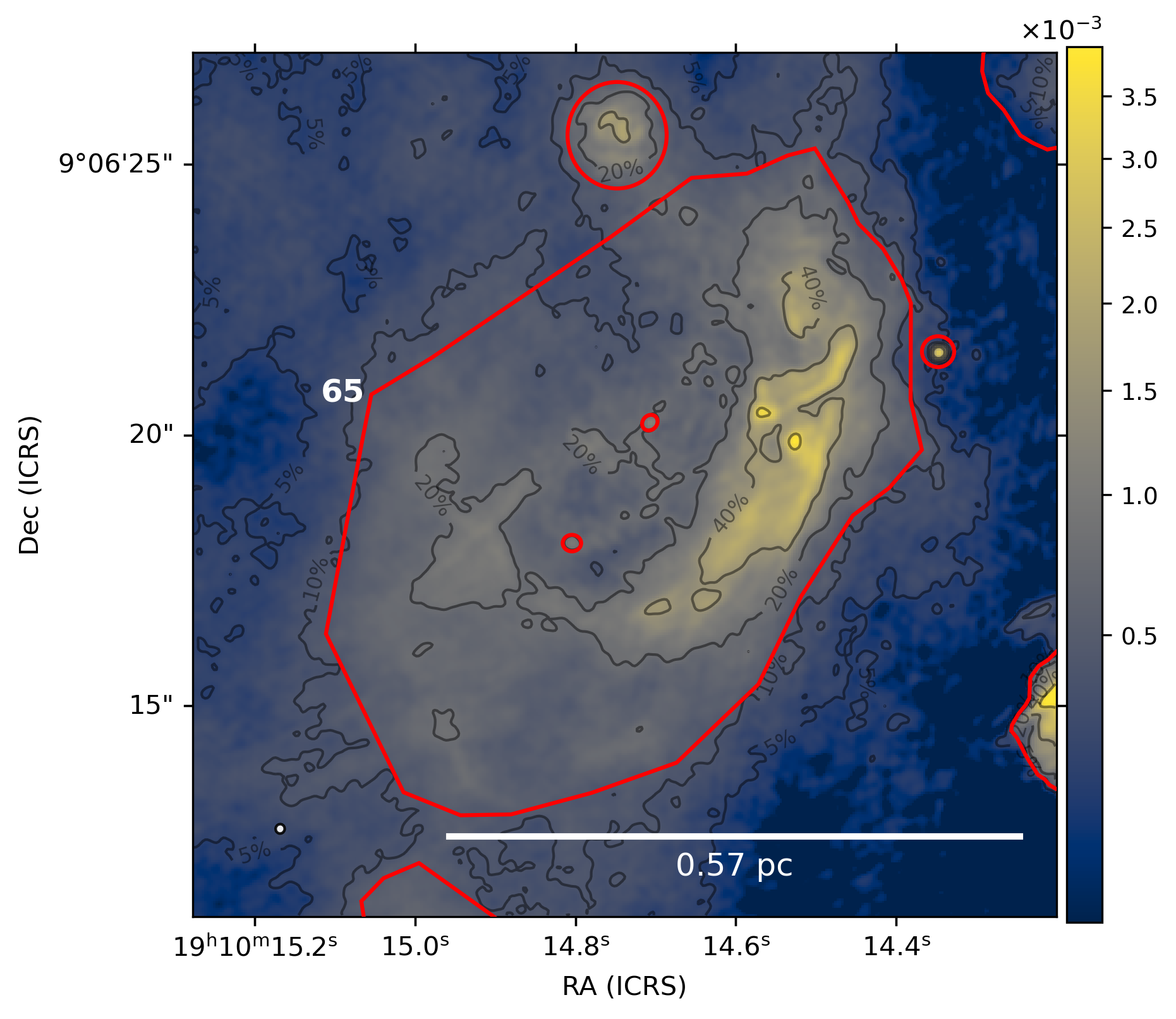}
    \includegraphics[width=0.32\textwidth]{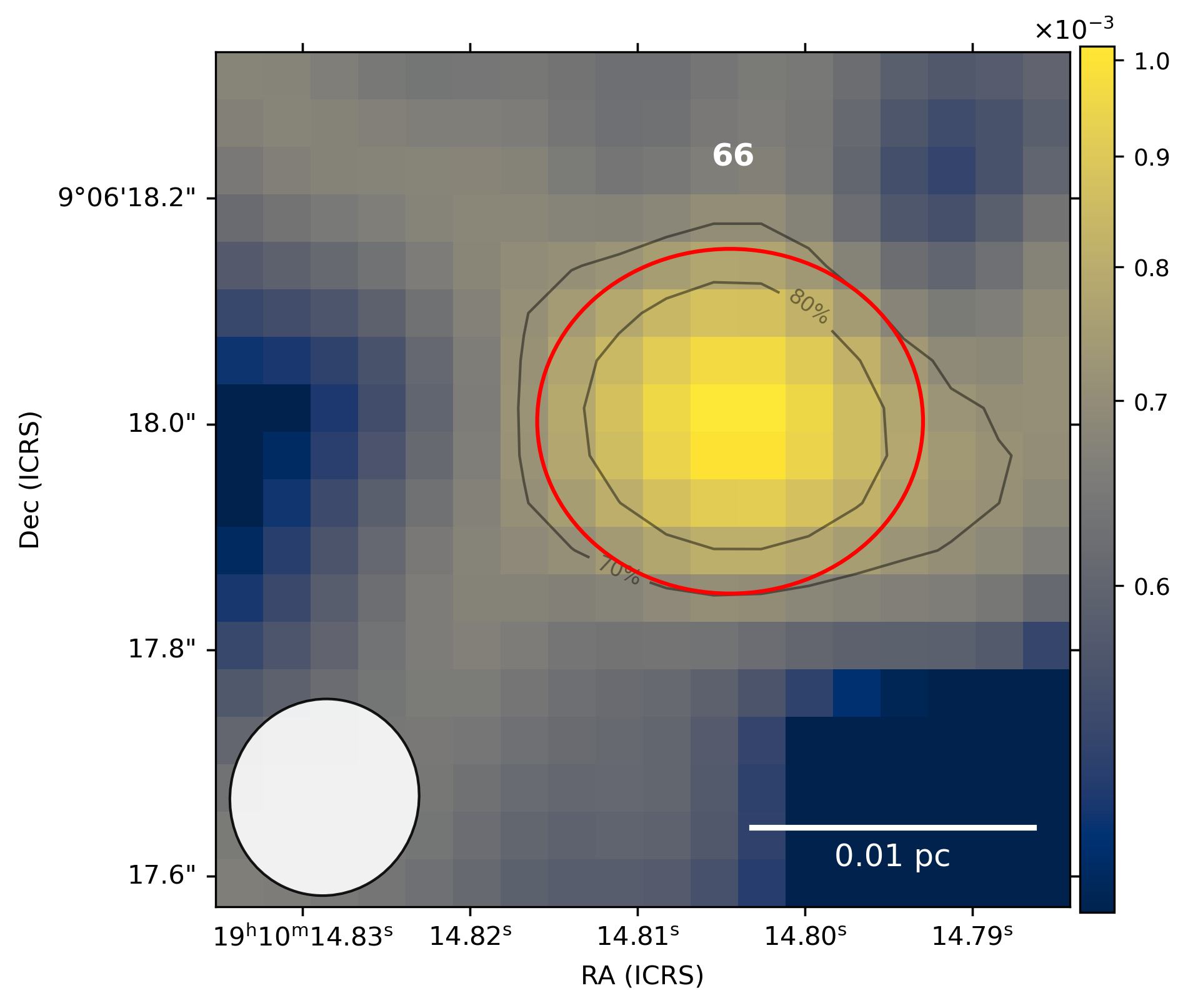}
    \includegraphics[width=0.32\textwidth]{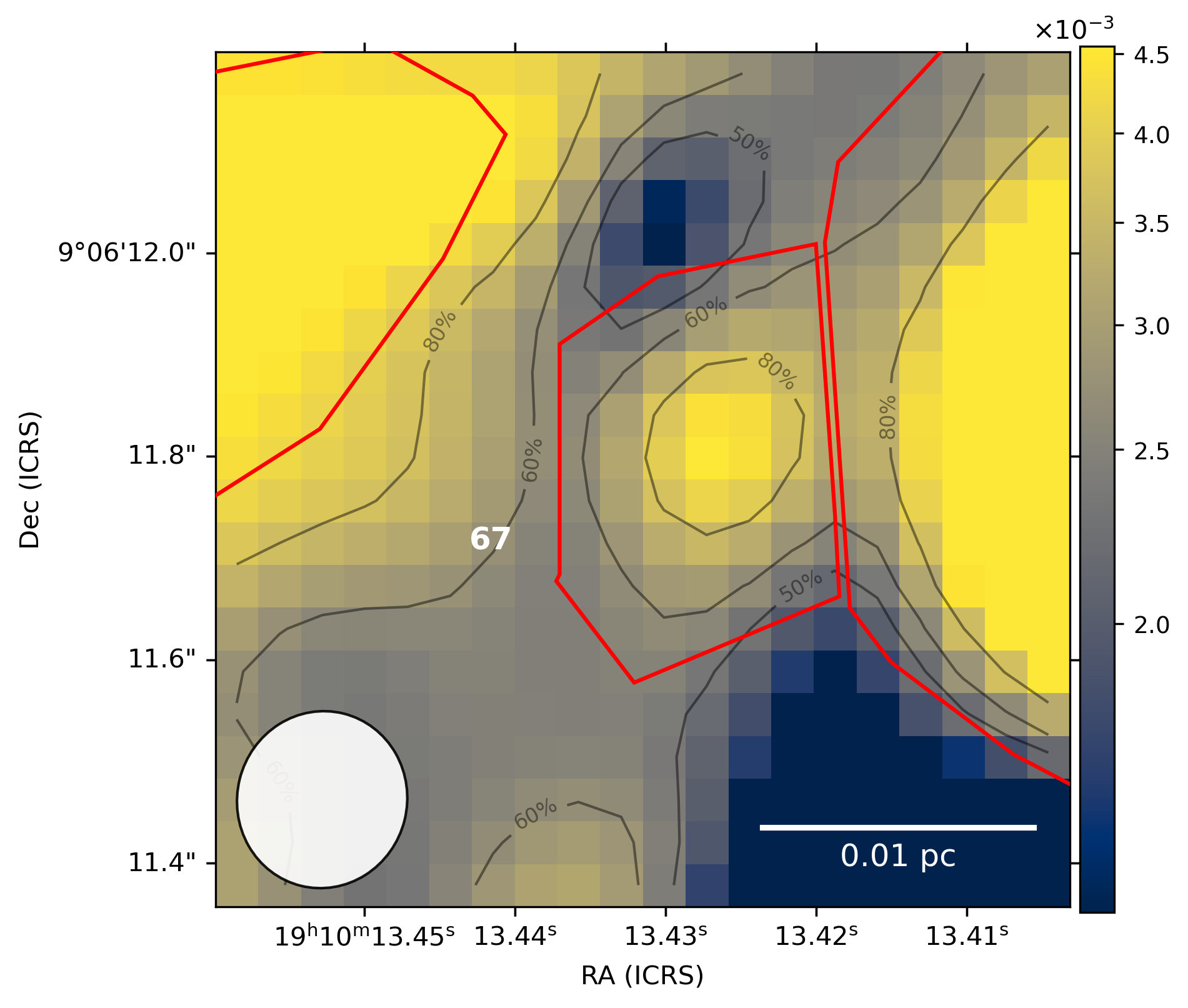}
    \includegraphics[width=0.32\textwidth]{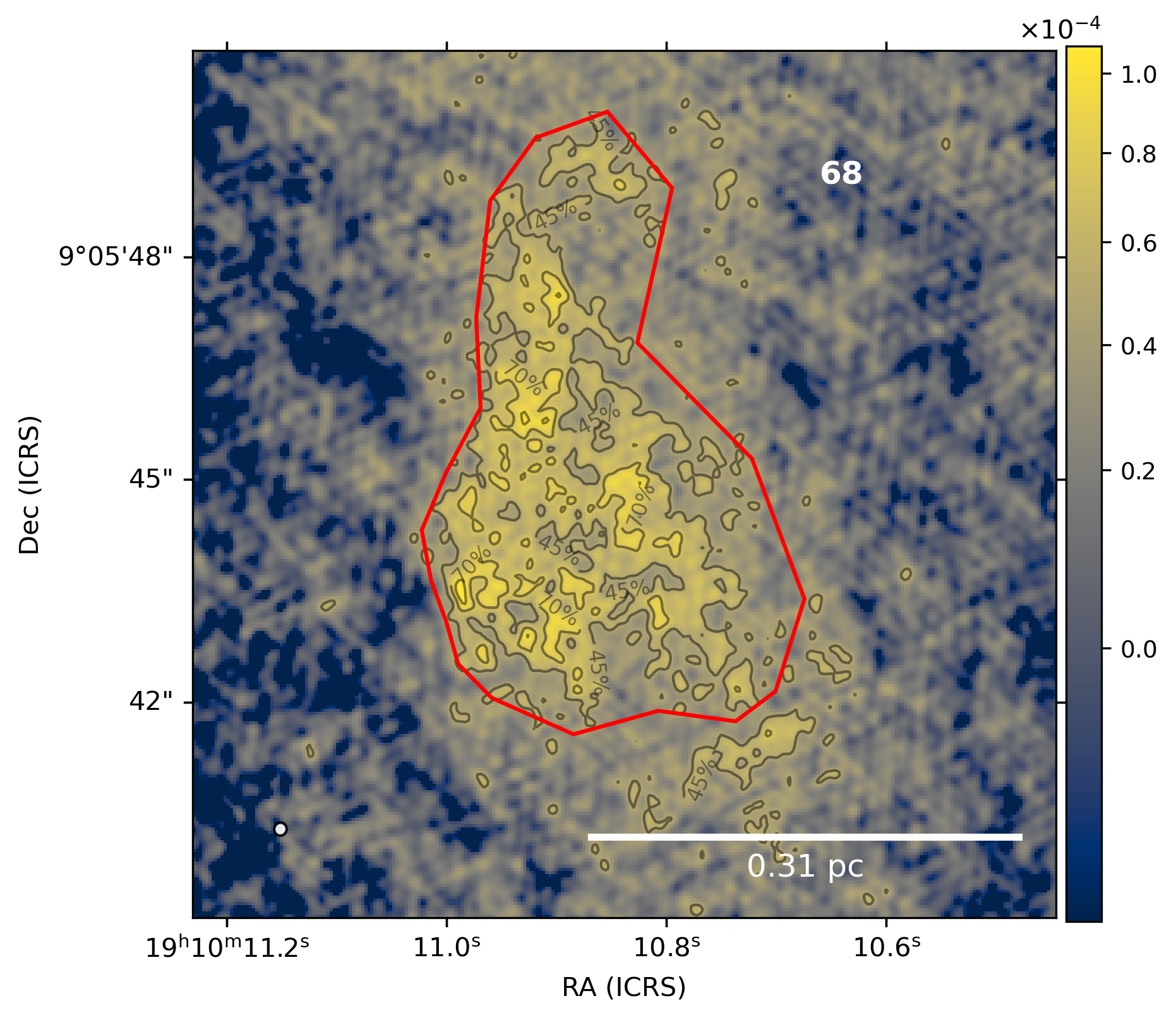}
    \includegraphics[width=0.32\textwidth]{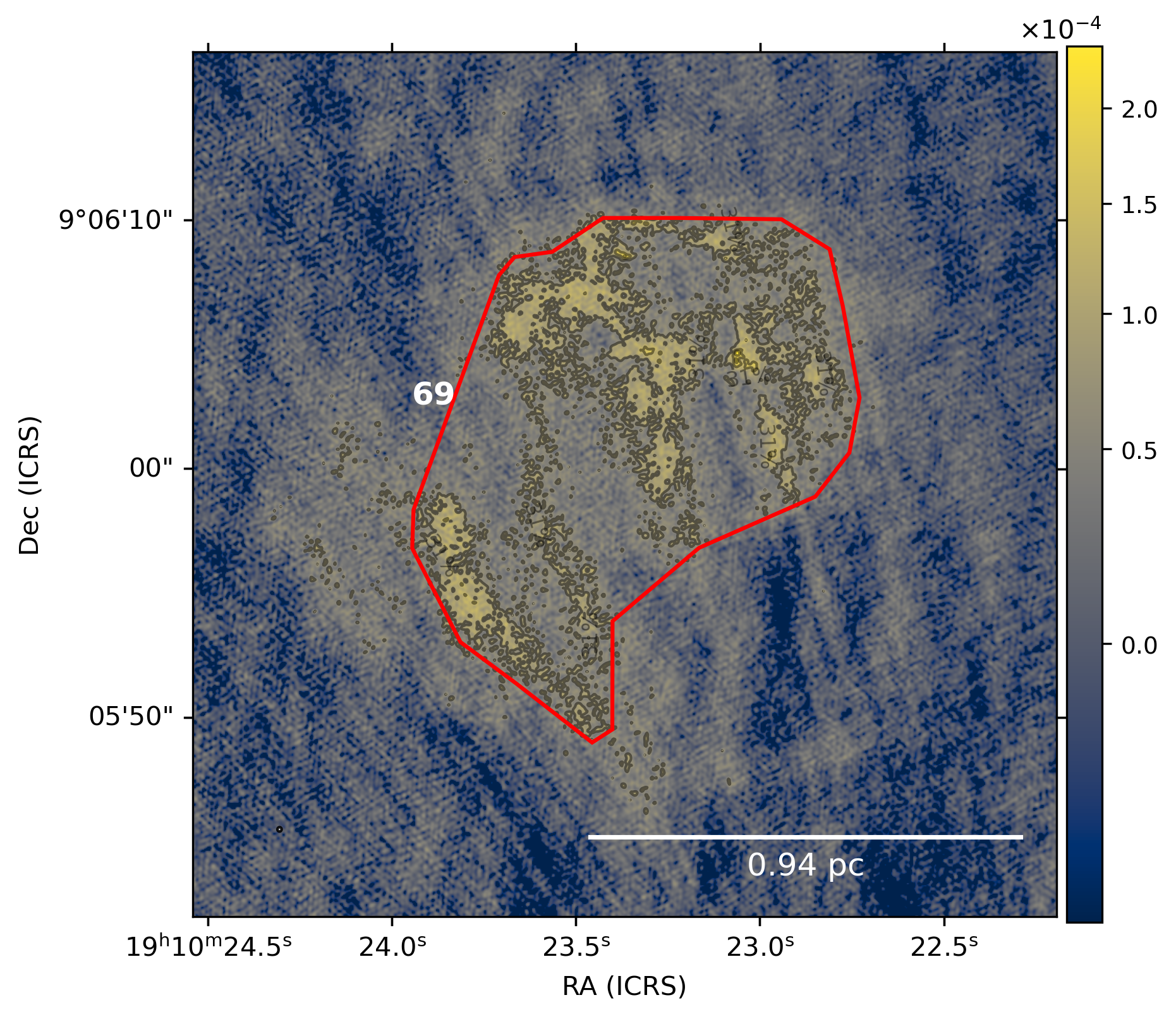}
    \includegraphics[width=0.32\textwidth]{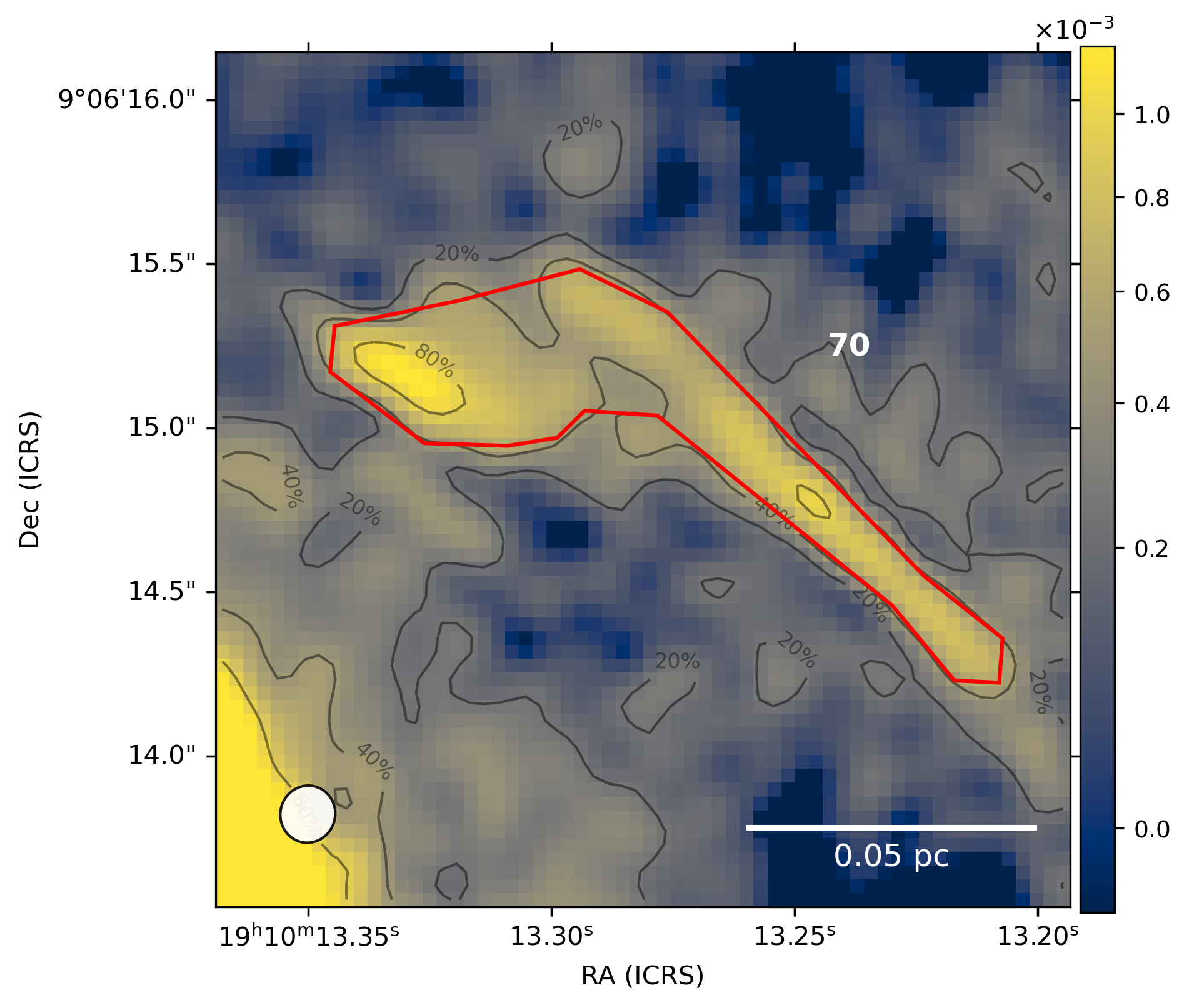}
    \includegraphics[width=0.32\textwidth]{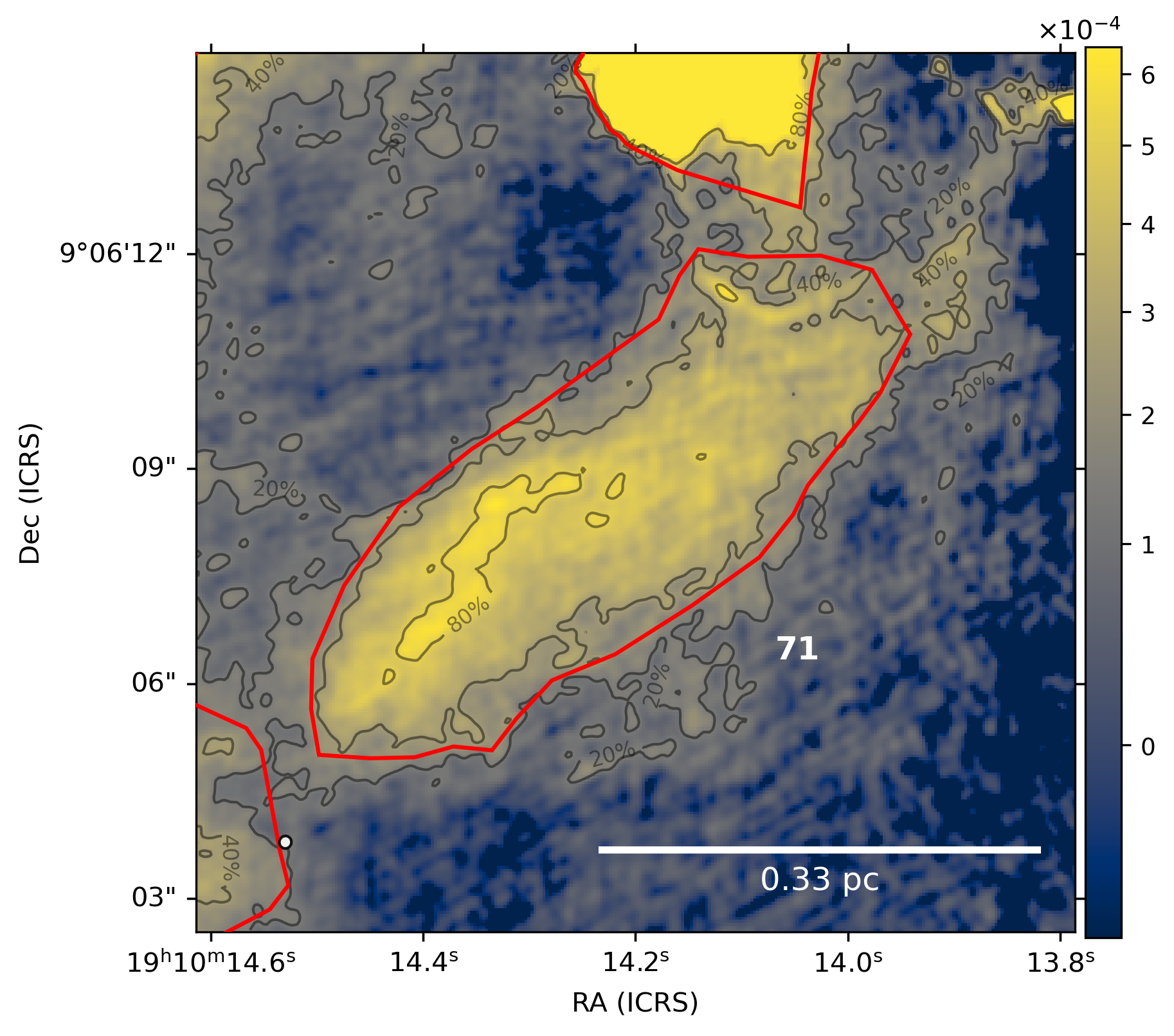}
    \includegraphics[width=0.32\textwidth]{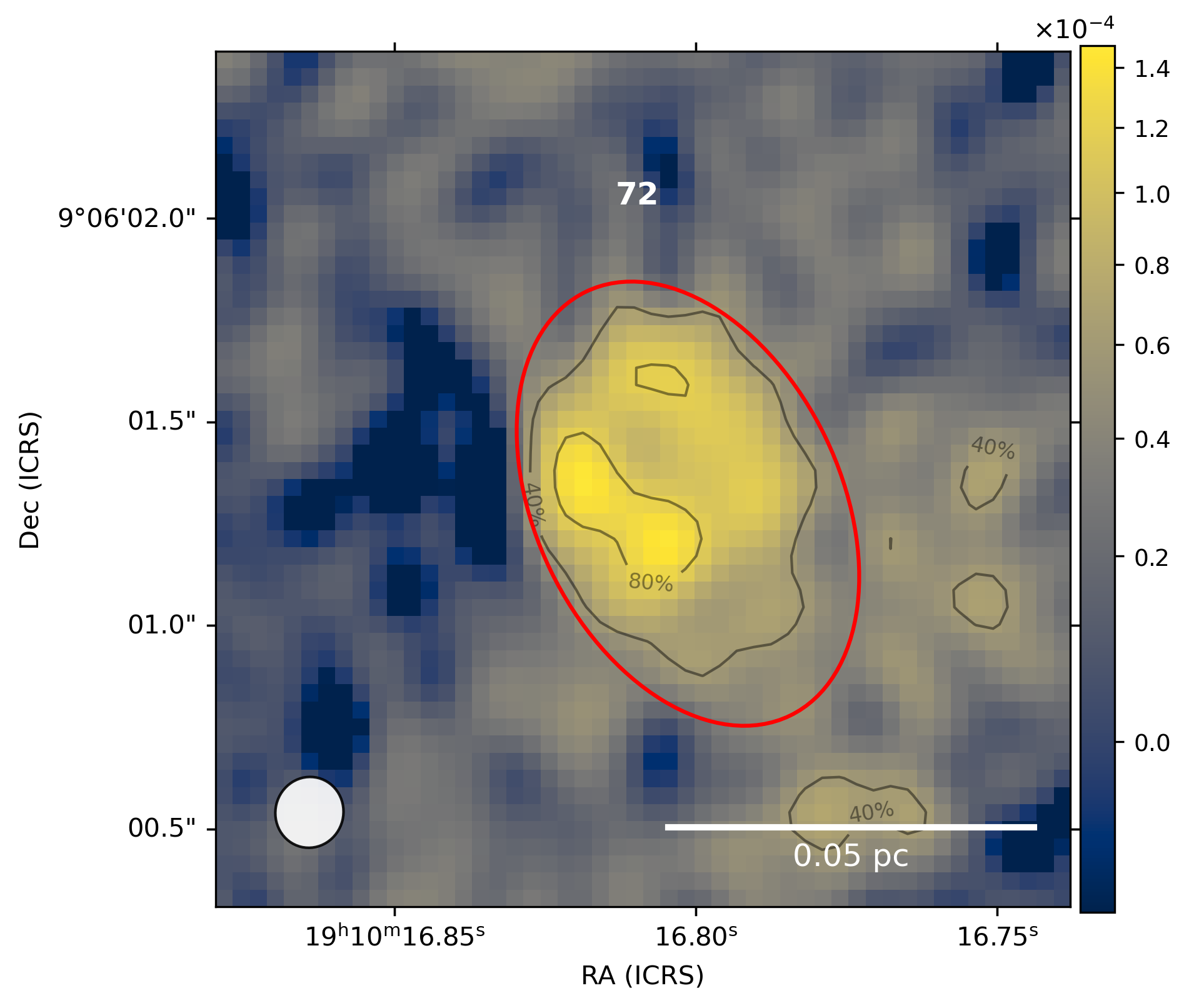}
    \includegraphics[width=0.32\textwidth]{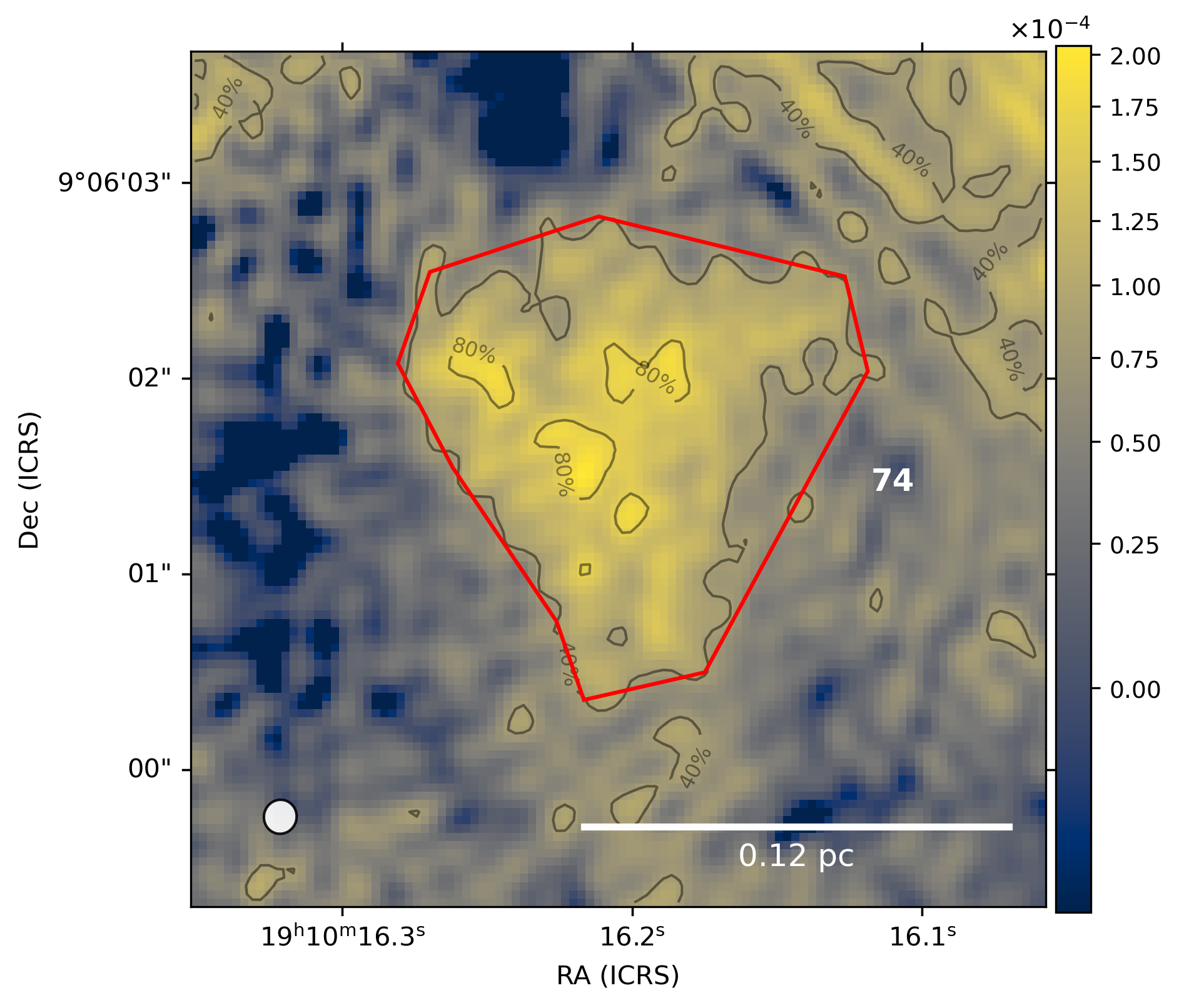}
    \includegraphics[width=0.32\textwidth]{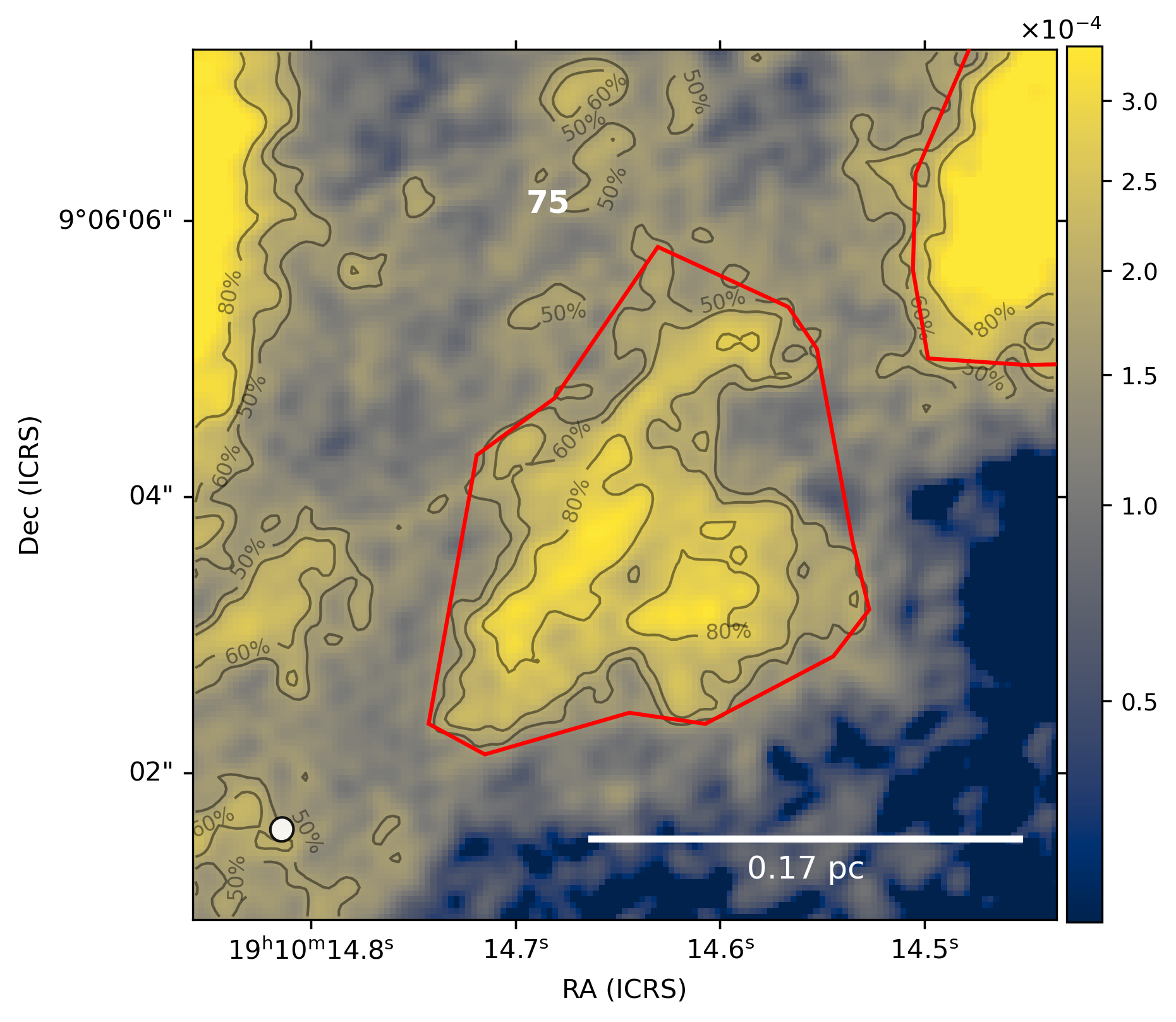}
  \caption{Zoomed-in views (continued).}
\end{figure*} 
\begin{figure*}
    \centering
    \includegraphics[width=0.32\textwidth]{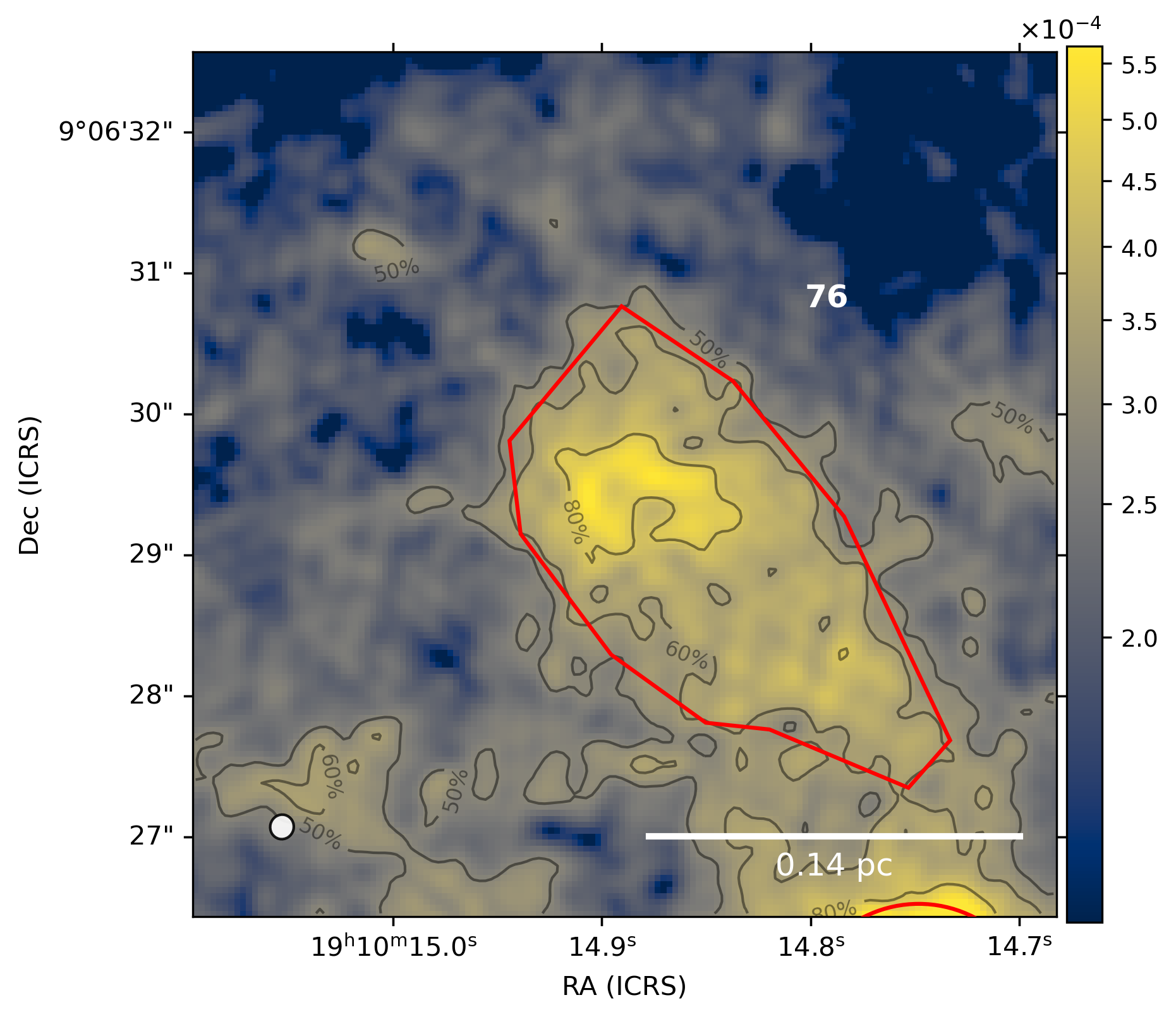}
    \includegraphics[width=0.32\textwidth]{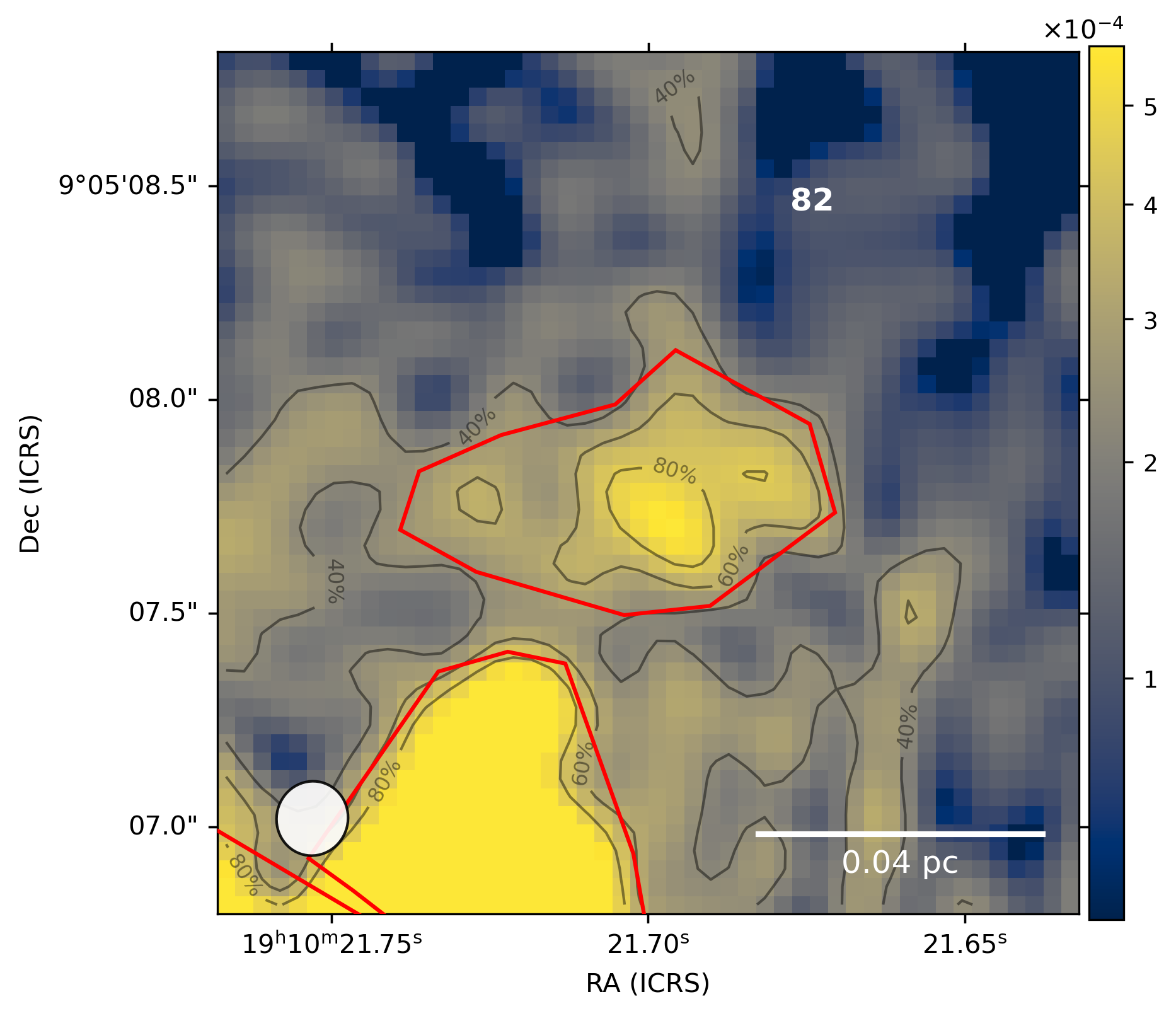}
    \includegraphics[width=0.32\textwidth]{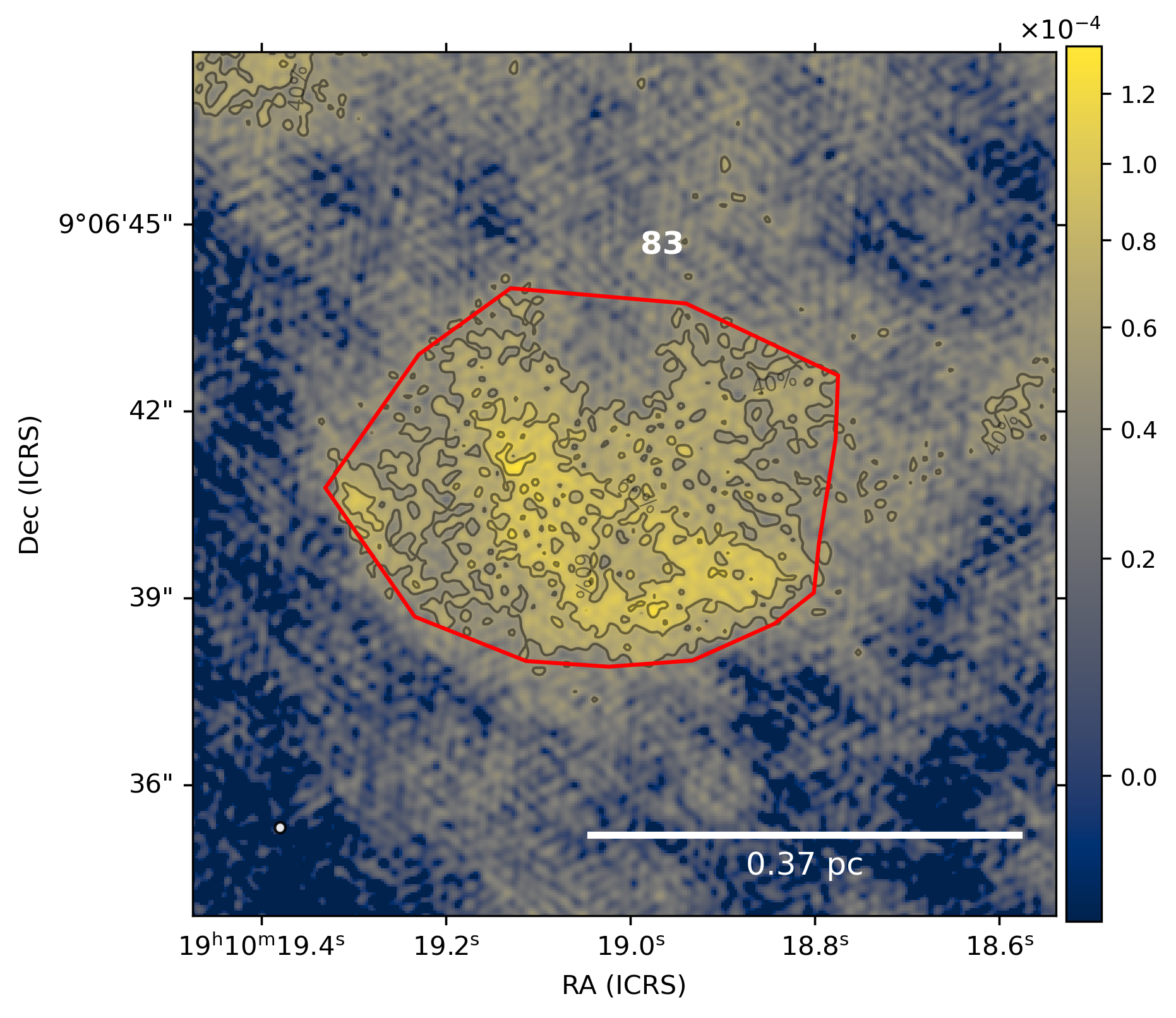}
    \includegraphics[width=0.32\textwidth]{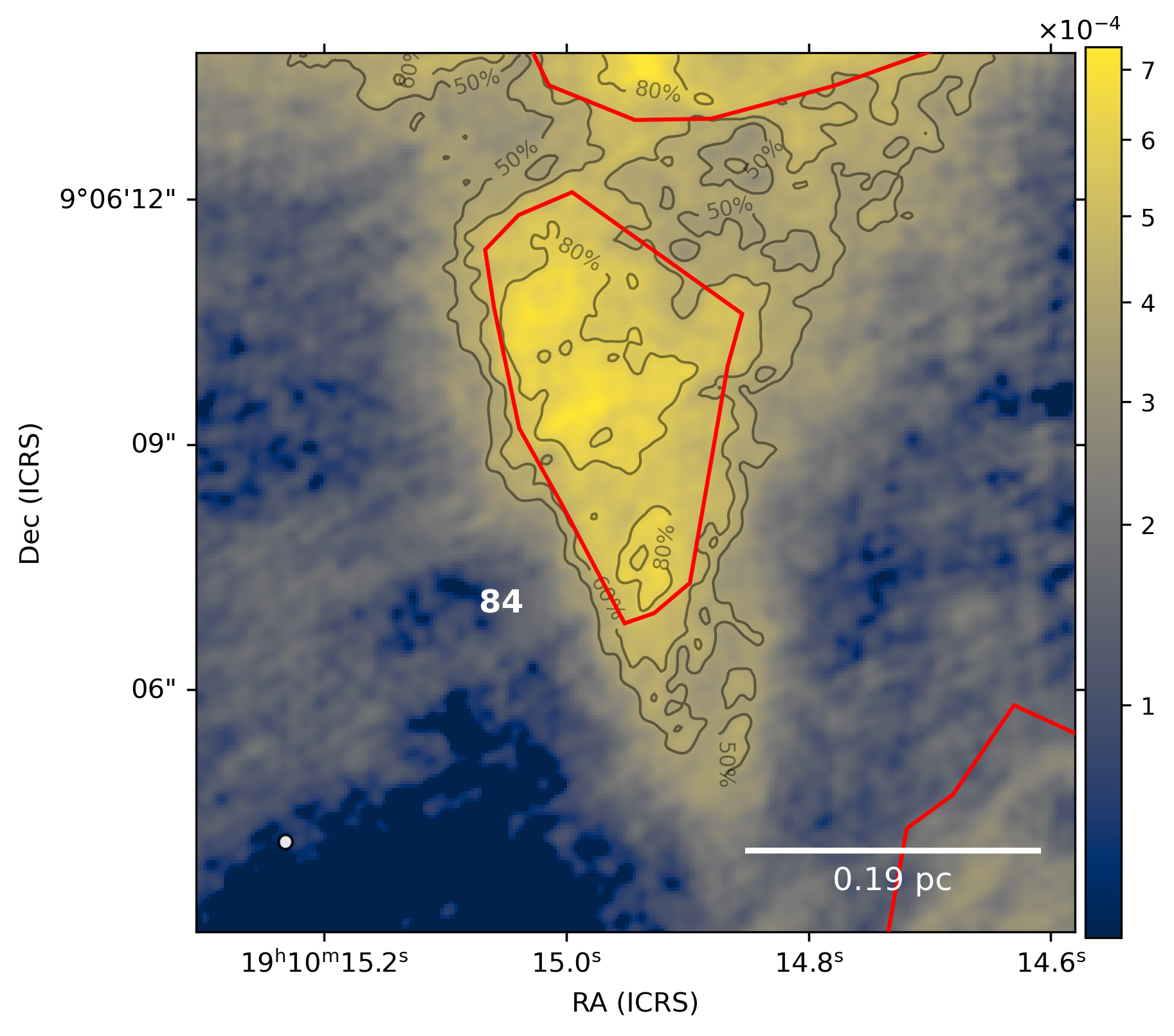}
    \includegraphics[width=0.32\textwidth]{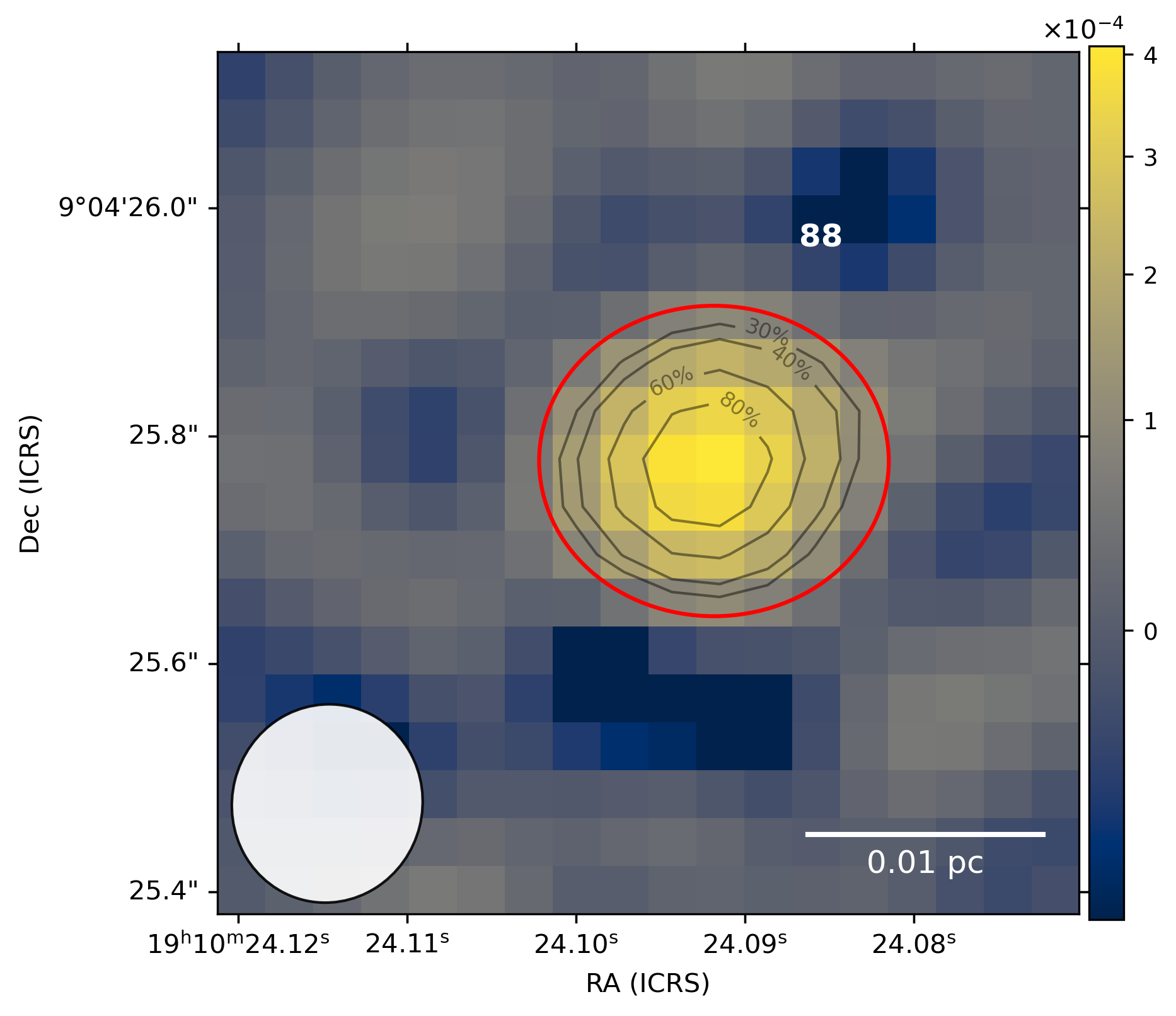}
    \includegraphics[width=0.32\textwidth]{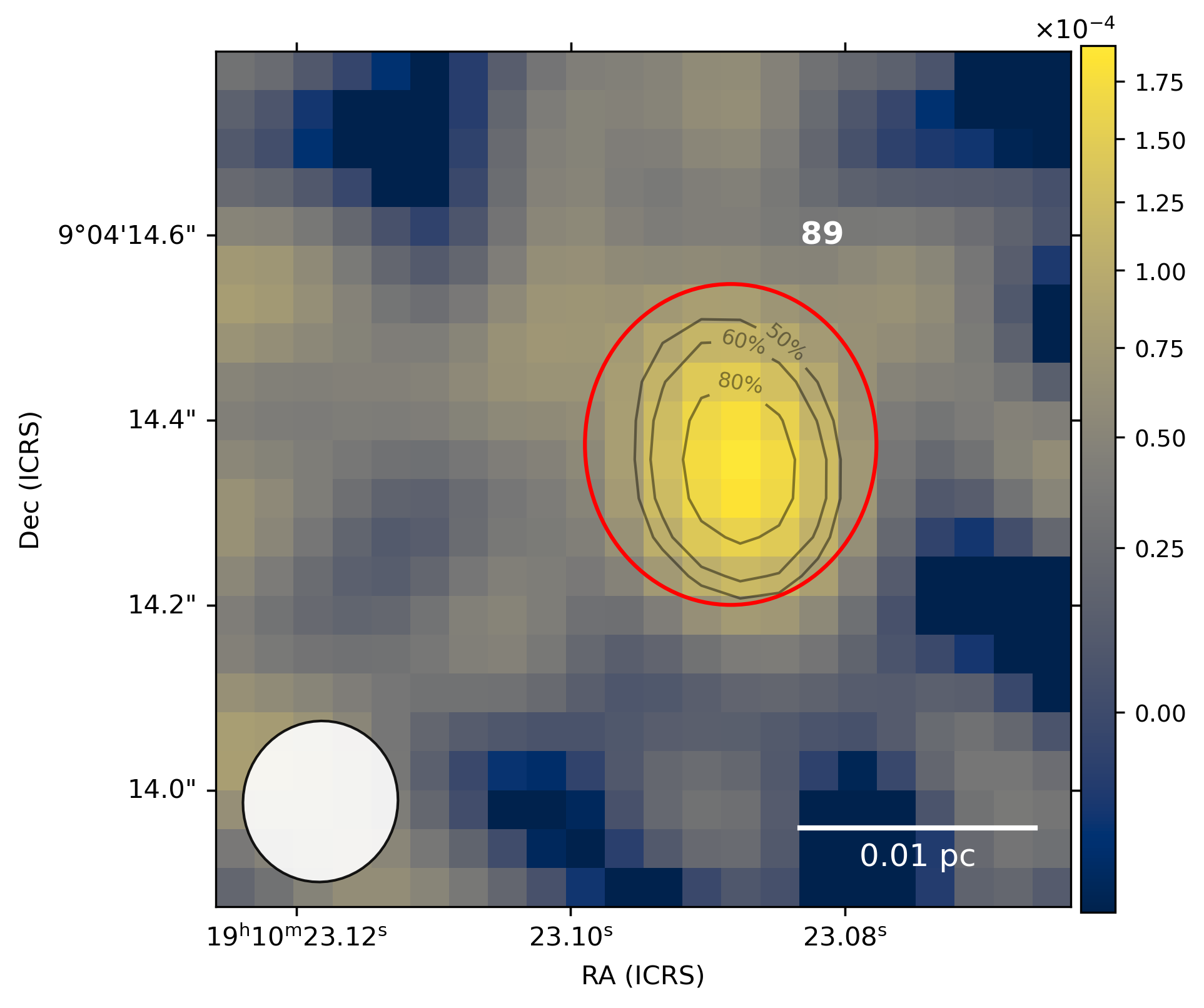}
    \includegraphics[width=0.32\textwidth]{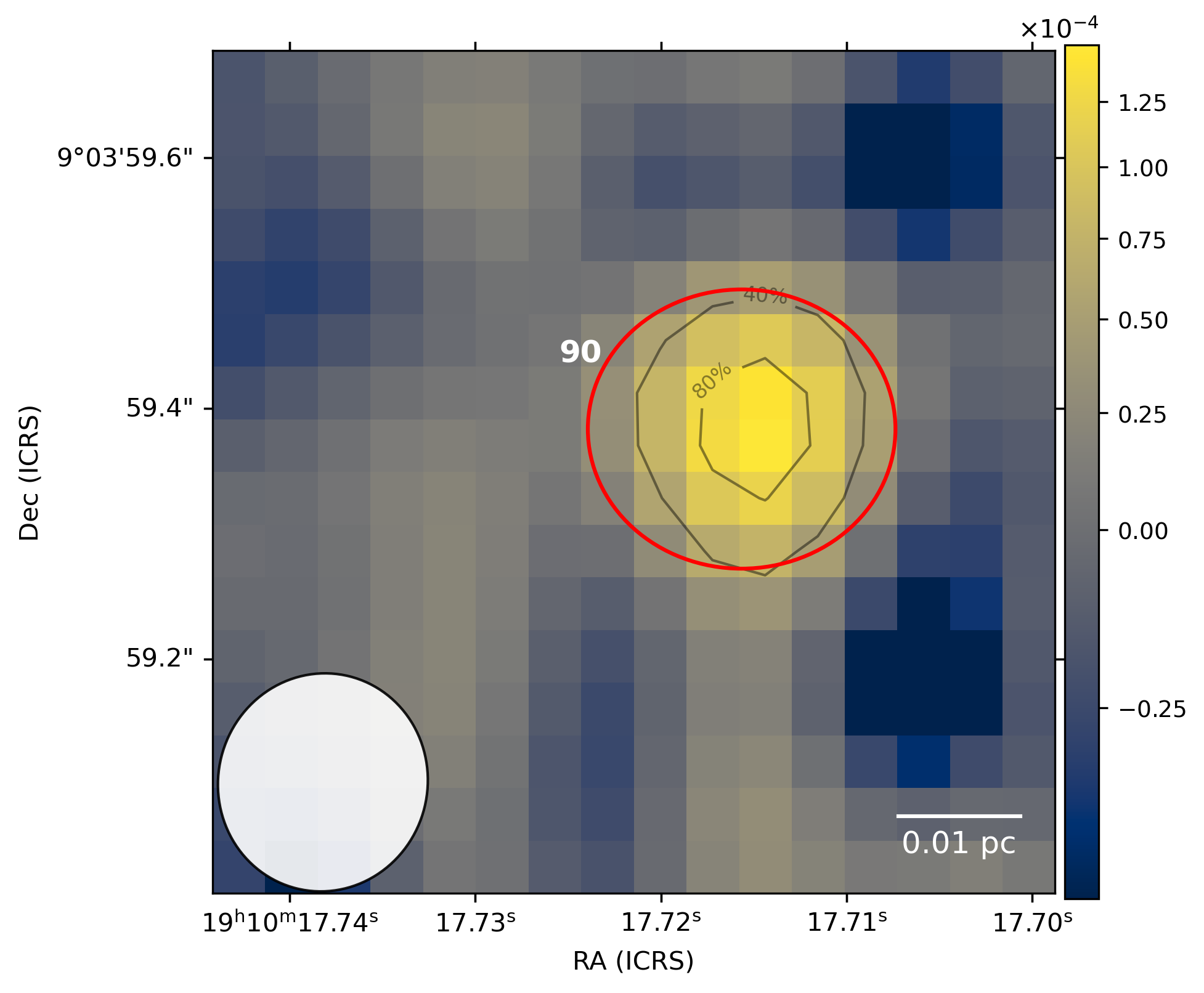}
    \includegraphics[width=0.32\textwidth]{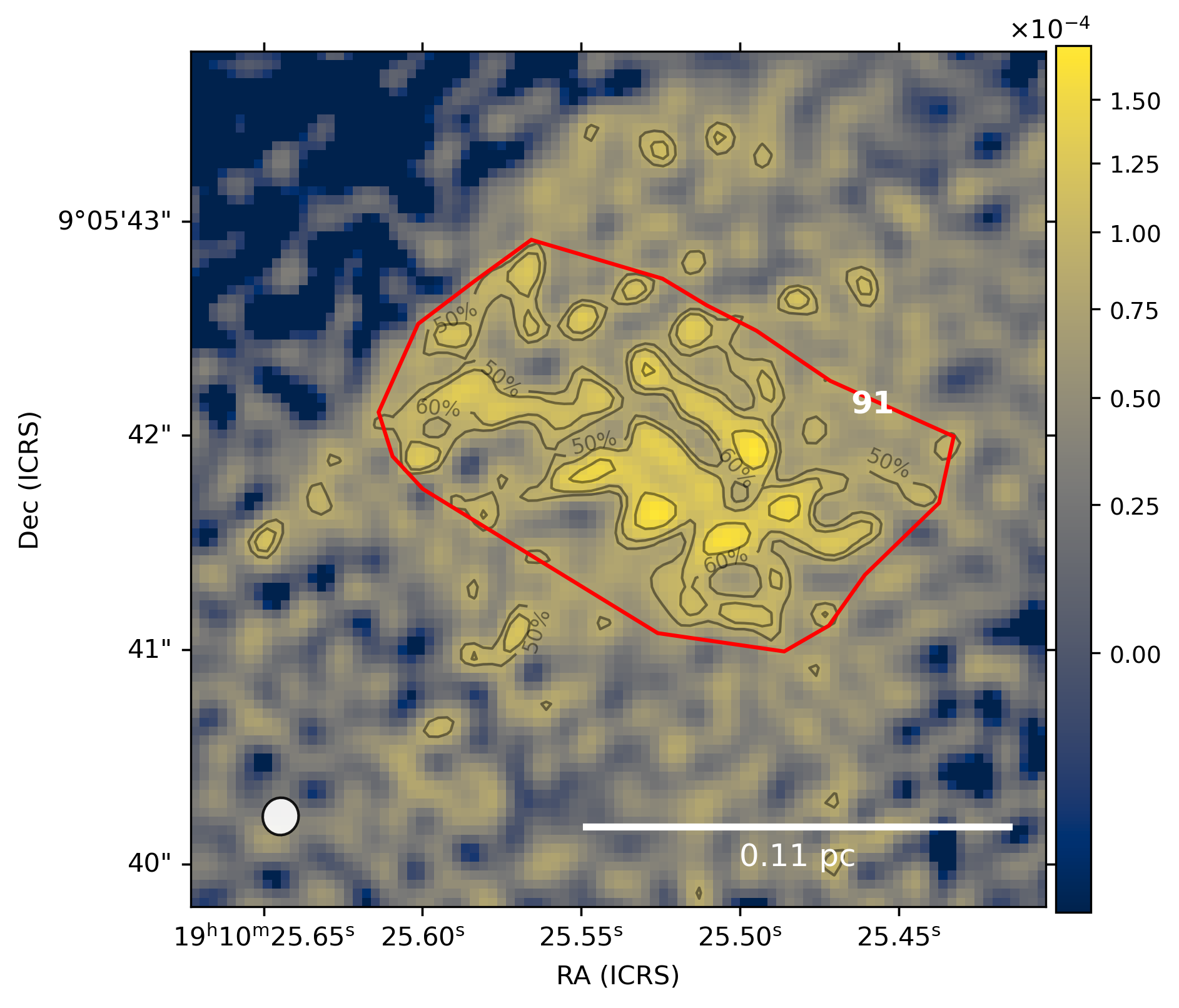}
    \includegraphics[width=0.32\textwidth]{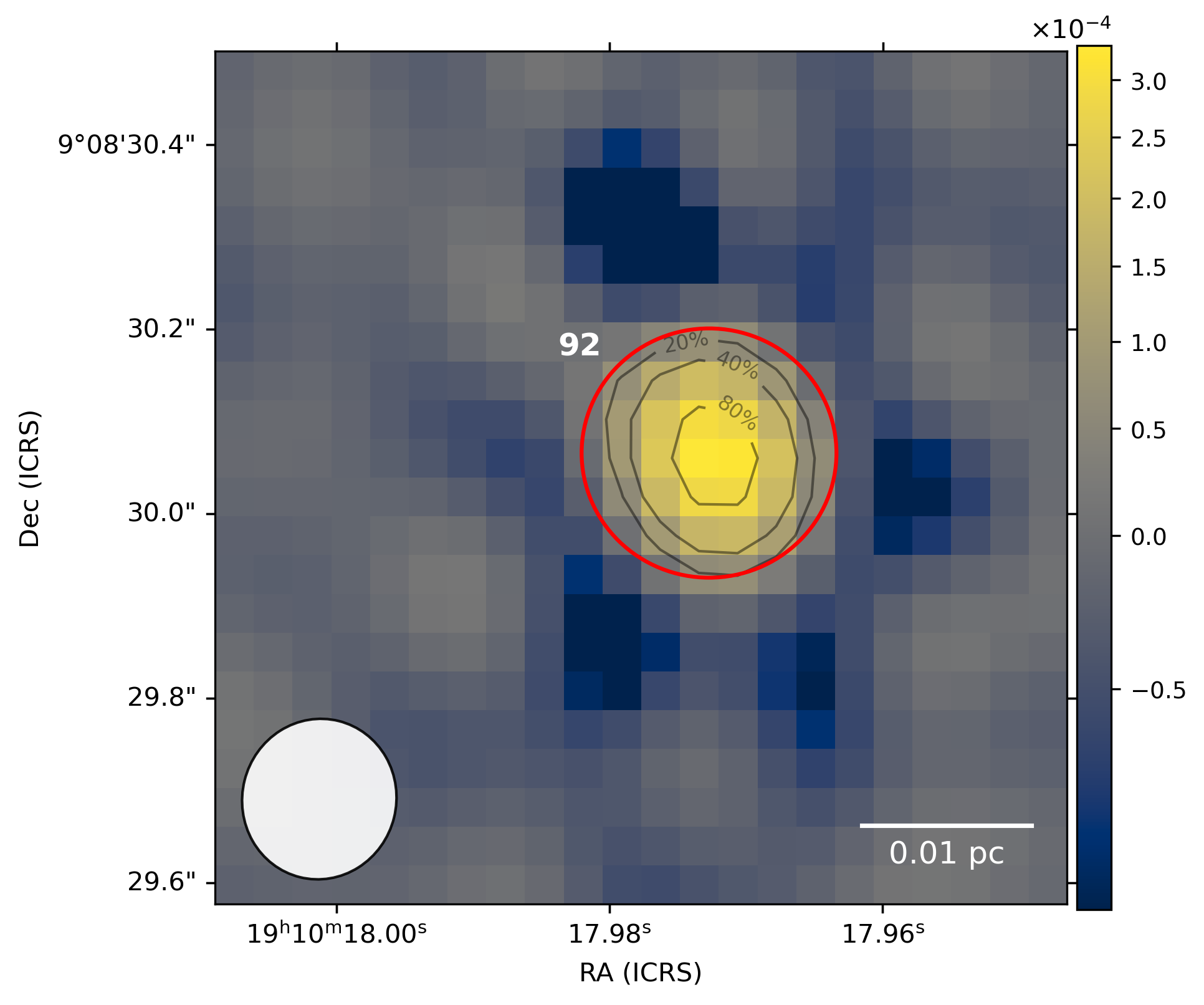}
    \includegraphics[width=0.32\textwidth]{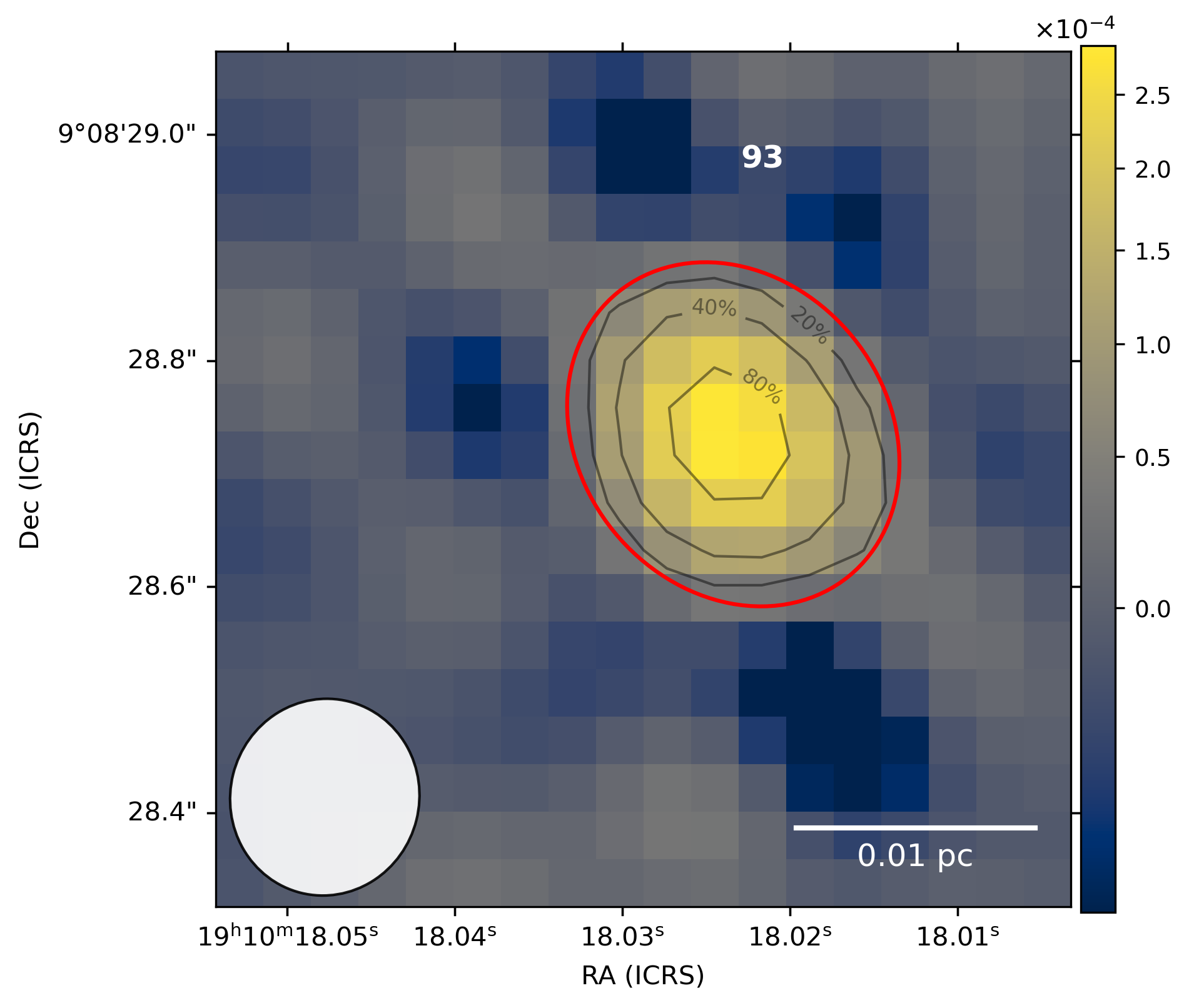}
    \includegraphics[width=0.32\textwidth]{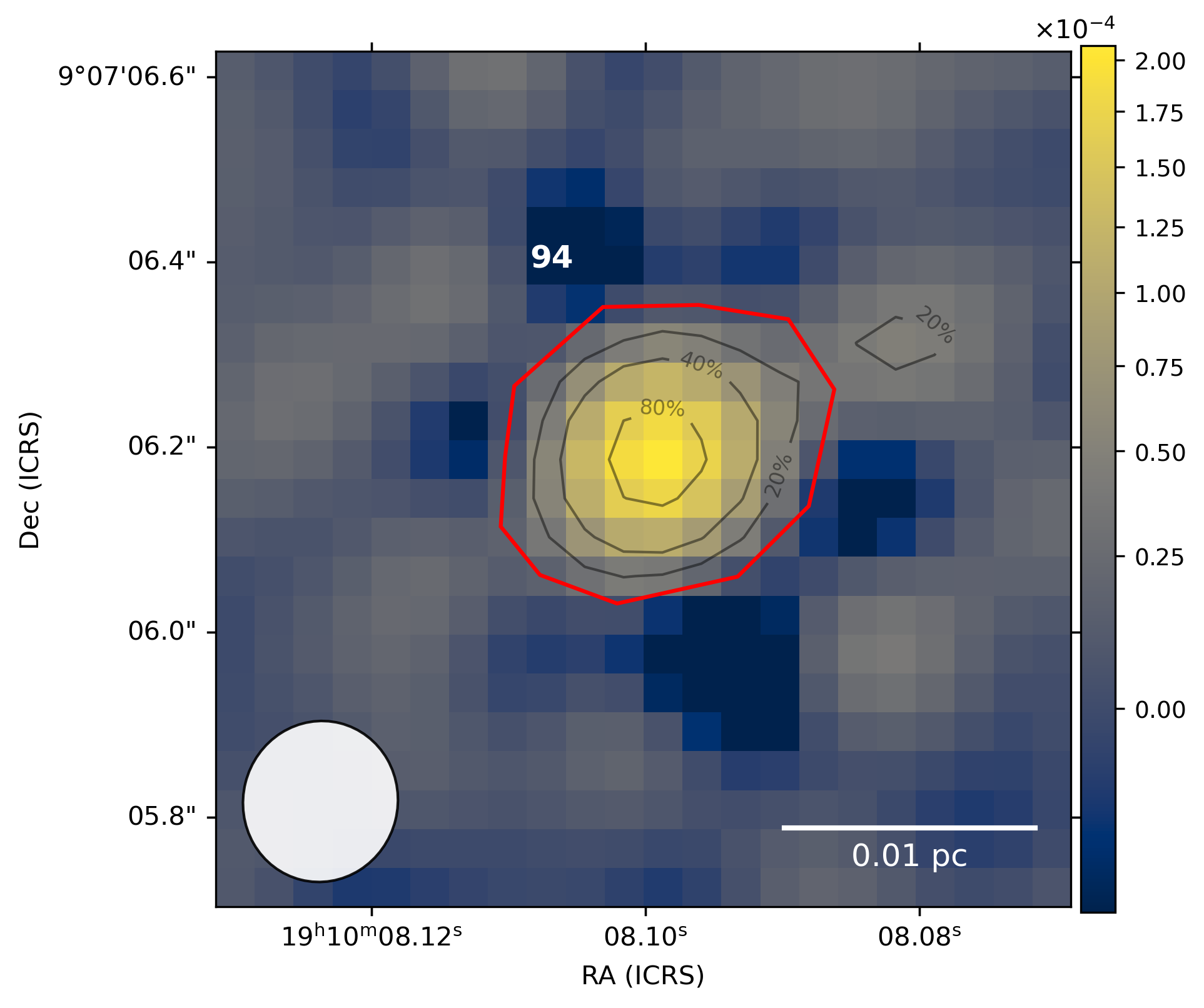}
    \includegraphics[width=0.32\textwidth]{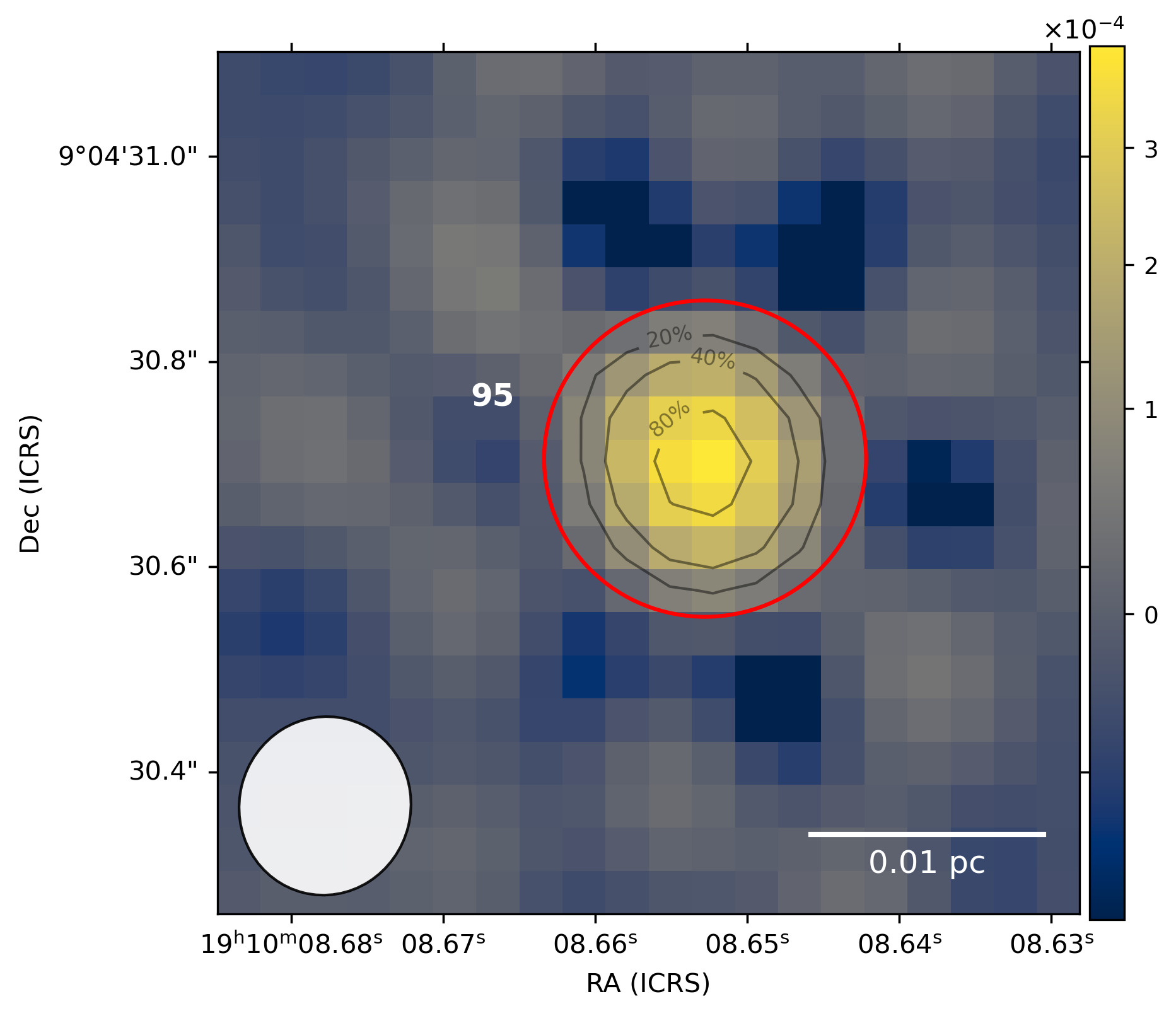}
  \caption{Zoomed-in views (continued).}
\end{figure*}

\end{document}